\RequirePackage{fix-cm}
\documentclass[aps,prx,10pt,twocolumn,a4paper,superscriptaddress,longbibliography,nofootinbib,floatfix]{revtex4-2}
\usepackage[a4paper,top=1.8cm,bottom=2cm,left=2cm,right=2cm,columnsep=0.5cm]{geometry}

\usepackage[utf8]{inputenc}
\usepackage[T1]{fontenc}
\usepackage{microtype}
\usepackage[british]{babel}
\usepackage{amsmath,amssymb,amsfonts,amsthm}
\usepackage{dsfont}
\ifdefined\pdfmapfile\pdfmapfile{+dstroke.map}\fi
\usepackage{xfrac}
\usepackage{physics}
\usepackage{braket}
\usepackage{booktabs}
\usepackage{graphicx}
\usepackage{float}
\usepackage{enumitem}
\usepackage{xcolor}
\usepackage{pgfplots}
\usetikzlibrary{arrows.meta,positioning}
\pgfplotsset{compat=1.18}
\usepgfplotslibrary{groupplots}
\newtheorem{lemma}{Lemma}
\newtheorem{corollary}{Corollary}
\newtheorem{proposition}{Proposition}

\usepackage[framemethod=TikZ]{mdframed}
\mdfdefinestyle{stmt}{%
  linewidth=0.5pt, linecolor=blue!22!black!25, backgroundcolor=blue!3,
  roundcorner=4pt,
  innertopmargin=8pt, innerbottommargin=9pt, innerleftmargin=9pt, innerrightmargin=9pt,
  skipabove=9pt, skipbelow=9pt, splittopskip=10pt}
\surroundwithmdframed[style=stmt]{theorem}
\surroundwithmdframed[style=stmt,nobreak=true]{lemma}
\surroundwithmdframed[style=stmt,nobreak=true]{corollary}
\surroundwithmdframed[style=stmt]{proposition}

\newcommand{\qdrift}{\textup{\textsc{qDRIFT}}}
\newcommand{\statedrift}{\textup{\textsc{StateDRIFT}}}
\newcommand{\ssub}[1]{\raisebox{-0.45ex}{$\scriptscriptstyle #1$}}

\newcommand{\Var}{\mathrm{Var}}
\newcommand{\Cov}{\mathrm{Cov}}
\newcommand{\Expect}{\mathbb{E}}
\newcommand{\bq}{\boldsymbol{q}}
\newcommand{\bgamma}{\boldsymbol{\gamma}}
\newcommand{\bbeta}{\boldsymbol{\beta}}
\newcommand{\bxi}{\boldsymbol{\xi}}
\allowdisplaybreaks

\usepackage[bookmarksdepth=1]{hyperref}
\hypersetup{colorlinks=true,linkcolor={red!40!black},
            citecolor={blue!60!black},urlcolor={blue!50!black}}
\begin{document}

\title{Problem-aware sampling for bias and variance reduction in randomised Hamiltonian simulation}

\author{Brian~Coyle}
\email{brian.coyle@fujitsu.com}
\affiliation{Fujitsu Research of Europe Ltd., Slough SL1 2BE, UK}
\affiliation{School of Informatics, University of Edinburgh, Edinburgh, UK}
\author{Jonathan~H.~Fetherolf}
\affiliation{Fujitsu Research of Europe Ltd., Slough SL1 2BE, UK}
\author{Antonio~M\'{a}rquez~Romero}
\affiliation{Fujitsu Research of Europe Ltd., Pozuelo de Alarc\'{o}n, 28224 Madrid, Spain}
\author{Stefano~Scali}
\affiliation{Fujitsu Research of Europe Ltd., Slough SL1 2BE, UK}
\affiliation{Department of Physics and Astronomy, University of Exeter, Exeter EX4 4QL, UK}
\author{Giuseppe~Buonaiuto}
\affiliation{Fujitsu Research of Europe Ltd., Pozuelo de Alarc\'{o}n, 28224 Madrid, Spain}
\author{Josh~Kirsopp}
\affiliation{Fujitsu Research of Europe Ltd., Slough SL1 2BE, UK}

\date{\today}

\begin{abstract}
Randomised quantum simulation algorithms such as qDRIFT reduce circuit depth by replacing deterministic product formulas with stochastic compilations, but achieving high accuracy can require many steps.
In vanilla qDRIFT, Hamiltonian terms are sampled with probabilities proportional to their coefficient magnitudes. We propose StateDRIFT, a variant of qDRIFT which optimises its sampling distribution to estimate time-evolved quantities using knowledge of the initial state and target observable, with a provable truncation bound that keeps qDRIFT's guarantees.  The distribution at each depth is chosen by solving an optimisation problem targeting bias and/or variance.  StateDRIFT can significantly reduce the bias and variance of estimators, and hence circuit depth, which we test on short-time dynamics of lattice models and molecules under Jordan--Wigner up to 18 qubits.  For example, with BeH$_2$ StateDRIFT can reduce the number of two-qubit gates required to achieve a given accuracy from $11{,}201$ to $983$.
StateDRIFT involves measuring observables and commutators on the initial state, so we propose two practical ways to estimate these: classical surrogates such as tensor networks, and classical shadows, which recover most or all of the saving from exact oracles in the systems tested.
\end{abstract}

\maketitle
\interdisplaylinepenalty=10000
\setlength{\parskip}{4pt plus 1pt minus 1pt}

\begin{figure*}[!t]
\centering
\includegraphics[width=\textwidth]{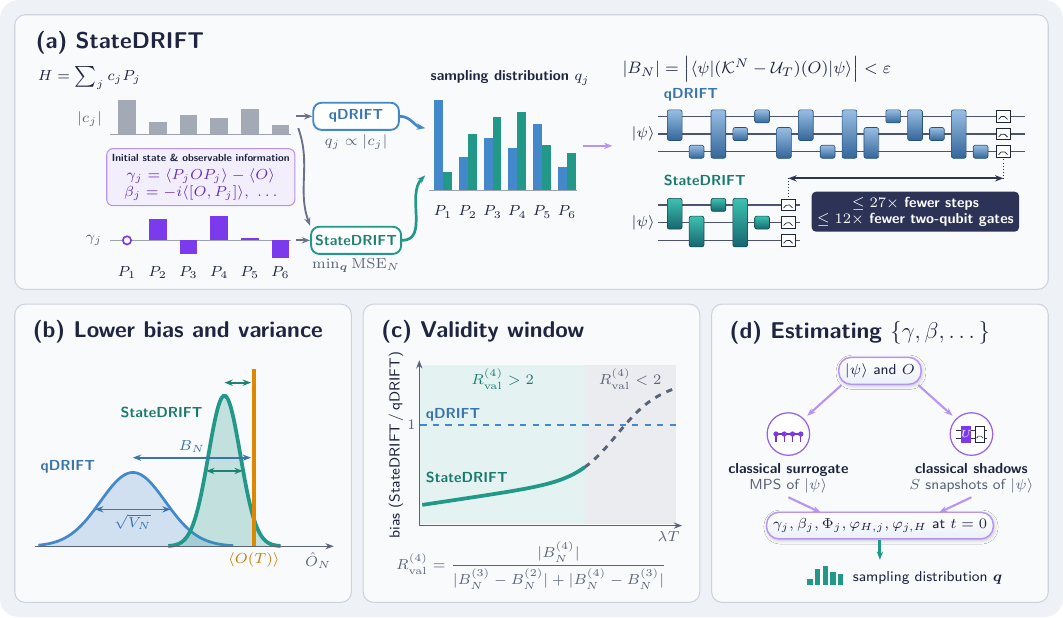}
\caption{\textbf{StateDRIFT.}
\textbf{(a)}~Vanilla \qdrift{} samples according to a distribution $\bq$ which is proportional to the weight of the
Hamiltonian terms $|c_j|$.  \statedrift{} computes state- and observable-dependent quantities, $\gamma_j$, $\beta_j$,
$\varphi_{H,j}$, $\varphi_{j,H}$ and $\Phi_j$, and reweights the distribution according to them to \textbf{(b)}~reduce
the bias and/or variance.  The bias is given by $B_N$ between the sampled \qdrift{} channel, $\mathcal{K}$, and the
true channel $\mathcal{U}_T$.  As a result, the circuit depths required to achieve a certain accuracy $\varepsilon$ can be
significantly lower than vanilla \qdrift{}, with up to $11.4\times$ fewer two-qubit gates in our largest margin
example.  \textbf{(c)}~\statedrift{} uses truncated formulae to 4th order in $dt$, e.g.\
$B^{(4)}_N$ for the bias, which can give inaccurate results due to truncation error.  A validity metric, $R_{\rm val}^{(4)}$,
computed on this truncated formula checks the validity of the distribution allocation in reducing bias and variance over
\qdrift{}.  We find this metric accurate in all cases we test in this work.  \textbf{(d)}~Classical surrogates or classical shadows can be used to
compute the $\gamma_j$, $\beta_j$, etc.\ quantities for \statedrift{}.}
\label{fig:hero}
\end{figure*}

\section{Introduction}
\label{sec:intro}
The original proposed use for quantum computers was in simulating quantum systems
themselves~\cite{Feynman1982,Lloyd1996}.  Quantum computers may be most useful for simulating strongly correlated
molecules and materials, whose exact simulation cost grows exponentially with system size for classical computers, and
for evolving such systems dynamically through time and evaluating their real-time
response~\cite{Bauer2020}.

Subsequent decades of research have developed a suite of tools and algorithms for the purpose of approximating
time evolution on quantum computers, via the operator $e^{-iHT}$, including Suzuki--Trotter methods~\cite{Lloyd1996,Suzuki1991,Childs2021}, LCUs~\cite{ChildsWiebe2012,Berry2015Taylor},
qubitisation and QSP based approaches~\cite{LowChuang2017,LowChuang2019}, along with a family of randomised algorithms
including random-ordering product formulas~\cite{Childs2019Randomization}, \qdrift{}~\cite{Campbell2019} and
qSWIFT~\cite{Nakaji2024qSWIFT}.  These are evolution methods which can be used, for example, in phase
estimation routines such as QPE~\cite{Kitaev1995,AbramsLloyd1999}, early fault-tolerant single-ancilla methods such as
QCELS~\cite{DingLin2023QCELS} or QMEGS~\cite{Ding2024QMEGS}, or in ground-state preparation methods such as imaginary time
evolution~\cite{McArdle2019VarQITE,Motta2020QITE} and its near-term QC-QMC
version~\cite{Huggins2022QCQMC,Zhang2025QCQMC,Kanno2024QCQMC,Buonaiuto2026QCQMC}.  Here we focus on the common primitive of estimating a time-evolved expectation value $\braket{O(T)}$.

In these applications, the dominant cost is the expense of decomposing $e^{-iHT}$ into a
series of elementary gates.  Deterministic product formulae incur a Trotter error, which is controlled by the commutators
of the $L$ Hamiltonian terms~\cite{Childs2021}, and the complexity scales with $L$.  Randomised methods such as
\qdrift{}~\cite{Campbell2019} replace this Trotter error with one which depends on the total weight of all Hamiltonian
terms, $\lambda=\sum_j|c_j|$ rather than $L$.  In \qdrift{} each step applies a single rotation
$e^{-i\theta_jP_j}$ whose generator is sampled from the terms of $H=\sum_jc_jP_j$ with probability $q_j\propto|c_j|$.
However, being non-deterministic, these \qdrift{} and adjacent methods produce estimators with potentially large bias
and variance~\cite{Campbell2019,Chen2021Concentration}.  These errors translate into the number of steps and evaluation
trajectories required to achieve a certain target accuracy, $\varepsilon$.  Compared with methods whose complexity scales
optimally (such as qubitisation with an additive $\log(\sfrac{1}{\varepsilon})$ factor~\cite{LowChuang2019}),
product formulae and randomised methods trade a sub-optimal scaling ($\mathcal{O}(\varepsilon^{-1})$ for
first-order formulae and \qdrift{}) for an ancilla-free implementation with small constant overheads at the sizes of
interest~\cite{Childs2018PNAS,Campbell2019}.  Given the minimal qubit requirements, if one
could reduce the bias and variance sufficiently, it would make randomised methods among the most
promising candidates for near and mid-term quantum implementations, where qubits and circuit depth are scarce.

To address this error reduction, several works have been proposed, and we focus on those
specifically related to \qdrift{}.  Ref.~\cite{KGR2023} showed that the underlying distribution could be made arbitrary,
while retaining asymptotic unbiasedness using corrections to each rotation.  To target the circuit depth specifically,
Ref.~\cite{KGR2023} adds gate-cost aware reweightings to the distribution, which reduces circuit depth.  Ref.~\cite{CAS2026} instead reweights measurement outcomes to minimise a gate-cost weighted variance.
For more complex per-step operations, qSWIFT~\cite{Nakaji2024qSWIFT} uses higher-order terms
of the \qdrift{} error expansion to cancel the leading order error but introduces an extra
ancilla qubit, while qSHIFT~\cite{Lee2026} includes higher orders but without the ancilla, using adaptive sampling and
classical computation.  On the other hand, MLMC-\qdrift{}~\cite{Mohammadipour2026} proposes multilevel estimators to reduce
variance.
Since both deterministic and randomised approaches have their respective advantages, several works
propose to combine them to tackle the relevant parts of the problem where each excels.  For example, composite and
partially randomised schemes apply a product formula to the terms with the largest coefficients
and sample from the remainder~\cite{Ouyang2020,JinLi2021PartiallyRandom,Hagan2023Composite,Pocrnic2024}. Ref.~\cite{Paganelli2026When} shows that treating the dominant terms deterministically and sampling the rest is useful for Hamiltonians with many terms and unevenly weighted coefficients, at moderate precision.  Partially randomised evolution has been used for phase estimation~\cite{Gunther2026PartialRandom}, and combining it with
the partially fault-tolerant STAR architecture~\cite{Akahoshi2023} brings chemically accurate molecular simulation into the early
fault-tolerant regime~\cite{Kanasugi2026EarlyFTQPE}.  Other ways of reducing the bias include using multiple increasing depth \qdrift{} runs combined
with Richardson extrapolation~\cite{Watson2024RandomlyCompiled} (multi-product formulas~\cite{Zhuk2024DynamicMPF} and Richardson extrapolation~\cite{Watson2024TrotterMitigation} do the same for deterministic product formulas), at the cost of a
larger variance, or removing the bias entirely with unbiased compilers, which write $e^{-iHT}$ as a quasi-probability mixture of circuits
and need correspondingly more samples~\cite{Kiumi2025TEPAI,Granet2024Hamiltonian}.  For unbiased compilers,
Ref.~\cite{Dai2026structure} partly reduces this sampling overhead using classical variance-reduction techniques.  However, all of these methods define the \qdrift{} sampling distribution from the Hamiltonian alone. Product-formula errors can also be bounded more tightly using the initial state~\cite{Sahinoglu2021LowEnergy,Chen2024AverageCase} or the observable~\cite{Yu2025ObservableDriven,Li2024Principal}, but these works use that information in deterministic formulas, not in a sampling distribution.
Refs.~\cite{Wu2025fluctuation,Fan2025adaptive} update the sampling rule using problem-aware information and truncated
expansions, but use only the state, re-measured at every step, and minimise the distance between the
approximate and exact evolved states rather than the error in a chosen observable.  For the energy as a targeted observable, Refs.~\cite{Kivlichan2019PhaseEstimation,Poulin2015TrotterStep} have used
classically computed approximate ground states to allocate Hamiltonian term frequencies, and Ref.~\cite{Li2024Principal} uses
observable dependence in error bounds for Trotterised methods.

Here we introduce \statedrift{}, which chooses the \qdrift{} sampling distribution from the initial state, $\ket{\psi}$,
and the observable, $O$, as well as from the Hamiltonian (Figure~\ref{fig:hero}) to reduce the error in estimating
time evolved quantities, $\braket{O(T)}$.  It involves optimising over a parameterised distribution which can be used
to target the bias and/or the variance of an estimator, or a combination of the two to reduce both simultaneously.  We
use a fourth-order expansion to give an objective function to estimate, while the quantities required for the probability allocation
are low-order disturbances and commutators involving $\ket{\psi}$, $O$ and the Hamiltonian terms $P_j$.  We also
provide rigorous error bounds on the truncation.  These quantities can be effectively estimated on classical surrogates
such as tensor networks, or measured with classical shadows~\cite{Huang2020Shadows,Zhao2021Fermionic,Wan2023Matchgate}, and we show that these optimised distributions can
significantly reduce the circuit complexity and number of trajectories required to estimate observables.  \statedrift{} only requires information on $t=0$ quantities, and the default is to fix the distribution once before any simulation. It is also compatible with many of the adaptive or hybrid strategies mentioned above, and as such can be generalised to longer time evolution at the expense of more quantum resources.

\begin{figure}[!t]
\centering
\includegraphics[width=\columnwidth]{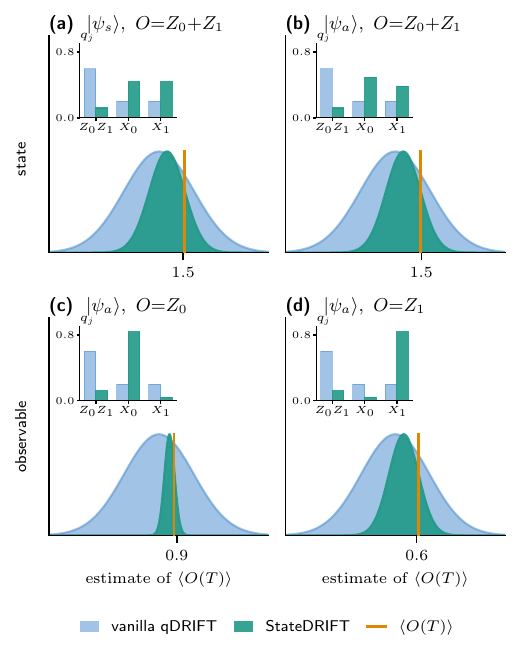}
\caption{\textbf{Toy example: two-site transverse-field Ising quench
$H=3Z_0Z_1+X_0+X_1$, estimating $\braket{O(T)}$ with one \qdrift{} step at $T=0.2$.}
Insets: \qdrift{} probability allocations over the three terms (vanilla \qdrift{} blue,
\statedrift{} teal throughout).
Main panels: each estimator as a band about its mean of width $\sqrt{\mathrm{MSE}}$~\eqref{eq:mse-obj}
at $M=1$ trajectory, against the exact value of $\braket{O(T)}$ (orange).  Both estimators undershoot $\braket{O(T)}$, but at fixed $N=1$ \statedrift{}
has both the smaller bias and the smaller variance. \textbf{(a,b)~State dependence} at fixed $O=Z_0+Z_1$, for (a) a symmetric
state $\ket{\psi_s}$ and (b) an asymmetric $\ket{\psi_a}$. \textbf{(c,d)~Observable
dependence} at fixed $\ket{\psi_a}$, for (c) $O=Z_0$ and (d) $O=Z_1$.}
\label{fig:toy-qubit-main}
\end{figure}

\section{State-aware bias and variance}
\label{sec:state-aware}

We estimate the time-evolved expectation value
\begin{equation}
  \braket{O(T)} = \braket{\psi|e^{iHT}Oe^{-iHT}|\psi}
  \label{eq:estimand}
\end{equation}
for a Hamiltonian $H=\sum_{j=1}^{L}c_jP_j$ with real coefficients $c_j$ and Pauli strings $P_j$, a
reference state $\ket{\psi}$, and an observable $O$.  An identity term only adds a global phase,
which cancels in $\braket{O(T)}$, so we drop it.  The total evolution time $T$ is fixed, and the number of \qdrift{} steps
$N$ sets the step size $dt=T/N$.  Unless stated otherwise, $[H,O]\neq0$; we also treat
$[H,O]=0$, including energy estimation, as an important simplifying case.  Throughout, the
baseline is vanilla \qdrift{},
$q_j^c=|c_j|/\lambda$ with $\lambda=\sum_j|c_j|$~\cite{Campbell2019}.

Our aim is to choose the sampling distribution $\bq$ using information about $\ket{\psi}$
and $O$ at $t=0$.  Relevant to our results is Ref.~\cite{Chen2021Concentration}, which showed that for a fixed input
state, a single realisation concentrates about the exact evolution, and the gate count needed to reach
a target accuracy is lower by a factor of the qubit number than the state-independent bound that
must hold for all inputs.  This saving requires only that the state be fixed, not that it be known,
and so does not exploit state-specific structure.

\subsubsection*{Estimator and error decomposition}

A \qdrift{} trajectory samples an index $j_k$ independently from $\bq$ at each step and
applies
\begin{equation}
  V_{j_k}:=e^{-i\theta_{j_k}P_{j_k}}, \qquad \theta_j:=\frac{c_jdt}{q_j}.
\end{equation}
The corresponding estimator is
\begin{equation}
  \hat O_N:=\braket{\psi|U_N^\dagger O U_N|\psi}, \qquad U_N:=\prod_{k=1}^{N}V_{j_k}.
  \label{eq:qdrift-est}
\end{equation}
For $M$ independent trajectories, the mean squared error (MSE) between the estimate and the true evolution is given by
\begin{equation}
\begin{split}
\mathrm{MSE}_N&:=\mathbb{E}_{\bq}\Bigl[\Bigl(\tfrac1M\textstyle\sum_{m=1}^{M}\hat{O}_N^{(m)}
  - \braket{O(T)}\Bigr)^{\!2}\Bigr] \\&=B_N^2+\frac{V_N}{M},
  \\
  B_N&:=\mathbb E_{\bq}[\hat O_N]-\braket{O(T)},
  \\
  V_N&:=\operatorname{Var}_{\bq}[\hat O_N]= \mathbb{E}_{\bq}[\hat{O}_N^2] - \mathbb{E}_{\bq}[\hat{O}_N]^2.
  \label{eq:mse-obj}
\end{split}
\end{equation}
Thus, additional trajectories suppress the variance $V_N$ contribution but not the bias $B_N$, and so reducing the bias presents
a useful lever to reduce the overall circuit depth required to reach a given accuracy in the
simulation.

\subsubsection*{Single-step illustration}

For one step, conjugation by a Pauli rotation gives the exact expression
\begin{equation}
\begin{split}
  \hat O_1(j)&=\braket{O}
  +\sin^2\!\theta_j\,\gamma_j
  +\tfrac12\sin(2\theta_j)\,\beta_j,\\
  \gamma_j&:=\braket{\psi|P_jOP_j|\psi}-\braket{O},
  \\
  \beta_j&:=-i\braket{\psi|[O,P_j]|\psi} .
  \label{eq:gamma-beta-def}
\end{split}
\end{equation}
Writing
$s_j:=\sin^2\!\theta_j\,\gamma_j+\tfrac12\sin(2\theta_j)\,\beta_j$ for the shift term $j$ produces in
the estimator, the exact single-step bias and variance are
\begin{align}
  B_1&=\braket{O}+\sum_jq_js_j-\braket{O(T)},
  \label{eq:toy-bias}\\
  V_1&=\sum_{j<k}q_jq_k(s_j-s_k)^2.
  \label{eq:toy-var}
\end{align}
The state- and observable-dependent responses $\gamma_j$ and $\beta_j$ will determine how
the Hamiltonian terms should be sampled.

The mechanism can be illustrated in a two-site transverse-field Ising model,
\begin{equation}
  H=3Z_0Z_1+X_0+X_1,
  \label{eq:toy-H}
\end{equation}
with the real product states
$\ket{\psi_s}=\ket{\sfrac{\pi}{4}}^{\otimes2}$ and
$\ket{\psi_a}=\ket{\sfrac{\pi}{6}}\otimes\ket{\sfrac{\pi}{3}}$, where
$\ket{\theta}=\cos(\theta/2)\ket{0}+\sin(\theta/2)\ket{1}$.  We estimate either the
total magnetisation $Z_0+Z_1$ or one of the local magnetisations $Z_0$ and $Z_1$.
Vanilla \qdrift{} assigns the distribution
$\bq^c=(\sfrac35,\sfrac15,\sfrac15)$ to $(Z_0Z_1,X_0,X_1)$, following the Hamiltonian
coefficients alone.

Because the states and operators are real while
$[O,P_j]$ is anti-Hermitian, $\beta_j=0$ for every term.  For small angles,
$s_j=(c_jT/q_j)^2\gamma_j+\mathcal O(T^4)$, and the leading variance is minimised by
\begin{equation}
  q_j\propto|c_j|\sqrt{|\gamma_j|}.
\end{equation}
This distribution equalises the leading shifts when the non-zero $\gamma_j$ have a common
sign.  Here $Z_0Z_1$ commutes with every observable considered, so
$\gamma_{Z_0Z_1}=0$, whereas conjugation by $X_i$ flips $Z_i$ and gives
$\gamma_{X_i}=-2\braket{Z_i}$ whenever site $i$ occurs in $O$.  Vanilla \qdrift{}
therefore gives most of its probability to a term that does not directly change the selected
observable, while the state-aware rule moves that probability to the transverse terms.

Figure~\ref{fig:toy-qubit-main} shows both kinds of conditioning.  For fixed
$O=Z_0+Z_1$, the symmetric state assigns equal weight to $X_0$ and $X_1$, whereas the
asymmetric state favours the site with larger $|\braket{Z_i}|$.  For fixed
$\ket{\psi_a}$, choosing $O=Z_0$ or $O=Z_1$ concentrates the state-aware weight on $X_0$
or $X_1$, respectively.  At $T=0.2$, the state-aware estimator has both lower bias and lower
variance in all four cases and reduces $\sqrt{\mathrm{MSE}}$ by $2$--$7\times$.  The plotted
distributions retain $20\%$ of the vanilla distribution through the minimum-weight mixture
introduced below.

\subsubsection*{Generalisation to multiple steps}

The single-step toy example depends only on $\gamma_j$ and $\beta_j$.  At $N>1$ extra cross-step terms
enter the expansion, because we get contributions from pairs of sampled Pauli terms
$P_{j_k}$ and $P_{j_{k'}}$ with $k\neq k'$, so we need more flexible definitions.  We define the following terms, in
which $A$ and $B$ stand for either a Hamiltonian term $P_j$ or $H$ itself,
\begin{equation}
\begin{split}
  \gamma_A(X)&:=\braket{\psi|AXA|\psi}-\braket{X},\\
  \beta_A(X)&:=-i\braket{\psi|[X,A]|\psi},\\
  \varphi_{A,B}(X)&:=\braket{\psi|[A,[X,B]]|\psi} .
  \label{eq:cross-step-defs}
\end{split}
\end{equation}
We also use the contractions
\begin{equation}
\begin{split}
  \beta_H(X)&:=\sum_jc_j\beta_{P_j}(X),\\
  \varphi_{H,P_j}(X)&:=\sum_bc_b\varphi_{P_b,P_j}(X),\\
  \varphi_{P_j,H}(X)&:=\sum_bc_b\varphi_{P_j,P_b}(X),\\
  \varphi_{H,H}(X)&:=\sum_{a,b}c_ac_b\varphi_{P_a,P_b}(X)
    =\bra{\psi}[H,[X,H]]\ket{\psi}.
  \label{eq:H-sums}
\end{split}
\end{equation}
When $X=O$, we suppress the observable and write $\gamma_j := \gamma_{P_j}(O)$, $\beta_j := \beta_{P_j}(O)$,
$\varphi_{a,b} := \varphi_{P_a,P_b}(O)$, $\varphi_{H,j} := \varphi_{H,P_j}(O)$ and
$\varphi_{j,H} := \varphi_{P_j,H}(O)$.  In particular,
\begin{equation}
  \beta_H=\frac{d\braket{O(t)}}{dt}\bigg|_0,
  \qquad
  \varphi_{H,H}=\frac{d^2\braket{O(t)}}{dt^2}\bigg|_0.
  \label{eq:target-derivatives}
\end{equation}

The above quantities are all independent of $\bq$.  The general bias and variance we will derive depend on $\bq$ only
through density terms that weight these quantities by powers of $c_j/q_j$, the first of which
are
\begin{align}
  b_{\gamma,2}&:=\sum_j\frac{c_j^2\gamma_j}{q_j}
    =\mathbb E_{\bq}\!\left[\left(\frac{c_j}{q_j}\right)^2\gamma_j\right],\notag\\
  v_{\gamma^2,4}&:=\sum_j\frac{c_j^4\gamma_j^2}{q_j^3}
    =\mathbb E_{\bq}\!\left[\left(\frac{c_j}{q_j}\right)^4\gamma_j^2\right],\notag\\
  v_{\beta^2,2}&:=\sum_j\frac{c_j^2\beta_j^2}{q_j}-\beta_H^2
    =\operatorname{Var}_{\bq}\!\left[\left(\frac{c_j}{q_j}\right)\beta_j\right].
  \label{eq:per-step-defs}
\end{align}
The expectation is over the term $j$ sampled from $\bq$ itself, and each subscript names what is
summed and the power of the weight $c_j^{\,p}/q_j^{\,p-1}$, a convention every density below follows.
The last is a variance rather than a raw moment because $\mathbb{E}_{\bq}[(c_j/q_j)\beta_j] =
\beta_H$, and every density built from $\beta_j$ below subtracts this mean.  Combining $v_{\gamma^2,4}$ and $b_{\gamma,2}$ gives the variance of the shift a single sample
produces,
\begin{equation}
  v_\gamma:=v_{\gamma^2,4}-b_{\gamma,2}^2
  =\operatorname{Var}_{\bq}\!\left[\left(\frac{c_j}{q_j}\right)^2\gamma_j\right].
  \label{eq:Vgamma-def}
\end{equation}
Along with the terms above, cross-step densities couple pairs of steps, for example
\begin{equation}
  v_\varphi:=\sum_{a,b}\frac{c_a^2c_b^2\varphi_{a,b}^2}{q_aq_b},
  \label{eq:cross-step-densities}
\end{equation}
with the rest defined in Appendix~\ref{app:densities}.
Expanding to fourth order in $dt$ gives (Appendix~\ref{app:S2})
\begin{equation}
\begin{split}
  &B_N =N\bigl(b_{\gamma,2}-\tfrac12\varphi_{H,H}\bigr)dt^2 \\
  &+N\bigl[b_{\beta,3}+(N{-}1)B_{3,2}\bigr]dt^3 \\
  &+N\bigl[b_{\gamma,4}+(N{-}1)B_{4,2}+(N{-}1)(N{-}2)B_{4,3}\bigr]dt^4 \\
  &+\mathcal O(dt^5),
\end{split}
\label{eq:BN_full}
\end{equation}
and
\begin{equation}
\begin{split}
  &V_N =Nv_{\beta^2,2}dt^2 \\
  &+N\bigl[v_{\beta\gamma,3}+(N{-}1)V_{3,2}\bigr]dt^3 \\
  &+N\bigl[v_\gamma-v_{\beta^2,4}+(N{-}1)V_{4,2}+(N{-}1)(N{-}2)V_{4,3}\bigr]dt^4 \\
  &+\mathcal O(dt^5).
\end{split}
\label{eq:VN_full}
\end{equation}
The single-step terms $b_{\beta,3}$, $b_{\gamma,4}$,
$v_{\beta\gamma,3}$ and $v_{\beta^2,4}$, together with the remaining cross-step
densities, are defined in Appendix~\ref{app:densities}.
The leading bias is proportional to the distance between $b_{\gamma,2}$, from the distribution, and the target
$\tfrac12\varphi_{H,H}$.  When the contributions to
$b_{\gamma,2}$ have mixed signs, a distribution allocation of qDRIFT probabilities $\bq$ can tune these two
quantities to cancel and remove the $dt^2$ bias contribution, leaving the terms of order $dt^3$ and higher.  Since $dt = \sfrac{T}{N}$, the cross-step terms $N(N{-}1)B_{3,2}\,dt^3$ and $N(N{-}1)(N{-}2)B_{4,3}\,dt^4$ also fall as $\sfrac{1}{N}$, so the bias keeps its $\sfrac{1}{N}$ scaling with a smaller coefficient.  The variance separates into single-step terms proportional to $N$
and cross-step terms with higher powers of $N$.

\subsubsection*{Symmetry and observable-dependent simplifications}

The $\beta_j$ term is computed via the gradient as follows:
\begin{equation}
  \beta_j=\frac{d}{d\theta}
  \braket{\psi|e^{i\theta P_j}Oe^{-i\theta P_j}|\psi}\bigg|_{\theta=0}
  =-i\braket{\psi|[O,P_j]|\psi} .
  \label{eq:beta-slope}
\end{equation}

\begin{lemma}\label{lem:beta}
If a basis exists in which the amplitudes of $\ket{\psi}$ and the matrix entries of $O$
and $P_j$ are all real, then $\beta_j=0$.
\end{lemma}
\begin{proof}
Hermitian operators with real entries are symmetric, so $[O,P_j]$ is real and
antisymmetric and its expectation in a state with real amplitudes vanishes.  Equivalently,
time-reversal symmetry~\cite[Ch.~4]{SakuraiNapolitano2020} makes the expectation in
Eq.~\eqref{eq:beta-slope} even in $\theta$, and its slope at the origin is zero.
\end{proof}

\begin{corollary}\label{cor:beta-pauli}
If $\ket{\psi}$ has real amplitudes in the computational basis and $P_j$ and every Pauli string of $O$
contain an even number of $Y$ operators, then $\beta_j=0$.
\end{corollary}
The proof is given in Appendix~\ref{app:S3}.  Electronic-structure Hamiltonians over
real molecular orbitals, real observables, and Hartree--Fock or other real-amplitude ansatz states
naturally satisfy this condition.  Complex reference states or Hamiltonian terms with an odd
number of $Y$ operators need not; our lattice benchmarks include this $\beta_j\neq0$ regime.

Under the reality condition of Lemma~\ref{lem:beta}, $\beta_j=0$ and the bias and variance are even in $dt$, so the odd orders vanish and, to fourth order, the expansion reduces to
\begin{align}
  &B_N =N\bigl(b_{\gamma,2}-\tfrac12\varphi_{H,H}\bigr)dt^2 \notag\\*
  &+N\bigl[b_{\gamma,4}+(N{-}1)B_{4,2}+(N{-}1)(N{-}2)B_{4,3}\bigr]dt^4 \notag\\*
  &+\mathcal{O}(dt^6), \notag\\*
  &V_N =N\bigl[v_\gamma+(N{-}1)V_{4,2}+(N{-}1)(N{-}2)V_{4,3}\bigr]dt^4 \notag\\*
  &+\mathcal{O}(dt^6).
  \label{eq:VN_beta0}
\end{align}
In Appendix~\ref{app:S3}, we detail which terms survive under each condition.
For $N=1$ this becomes
\begin{equation}
\begin{split}
  B_1&=\bigl(b_{\gamma,2}-\tfrac12\varphi_{H,H}\bigr)dt^2
       +b_{\gamma,4}dt^4+\mathcal{O}(dt^6),\\
  V_1&=v_\gamma dt^4+\mathcal{O}(dt^6),
  \label{eq:V1B1-beta0}
\end{split}
\end{equation}
whereas without the reality condition
\begin{align}
  B_1&=\bigl(b_{\gamma,2}-\tfrac12\varphi_{H,H}\bigr)dt^2       +b_{\beta,3}dt^3+b_{\gamma,4}dt^4\notag\\
  &\quad+\mathcal{O}(dt^5),\notag\\
V_1&=v_{\beta^2,2}dt^2+v_{\beta\gamma,3}dt^3+(v_\gamma-v_{\beta^2,4})dt^4\notag\\
  &\quad+\mathcal{O}(dt^5).
  \label{eq:V1B1-general}
\end{align}
For energy estimation, $O=H$, the quantities $\varphi_{j,H}$ and
$\varphi_{H,H}$ vanish.  In the real-amplitude regime,
\begin{equation}
\begin{split}
  &B_N =Nb_{\gamma,2}dt^2 \\
  &+N\bigl[b_{\gamma,4}+(N{-}1)B_{4,2}+(N{-}1)(N{-}2)B_{4,3}\bigr]dt^4 \\
  &+\mathcal{O}(dt^6),\\
  &V_N =N\bigl[v_\gamma+(N{-}1)(v_{\gamma\cdot\varphi_{H,j},3}+\tfrac12v_\varphi) \\
  &\quad +\tfrac13(N{-}1)(N{-}2)v_{\varphi_{H,j}^2,2}\bigr]dt^4 \\
  &+\mathcal{O}(dt^6).
\end{split}
\label{eq:V1B1-energy}
\end{equation}
Away from an eigenstate (Lemma~\ref{lem:collapse}, Appendix~\ref{app:S3}), the
$\varphi_{H,\cdot}$ densities need not vanish.  Their higher powers of $N$ make them
increasingly important when the distribution is deployed at greater circuit depth.

\subsubsection*{Vanilla \qdrift{} mixing}

A distribution optimised for the reference state and observable may assign zero weight to a
term with $\gamma_j=0$.  Such a term can nevertheless drive the later dynamics when $N>1$.
We therefore mix every state-aware rule with the vanilla distribution,
\begin{equation}
  \bq=\tau\bq^c+(1-\tau)\bq^{\mathsf{SD}},
  \qquad \tau\in(0,1].
  \label{eq:tau-mixing}
\end{equation}
Besides preserving the dynamics, $\tau$ controls the largest reweighting relative to
vanilla \qdrift{}.

\begin{lemma}[Bounded reweighting]\label{lem:minweight}
Let $\bq$ be given by~\eqref{eq:tau-mixing} for any $\bq^{\mathsf{SD}}$ with
$q^{\mathsf{SD}}_j \ge 0$, and let $\omega_j := q^c_j/q_j$ be the probability ratio relative to
vanilla \qdrift{}.  Then
\begin{equation}
  \max_j \omega_j \le \frac{1}{\tau},
  \qquad
  \Expect_{\bq^c}[\omega]=\sum_j\frac{(q_j^c)^2}{q_j}\le\frac{1}{\tau}.
  \label{eq:omega-bound}
\end{equation}
\end{lemma}

We prove this in Appendix~\ref{app:S4}.  Through the rest of the paper, we set $\tau=0.2$ as a
minimum value.

\section{\statedrift{} distributions}
\label{sec:allocations}

Next, given the analytical bias and variance formulae, we derive the corresponding
\statedrift{} distribution allocations for the probabilities $\bq$.  The bias carries the distribution
through $b_{\gamma,2}$ and the variance through $v_\gamma$ or $v_{\beta^2,2}$, and in general
these are minimised by different $\bq$, so no single probability allocation is optimal for both.  We first
give the variance-optimal distribution allocations, which are available in closed form, and then the bias,
which is not.  For simplicity and to build intuition, we begin with $N=1$ so only $\gamma_j$ and
$\beta_j$, and not the cross-step densities, contribute to the bias and variance.  We then
generalise to $N>1$.

\subsubsection*{Minimum-variance distribution allocation}

\paragraph{\textbf{$\gamma$-driven regime ($N=1$, $\beta_j = 0$).}}
At $\beta_j = 0$ the single-step variance is $v_\gamma\,dt^4$~\eqref{eq:V1B1-beta0}.
When $\mathrm{sign}(\gamma_i)=\mathrm{sign}(\gamma_j)\;\;\forall\,\gamma_i,\gamma_j \neq 0$,
it is minimised by
\begin{equation}
  \bq^\star_\gamma \;:\; q_j \;\propto\; |c_j|\sqrt{|\gamma_j|}
  \label{eq:gamma-neyman}
\end{equation}
because $v_\gamma = \Var_{\bq}[X_j]$ for the random variable
$X_j = (c_j/q_j)^2\gamma_j$~\eqref{eq:Vgamma-def}.  A variance vanishes when its argument takes a
common value $x$ for every $j$, and solving $(c_j/q_j)^2\gamma_j = x$ with
$\sum_j q_j = 1$ fixes $\sqrt{|x|} = \sum_j|c_j|\sqrt{|\gamma_j|}$, leaving the
distribution~\eqref{eq:gamma-neyman} (Appendix~\ref{app:S4}).

\paragraph{\textbf{$\beta$-driven regime ($N=1$, $\beta_j \ne 0$).}}
With $\beta_j \ne 0$ the leading order of the variance is $v_{\beta^2,2}\,dt^2$, with
$v_{\beta^2,2} = \sum_j c_j^2\beta_j^2/q_j - \beta_H^2$~\eqref{eq:per-step-defs}.  The weights
$c_j^2\beta_j^2$ are non-negative and $\beta_H$ does not depend on $\bq$, so Cauchy--Schwarz
(Appendix~\ref{app:S4}) minimises the variance, for $c_j\beta_j\neq0$ for every $j$, at
\begin{equation}
\begin{aligned}
  \bq^\star_\beta \;&:\; q_j \;\propto\; |c_j||\beta_j|, \\
  V_1(\bq^\star_\beta) \;&=\; \biggl[\Bigl(\sum_j |c_j||\beta_j|\Bigr)^2
    - \beta_H^2\biggr] dt^2 + \mathcal{O}(dt^3).
\end{aligned}
  \label{eq:beta-neyman}
\end{equation}
A density carrying a single power of $1/q_j$ is minimised by weights proportional to the
square root of its per-term coefficient, which is the Neyman
allocation~\cite{Neyman1934,Owen2013}.  Similar problem-specific information appears also in
Refs.~\cite{Wecker2015,Crawford2021,Arrasmith2020}, where it is used to allocate measurements across Pauli groups
in variational energy estimation, or operator sampling in adaptive optimisers.

\begin{figure*}[t!]
\centering
\includegraphics[width=\textwidth]{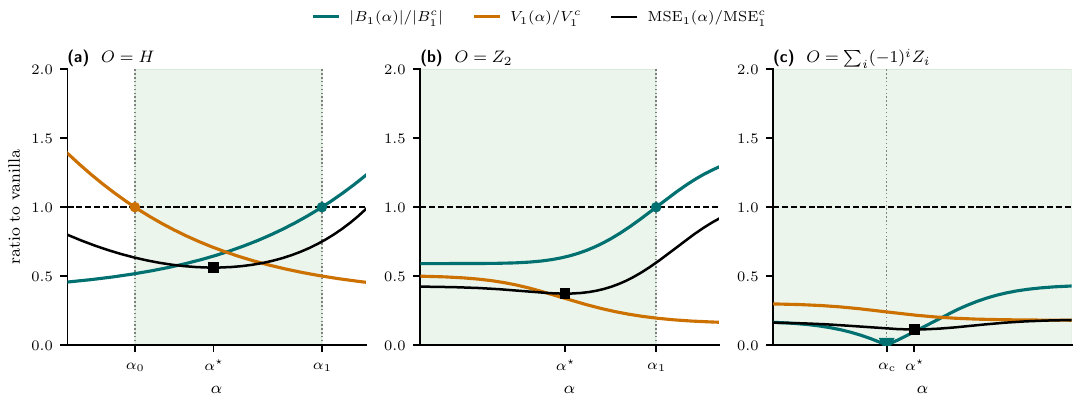}
\caption{\textbf{Joint improvement across the interpolating family.}
Ratios to the vanilla \qdrift{} baseline for the XY chain
$H = \sum_i (X_iX_{i+1} + Y_iY_{i+1}) + \sum_i Z_i$ at $n=3$, with $\tau$-mixing $\tau=0.20$ and
$dt = 10^{-3}$.  The reference state is
$\ket{\psi} = e^{-i\theta_X\sum_i Z_i}\,e^{-i\theta_{XY}\sum_i (X_iX_{i+1}+Y_iY_{i+1})}\ket{011}$
with $\theta_{XY} = 0.2$ and $\theta_X = 1.3$, one Hamiltonian-variational layer per term group.
Teal and orange are the exact single-step $|B_1|$ and $V_1$, black the
mean-squared error~\eqref{eq:mse} with the square at its minimum $\alpha^\star$.  The shaded region
is where both bias and variance are reduced below vanilla \qdrift{}, and tends to the interval
$(\alpha_0,\alpha_1)$ of~\eqref{eq:interval-boundaries} as $dt\to0$.  In each panel we take the
number of \qdrift{} runs, $M$, to be the value at which the two contributions to the mean-squared
error are equal at the vanilla distribution, $M = V_1(\bq^c)/B_1(\bq^c)^2$, giving
$4.4\times10^4$, $1.6\times10^4$ and $1.4\times10^5$ for (a), (b) and (c) respectively.
The subplots show different observables, each exhibiting different behaviour.
\textbf{(a)} shows the energy, $O=H$, where $\tfrac12\varphi_{H,H}=0$ and $\bq^\star_\gamma$
costs variance, so both $\alpha_0$ and $\alpha_1$ of~\eqref{eq:interval-boundaries} exist.
\textbf{(b)} is a local observable ($I\otimes Z_2\otimes I$) with $\tfrac12\varphi_{H,H}\neq0$, where
$\bq^\star_\gamma$ already improves the variance, so only $\alpha_1$ exists.
\textbf{(c)}, the staggered magnetisation $O=\sum_i(-1)^iZ_i$, shows a regime where the $\gamma$ values can be mixed sign, and hence a cancellation
$b_{\gamma,2}\to\tfrac12\varphi_{H,H}$ can occur in the bias, at $\alpha_{\mathrm c}$, where
$|B_1|$ falls to $10^{-4}$ of vanilla.  The latter is outside the regime of
Corollary~\ref{prop:joint-main} because sign-definite $\bgamma$ is what makes $b_{\gamma,2}$
non-decreasing in $\alpha$, and without it neither endpoint exists.}
\label{fig:joint-crossover}
\end{figure*}

\subsubsection*{Bias-cancelling distribution allocation}

While the two distributions above minimise the variance in each case, they do not
minimise the bias of the estimator.  Fixing $\bq$ by~\eqref{eq:gamma-neyman} (under Lemma~\ref{lem:beta})
or~\eqref{eq:beta-neyman} leaves the single-step biases as:
\begin{align}
  B_1(\bq^\star_\gamma) &= \bigl(b^\star_\gamma - \tfrac12\varphi_{H,H}\bigr)dt^2
    + b_{\gamma,4}(\bq^\star_\gamma)\,dt^4 + \mathcal{O}(dt^6), \notag\\
  B_1(\bq^\star_\beta) &= \biggl[\Bigl(\sum_k|c_k||\beta_k|\Bigr)
    \Bigl(\sum_j \frac{|c_j|\gamma_j}{|\beta_j|}\Bigr) \notag\\
    &\qquad - \tfrac12\varphi_{H,H}\biggr]dt^2 + \mathcal{O}(dt^3)
\end{align}
with $b^\star_\gamma := b_{\gamma,2}(\bq^\star_\gamma)$ and
$|b^\star_\gamma| = \bigl(\sum_j|c_j|\sqrt{|\gamma_j|}\bigr)^2$.
The shift $\tfrac12\varphi_{H,H}$ does not depend on $\bq$, so a distribution which sends
$b_{\gamma,2}$ to $\tfrac12\varphi_{H,H}$ removes the bias at order $dt^2$.  Since
$\bq^\star_\gamma$ sends $b_{\gamma,2}$ to $b^\star_\gamma$ instead, minimising the density
and cancelling the bias coincide only when we have either 1)~$\tfrac12\varphi_{H,H}\le b^\star_\gamma$ when every $\gamma_j>0$ or 2)~$\tfrac12\varphi_{H,H}\ge b^\star_\gamma$ when every $\gamma_j<0$.  Then the bias is minimised by~\eqref{eq:gamma-neyman} but not cancelled.  This is the case for the energy observable $O=H$, where $\varphi_{H,H}=0$ and the $dt^2$ bias is $b_{\gamma,2}\,dt^2$.  The two orders pull in
different directions even then, since the $\bq$-dependent part of $b_{\gamma,4}$, $\sum_j\sfrac{c_j^4\gamma_j}{q_j^3}$, has its smallest magnitude at $q_j \propto |c_j||\gamma_j|^{1/4}$
and $b_{\gamma,2}$ at~\eqref{eq:gamma-neyman}, so the total bias has no closed-form
minimiser.

The second bias is not minimised at all.  Built from $\bbeta$ alone,
$\bq^\star_\beta$ minimises $b_{\gamma,2}$ only when $|\gamma_j| \propto \beta_j^2$, and the bias
it leaves weights each term by $\gamma_j/|\beta_j|$, so minimising the variance alone can increase the bias: $\bq^\star_\beta$ rarely
samples terms with small $|\beta_j|$, which barely affect the variance but still contribute to the bias through $\gamma_j$.  At small $dt$ the $dt^2$ variance still dominates the $dt^4$ bias, so
$\bq^\star_\beta$ minimises $\mathrm{MSE}_1$ at small $M$, while increasing $M$ suppresses
$V_1/M$ and leaves $B_1^2$.

\paragraph{\textbf{Joint distribution allocation ($N=1$).}}
In the above, we had special cases with closed-form variance-optimal distributions, which
were not bias-optimal, and minimising the bias or the variance individually leads to a worsening
of the other.  In this section, we propose a distribution which interpolates between these
regimes, so that an optimal trade-off between the two quantities can be found instead.  We
therefore optimise over a parameterised family of distributions.

A density $\sum_j c_j^2 f_j/q_j$ with $f_j > 0$ is minimised at
$q_j \propto |c_j|\sqrt{f_j}$, and both $|b_{\gamma,2}|$ ($f_j = |\gamma_j|$ for sign-definite $\bgamma$) and
$v_{\beta^2,2}$ ($f_j = \beta_j^2$) take that form up to the $\bq$-independent
$\beta_H^2$, as does any weighted sum of the two.
With a hyperparameter $\alpha \ge 0$ we therefore take
\begin{equation}
  J_\alpha(\bq)
  \;:=\; \sum_j \frac{c_j^2(|\gamma_j| + \alpha\beta_j^2)}{q_j}
  \label{eq:combined_obj}
\end{equation}
and the same Cauchy--Schwarz argument yields the combined
distribution allocation (Appendix~\ref{app:S4}, Lemma~\ref{lem:combined}):
\begin{equation}
  \bq^{(\alpha)} \;:\; q_j \;\propto\; |c_j|\sqrt{|\gamma_j| + \alpha\beta_j^2}
  \label{eq:combined-neyman}
\end{equation}
At $\alpha = 0$ this recovers $\bq^\star_\gamma$; as $\alpha\to\infty$
the $\beta_j^2$ term dominates and it recovers $\bq^\star_\beta$, so the two variance optima are
its ends.
To choose $\alpha$ at a given sample budget, we return to the observable estimate
itself. The target $\braket{O(T)}$ is estimated by averaging the
single-step ($N=1$) qDRIFT estimator $\hat{O}_1$ over $M$ independent samples, where
throughout, $M$ counts independent random-compilation trajectories rather than measurement shots
on a fixed circuit. We fix $\alpha$ by minimising the mean-squared error of this
$M$-sample estimate,
\begin{equation}
\begin{split}
  \mathrm{MSE}_1(\alpha) \;:&=\; \mathbb{E}\!\left[
    \Bigl(\tfrac{1}{M}\textstyle\sum_{k=1}^M \hat{O}_1^{(k)} - \braket{O(T)}\Bigr)^{\!2}
  \right] \\
  \;&=\; B_1(\alpha)^2 + \frac{V_1(\alpha)}{M}.
\end{split}
  \label{eq:mse-def}
\end{equation}
To $O(dt^4)$, the bias contributes $B_1^2 = (b_{\gamma,2} - \tfrac12\varphi_{H,H})^2 dt^4$, and its higher terms enter only at $dt^5$ and beyond. $V_1$ is the full single-step
variance~\eqref{eq:VN_full} at $N=1$, so
\begin{equation}
\begin{split}
  \mathrm{MSE}_1(\alpha) &\;=\; (b_{\gamma,2} - \tfrac12\varphi_{H,H})^2\,dt^4 \\
  &\quad + \frac{1}{M}\Bigl[ v_{\beta^2,2}\,dt^2 + v_{\beta\gamma,3}\,dt^3 \\
  &\qquad + \bigl(v_\gamma - v_{\beta^2,4}\bigr)dt^4 \Bigr],
\end{split}
\label{eq:mse}
\end{equation}
where each density is a function of $\alpha$ through the combined
distribution~\eqref{eq:combined-neyman}. This is a smooth scalar function of
$\alpha$, minimised by gradient descent with the closed-form derivative
$d\,\mathrm{MSE}_1/d\alpha$ from Appendix~\ref{app:S5}; the minimum
$\alpha^\star = \arg\min_\alpha \mathrm{MSE}_1(\alpha)$ is the distribution allocation we deploy.
We can use this joint distribution allocation to prove the existence of certain regimes where both
bias and variance can be improved simultaneously over the vanilla \qdrift{} distribution.  The mechanism behind this is
monotonicity, as the following lemma shows.

\begin{lemma}[Monotone distribution allocation densities]\label{lem:monotone}
Let $\gamma_j \neq 0$ and
$\mathrm{sign}(\gamma_i) = \mathrm{sign}(\gamma_j)\;\;\forall\,i,j$.  Then along the
family~\eqref{eq:combined-neyman}, $|b_{\gamma,2}(\alpha)|$ is non-decreasing and
$v_{\beta^2,2}(\alpha)$ is non-increasing in $\alpha$.
\end{lemma}

\begin{corollary}[Joint improvement]\label{prop:joint-main}
Let the hypotheses of Lemma~\ref{lem:monotone} hold, and let
$\tfrac12\varphi_{H,H}\le b_{\gamma,2}(\bq^\star_\gamma)$ when every $\gamma_j>0$, or
$b_{\gamma,2}(\bq^\star_\gamma)\le\tfrac12\varphi_{H,H}$ when every $\gamma_j<0$, and let the $|\beta_j|$ not all be equal.
If the leading bias and variance coefficients of vanilla \qdrift{}, $q_j^c\propto|c_j|$, are non-zero, then some
$\alpha\ge0$ has
\begin{equation}
  \lim_{dt\to0}\frac{\bigl|B_1(\bq^{(\alpha)})\bigr|}{\bigl|B_1(\bq^c)\bigr|} \le 1
  \quad\text{and}\quad
  \lim_{dt\to0}\frac{V_1\bigl(\bq^{(\alpha)}\bigr)}{V_1(\bq^c)} \le 1 ,
  \label{eq:joint-weak-main}
\end{equation}
with $B_1$ and $V_1$ the full single-step
bias and variance, whose expansions are given in~\eqref{eq:V1B1-general}.
\end{corollary}

\noindent The proofs of
Lemma~\ref{lem:monotone} and Corollary~\ref{prop:joint-main} are given in Appendix~\ref{app:S4}.  Increasing $\alpha$ transfers
weight from the terms with large $|\gamma_j|$ to those with large $|\beta_j|$, which raises
$b_{\gamma,2}$ and lowers $v_{\beta^2,2}$ at every $\alpha$, so the bias density is smaller than vanilla for $\alpha<\alpha_1$, and the variance density is smaller for $\alpha>\alpha_0$.  Between the two crossings both remain
below vanilla.  They satisfy
\begin{equation}
\begin{aligned}
  |b_{\gamma,2}(\bq^{(\alpha_1)})-\tfrac12\varphi_{H,H}|
    &= |b_{\gamma,2}(\bq^c)-\tfrac12\varphi_{H,H}|, \\
  v_{\beta^2,2}\!\bigl(\bq^{(\alpha_0)}\bigr) &= v_{\beta^2,2}(\bq^c).
\end{aligned}
  \label{eq:interval-boundaries}
\end{equation}
When $\mathrm{sign}(\gamma_i)=\mathrm{sign}(\gamma_j)\;\forall\,i,j$, every distribution in the parameterised
family~\eqref{eq:combined-neyman} is the optimal distribution for some balance of $|b_{\gamma,2}|$ (leading bias) and
$v_{\beta^2,2}$ (leading variance).  Each member of the parameterised family $\bq^{(\alpha)}$~\eqref{eq:combined-neyman} therefore has the smallest
possible leading variance for its leading bias.  As a result, at least one value of $\alpha$ is no worse than vanilla
\qdrift{} with respect to both leading bias and variance.  The corollary can be tightened to a strict inequality with respect to the
vanilla \qdrift{} distribution with further assumptions (Corollary~\ref{cor:joint-sm}).  The ordering assumption ($b_{\gamma,2}(\bq^\star_\gamma)\ge\tfrac12\varphi_{H,H}$ or
$b_{\gamma,2}(\bq^\star_\gamma)\le\tfrac12\varphi_{H,H}$) in the corollary precludes cases where cancellation can occur,
so where such cancellation exists, we find it numerically as in Figure~\ref{fig:joint-crossover}c.

Since Lemma~\ref{lem:monotone} controls $b_{\gamma,2}$ and $v_{\beta^2,2}$ only, at fixed $T$ the contribution of other terms for $N>1$ means we cannot guarantee a region of joint
improvement, so we optimise the distribution heuristically instead.

Figure~\ref{fig:joint-crossover} illustrates the family on a three-site XY chain
for three observables, the energy, a local magnetisation and the staggered magnetisation.  Within
the interval either quantity may be traded against the other without losing on both, and
$\alpha^\star$ lies inside $(\alpha_0,\alpha_1)$ in (a) and (b).  Appendix~\ref{app:S4} treats mixed-sign $\bgamma$ (Fig.~\ref{fig:joint-crossover}c), one of the two ways the assumptions of Lemma~\ref{lem:monotone} can fail, the other being a vanishing $\gamma_j$.  In all three panels the $Z_i$ terms have $\gamma_j=0$, since they commute with $O$ in (b) and (c), and in (a) $\braket{X_iX_{i+1}+Y_iY_{i+1}}=0$ on every bond for this reference state.

\paragraph{\textbf{Depth ($N>1$).}}
Next we generalise the above to $N>1$ \qdrift{} steps, using
$B_N$~\eqref{eq:BN_full} and $V_N$~\eqref{eq:VN_full} in full.  At fixed $T = N\,dt$, the fourth-order terms
of $B_N$ and $V_N$ scale differently with depth.  The single-step
parts, $N\,b_{\gamma,4}$ and $N(v_\gamma - v_{\beta^2,4})$, fall as $T^4/N^3$, the $N(N{-}1)$
groups $B_{4,2}$ and $V_{4,2}$ as $T^4/N^2$, and the $N(N{-}1)(N{-}2)$ groups $B_{4,3}$ and
$V_{4,3}$ only as $T^4/N$, so these terms dominate the bias and the variance alike as the circuit
deepens.
Every retained group therefore falls at least as $\sfrac{1}{N}$ at fixed $T$.  Note, Ref.~\cite{Watson2024RandomlyCompiled} also does an order expansion of the
bias as powers of $N^{-1}$, at a fixed $T$, with coefficients bounded by powers of $\lambda T$.  Our expansion differs by
instead expanding as powers of $dt$ at fixed $N$.  In their variable, \eqref{eq:BN_full} keeps each coefficient only to
order $T^4$, so its accuracy is set by $\lambda T$ rather than by the step size, and added depth does not improve it.  In Appendix~\ref{app:S7} we provide an error bound for our
truncation, demonstrated numerically in Appendix~\ref{app:S9} (Figure~\ref{fig:supp-dt-check}).
The mean-squared error is
\begin{equation}
  \mathrm{MSE}_N \;=\; B_N^2 + \frac{V_N}{M}.
  \label{eq:mseN}
\end{equation}

The cross-step terms (e.g.\ those that involve terms like $N(N{-}1)$, etc.) bring in
three further per-term quantities: the coupling
$\Phi_j := \tfrac12\sum_b c_b^2\bigl(\varphi_{j,b}^2+\varphi_{b,j}^2\bigr)$ from $v_\varphi$, and the shifts $\varphi_{H,j}$ and $\varphi_{j,H}$ from
$v_{\varphi_{H,j}^2,2}$ and $v_{\varphi_{j,H}^2,2}$ (Appendix~\ref{app:densities}), with both shifts
from~\eqref{eq:H-sums}.  Generalising~\eqref{eq:combined-neyman}, the distribution we use is
\begin{equation}
\begin{split}
  q^{(\bxi)}_j &\propto
  |c_j|\sqrt{\rho_j|\gamma_j| + \alpha_j\beta_j^2 + \mu_j\Phi_j + \nu_j\varphi_{H,j}^2
    + \nu'_j\varphi_{j,H}^2\,}, \\
  &\quad \rho_j>0,\quad \alpha_j,\mu_j,\nu_j,\nu'_j \ge 0
\end{split}
\label{eq:joint-N}
\end{equation}
with one weight per term for each density, collected as
$\bxi:=(\rho_j,\alpha_j,\mu_j,\nu_j,\nu'_j)_{j=1}^{L}$ and defined up to a common positive scale.
Each $q_j$ depends on its weights only through the bracket, which free positive weights can set to
any positive value, so the family is a reparameterisation of the interior of the probability
simplex rather than a restriction of it.  The
weights $\bxi$ are set by minimising $\mathrm{MSE}_N$ subject to the bias and variance
each staying within a margin $\eta$ of vanilla \qdrift{},
$|B_N| \le (1+\eta)|B_N^{\mathrm{van}}|$ and $V_N \le (1+\eta)V_N^{\mathrm{van}}$, fixed at
$\eta=5\%$ throughout (Appendix~\ref{app:S5}).  We can also reduce the complexity of the optimisation by
setting $\rho_j=1$, $\alpha_j=\alpha$, $\mu_j=\mu$, $\nu_j=\nu$ and $\nu'_j=\nu'$ for every $j$, which reduces
the $5L$ per-term weights to four shared scalars $(\alpha,\mu,\nu,\nu')$ (see Appendix~\ref{app:S9}).
Note that the $5L$ per-Pauli distribution we optimise is overparameterised relative to the number of
parameters required to specify $\bq$ ($L-1$).  Optimising these weights gives the same distributions as
optimising $\bq$ directly, but in the parameterised case each parameter multiplies a term whose effect on the bias
or variance is known.  As a result, the optimisation can be warm-started from the above closed-form optimal rules,
or from some other setting of the parameters (e.g.\ shared across each $j$) instead of from an arbitrary $\bq$.
Every rule is mixed with vanilla \qdrift{} as in~\eqref{eq:tau-mixing} so that each $q_j>0$.

Note that Refs.~\cite{Wu2025fluctuation,Fan2025adaptive} also introduce state-aware \qdrift{} distributions, but are agnostic to the target operator.
Ref.~\cite{Wu2025fluctuation} weights each term by its fluctuation in the reference state,
$q_j\propto\Delta H_j:=|c_j|\sqrt{1-\braket{P_j}^2}$, to minimise the distance between the approximate and exact
evolved states.  At their first step, both rules are members of our general distribution allocation (Eq.~\eqref{eq:joint-N})
with $O=\ket{\psi}\bra{\psi}$, such that $\braket{O(T)}=|\bra{\psi}e^{-iHT}\ket{\psi}|^2$.  In this case, $\gamma_j$ evaluates to $\braket{P_j}^2-1$, so the weights $\rho_j=1$ give $|c_j|\sqrt{|\gamma_j|}=\Delta H_j$,
the rule of Ref.~\cite{Wu2025fluctuation}, and $\rho_j=|\gamma_j|^{-1/2}$ gives $|c_j||\gamma_j|^{1/4}$, that of Ref.~\cite{Fan2025adaptive}.  As a result, however, bias
cancellation cannot be achieved as it can in our method.  Since $\gamma_j\le0$ for all $j$, $b_{\gamma,2}$ is single sign.  As a
result:
\begin{equation}
  \bigl|b_{\gamma,2}-\tfrac12\varphi_{H,H}\bigr|
   \;\ge\; \Bigl(\sum_j \Delta H_j\Bigr)^{\!2} - \Var_\psi(H) \;\ge\; 0,
  \label{eq:main-nocancel}
\end{equation}
a gap that does not depend on $\bq$, so it cannot be closed by a more optimal choice of distribution (Appendix~\ref{app:S4}).

The new terms in the
distribution~\eqref{eq:joint-N}, specifically $\Phi_j$, $\varphi_{H,j}$ and $\varphi_{j,H}$, target only the variance
via the mean-squared error as written in~\eqref{eq:mseN}, and \emph{not} the bias.  As in the
$N=1$ example, the bias is only targeted via the $\gamma_j$ terms in the
$\mathrm{MSE}$ optimisation.  This is because to order $dt^4$, none of the three new terms appears in the
bias squared, $B_N^2$, although $\varphi_{H,j}$ and $\varphi_{j,H}$ do appear in $B_N$.  As such, if one wished to allocate more
resources to bias reduction in our approach, the $\mathrm{MSE}$ would need to be expanded beyond
$dt^4$.  Alternatively, simply removing the $\Phi_j$, $\varphi_{H,j}$ and $\varphi_{j,H}$ terms would still
correctly target bias reduction to $\mathcal{O}(dt^4)$.  However, for consistency in this work,
we retain only the $\mathcal{O}(dt^4)$ orders.

\begin{figure}[!t]
\centering
\includegraphics[width=\columnwidth]{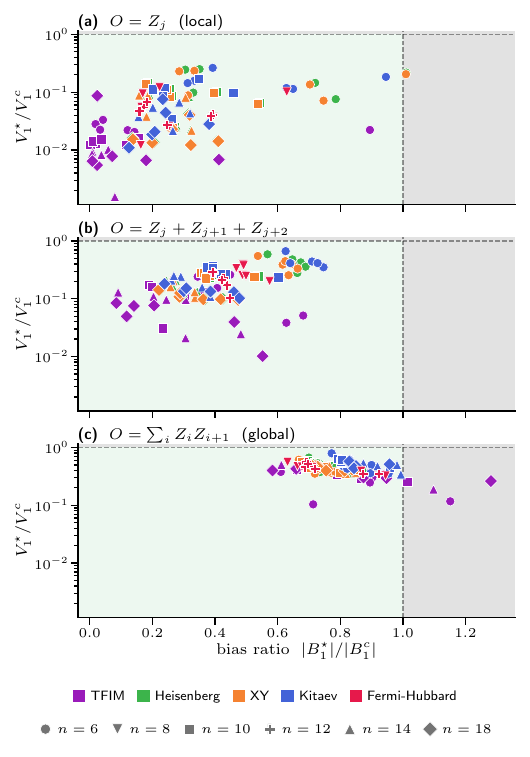}
\caption{\textbf{\statedrift{} joint bias and variance reduction for lattice quenches
($\beta_j\neq0$) on a time-evolved observable $\braket{O(T)}$, for a single \qdrift{} step
($N=1$).} We evolve each initial state of the form
$\bigotimes_{i=1}^{n}\bigl(\cos\tfrac{a_i}{2}\,\ket{0}
+ e^{\mathrm{i}\phi_i}\sin\tfrac{a_i}{2}\,\ket{1}\bigr)$, with
$a_i\sim\mathcal{U}(0.3,1.6)$ and $\phi_i\sim\mathcal{U}(0,2\pi)$, according to the
transverse-field Ising, Heisenberg, XY and Kitaev chain Hamiltonians and the
Fermi--Hubbard model on $2\!\times\!2$ and $2\!\times\!3$ lattices (Eqs.~\eqref{eq:lattice-hams}--\eqref{eq:lattice-hams-fh}), on $n=6$ to $18$
qubits.
We measure observables with decreasing locality, (a) a single-site $Z_j$,
(b) a three-site region sum $Z_j+Z_{j+1}+Z_{j+2}$ and (c) the whole-chain
correlator $\sum_i Z_iZ_{i+1}$, each non-conserved so that all five models evolve, at
minimum weight $\tau=0.2$.  We take a single step with $\lambda\,dt=0.02$, so $T=dt$, and choose $\alpha^\star$ to minimise $B_1^2+\sfrac{V_1}{M}$ at $M=300$.
In each case, we
compare the bias and variance of the \statedrift{} distribution to that of vanilla
\qdrift{} ($q_j\propto|c_j|$) as a ratio, so $<1$ is better and the green region improves
both bias and variance.  Every instance improves at least one of the two, and the vast
majority improve both.}
\label{fig:statedrift}
\end{figure}

As above, we can make three simplifications to this probability allocation.  First, when
$\beta_j = 0$ we set $\alpha_j = 0$.  Secondly, at an eigenstate of $H$, where
$\varphi_{H,j} = 0$ (Lemma~\ref{lem:collapse}), we can remove the $\nu_j$ terms also.  We are left
with the three-density family
\begin{equation}
  q^{(\bxi)}_j \;\propto\; |c_j|\sqrt{\,\rho_j|\gamma_j| + \mu_j\Phi_j + \nu'_j\varphi_{j,H}^2\,}
  \label{eq:joint-beta0}
\end{equation}
The third simplification is for the energy, $O=H$, where the conjugate shift
$\varphi_{j,H}$ vanishes as well and the mean-squared error becomes
\begin{equation}
\begin{split}
  \mathrm{MSE}_N(\bxi) &\;=\; N^2 b_{\gamma,2}(\bxi)^2\,dt^4 \\
  &\quad + \frac{1}{M}\Bigl[\, N\,v_\gamma(\bxi)
  + \tfrac{N(N-1)}{2}\,v_\varphi(\bxi) \,\Bigr]dt^4
\end{split}
\label{eq:mse-beta0}
\end{equation}
with the distribution
\begin{equation}
  q^{(\bxi^\star)}_j \;\propto\; |c_j|\sqrt{\,\rho^\star_j\gamma_j + \mu^\star_j\Phi_j\,},
  \quad
  \bxi^\star = \operatorname*{arg\,min}_{\bxi}\mathrm{MSE}_N(\bxi)
  \label{eq:joint-energy}
\end{equation}
minimised at the target depth.  The modulus
of~\eqref{eq:joint-beta0} is dropped here because at the ground state
$\gamma_j = \braket{\psi_0|P_j H P_j|\psi_0} - E_0 \ge 0$ by the variational principle.

Besides these analytic simplifications, we can also use heuristic simplifications, for example simply
optimising $q_j \propto |c_j|\sqrt{\rho_j|\gamma_j|}$.  We demonstrate in Appendix~\ref{app:S9}
(Figure~\ref{fig:simplified-alloc}) that this is effective at achieving the majority of bias reduction, yet still a
gap remains relative to optimising the most general distribution.

\paragraph{\textbf{Error and validity of the distribution.}}
In Proposition~\ref{prop:main-RN}, we compute the residual error of the \statedrift{} distribution relative to the true
bias (proof in Appendix~\ref{app:S7}).

\begin{proposition}[Truncation error]\label{prop:main-RN}
The error $R_N=B_N-B_N^{(4)}$ of the fourth-order bias is bounded by
\begin{equation}
  |R_N| \;\le\; \|\delta_2\|\,\frac{(2\lambda T)^3}{3!}\,\frac{T^2}{N}\,\|O\| + \mathcal{O}\bigl(N^{-2}\bigr) ,
  \label{eq:main-RN}
\end{equation}
where $\delta_2$ is the $dt^2$ coefficient of the error of a single \qdrift{} step
(Lemma~\ref{lem:s2-defect}), with $\|\delta_2\|\le2\sum_j\sfrac{c_j^2}{q_j}+2\lambda^2$.
\end{proposition}

\begin{figure*}[!t]
\centering
\includegraphics[width=\linewidth]{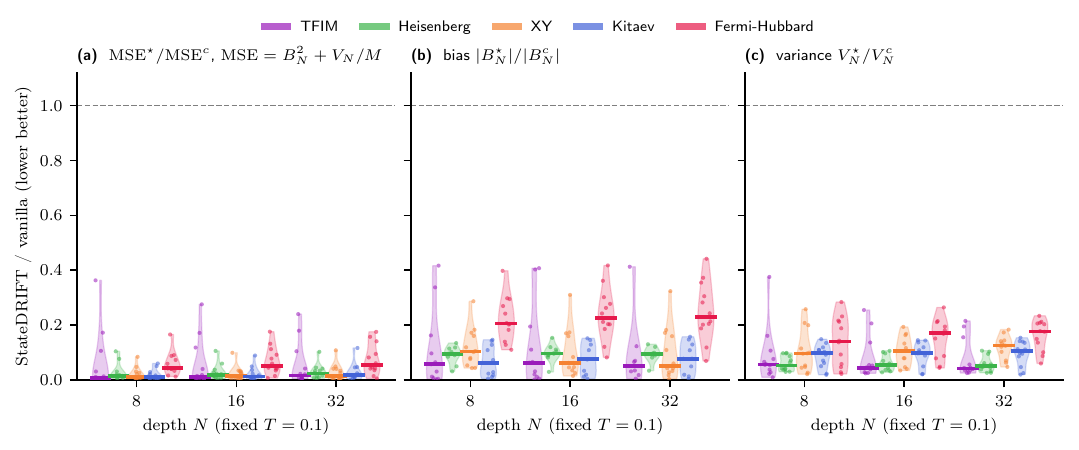}
\caption{\textbf{StateDRIFT on time-evolved lattice models.}  Ratio of \statedrift{} to vanilla
\qdrift{} for \textbf{(a)}~the mean-squared error at $M=1000$, \textbf{(b)}~the bias $|B_N|$ and \textbf{(c)}~the variance
$V_N$ (lower is better), against depth $N=8$, $16$ and $32$ at fixed $T=0.1$.  We test on the transverse-field Ising, Heisenberg, XY and Kitaev
chains and the $2\!\times\!2$ Fermi--Hubbard model at $n=8$ qubits, with $O=Z_0$, over $12$ random product states for each model.}
\label{fig:nsweep}
\end{figure*}

Since the leading error is of the same $\mathcal{O}(N^{-1})$ as the leading bias for which we optimise, the dominant
error scales with $\lambda T$.  However, we can still use the truncated expression to effectively \emph{rank}
distributions.  The deployed $\bq^\star$ has a smaller exact bias than vanilla $\bq^c$
whenever (Appendix~\ref{app:S8})
\begin{equation}
  \bigl|B_N^{(4)}(\bq^c)\bigr|-\bigl|B_N^{(4)}(\bq^\star)\bigr| \;>\; |R_N(\bq^\star)|+|R_N(\bq^c)| .
  \label{eq:main-rank}
\end{equation}
The bound~\eqref{eq:main-RN} is too loose to determine when the criterion~\eqref{eq:main-rank} is satisfied, so instead we can
define a heuristic `validity' metric based on subsequent error truncation to determine when \statedrift{} may be useful or
not.  We define $g_p:=B_N^{(p)}-B_N^{(p-1)}$, where $B_N^{(p)}$ is the bias truncated at
order $dt^p$ ($B_N^{(1)}=0$), and the metric compares the last kept terms, $g_3$ and $g_4$, with the total truncated bias itself
($B_N^{(4)}$).  By construction we optimise $B_N^{(4)}(\bq^\star)\approx0$.  We assume that the truncated terms also decrease with increasing order, and also decay fast enough that the
first remainder error of each parity dominates the rest, i.e.:
\begin{equation}
  \begin{aligned}
  |R_N(\bq^c)| \;&=\; \Bigl|\sum_{p\ge5}g_p(\bq^c)\Bigr| \;\approx\; |g_5(\bq^c)+g_6(\bq^c)|\\
  &\le\; |g_3(\bq^c)|+|g_4(\bq^c)| .
  \end{aligned}
  \label{eq:main-Rc}
\end{equation}%
Secondly, since the \statedrift{} distribution contains $\tau$-mixing, it stays within $\sfrac{1}{\tau}$ of vanilla \qdrift{}
(Lemma~\ref{lem:minweight}), and we assume $|R_N(\bq^\star)|\approx|R_N(\bq^c)|$.  Then criterion~\eqref{eq:main-rank}
becomes $|B_N^{(4)}(\bq^c)|>2(|g_3(\bq^c)|+|g_4(\bq^c)|)$ and as such we define the validity metric as:
\begin{equation}
  R_{\rm val}^{(4)} \;:=\; \frac{\bigl|B_N^{(4)}(\bq^c)\bigr|}{\bigl|g_3(\bq^c)\bigr|+\bigl|g_4(\bq^c)\bigr|} \;>\; 2 .
  \label{eq:main-rvan}
\end{equation}%
If~\eqref{eq:main-rvan} holds on the target problem and the assumptions above are valid as bounds,
$B_N^{(4)}(\bq^\star)=0$ and $|R_N(\bq^\star)|,\,|R_N(\bq^c)|\le|g_3(\bq^c)|+|g_4(\bq^c)|$, then \eqref{eq:main-rank} holds
and we have a guarantee that \statedrift{} will outperform vanilla \qdrift{} in achieving a lower bias.
In all cases tested in this work (Appendix~\ref{app:S8}), every depth satisfying~\eqref{eq:main-rvan} is a \statedrift{} win or tie, and every
depth at which vanilla \qdrift{} wins fails~\eqref{eq:main-rvan}.  Note that the criterion is conservative in favour of
\qdrift{}, and thus we observe many cases where the \statedrift{} still has a lower bias, yet~\eqref{eq:main-rvan} rules them out.

\section{StateDRIFT benchmarks}
\label{sec:benchmarks}

Next, we demonstrate the effectiveness of \statedrift{} in minimising either the bias or the
variance, or both, of an evolved expectation value for several examples.  We first assume exact,
oracular access to densities $\gamma_j$, $\beta_j$ and the cross-step $\Phi_j$,
$\varphi_{H,j}$, $\varphi_{j,H}$, supported by the theoretical framework of the previous sections.  This is not realistic in practice,
as we do not have access to these quantities exactly,
since that would require computing observables on the true target state $\ket{\psi}$.  In the following
sections we describe two practical deployments of \statedrift{} that approximate or measure the
distribution densities $\hat\gamma_j$, $\hat\beta_j$, $\hat\Phi_j$, $\hat\varphi_{H,j}$, $\hat\varphi_{j,H}$ without the
oracle.  The first approximates them from a classically simulable surrogate of the reference state,
the second measures them by a classical-shadow approach.

All simulations use the open-source OpenQARP library~\cite{OpenQARP}
(\url{https://github.com/OpenQARP/openqarp}), with the molecular Hamiltonians built using PySCF~\cite{Sun2020}.
We first benchmark \statedrift{} on five lattice models, the transverse-field Ising,
Heisenberg, XY and Kitaev chains and the Fermi--Hubbard model on $2\!\times\!2$ and $2\!\times\!3$ lattices.  The spin chains are the transverse-field
Ising, the isotropic Heisenberg and the XY model:
\begin{align}
  H_{\mathrm{TFI}}  &= -J\sum_{i=1}^{n-1} Z_i Z_{i+1} - h\sum_{i=1}^{n} X_i, \label{eq:lattice-hams}\\
  H_{\mathrm{Heis}} &= \frac{J}{4}\sum_{i=1}^{n-1}\bigl(X_i X_{i+1} + Y_i Y_{i+1} + Z_i Z_{i+1}\bigr), \\
  H_{\mathrm{XY}} &= \sum_{i=1}^{n-1}\bigl(J_x X_i X_{i+1}+J_y Y_i Y_{i+1}\bigr)+h\sum_{i=1}^{n} Z_i.
\end{align}
The fermionic models are the spinless Kitaev ($p$-wave) chain and the
Fermi--Hubbard model,
\begin{align}
  H_{\mathrm{Kit}} &= \sum_{i=1}^{n-1}\bigl(-t_{\rm h}\,a_i^\dagger a_{i+1}
                      + \Delta\,a_i^\dagger a_{i+1}^\dagger + \mathrm{h.c.}\bigr) \notag\\
                   &\qquad - \mu_{\rm c}\sum_{i=1}^{n} a_i^\dagger a_i, \label{eq:lattice-hams-ferm}\\
  H_{\mathrm{FH}}  &= -t_{\rm h}\!\!\sum_{\braket{i,j},\,\sigma}\!\bigl(a_{i\sigma}^\dagger a_{j\sigma}
                      + \mathrm{h.c.}\bigr) + U\sum_i n_{i\uparrow} n_{i\downarrow} , \label{eq:lattice-hams-fh}
\end{align}
with Fermi--Hubbard on an $L_x\times L_y$ lattice, mapped to $2L_xL_y$ qubits by the Jordan--Wigner
transformation.  We use $J=h=1$ for the transverse-field Ising model, $J=1$ for the Heisenberg model, $J_x=J_y=h=1$ for the XY model, $t_{\rm h}=1$, $\mu_{\rm c}=0.5$ and $\Delta=0.6$ for the Kitaev chain, and $t_{\rm h}=1$ and $U=2$ with open boundaries for Fermi--Hubbard.  We randomise the initial state, $\ket{\psi}$, for each model as a different random product state, with varying particle number.

At a single
\qdrift{} step (Figure~\ref{fig:statedrift}), evolving product-state references under each model so
that $\beta_j\neq0$ generically (the $\beta$-driven regime of Appendix~\ref{app:S4}), a distribution minimising
the variance alone lowers it but tends to raise the bias above vanilla \qdrift{} for the majority of
references, and vice versa.  The combined family $q_j\propto|c_j|\sqrt{|\gamma_j|+\alpha^\star\beta_j^2}$~\eqref{eq:combined-neyman},
with $\alpha^\star$ the MSE-minimising weight~\eqref{eq:mse}, reconciles the two: for the vast majority of references it improves both the bias and the
variance.  The size of the improvement grows with
how unevenly the per-term densities are distributed across the Hamiltonian terms, and vanishes when
they are uniform, where vanilla \qdrift{} is already optimal.

\begin{figure*}[!t]
\centering
\includegraphics[width=\linewidth]{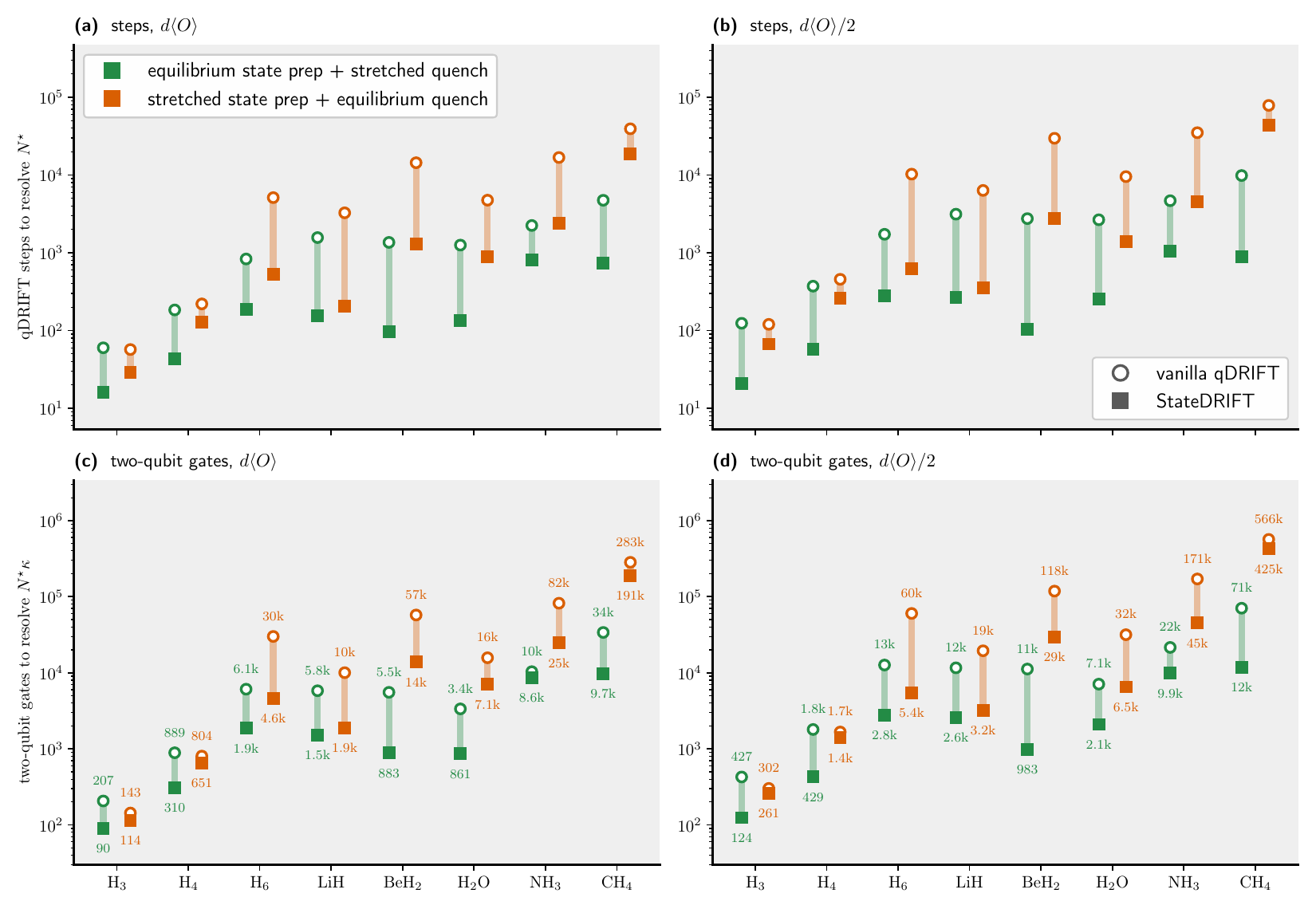}
\caption{\textbf{\statedrift{} resolves the dynamics of molecular Hamiltonians in
fewer steps and fewer two-qubit gates.}  \textbf{(a,b)}~\qdrift{} steps $N^\star$ and \textbf{(c,d)}~two-qubit
gates $N^\star\kappa$ required such that $N^\star$ sufficiently \emph{resolves} the bias, i.e.\ the minimal $N$
such that $|\widehat B_{N'}|+2\hat\sigma_{N'}<\varepsilon$ for all $N'\ge N$, with $\hat\sigma_{N'}=\sqrt{\sfrac{\widehat V_{N'}}{M}}$ the standard error of the sampled bias $\widehat B_{N'}$ and $\widehat V_{N'}$ the sample variance of the $M$ trajectory estimates (Appendix~\ref{app:S9}, Resolving depth), at
$\varepsilon=d\braket{O}$ (\textbf{a,c}) and $\varepsilon=\sfrac{d\braket{O}}{2}$ (\textbf{b,d}), for H$_3$
($n=6$) to CH$_4$ ($n=18$) with $O=-\sfrac{1}{2}\sum_p(p{+}1)Z_p$.  Green: the equilibrium ground state evolved
under the stretched-bond Hamiltonian.  Orange: the stretched ground state evolved under the equilibrium
Hamiltonian.  We run all molecules for time $T=1$ (atomic units) except H$_2$O, which runs for $T=\sfrac{1}{2}$.  Each depth is sampled with $M=5000$ trajectories
($M=500$ for CH$_4$).  Open circles vanilla \qdrift{}, filled squares
\statedrift{}.  The labels in (c,d) are the gate counts.}
\label{fig:chem}
\end{figure*}

Next, Figure~\ref{fig:nsweep} increases the depth $N$ for the same lattice families as
Figure~\ref{fig:statedrift}, and at each depth we solve the optimisation problem
\begin{equation}
\begin{split}
  \bxi^\star &\;=\; \arg\min_{\bxi}\ \mathrm{MSE}_N\big(\bq^{(\bxi)}\big) \\
  &\quad \text{s.t.}\quad |B_N| \le (1+\eta)\,|B_N^{\mathrm{van}}|,\quad
  V_N \le (1+\eta)\,V_N^{\mathrm{van}},
\end{split}
\label{eq:deploy-opt}
\end{equation}
over the $5L$ per-term weights of the general distribution family $\bq^{(\bxi)}$~\eqref{eq:joint-N}, using the
$dt^4$ expansion for the MSE~\eqref{eq:mseN} at a fixed shot budget $M$ and margin $\eta=5\%$.
We keep this margin to allow a slightly worse bias or variance where significant savings are available
in the other, while the ceiling relative to vanilla ensures neither is completely sacrificed for the
other.  The figure shows the resulting total error and its bias and variance parts across depth for
all five models, with the state-aware distribution achieving significantly lower bias and variance than vanilla \qdrift{} at every simulated $N$.

Quantum chemistry Hamiltonians are instead $\gamma$-driven ($\beta_j=0$, Lemma~\ref{lem:beta}),
where we optimise the simpler distribution
$q^{(\bxi)}_j\propto|c_j|\sqrt{\,\rho_j|\gamma_j|+\mu_j\Phi_j+\nu_j\varphi_{H,j}^2+\nu'_j\varphi_{j,H}^2\,}$, the $\beta_j=0$
($\alpha_j=0$) case of Eq.~\eqref{eq:joint-N}. Because the resolving depth is set by the bias alone,
here the weights minimise the fourth-order bias rather than the mean-squared error,
$\bxi^\star=\arg\min_{\bxi}|B_N(\bq^{(\bxi)})|$ subject to $V_N\le(1+\eta)V_N^{\mathrm{van}}$ with
$\eta=5\%$, re-solved at every depth.  Our molecular benchmarks are numerical quenches between the qubit Hamiltonians of two geometries (Appendix~\ref{app:S9}).  On a sudden quench we locate the depth $N^\star$ at which
the dynamics resolves, where the compilation bias falls below a threshold set by the observable's
motion, $|B_N|<\varepsilon$ with $\varepsilon = f(d\braket{O})$ and
$d\braket{O} := |\braket{O(T)} - \braket{O(0)}|$.
  As such, \statedrift{} resolves the dynamics
with fewer steps than vanilla \qdrift{}.  We show this for H$_3$, H$_4$, H$_6$, LiH, BeH$_2$, H$_2$O, NH$_3$ and CH$_4$,
spanning $6$ to $18$ qubits, in Figure~\ref{fig:chem}, resolving in $1.7$ to $26\times$ fewer steps. Here, we choose the weighted density $O=-\sfrac{1}{2}\sum_p(p{+}1)Z_p$, $=\sum_p(p{+}1)n_p$ up to a constant.  This is the spin-orbital occupations weighted by their index, so it changes whenever electrons move between orbitals.

As mentioned above, the primary resource we count is the number of \qdrift{} steps, $N$.  However, \statedrift{}
optimised distributions can reweight towards heavier Pauli terms that are more expensive to implement in hardware. As the two-qubit gate cost is a more faithful measure of practical
complexity, we also report it. Each term $P_j$ is exponentiated as $e^{-i\theta_j P_j}$ and compiled
by a CNOT ladder over its support, so a weight-$w_j$ Pauli string costs $C_j = 2(w_j-1)$ two-qubit
gates~\cite{Whitfield2011Simulation,Seeley2012BravyiKitaev,CAS2026}. A depth-$N$ trajectory then costs
$N\kappa(\bq)$ gates on average, with per-step cost
\begin{equation}
  \kappa(\bq) \;=\; \sum_j q_j C_j .
  \label{eq:kappa}
\end{equation}
This expected gate count $N\kappa$ is the figure of merit used by
Ref.~\cite{KGR2023}, which reduces it by sampling at $q_j\propto|c_j|/C_j$.
Reference~\cite{CAS2026} instead minimises a net cost, the expected circuit cost
multiplied by the second moment of the importance weight, and reweights the measured
outcome rather than the channel.  We attempt to adopt a \emph{cost-aware} version of our \statedrift{}
distribution, but find it is not helpful in reducing the gate count further (Appendix~\ref{app:S6}).

Both methods use the same $M$ trajectories, each starting from the same prepared state, so the ratio of $N^\star\kappa$ is also the ratio of the total two-qubit gate count of the evolution over all $M$ trajectories.  As discussed above, this step advantage does not automatically guarantee a reduction in
the quantum circuit gate count.  Nevertheless, we also see from Figure~\ref{fig:chem}(c,d) that the gate
count is reduced over vanilla by $1.2$--$11\times$, because the step saving outweighs the higher
per-step cost of the state-aware distribution.  The most dramatic saving we find is $11.4\times$ for BeH$_2$ prepared
at equilibrium, requiring $983$ gates to resolve the dynamics to half the threshold, $|B_N|<\sfrac{d\braket{O}}{2}$,
relative to vanilla \qdrift{}, which takes $11{,}201$.  Also, compared to the cost-aware (not optimised) distribution of Ref.~\cite{KGR2023}, evaluated, as in that work, with a cost of $0.1$ for each single-qubit term, \statedrift{} requires
$2.4$--$36\times$ fewer gates in that cost model on the problems tested here, and $1.1$--$11\times$ fewer than the fluctuation-guided distribution of Ref.~\cite{Wu2025fluctuation} (Figure~\ref{fig:cost-compare}).  We use $2(w_j-1)$ as the cost to implement $P_j$ with weight $w_j$ using a CNOT ladder~\cite{Whitfield2011Simulation,Seeley2012BravyiKitaev,CAS2026}.  In Figure~\ref{fig:target-sweep} we show that, for LiH, the gap in two-qubit gates between \statedrift{} and vanilla \qdrift{} grows with
the tightening of the required precision.  The example of H$_2$O shows the cancellation effect,
where a large fraction of negative $\gamma_j$ terms cause the
optimisation to drive $b_{\gamma,2}\to\tfrac12\varphi_{H,H}$, cancelling the bias at
$\mathcal{O}(dt^2)$.  As Appendix~\ref{app:S8} shows, this cancellation is also what limits the $t=0$ initialised \statedrift{}.  CH$_4$ is the
largest molecule we test, with $\lambda T=59$--$65$, yet \statedrift{} can find shorter circuits than \qdrift{} there.
H$_2$O has a comparable $\lambda T$ ($68$--$72$ at $T=1$) but fails for $T=1$ and H$_2$O has the largest observed leading
order bias cancellation effect.  This bias cancellation results in the removal of the leading $dt^2$ bias, and the remaining bias is
controlled by the truncated remainder (Proposition~\ref{prop:main-RN}).  These terms are not targeted by our distribution
and grow with $\lambda T$.  This could be mitigated in future works by higher-order truncations, or alternative
expansions which target terms grouped by $\sfrac{1}{N}$ directly (e.g.\ Ref.~\cite{Watson2024RandomlyCompiled}).  Nevertheless, this observed failure at $T=1$ for water is predicted by $R_{\rm val}^{(4)}$~\eqref{eq:main-rvan}, so we reduce to $T=\sfrac12$ for
Figure~\ref{fig:chem}.  We discuss this further in Appendix~\ref{app:S8}.

\section{Deployment from classical surrogates}
\label{sec:surrogates}

The first practical deployment of \statedrift{} builds the distribution from a classical
surrogate, replacing the oracle densities $\gamma_j,\beta_j$ and the cross-step terms
$\Phi_j,\varphi_{H,j},\varphi_{j,H}$ with surrogate estimates $\tilde\gamma_j,\tilde\beta_j,\tilde\Phi_j,
\tilde\varphi_{H,j},\tilde\varphi_{j,H}$ evaluated exactly on a classically simulable surrogate
$\ket{\tilde\psi}\approx\ket{\psi}$, so that $\tilde\gamma_j\approx\gamma_j$, and feeding them into the
same optimised family~\eqref{eq:joint-N} at no quantum cost.  To compute the distribution and solve the optimisation to fourth order, we need terms up to $\braket{\tilde\psi|[H,[H,[H,[H,O]]]]|\tilde\psi}$.  We assume $\ket{\tilde\psi}$ is a matrix-product state (MPS) of bond dimension $\chi$, and $H$ and $O$ are matrix-product operators (MPOs) of bond dimensions $D_H$ and $D_O$ respectively.  Consider a single commutator $\braket{\tilde\psi|[H,O]|\tilde\psi}=\braket{\tilde\psi|HO|\tilde\psi}-\braket{\tilde\psi|OH|\tilde\psi}$.  Each of $\braket{\tilde\psi|HO|\tilde\psi}$ and $\braket{\tilde\psi|OH|\tilde\psi}$ is the expectation value of an MPO of bond dimension $D_HD_O$, which a site-by-site contraction evaluates in $\mathcal{O}\bigl(n(\chi^3D_HD_O+\chi^2D_H^2D_O^2)\bigr)$~\cite{Schollwock2011}.  Each further $H$ (e.g.\ evaluating $[H,[H,O]]$) multiplies the MPO bond dimension by $D_H$.  Evaluating them on
$\ket{\tilde\psi}$ adds no shot noise, so the surrogate mismatch $\ket{\tilde\psi}\neq\ket{\psi}$ is the only source
of error in these estimated inputs.

Fixing the distribution from a classical surrogate does not make the problem classically easy.
The surrogate is used only to compute quantities at $t=0$, while the evolved $\braket{O(T)}$ still
comes from the quantum computation, and surrogate performance will necessarily not be able to track the
evolution accurately in general.  Where the reference is too entangled for any surrogate, we use
classical shadows as described in the next section.

The surrogate enters only through the densities it supplies.  Evaluating the disturbance
density~\eqref{eq:gamma-beta-def} on the surrogate gives
\begin{equation}
  \tilde\gamma_j = \braket{\tilde\psi|P_j O P_j|\tilde\psi} - \braket{\tilde\psi|O|\tilde\psi},
  \label{eq:gamma-surrogate}
\end{equation}
and $\tilde\beta_j,\tilde\Phi_j,\tilde\varphi_{H,j},\tilde\varphi_{j,H}$ follow by the same substitution
$\ket{\psi}\!\to\!\ket{\tilde\psi}$ in their definitions, so the deployment is fixed once $\ket{\tilde\psi}$ is
chosen.  The choice of surrogate model is problem-specific and resource-dependent.

The
Hartree--Fock determinant~\cite{SzaboOstlund}
$\ket{\tilde\psi_{\mathrm{HF}}}$ is the cheapest surrogate, a single Slater determinant that maps under Jordan--Wigner to a
computational-basis product state, so each density in~\eqref{eq:gamma-surrogate} is a product-state
expectation evaluated in closed form.  It is faithful at weak correlation but its estimate degrades once
correlation is strong, as at stretched bonds or near-degeneracies.  Tensor-network surrogates offer a
systematically improvable alternative, including the matrix-product
state~\cite{White1992,Schollwock2011,Orus2014,Garcia2024TensorSurvey} of bond dimension $\chi$,
\begin{equation}
  \ket{\tilde\psi_\chi} = \sum_{s_1\cdots s_n}\ \sum_{a_1\cdots a_{n-1}=1}^{\chi}
  A^{s_1}_{a_1} A^{s_2}_{a_1 a_2}\cdots A^{s_n}_{a_{n-1}}\,\ket{s_1\cdots s_n},
  \label{eq:mps}
\end{equation}
whose densities are evaluated by contracting its tensor legs.  The cost of manipulating a
matrix-product state, contraction included, is dictated by $\chi$, so fixing $\chi$ to a constant, as is
common practice, makes contractions efficient in system size.  It refines toward
$\ket{\psi}$ as $\chi$ grows, though the $\chi$ needed grows with entanglement and becomes
prohibitive for volume-law states.  Where a system admits a tailored classical method, an accurate
surrogate is available at no quantum cost.  For strongly correlated chemistry, orbital-optimised pair
coupled-cluster doubles (oo-pCCD)~\cite{oopccd} provides another candidate surrogate, whose accuracy must be assessed for the
system of interest.

Each surrogate is deployed through the same machinery as Figure~\ref{fig:chem}: the full
optimised family~\eqref{eq:joint-N}, built from the surrogate densities and mixed with vanilla at a
trainable weight $\tau$ restricted to a minimum value $\tau\!\ge\!0.2$.  A faithful surrogate remains free to concentrate.

\begin{figure}[!t]
\centering
\includegraphics[width=\columnwidth]{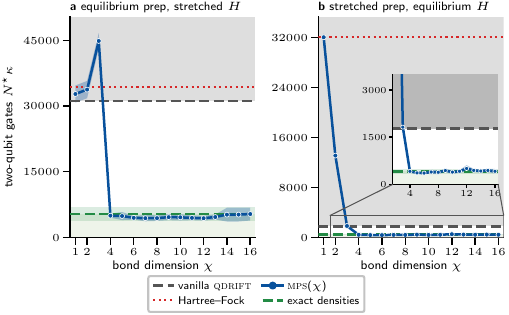}
\caption{\textbf{Deployment from classical surrogates.}  We measure the electronic dipole of LiH,
$O = -\sum_{pq} d_{pq}\,a_p^\dagger a_q$, with $d_{pq}=\braket{p|\hat z|q}$ the dipole
integrals along the bond axis.  Two-qubit gate counts $N^* \kappa$ are shown for the ground state of LiH \textbf{(a)} prepared at equilibrium
bond length and evolved under the Hamiltonian for the $2\times$ stretched bond length and \textbf{(b)}~prepared at the $2\times$ stretched
bond length and evolved under the Hamiltonian for the equilibrium bond length.
In each we use as surrogates, 1)~the Hartree--Fock single-determinant initial state and
2)~matrix product states with increasing bond dimension $\chi$.  Steps $N^\star$ are defined as in
Figure~\ref{fig:chem}, at $\varepsilon=d\braket{O}$.}
\label{fig:surrogate}
\end{figure}

Figure~\ref{fig:surrogate} benchmarks the Hartree--Fock and matrix-product state (MPS) surrogates on
simulating LiH with \statedrift{}, computing the non-commuting electronic dipole observable
on the equilibrium and stretched preparations of Figure~\ref{fig:chem}.  For each surrogate we compute the depth required
to resolve the change in dynamics, $d\braket{O}$, as in Figure~\ref{fig:chem}.  Hartree--Fock, whose product-state densities misjudge the
correlated ground state, deploys a distribution worse than vanilla.  In Figure~\ref{fig:surrogate}, we show that a bond dimension $\chi=4$ MPS is sufficiently expressive to capture the
relevant information for \statedrift{}, while a lower bond dimension can perform worse than vanilla \qdrift{} or even
the HF initial state.  Raising the bond dimension until the probability allocation optimisation converges to a minimum bias is also
an effective proxy to find the minimum expressivity required for \statedrift{}.

\begin{figure*}[!t]
\centering
\includegraphics[width=\textwidth]{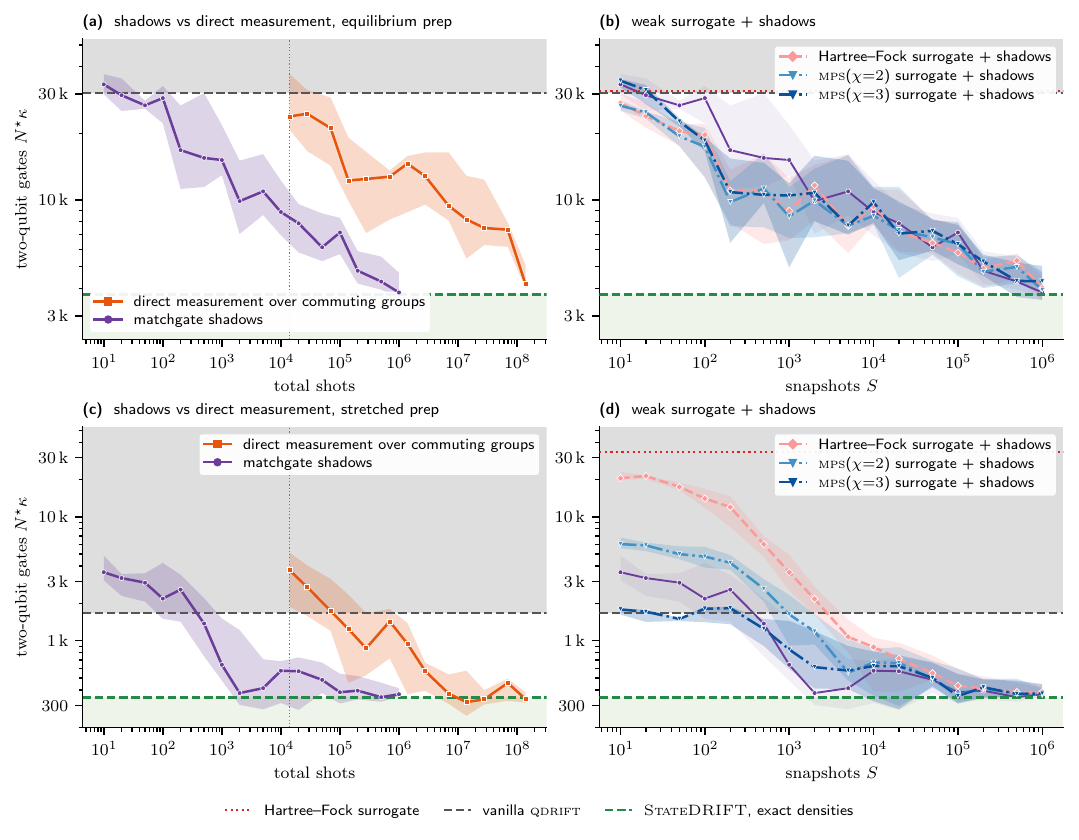}
\caption{\textbf{Deployment with classical shadows.}  Matchgate shadows are used to resolve the densities for
the same LiH evolved dipole as in Figure~\ref{fig:surrogate} with equilibrium \textbf{(a,b)} and stretched
\textbf{(c,d)} geometry initial states.  \textbf{(a,c)}~Matchgate shadows compared to direct measurements of grouped
commuting observables needed for each observable in $\gamma$, $\beta$, etc.  \textbf{(b,d)}~Combining a poor quality
classical surrogate with shadow-computed observables reduces measurements to achieve \statedrift{} accuracy over a
cold-started shadow probability allocation.  We use $M=1000$ trajectories to estimate the exact biases (and hence compute
$N^\star$).  Error bars show median and interquartile range over $10$ seeds.}
\label{fig:warmdrift}
\end{figure*}

\section{Deployment with classical shadows}
\label{sec:shadows}

When no classical surrogate is accurate enough, the distribution can instead be built by
measuring the reference state $\ket{\psi}$ itself.  A naive approach measures the expectation for each $\gamma_j$,
$\beta_j$ and cross-step term individually, which is wasteful.  Instead, we can use classical shadows~\cite{Aaronson2020ShadowTomography,Huang2020Shadows,Elben2023Toolbox} to estimate the observables we need to
evaluate on $\ket{\psi}$.  We then rerun the optimisation using these estimated quantities.  For fermionic
Hamiltonians, all the quantities we need to estimate ($\Phi_j$, $\varphi_{H,j}$, etc.) are functionals of the reference
state's reduced density matrices (RDMs), and we can use matchgate shadows~\cite{Zhao2021Fermionic,Wan2023Matchgate}
instead.  For more
general, non-fermionic systems with $\beta_j\neq0$ we instead use general Pauli
shadows~\cite{Huang2020Shadows,Huang2021Derandomization,Hadfield2022LocallyBiased}.

A classical shadow~\cite{Huang2020Shadows} samples a random unitary $W$ from an ensemble
$\mathcal{E}$, applies it to $\ket{\psi}$ and measures the rotated state in the computational basis to
obtain an outcome $\ket{b}$,
and forms the single-snapshot estimator
$\hat\sigma=\mathcal{M}_\mathcal{E}^{-1}\!\big(W^\dagger\ket{b}\bra{b}W\big)$, where
$\mathcal{M}_\mathcal{E}$ is the ensemble's measurement channel.  By construction
$\mathbb{E}[\hat\sigma]=\ket{\psi}\bra{\psi}$, so any density $\braket{X}$ is estimated by
averaging $\mathrm{tr}(X\hat\sigma)$ over $S$ snapshots.  A local-Clifford ensemble factorises,
$W=\bigotimes_i W_i$ and
$\hat\sigma=\bigotimes_i\big(3\,W_i^\dagger\ket{b_i}\bra{b_i}W_i-\mathds{1}\big)$, and the
single-snapshot estimator of a weight-$w$ Pauli has variance $\sim3^{w}$, so the number of snapshots to
reach a fixed precision grows exponentially in the Pauli weight.  In matchgate shadows~\cite{Zhao2021Fermionic,Wan2023Matchgate}, the ensemble of random unitaries is instead
fermionic Gaussian unitaries~\cite{Zhao2021Fermionic,Wan2023Matchgate}.  These matchgate unitaries rotate
the $2n$ Majorana operators among themselves as a transformation in the orthogonal group of dimension
$2n$, $O(2n)$.  Matchgate shadows can sample this transformation uniformly from the even permutations of the
Majorana operators~\cite{Zhao2021Fermionic}, from the full group $O(2n)$~\cite{Wan2023Matchgate}, or from other ensembles.
Here we use the permutation ensemble of Ref.~\cite{Zhao2021Fermionic}, because every input we estimate is a linear combination of products of the
Jordan--Wigner Majorana operators, and for each such product it has the same measurement channel and the same variance
bound,~\eqref{eq:matchgate-variance}, as the full group~\cite{Wan2023Matchgate}.  We therefore only need this simpler version, whose circuits are Clifford, built from
nearest-neighbour $S^\dagger$, $\exp\bigl(\sfrac{i\pi}{4}\,X_jX_{j+1}\bigr)$ and single-qubit Pauli gates.  Matchgate shadows give unbiased estimates of $k$-RDMs (reduced density matrices), or
equivalently every Majorana correlator of degree up to $2k$.  While the scaling of the variance (and hence the
number of samples to reach a desired accuracy) of standard classical shadows scales exponentially with the
operator weight, matchgate shadows have a variance that scales as:
\begin{equation}
  \mathrm{Var}_{\rm snapshot} \;\le\; \binom{2n}{2k}\Big/\binom{n}{k}
  \;=\; \frac{(2n)^k}{(2k-1)!!} + \mathcal{O}\bigl(n^{k-1}\bigr)
  \label{eq:matchgate-variance}
\end{equation}
for a degree-$2k$ correlator, which for a fixed $k$, is polynomial in $n$.  The inputs that are linear in the
state are therefore estimated without bias.  The squares in $\Phi_j$, $\varphi_{H,j}^2$ and $\varphi_{j,H}^2$ are formed
from the estimates directly, a plug-in estimate biased upward by the estimator variance, which falls as $1/S$.

Every term we need to compute for the distribution is composed of expectation values of simple operators or nested
commutators with the Hamiltonian terms.  For example, $\gamma_j=\braket{P_jOP_j}-\braket{O}$ and
$\varphi_{j,b}=\braket{[P_j,[O,P_b]]}$.  Since each of the Hamiltonian terms comes from fermionic operators, $a_m$ and $a_m^\dagger$, under Jordan--Wigner, we
can write them in terms of the $2n$ Majorana operators $\zeta_{2m}=a_m+a_m^\dagger$ and $\zeta_{2m+1}=-i(a_m-a_m^\dagger)$
as follows~\cite{JordanWigner1928,Zhao2021Fermionic}:
\begin{equation}
  P_j\propto\zeta_{\mathcal{I}_j}:=\prod_{\mu\in \mathcal{I}_j}\zeta_\mu ,
  \quad
  O=o_0+\frac{i}{2}\sum_{\mu,\nu}A_{\mu\nu}\,\zeta_\mu\zeta_\nu ,
  \label{eq:majorana-forms}
\end{equation}
where $|\mathcal{I}_j|\in\{2,4\}$ is the number of Majorana operators in term $j$. This is because electronic-structure Hamiltonians in the Born--Oppenheimer approximation and a finite 
basis~\cite{Bauer2020} contain only one-body terms $a_p^\dagger a_q$ and two-body terms $a_p^\dagger a_q^\dagger a_r a_s$, which map to two and
four Majorana operators respectively, and $A$ is real and antisymmetric, with $o_0$ the constant (identity) part of $O$.
To see what order, $k$, we require for \statedrift{}, we first note that a Majorana monomial
$\zeta_{\mathcal{I}}$ commutes with any $\zeta_{\mathcal{I}'}$ of even degree $|\mathcal{I}'|$, such as a Hamiltonian
term, unless $\mathcal{I}$ and $\mathcal{I}'$ share an odd number of indices, i.e.\ $|\mathcal{I}\cap\mathcal{I}'|$ is odd, in
which case they anticommute and their commutator is
$[\zeta_{\mathcal{I}},\zeta_{\mathcal{I}'}]=\zeta_{\mathcal{I}}\zeta_{\mathcal{I}'}-\zeta_{\mathcal{I}'}\zeta_{\mathcal{I}}=2\,\zeta_{\mathcal{I}}\zeta_{\mathcal{I}'}$.  Expanding $\zeta_{\mathcal{I}}\zeta_{\mathcal{I}'}$ using~\eqref{eq:majorana-forms}, we get
\begin{equation}
  \zeta_{\mathcal{I}}\zeta_{\mathcal{I}'}
    =\prod_{\mu\in\mathcal{I}}\zeta_\mu\prod_{\nu\in\mathcal{I}'}\zeta_\nu
    =\pm\prod_{\mu\in\mathcal{I}\cap\mathcal{I}'}\zeta_\mu^2\prod_{\nu\in\mathcal{S}}\zeta_\nu
    =\pm\,\zeta_{\mathcal{S}},
  \label{eq:majorana-product}
\end{equation}
where $\mathcal{S}=(\mathcal{I}\cup\mathcal{I}')\setminus(\mathcal{I}\cap\mathcal{I}')$ is the set of
indices which appear in exactly one of $\mathcal{I}$ and $\mathcal{I}'$, the sign collects a factor $-1$ for each
swap of two distinct Majorana operators, $\zeta_\mu\zeta_\nu=-\zeta_\nu\zeta_\mu$ for $\mu\neq\nu$, made to bring the two
copies of each shared $\zeta_\mu$, one from $\zeta_{\mathcal{I}}$ and one from $\zeta_{\mathcal{I}'}$, together.  We also have $\zeta_\mu^2=\mathds{1}$ and as
such the commutator $[\zeta_{\mathcal{I}},\zeta_{\mathcal{I}'}]$ is a single Majorana monomial of order
$|\mathcal{S}|=|\mathcal{I}|+|\mathcal{I}'|-2|\mathcal{I}\cap\mathcal{I}'|\le|\mathcal{I}|+|\mathcal{I}'|-2$.  Then, for a Hamiltonian term where $|\mathcal{I}'|\le4$, the degree of the commutator
$[\zeta_{\mathcal{I}},\zeta_{\mathcal{I}'}]$ is larger than that of $\zeta_{\mathcal{I}}$ by at most 2, since
$|\mathcal{S}|\le|\mathcal{I}|+4-2=|\mathcal{I}|+2$.  Starting from $O$, which has degree two
by~\eqref{eq:majorana-forms}, $[O,P_b]$ therefore has degree at most four and $\varphi_{j,b}=\braket{[P_j,[O,P_b]]}$ at
most six.  To compute $\gamma_j$, we note that conjugation of $O$ with $P_j$ flips the sign of the $\zeta_\mu$
with $\mu\in\mathcal{I}_j$, so
\begin{equation}
  \begin{aligned}
  \gamma_j&=\braket{P_jOP_j}-\braket{O}
    =\frac{i}{2}\sum_{\mu,\nu}A_{\mu\nu}\,(s_\mu s_\nu-1)\braket{\zeta_\mu\zeta_\nu}\\
  &=-2i\sum_{\mu\in\mathcal{I}_j}\sum_{\nu\notin\mathcal{I}_j}A_{\mu\nu}\braket{\zeta_\mu\zeta_\nu},
  \end{aligned}
  \label{eq:gamma-majorana}
\end{equation}
with $s_\mu=-1$ for $\mu\in\mathcal{I}_j$ and $s_\mu=+1$ otherwise: only pairs with exactly one index in
$\mathcal{I}_j$ change sign, and $A_{\nu\mu}\zeta_\nu\zeta_\mu=A_{\mu\nu}\zeta_\mu\zeta_\nu$ doubles each.
This is a product of 2 Majorana operators, so it requires the 1-RDM.  Similarly $\beta_j$ is written
as at most products of 4 Majorana operators so needs the 2-RDM.  The terms $\varphi_{j,b}$, $\Phi_j$, $\varphi_{H,j}$ and
$\varphi_{j,H}$ have at most 6 and require the 3-RDM, while the term from the fourth time
derivative of $\braket{O(t)}$ at $t=0$,
\begin{equation}
  \frac{d^4}{dt^4}\braket{O(t)}\Big|_{t=0}=\braket{[H,[H,[H,[H,O]]]]},
  \label{eq:fourth-derivative}
\end{equation}
is a product of at most 10 Majorana operators, hence requiring the 5-RDM.  Every term that depends on the distribution therefore needs at most the 3-RDM.  Only the fourth-derivative term, which does not depend on the distribution, needs the 5-RDM, which requires

\begin{equation}
  S=\mathcal{O}\!\left(\frac{n^5\log n}{\varepsilon_{\rm s}^2}\right)
  \label{eq:snapshot-count}
\end{equation}
matchgate snapshots to estimate every element of the 5-RDM to accuracy $\varepsilon_{\rm s}$~\cite{Zhao2021Fermionic}.
In Figure~\ref{fig:warmdrift} we demonstrate this for the same LiH dipole setup that was used to test the classical surrogates, which is
non-trivial due to the non-computational-basis nature of the dipole operator.  In Figure~\ref{fig:warmdrift}(a,c) we
show that matchgate shadows reach oracle-accuracy with far fewer shots than a strategy which distributes a fixed number
of shots across all commuting groups of operators to be estimated, and also in (b,d) we show that shadows can be
combined with non-converged surrogates to improve both.  The shadow snapshots are computed once, on the prepared state alone. As such, their cost is shared by every trajectory, depth, evolution time and accuracy target that uses the distribution.

\section{Discussion}
\label{sec:discussion}

In this work, we introduced \statedrift{}, which uses problem-specific information to optimise the \qdrift{}
sampling distribution.  \statedrift{} is capable of dramatically reducing circuit depths and trajectories in estimating
time-evolved observables.  We prove that, to fourth order, the bias and variance depend on the distribution
only through low order densities at $t=0$ and bound the error of this truncation. Furthermore, we show that at a single step, for
sign-definite $\gamma_j$, a probability allocation no worse than vanilla \qdrift{} in both bias and variance exists.  We demonstrate \statedrift{}
on lattice and molecular Hamiltonian evolutions and show that in practice, the bias and variance can be reduced by orders of magnitude on lattice models.
Classical surrogates or classical shadows remove the need for exact inputs, at no quantum cost for a surrogate and with a number of snapshots polynomial in the number of qubits for shadows,
and the method requires no extra ancillae over vanilla \qdrift{} since we only alter the sampling distribution.
Finally, we also have a hyperparameter $\tau$ which controls the deviation from vanilla \qdrift{} and ensures
\statedrift{} keeps the worst-case error bounds of \qdrift{}, loosened by at most a $\tau$-dependent factor (Lemma~\ref{lem:minweight}), which matters
at long propagation times where the fourth-order expansion no longer holds.

There are several possible avenues for future work.  The first is to tighten the error bounds.  Our bound on the
truncation (Proposition~\ref{prop:main-RN}) is set by $\lambda T$ rather than by the
depth~\cite{Watson2024RandomlyCompiled}, and a tighter bound, together with a bound on how errors propagate into the resolving depth from the estimated
densities, would make the probability allocations rigorous at the depths used in practice.
The second is to utilise \statedrift{} for more challenging applications, beyond the classically tractable systems of
$6$ to $18$ qubits benchmarked here, at the values of $\lambda T$ that the application requires.  Higher-order
truncations, or expansions that target the terms grouped by $\sfrac1N$ directly (as Richardson
extrapolation~\cite{Watson2024RandomlyCompiled,Watson2024TrotterMitigation} does), would extend the distribution to larger
$\lambda T$, including cases such as H$_2$O at $T=1$.  The third is to combine \statedrift{} with composite and partially randomised schemes,
which apply a product formula to the terms with the largest coefficients and sample from the
remainder~\cite{Ouyang2020,JinLi2021PartiallyRandom,Hagan2023Composite,Pocrnic2024}. This could include partially randomised
phase estimation~\cite{Gunther2026PartialRandom}, which, combined with the STAR architecture~\cite{Akahoshi2023,Kanasugi2026EarlyFTQPE},
would bring several simulation targets into the accessible regime on early fault-tolerant devices.

\begin{acknowledgments}
The authors acknowledge support from Fujitsu Research of Europe and Fujitsu Limited.

\textit{Use of large language models.}---The authors used large language model tools (Claude, Anthropic;
Codex, OpenAI) to assist with software development, code review and derivations, and editing of the
manuscript text and its \LaTeX{} structure.  The authors reviewed and verified all outputs and take full
responsibility for the content of this work.

\textit{Competing interests.}---The authors declare no competing interests.
\end{acknowledgments}

\clearpage
\onecolumngrid
\appendix
\interdisplaylinepenalty=0

\section{Bias and variance densities}
\label{app:densities}

\subsection{Depth groups of the bias and variance}

The expressions~\eqref{eq:BN_full} and~\eqref{eq:VN_full} are grouped by order in $dt$
and then by power of $N$: $B_{p,k}$ and $V_{p,k}$ collect everything entering at order $dt^p$ with
$k$ factors of $N$, with the leading terms of each written out in the main text.  We use upper case
for these to denote that they are linear combinations of simpler densities (below).  For the
bias, we have the extra terms:
\begin{equation}
\begin{gathered}
  B_{3,2} = \tfrac12\bigl(b_{\beta\Delta,2} + b_{\gamma H,2}\bigr), \qquad
  B_{4,3} = \tfrac16\bigl(b_{\varphi\Delta,2} + b_{\gamma HH,2} + b_{\beta\Delta H,2}\bigr), \\
  B_{4,2} = \tfrac14\bigl(b_{\varphi\Delta,2} + b_{\gamma HH,2}\bigr)
    + \tfrac12\bigl(b_{\varphi_{H,j},3} + b_{\varphi_{j,H},3}\bigr) + \tfrac12 b_{\gamma\gamma,2} .
\end{gathered}
  \label{eq:methods-bias-groups}
\end{equation}
For the variance, we have the extra terms:
\begin{equation}
\begin{gathered}
  V_{3,2} = v_{\beta\cdot\varphi_{H,j},2} + v_{\beta\cdot\varphi_{j,H},2}, \qquad
  V_{4,3} = \tfrac13\bigl(v_{\varphi_{H,j}^2,2} + v_{\varphi_{j,H}^2,2}
    + v_{\varphi_{\ssub{H,j}}\varphi_{\ssub{j,H}},2}\bigr) + v_{\beta\cdot\psi} - \varphi_{H,H}^2, \\
  V_{4,2} = v_{\gamma\cdot\varphi_{H,j},3} + v_{\gamma\cdot\varphi_{j,H},3}
    + \tfrac12 v_\varphi + v_{\beta\cdot\varphi} - \tfrac12\varphi_{H,H}^2 - 2b_{\gamma,2}\varphi_{H,H} .
\end{gathered}
  \label{eq:methods-variance-groups}
\end{equation}
The densities themselves are defined below.

\subsection{Density definitions}

Every density in~\eqref{eq:BN_full} and~\eqref{eq:VN_full} is a weighted sum of the
quantities defined in~\eqref{eq:cross-step-defs} over one or two terms.  As a reminder, we have the terms (where $A$, $B$, $X$ can be any
operator):
\begin{equation}
  \gamma_A(X):=\braket{\psi|AXA|\psi}-\braket{X}, \qquad
  \beta_A(X):=-i\braket{\psi|[X,A]|\psi}, \qquad
  \varphi_{A,B}(X):=\braket{\psi|[A,[X,B]]|\psi},
  \label{eq:methods-defs-gamma}
\end{equation}
with the contractions over the terms of $H$, each taking the same argument $X$,
\begin{equation}
  \beta_H:=\sum_jc_j\beta_{P_j}, \qquad
  \varphi_{H,H}:=\sum_{a,b}c_ac_b\varphi_{P_a,P_b}, \qquad
  \varphi_{H,P_j}:=\sum_bc_b\varphi_{P_b,P_j}, \qquad
  \varphi_{P_j,H}:=\sum_bc_b\varphi_{P_j,P_b},
  \label{eq:methods-defs-betaH}
\end{equation}
and the disturbance $\Delta_j(X):=P_jXP_j-X$.  For compactness, we use the shorthand, $\gamma_j:=\gamma_{P_j}(O)$,
$\beta_j:=\beta_{P_j}(O)$, $\varphi_{a,b}:=\varphi_{P_a,P_b}(O)$, $\varphi_{H,j}:=\varphi_{H,P_j}(O)$ and
$\varphi_{j,H}:=\varphi_{P_j,H}(O)$.  We use the subscript notation to identify each term as
follows.  For example, $v_{X,p}$ indicates that the quantity $X$ appears in the sum and $p$ is the power
of the weight, $\sfrac{c_j^{\,p}}{q_j^{\,p-1}}$.  The full list of terms, including those already
given in~\eqref{eq:per-step-defs}--\eqref{eq:cross-step-densities}, is
\begin{equation}
\begin{alignedat}{2}
  v_{\beta^2,2} &= \sum_j \frac{c_j^2\beta_j^2}{q_j} - \beta_H^2, &\qquad
  v_\gamma &= \sum_j \frac{c_j^4\gamma_j^2}{q_j^3} - b_{\gamma,2}^2, \\
  v_{\beta\gamma,3} &= 2\Bigl(\sum_j \frac{c_j^3\beta_j\gamma_j}{q_j^2} - \beta_H\,b_{\gamma,2}\Bigr), &\qquad
  v_{\beta^2,4} &= \tfrac43\Bigl(\sum_j \frac{c_j^4\beta_j^2}{q_j^3} - \beta_H\sum_j \frac{c_j^3\beta_j}{q_j^2}\Bigr), \\
  v_{\beta\cdot\varphi_{H,j},2} &= \sum_j \frac{c_j^2\beta_j\cdot\varphi_{H,j}}{q_j} - \beta_H\cdot\varphi_{H,H}, &\qquad
  v_{\beta\cdot\varphi_{j,H},2} &= \sum_j \frac{c_j^2\beta_j\cdot\varphi_{j,H}}{q_j} - \beta_H\cdot\varphi_{H,H},
\end{alignedat}
  \label{eq:methods-single-sum-densities}
\end{equation}
and
\begin{equation}
\begin{alignedat}{2}
  v_{\gamma\cdot\varphi_{H,j},3} &= \sum_j \frac{c_j^3\gamma_j\cdot\varphi_{H,j}}{q_j^2}, &\qquad
  v_{\gamma\cdot\varphi_{j,H},3} &= \sum_j \frac{c_j^3\gamma_j\cdot\varphi_{j,H}}{q_j^2}, \\
  v_{\varphi_{H,j}^2,2} &= \sum_j \frac{c_j^2\varphi_{H,j}^2}{q_j}, &\qquad
  v_{\varphi_{j,H}^2,2} &= \sum_j \frac{c_j^2\varphi_{j,H}^2}{q_j}, \\
  v_{\varphi_{\ssub{H,j}}\varphi_{\ssub{j,H}},2} &= \sum_j \frac{c_j^2\varphi_{H,j}\varphi_{j,H}}{q_j}, &\qquad
  v_\varphi &= \sum_{a,b} \frac{c_a^2c_b^2\varphi_{a,b}^2}{q_aq_b} .
\end{alignedat}
  \label{eq:methods-variance-densities}
\end{equation}
The two fourth-order cross-step densities $v_{\beta\cdot\varphi}$ and $v_{\beta\cdot\psi}$, which
carry $\beta_j$ against the coupling and against a nested double commutator with $H$, are given in
Appendix~\ref{app:S2}.

The corresponding bias densities are:
\begin{equation}
\begin{aligned}
  b_{\gamma,2} &= \sum_j \frac{c_j^2\gamma_j}{q_j}, \qquad
  b_{\gamma,4} = -\tfrac13\sum_j \frac{c_j^4\gamma_j}{q_j^3} - \tfrac1{24}\varphi_{H,H}\bigl(-[H,[H,O]]\bigr) , \\
  b_{\beta\Delta,2} &= \sum_j \frac{c_j^2}{q_j}\,\beta_H\bigl(\Delta_j(O)\bigr) - \tfrac12 \beta_H\bigl(-[H,[H,O]]\bigr), \qquad
  b_{\gamma H,2} = \sum_j \frac{c_j^2}{q_j}\,\gamma_j\bigl(i[H,O]\bigr) - \tfrac12 \beta_H\bigl(-[H,[H,O]]\bigr), \\
  b_{\beta,3} &= -\tfrac23\sum_j \frac{c_j^3\beta_j}{q_j^2} - \tfrac16 \beta_H\bigl(-[H,[H,O]]\bigr), \\
  b_{\varphi_{H,j},3} &= -\tfrac23\sum_j \frac{c_j^3\varphi_{H,j}}{q_j^2} - \tfrac16 \varphi_{H,H}\bigl(-[H,[H,O]]\bigr), \qquad
  b_{\varphi_{j,H},3} = -\tfrac23\sum_j \frac{c_j^3\varphi_{j,H}}{q_j^2} - \tfrac16 \varphi_{H,H}\bigl(-[H,[H,O]]\bigr),
\end{aligned}
  \label{eq:methods-bias-densities}
\end{equation}
with $b_{\gamma,4}$ the fourth-order counterpart of $b_{\gamma,2}$, and the three densities
carrying two factors of $H$ follow the same pattern one order higher,
\begin{equation}
\begin{aligned}
  b_{\varphi\Delta,2} &= \sum_j \frac{c_j^2}{q_j}\,\varphi_{H,H}\bigl(\Delta_j(O)\bigr) - \tfrac12 \varphi_{H,H}\bigl(-[H,[H,O]]\bigr), \\
  b_{\gamma HH,2} &= \sum_j \frac{c_j^2}{q_j}\,\gamma_j\bigl(-[H,[H,O]]\bigr) - \tfrac12 \varphi_{H,H}\bigl(-[H,[H,O]]\bigr), \\
  b_{\beta\Delta H,2} &= \sum_j \frac{c_j^2}{q_j}\,\beta_H\bigl(\Delta_j(i[H,O])\bigr) - \tfrac12 \varphi_{H,H}\bigl(-[H,[H,O]]\bigr) ,
\end{aligned}
  \label{eq:methods-bias-densities-4}
\end{equation}
while $b_{\gamma\gamma,2}$, from two sampled terms, is
\begin{equation}
  b_{\gamma\gamma,2} = \sum_{j,k} \frac{c_j^2c_k^2}{q_jq_k}\,\gamma_{j,k}
    \;-\; \tfrac12\bigl(b_{\varphi\Delta,2} + b_{\gamma HH,2}\bigr) \;-\; \tfrac14\,\varphi_{H,H}\bigl(-[H,[H,O]]\bigr),
  \label{eq:methods-pair-density}
\end{equation}
with
\begin{equation}
  \gamma_{j,k} := \bra{\psi}P_jP_kOP_kP_j\ket{\psi} - \gamma_j - \gamma_k - \braket{O} .
  \label{eq:methods-pair-gamma}
\end{equation}
All of these are algebraic in $\bq$ with elementary
gradients, which is what makes the distribution solvable at each depth.

\section{A toy example}\label{app:S1}

In this section, we derive the single-step quantities of the main-text toy
(Fig.~\ref{fig:toy-qubit-main}) and evaluate them for the four cases plotted there.

\paragraph{Setup.}
The generator is the two-site transverse-field Ising Hamiltonian
\begin{equation}
  H = 3Z_0Z_1 + X_0 + X_1 ,
  \qquad (P_1,P_2,P_3) = (Z_0Z_1,\,X_0,\,X_1),
  \qquad \boldsymbol{c} = (3,1,1)
  \label{eq:s1-H}
\end{equation}
with $\lambda = \sum_j|c_j| = 5$.  The reference states are the real product states
\begin{equation}
  \ket{\psi_s} = \ket{\sfrac{\pi}{4}}^{\otimes 2},
  \qquad
  \ket{\psi_a} = \ket{\sfrac{\pi}{6}}\otimes\ket{\sfrac{\pi}{3}},
  \qquad
  \ket{\theta} = \cos\tfrac\theta2\ket{0} + \sin\tfrac\theta2\ket{1}
  \label{eq:s1-states}
\end{equation}
with $\braket{\theta|Z|\theta} = \cos\theta$.  The observable is
$O\in\{Z_0+Z_1,\,Z_0,\,Z_1\}$ and the estimand is
$\braket{O(T)} = \braket{\psi|e^{iHT}Oe^{-iHT}|\psi}$ at $T=0.2$, taken
with a single step, $N=1$ and $dt=T$.

\paragraph{The single-step estimator.}
A \qdrift{} step samples $j$ with probability $q_j$ and applies
$V_j = e^{-i\theta_j P_j}$ at $\theta_j = \sfrac{c_j\,dt}{q_j}$.  Since $P_j^2 = \mathds{1}$,
\begin{equation}
  V_j = \cos\theta_j\,\mathds{1} - i\sin\theta_j\,P_j
  \label{eq:s1-Vj}
\end{equation}
Conjugating $O$ by \eqref{eq:s1-Vj},
\begin{align}
  V_j^\dagger O V_j
    &= \big(\cos\theta_j\,\mathds{1} + i\sin\theta_j\,P_j\big)\,O\,
       \big(\cos\theta_j\,\mathds{1} - i\sin\theta_j\,P_j\big) \nonumber \\
    &= \cos^2\!\theta_j\,O + \sin^2\!\theta_j\,P_j O P_j
       + i\sin\theta_j\cos\theta_j\,\big(P_j O - O P_j\big) \nonumber \\
    &= O + \sin^2\!\theta_j\,\big(P_j O P_j - O\big)
       - \tfrac{i}{2}\sin 2\theta_j\,[O,P_j]
  \label{eq:s1-conj}
\end{align}
the last line using $\cos^2\!\theta_j = 1-\sin^2\!\theta_j$ and
$2\sin\theta_j\cos\theta_j = \sin2\theta_j$.  With
$\hat O_1(j) := \braket{\psi|V_j^\dagger O V_j|\psi}$ and the definitions
\begin{equation}
  \gamma_j := \braket{\psi|P_j O P_j|\psi} - \braket{O} ,
  \qquad
  \beta_j := -i\braket{\psi|[O,P_j]|\psi}
  \label{eq:s1-gamma-beta}
\end{equation}
the expectation of \eqref{eq:s1-conj} in $\ket{\psi}$ is
\begin{equation}
  \hat O_1(j) = \braket{O} + \sin^2\!\theta_j\,\gamma_j
              + \tfrac12\sin2\theta_j\,\beta_j
  \label{eq:s1-est}
\end{equation}

\paragraph{The first-order term vanishes.}
$[O,P_j]^\dagger = -[O,P_j]$, so $\braket{\psi|[O,P_j]|\psi}$ is purely
imaginary.  Each $P_j$ in \eqref{eq:s1-H} and each $O$ is a real matrix in the
computational basis, and $\ket{\psi_s}$, $\ket{\psi_a}$ have real amplitudes,
so $\braket{\psi|[O,P_j]|\psi}$ is also real.  A number both real and purely
imaginary is zero, giving
\begin{equation}
  \beta_j = 0 ,
  \qquad
  \hat O_1(j) = \braket{O} + s_j ,
  \qquad
  s_j := \sin^2\!\theta_j\,\gamma_j
  \label{eq:s1-beta-zero}
\end{equation}

\paragraph{Bias and variance.}
The bias of \eqref{eq:s1-beta-zero} against the exact evolution is
\begin{equation}
  B_1 := \mathbb{E}_{\bq}[\hat O_1] - \braket{O(T)}
       = \braket{O} + \sum_j q_j s_j - \braket{O(T)}
  \label{eq:s1-bias}
\end{equation}
Since $\braket{O}$ is constant, $\Var_{\bq}[\hat O_1] = \Var_{\bq}[s]$.  With
$\sum_j q_j = 1$,
\begin{align}
  \tfrac12\sum_{j,k} q_j q_k (s_j-s_k)^2
    &= \tfrac12\Big[\sum_j q_j s_j^2 \sum_k q_k
       - 2\sum_j q_j s_j \sum_k q_k s_k
       + \sum_k q_k s_k^2 \sum_j q_j \Big] \nonumber \\
    &= \sum_j q_j s_j^2 - \Big(\sum_j q_j s_j\Big)^2 
  \label{eq:s1-var-expand}
\end{align}
The $j=k$ terms of the double sum vanish, leaving
\begin{equation}
  V_1 := \Var_{\bq}[\hat O_1] = \sum_{j<k} q_j q_k (s_j-s_k)^2 
  \label{eq:s1-var}
\end{equation}

\paragraph{Computing $\boldsymbol{\gamma}$.}
$\{X_i,Z_i\}=0$ gives $X_0(Z_0+Z_1)X_0 = -Z_0+Z_1$ and $X_1(Z_0+Z_1)X_1 = Z_0-Z_1$,
so for $O = Z_0+Z_1$
\begin{equation}
  \gamma_{X_0} = -2\braket{Z_0} ,
  \qquad
  \gamma_{X_1} = -2\braket{Z_1}
  \label{eq:s1-gamma-X}
\end{equation}
For $O = Z_0$ the same algebra gives $\gamma_{X_0} = -2\braket{Z_0}$ and
$\gamma_{X_1} = 0$, and the two exchange for $O = Z_1$.
Every observable here commutes with the coupling, $[O,Z_0Z_1] = 0$, hence
$Z_0Z_1\,O\,Z_0Z_1 = O$ and
\begin{equation}
  \gamma_{Z_0Z_1} = 0
  \label{eq:s1-gamma-ZZ}
\end{equation}
at every angle.  Because $Z_0Z_1$ commutes with $O$, its rotation leaves $O$ unchanged, so sampling
$Z_0Z_1$ returns $\hat O_1 = \braket{O}$ exactly for any $\theta_{Z_0Z_1}$, including the large angles that a
small $q_{Z_0Z_1}$ produces.  It drops out of $B_1$ and reduces \eqref{eq:s1-var} to
\begin{equation}
  V_1 = q_{Z_0Z_1} q_{X_0}\, s_{X_0}^2 + q_{Z_0Z_1} q_{X_1}\, s_{X_1}^2 + q_{X_0} q_{X_1}\,(s_{X_0}-s_{X_1})^2 
  \label{eq:s1-var-reduced}
\end{equation}
Vanilla \qdrift{} gives the $Z_0Z_1$ term the largest weight,
$\bq^c = (\sfrac35,\sfrac15,\sfrac15)$, and sets a common angle
$\theta_j = \sfrac{c_jT}{q_j} = T\lambda = 1$ for all three terms.

\paragraph{Small angles and the $\gamma$-driven rule.}
With the exact shifts $s_j = \gamma_j\sin^2\!\bigl(\sfrac{c_jT}{q_j}\bigr)$ there is no closed-form $\bq$ that minimises
$V_1$~\eqref{eq:s1-var}.  By expanding $s_j$ with a small angle approximation, we can find closed form
optimisers to certain orders.
For $\theta_j\ll1$, $\sin^2\!\theta_j = \theta_j^2 + \mathcal{O}(\theta_j^4)$ and
\begin{equation}
  s_j = \Big(\frac{c_j T}{q_j}\Big)^{\!2}\gamma_j + \mathcal{O}(\theta_j^4)
  \label{eq:s1-small-angle}
\end{equation}
Substituting $q_j = \sfrac{|c_j|\sqrt{|\gamma_j|}}{K}$ with
$K = \sum_k |c_k|\sqrt{|\gamma_k|}$ gives
$\sfrac{c_jT}{q_j} = \mathrm{sgn}(c_j)\,\sfrac{TK}{\sqrt{|\gamma_j|}}$ and
\begin{equation}
  s_j = T^2K^2\,\mathrm{sgn}(\gamma_j) + \mathcal{O}(\theta_j^4)
  \label{eq:s1-neyman-shift}
\end{equation}
The $s_j$ coincide when the $\gamma_j$ share a sign, and \eqref{eq:s1-var}
vanishes at leading order.  This is the $\gamma$-driven optimum
\begin{equation}
  q_j^\star \propto |c_j|\sqrt{|\gamma_j|}
  \label{eq:s1-qstar}
\end{equation}
of the main text~\eqref{eq:gamma-neyman}.  It is a closed-form minimiser only when the non-zero
$\gamma_j$ share a sign (Proposition~\ref{prop:real-neyman}).  In both states every non-zero
$\gamma_j$ is negative, so the sign condition holds.  Terms with $\gamma_j=0$, such as $Z_0Z_1$,
receive zero probability under~\eqref{eq:s1-qstar}.  We deal with this in the below paragraph.

\paragraph{Minimum weight.}
Equation~\eqref{eq:s1-gamma-ZZ} sends $q^\star_{Z_0Z_1}$ to zero and $\theta_{Z_0Z_1}$ to
infinity.  To avoid this, and to ensure \statedrift{} unbiasedness in a limit, we add a minimum weight
$\tau$ that mixes \eqref{eq:s1-qstar} with vanilla \qdrift{} allocation $\bq^c$,
\begin{equation}
  \bq = \tau\,\bq^c + (1-\tau)\,\bq^\star ,
  \qquad \tau = 0.2
  \label{eq:s1-tau}
\end{equation}
holding $q_{Z_0Z_1} = 0.12$ and $\theta_{Z_0Z_1} = 5$ finite.
Since $q^\star_{Z_0Z_1}=0$, the probability ratio
$\omega_{Z_0Z_1} = \sfrac{q^c_{Z_0Z_1}}{q_{Z_0Z_1}} = \sfrac{0.6}{0.12} = \sfrac{1}{\tau}$ equals the largest value allowed by
the bounded-reweighting lemma of the main text (Lemma~\ref{lem:minweight}).  The minimum
weight is the only source of $V_1$ at $N=1$ to leading order.  Under $\bq^\star$~\eqref{eq:s1-qstar}, $\gamma_{Z_0Z_1}=0$~\eqref{eq:s1-gamma-ZZ} gives $q^\star_{Z_0Z_1}=0$, so for every observable the first two terms of~\eqref{eq:s1-var-reduced} vanish,
\begin{equation}
  q^\star_{Z_0Z_1}q^\star_{X_0}\,s_{X_0}^2 \;=\; q^\star_{Z_0Z_1}q^\star_{X_1}\,s_{X_1}^2 \;=\; 0 .
\end{equation}
For $O=Z_0+Z_1$, $\gamma_{X_0}$ and $\gamma_{X_1}$ are both negative, so~\eqref{eq:s1-neyman-shift} gives
$s_{X_0}=s_{X_1}=-T^2K^2$ and
\begin{equation}
  q^\star_{X_0}q^\star_{X_1}\,(s_{X_0}-s_{X_1})^2
  \;=\; q^\star_{X_0}q^\star_{X_1}\,\bigl(-T^2K^2+T^2K^2\bigr)^2 \;=\; 0 .
\end{equation}
For $O=Z_0$, $\gamma_{X_1}=0$ gives $q^\star_{X_1}=0$, and for $O=Z_1$, $\gamma_{X_0}=0$ gives $q^\star_{X_0}=0$, so in both
cases the last term of~\eqref{eq:s1-var-reduced} also vanishes,
\begin{equation}
  q^\star_{X_0}q^\star_{X_1}\,(s_{X_0}-s_{X_1})^2 \;=\; 0 .
\end{equation}
Hence $V_1(\bq^\star)=0$ at leading order for all three observables.  The mixing~\eqref{eq:s1-tau} gives each
zero-probability term the weight $\tau q^c_j$, which produces the small $V_1$ of Table~\ref{tab:s1-alloc}.  At $N>1$ the $Z_0Z_1$ rotation generates
the dynamics of \eqref{eq:s1-H}, and $\tau$ keeps it in the compilation.

\begin{table}[H]
\centering
\caption{Single-step bias, variance and RMSE at $T=0.2$, $N=1$, under vanilla
\qdrift{} and \statedrift{}~\eqref{eq:s1-tau}, against $\braket{O(T)}$ from exact
diagonalisation.  $\mathrm{RMSE}=\sqrt{B_1^2+V_1}$ is the band width of
Fig.~\ref{fig:toy-qubit-main}, and $\bq$ lists $(q_{Z_0Z_1},q_{X_0},q_{X_1})$ under
\statedrift{}.}
\label{tab:s1-alloc}
\begin{tabular}{llrrrrcrrr}
\hline
 & & & \multicolumn{3}{c}{vanilla \qdrift{}} & \multicolumn{4}{c}{\statedrift{}} \\
\cline{4-6}\cline{7-10}
state & $O$ & $\braket{O(T)}$ & $B_1$ & $V_1$ & RMSE
      & $\bq$ & $B_1$ & $V_1$ & RMSE \\
\hline
$\ket{\psi_s}$ & $Z_0+Z_1$ & $1.5221$ & $-0.5084$ & $0.2407$ & $0.7065$ & $(0.120,0.440,0.440)$ & $-0.3478$ & $0.0078$ & $0.3589$ \\
$\ket{\psi_a}$ & $Z_0+Z_1$ & $1.4773$ & $-0.4982$ & $0.2514$ & $0.7068$ & $(0.120,0.495,0.385)$ & $-0.3387$ & $0.0072$ & $0.3491$ \\
$\ket{\psi_a}$ & $Z_0$     & $0.8572$ & $-0.2364$ & $0.2407$ & $0.5446$ & $(0.120,0.840,0.040)$ & $-0.0721$ & $0.0012$ & $0.0803$ \\
$\ket{\psi_a}$ & $Z_1$     & $0.6201$ & $-0.2617$ & $0.0802$ & $0.3856$ & $(0.120,0.040,0.840)$ & $-0.1668$ & $0.0004$ & $0.1681$ \\
\hline
\end{tabular}
\end{table}

Both allocations undershoot $\braket{O(T)}$ in all four rows.  \eqref{eq:s1-tau} lowers the RMSE
by $2.0$, $2.0$, $6.8$ and $2.3$ times across the four rows.  The gain is largest
for $O=Z_0$, where $\gamma_{X_1} = 0$ leaves one shifting term out of three and
vanilla \qdrift{} spends $80\%$ of its weight on terms that, when sampled, return
$\braket{O}$.  The same holds for $O=Z_1$, where the gain is smaller because there $\tfrac12\varphi_{H,H}=3.5$ is positive while every shift is negative, so $|b_{\gamma,2}-\tfrac12\varphi_{H,H}|\ge3.5$ for every allocation, against $\tfrac12\varphi_{H,H}=-0.23$ for $O=Z_0$.

\section{Variance and bias formulas}\label{app:S2}

In this section, we derive the single-step and $N$-step variance and bias of the
estimator, expanding both to $\mathcal{O}(dt^4)$, and identify every density that enters the
main-text expressions.

{%
\subsection*{Setup and notation}

We derive the master bias and variance formulas~\eqref{eq:BN_full} and~\eqref{eq:VN_full} for a general time-evolved
observable $O$, expanding the single-step estimator to $\mathcal{O}(dt^4)$ and composing
$N$ steps.  Two operators enter with different roles: the \emph{generators} are the Pauli
terms $P_j$ of $H = \sum_j c_j P_j$, which build each qDRIFT step
$V_j = e^{-i\theta_jP_j}$, while the \emph{measured} operator is the observable
$O$.  Setting $O = H$ recovers energy estimation, in which the two coincide.  Define the
single-step estimator $\hat O_1(j) := \braket{\psi|V_j^\dagger O V_j|\psi}$ and write
$\theta_j := \sfrac{c_j\,dt}{q_j}$ and $b_{\gamma,2} := \sum_j \sfrac{c_j^2\gamma_j}{q_j}$ throughout.  With
$\beta_j$ and $\gamma_j$ as in~\eqref{eq:s1-gamma-beta}, the cross-step scalar is
$\varphi_{a,b} := \braket{\psi|[P_a,[O,P_b]]|\psi}$, together with the two contractions of $\varphi$ against the generator coefficients, which
differ in general:
\begin{align}
  \varphi_{H,a} &:= \sum_b c_b\,\varphi_{b,a} = \braket{\psi|[H,[O,P_a]]|\psi}, \label{eq:s2-chi}\\
  \varphi_{a,H}    &:= \sum_b c_b\,\varphi_{a,b} = \braket{\psi|[P_a,[O,H]]|\psi}. \label{eq:s2-J}
\end{align}
All are real, since $P_jOP_j$, $-i[O,P_j]$ and $[P_a,[O,P_b]]$ are Hermitian.  Two further
quantities appear in the derivations, $\beta_H$ and $\varphi_{H,H}$, which are the first and
second derivatives of $\braket{O(t)}$ evaluated at $t=0$,
\begin{equation}
  \beta_H := \sum_j c_j\beta_j = \tfrac{d}{dt}\braket{O(t)}\big|_0,
  \qquad
  \tfrac12\varphi_{H,H} := \tfrac12\sum_a c_a\varphi_{H,a} = -\tfrac12\braket{[H,[H,O]]}
       = \tfrac12\tfrac{d^2}{dt^2}\braket{O(t)}\big|_0 
  \label{eq:s2-btot-kc}
\end{equation}
For energy estimation, $O=H$, both vanish,
$\beta_H\propto\braket{[H,H]}=0$
and $\tfrac12\varphi_{H,H}\propto\braket{[H,[H,H]]}=0$.  With $O=H$ we also have that $\varphi_{a,H} = \braket{[P_a,[H,H]]} = 0\;\;\forall a$,
while $\varphi_{H,a} = \braket{[H,[H,P_a]]} \neq 0$ in general.  The energy
formula~\eqref{eq:V1B1-energy} therefore retains $v_{\gamma\cdot\varphi_{H,j},3}$ and
$v_{\varphi_{H,j}^2,2}$ and drops every density built from $\varphi_{j,H}$.  Their weighted sums
still agree, $\sum_a c_a\varphi_{H,a} = \sum_a c_a\varphi_{a,H} = \varphi_{H,H}$.
}

The bias and the variance are built from the same maps.  Write $\mathcal{V}_j$ for the observable rotated by step
$j$, $\mathcal{K}$ for the mean single step and $\mathcal{U}_t$ for the exact evolution,
\begin{equation}
  \mathcal{V}_j(X) := V_j^\dagger X V_j, \qquad
  \mathcal{K}(X) := \sum_j q_j\,\mathcal{V}_j(X), \qquad
  \mathcal{U}_t(X) := e^{iHt} X e^{-iHt} .
  \label{eq:s2-meanchannel}
\end{equation}
We use $\mathcal{E}$ to denote \qdrift{} acting on the initial state, $\mathcal{E}(\rho)=\sum_jq_j\,V_j\rho V_j^\dagger$,
and $\mathcal{K}$ to denote it acting on the observable, $\mathcal{K}(O)=\sum_jq_j\,V_j^\dagger OV_j$~\eqref{eq:s2-meanchannel}.
The two give the same expectation,
\begin{equation}
  \Tr\bigl[O\,\mathcal{E}^N(\rho)\bigr] \;=\; \Tr\bigl[\mathcal{K}^N(O)\,\rho\bigr] ,
  \label{eq:s2-duality}
\end{equation}
$\mathcal{E}$ is more standard in the \qdrift{} literature~\cite{Campbell2019,Watson2024RandomlyCompiled}, and the two
are equivalent by~\eqref{eq:s2-duality}.  We work with $\mathcal{K}$, the Heisenberg picture, because it allows us to
naturally write the $\gamma_j$, $\beta_j$, etc.\ as expectation values of measurable operators (commutators) rather than
using commutators between states and operators.
A trajectory $(j_1,\dots,j_N)$ then gives
\begin{equation}
  \hat O_N = \Braket{\mathcal{V}_{j_1}\bigl(\mathcal{V}_{j_2}(\cdots\mathcal{V}_{j_N}(O))\bigr)} .
  \label{eq:s2-Vmap}
\end{equation}
The change a single step makes to an operator, which both expansions start from, is the following.

\begin{lemma}[One-step change]\label{lem:s2-onestep}
Write $r_j(X) := \mathcal{V}_j(X) - X$ for the change of $X$ under step $j$, at angle $\theta_j = \sfrac{c_j\,dt}{q_j}$.
Then
\begin{equation}
  r_j(X) = i[P_j,X]\,\frac{c_j}{q_j}\,dt + \Delta_j(X)\,\frac{c_j^2}{q_j^2}\,dt^2
   - \tfrac23\,i[P_j,X]\,\frac{c_j^3}{q_j^3}\,dt^3 - \tfrac13\,\Delta_j(X)\,\frac{c_j^4}{q_j^4}\,dt^4 + \mathcal{O}(dt^5).
  \label{eq:s2-rj}
\end{equation}
Averaging over the sampled term returns the mean step,
\begin{equation}
  \sum_j q_j\,r_j(X) = \mathcal{K}(X) - X .
  \label{eq:s2-rj-mean}
\end{equation}
\end{lemma}

\begin{proof}
Step $j$ applies $V_j = e^{-i\theta_jP_j}$ at $\theta_j = \sfrac{c_j\,dt}{q_j}$.  Since $P_j^2 = \mathds{1}$, we have:
\begin{equation}
  V_j = \cos\theta_j\,\mathds{1} - i\sin\theta_j\,P_j ,
\end{equation}
which is~\eqref{eq:s1-Vj}.  Conjugating $X$ by it
gives~\eqref{eq:s1-conj} with $O$ replaced by $X$, so that a step is exact in its angle,
\begin{equation}
  r_j(X) = V_j^\dagger XV_j - X = \sin^2\theta_j\,\Delta_j(X) + \tfrac12\sin(2\theta_j)\,i[P_j,X] .
  \label{eq:s2-rj-exact}
\end{equation}

The small-angle expansions are
\begin{equation}
  \sin^2\theta = \theta^2 - \tfrac13\theta^4 + \mathcal{O}(\theta^6), \qquad
  \tfrac12\sin2\theta = \theta - \tfrac23\theta^3 + \mathcal{O}(\theta^5),
\end{equation}
so that at $\theta_j = \sfrac{c_j\,dt}{q_j}$,
\begin{equation}
  \sin^2\theta_j = \frac{c_j^2}{q_j^2}\,dt^2 - \frac13\frac{c_j^4}{q_j^4}\,dt^4 + \mathcal{O}(dt^6), \qquad
  \tfrac12\sin2\theta_j = \frac{c_j}{q_j}\,dt - \frac23\frac{c_j^3}{q_j^3}\,dt^3 + \mathcal{O}(dt^5).
\end{equation}
Substituting these into~\eqref{eq:s2-rj-exact} and collecting powers of $dt$ gives~\eqref{eq:s2-rj}.

Finally, $\sum_jq_j\mathcal{V}_j = \mathcal{K}$ by~\eqref{eq:s2-meanchannel} and $\sum_jq_j = 1$, so summing $r_j(X)$
with weights $q_j$ gives~\eqref{eq:s2-rj-mean}.
\end{proof}

\subsection*{Bias}

We expand both the bias and variance to order $dt^4$.  Because the $N$ samples are
independent, the mean of the $N$-step estimator is the $N$-fold composition of the mean of a single
step, so it is enough to expand one step and then compose.  Averaging each $\mathcal{V}_j$ of~\eqref{eq:s2-Vmap} applies
$\mathcal{K}$, so
$\mathbb{E}_{\bq}[\hat O_N] = \bra{\psi}\mathcal{K}^N(O)\ket{\psi}$ with $\mathcal{K}^N := \mathcal{K}\circ\cdots\circ\mathcal{K}$ the $N$-fold
composition.  The exact value is $\braket{O(T)} = \bra{\psi}\mathcal{U}_T(O)\ket{\psi}$ with $T = N\,dt$, so the bias is
$B_N = \bra{\psi}(\mathcal{K}^N - \mathcal{U}_T)(O)\ket{\psi}$ exactly.

Proposition~\ref{prop:multistep_bias} at the end of this subsection writes $B_N$ in the following densities, with $\mathcal{L}_H:=i[H,\cdot\,]$, so that $\mathcal{L}_H(O)=i[H,O]$ and $\mathcal{L}_H^2(O)=-[H,[H,O]]$, $\Delta_j(X) := P_jXP_j - X$, $w_j := \sfrac{c_j^2}{q_j}$, and
$\gamma_j(X)$, $\beta_H(X)$ and $\varphi_{H,H}(X)$ are the quantities of~\eqref{eq:s1-gamma-beta} and~\eqref{eq:s2-btot-kc} with
$O$ replaced by $X$, and $\varphi_{H,j}$, $\varphi_{j,H}$ are given by~\eqref{eq:s2-chi}--\eqref{eq:s2-J}:
\begin{equation}
\begin{gathered}
\begin{aligned}
  b_{\gamma,2} &:= \sum_j w_j\gamma_j, &
  b_{\beta,3} &:= -\tfrac23\sum_j \frac{c_j^3\beta_j}{q_j^2} - \tfrac16\beta_H(\mathcal{L}_H^2(O)), \\
  b_{\gamma,4} &:= -\tfrac13\sum_j \frac{c_j^4\gamma_j}{q_j^3} - \tfrac1{24}\varphi_{H,H}(\mathcal{L}_H^2(O)), &
  b_{\beta\Delta H,2} &:= \sum_j w_j\beta_H(\Delta_j\mathcal{L}_H(O)) - \tfrac12\varphi_{H,H}(\mathcal{L}_H^2(O)), \\
  b_{\beta\Delta,2} &:= \sum_j w_j\beta_H(\Delta_jO) - \tfrac12\beta_H(\mathcal{L}_H^2(O)), &
  b_{\gamma H,2} &:= \sum_j w_j\gamma_j(\mathcal{L}_H(O)) - \tfrac12\beta_H(\mathcal{L}_H^2(O)), \\
  b_{\varphi_{H,j},3} &:= -\tfrac23\sum_j \frac{c_j^3\varphi_{H,j}}{q_j^2} - \tfrac16\varphi_{H,H}(\mathcal{L}_H^2(O)), &
  b_{\varphi_{j,H},3} &:= -\tfrac23\sum_j \frac{c_j^3\varphi_{j,H}}{q_j^2} - \tfrac16\varphi_{H,H}(\mathcal{L}_H^2(O)), \\
  b_{\varphi\Delta,2} &:= \sum_j w_j\varphi_{H,H}(\Delta_jO) - \tfrac12\varphi_{H,H}(\mathcal{L}_H^2(O)), &
  b_{\gamma HH,2} &:= \sum_j w_j\gamma_j(\mathcal{L}_H^2(O)) - \tfrac12\varphi_{H,H}(\mathcal{L}_H^2(O)),
\end{aligned}\\
  b_{\gamma\gamma,2} := \sum_{j,k} w_jw_k\,\gamma_{j,k}
    - \tfrac12\sum_j w_j\bigl[\gamma_j(\mathcal{L}_H^2(O)) + \varphi_{H,H}\bigl(\Delta_j(O)\bigr)\bigr]
    + \tfrac14\varphi_{H,H}(\mathcal{L}_H^2(O)),\\
  \gamma_{j,k} := \braket{P_jP_kOP_kP_j} - \gamma_j - \gamma_k - \braket{O}.
\end{gathered}
  \label{eq:s2-bias-densities}
\end{equation}
The subscript $p$ on each density is the $p$ of the $\delta_p$ inside it.

{%
We compute $B_N$ to order $dt^4$ in five steps, and we combine the steps into the expansion of $B_N$ in Proposition~\ref{prop:multistep_bias} at the end of this section.  Lemma~\ref{lem:s2-meanstep} averages the one-step
change of Lemma~\ref{lem:s2-onestep} to give the mean single step, Lemma~\ref{lem:s2-defect} subtracts the exact step
to give the map $\delta=\mathcal{K}-\mathcal{U}_{dt}$, Lemma~\ref{lem:s2-compose} composes $N$ steps, Lemma~\ref{lem:s2-bias-one} evaluates the terms with
one factor of $\delta$, and Lemma~\ref{lem:s2-pair} evaluates the
term with two factors of $\delta$.  Throughout, $w_j := \sfrac{c_j^2}{q_j}$.

\begin{lemma}[Mean single step]\label{lem:s2-meanstep}
For any operator $X$,
\begin{equation}
  \mathcal{K}(X) - X
    = \sum_j c_j\bigl(-i[X,P_j]\bigr)\,dt
    + \sum_j w_j\,\Delta_j(X)\,dt^2
    - \tfrac23\,\sum_j \frac{c_j^3}{q_j^2}\bigl(-i[X,P_j]\bigr)\,dt^3
    - \tfrac13\,\sum_j \frac{c_j^4}{q_j^3}\Delta_j(X)\,dt^4
    + \mathcal{O}(dt^5).
  \label{eq:s2-meanstep-exp}
\end{equation}
\end{lemma}

\begin{proof}
By~\eqref{eq:s2-rj-mean}, $\mathcal{K}(X) - X$ is the one-step change averaged over the sampled term,
\begin{equation*}
  \mathcal{K}(X) - X = \sum_j q_j\,r_j(X) .
\end{equation*}
Substituting the expansion~\eqref{eq:s2-rj} of $r_j(X)$ and carrying the weight $q_j$ onto each power of
$\sfrac{c_j}{q_j}$,
\begin{equation}
  q_j\frac{c_j}{q_j} = c_j, \qquad q_j\frac{c_j^2}{q_j^2} = w_j, \qquad
  q_j\frac{c_j^3}{q_j^3} = \frac{c_j^3}{q_j^2}, \qquad q_j\frac{c_j^4}{q_j^4} = \frac{c_j^4}{q_j^3},
  \label{eq:s2-uw}
\end{equation}
and using $i[P_j,X] = -i[X,P_j]$, the four terms become those of~\eqref{eq:s2-meanstep-exp}.
\end{proof}

\begin{lemma}[Difference from the exact step]\label{lem:s2-defect}
The map $\delta := \mathcal{K} - \mathcal{U}_{dt}$, which acts on operators and is the difference between the mean
step and the exact step, satisfies
\begin{equation}
  \delta = \delta_2\,dt^2 + \delta_3\,dt^3 + \delta_4\,dt^4 + \mathcal{O}(dt^5)
  \label{eq:s2-defect}
\end{equation}
with
\begin{align*}
  \delta_2 &= \sum_j w_j\,\Delta_j - \tfrac12\,\mathcal{L}_H^2, \\
  \delta_3 &= -\tfrac23\sum_j \frac{c_j^3}{q_j^2}\,\mathcal{L}_j - \tfrac16\,\mathcal{L}_H^3, \\
  \delta_4 &= -\tfrac13\sum_j \frac{c_j^4}{q_j^3}\,\Delta_j - \tfrac1{24}\,\mathcal{L}_H^4,
\end{align*}
where $\mathcal{L}_j:=i[P_j,\cdot\,]$, and, on the reference state,
\begin{equation}
  \braket{\delta_2(O)} = b_{\gamma,2} - \tfrac12\varphi_{H,H}, \qquad
  \braket{\delta_3(O)} = b_{\beta,3}, \qquad
  \braket{\delta_4(O)} = b_{\gamma,4}.
  \label{eq:s2-defect-eval}
\end{equation}
The $\mathcal{O}(dt)$ terms cancel for every $\bq$ with all $q_j>0$, and every $\delta_p$ vanishes for a single-term
Hamiltonian.
\end{lemma}

\begin{proof}
Given an operator, $X$, a \qdrift{} step, $j$, acts on it as follows, $\mathcal{V}_j(X)=V_j^\dagger XV_j=e^{i\theta_jP_j}Xe^{-i\theta_jP_j}$
with $\theta_j=\sfrac{c_j\,dt}{q_j}$. We use the notation of Ref.~\cite{Nakaji2024qSWIFT}, which writes expansions in
terms of $\mathcal{L}_j:=i[P_j,\cdot\,]$. In this notation we can write $\mathcal{V}_j=e^{\theta_j\mathcal{L}_j}$ and
$\mathcal{U}_{dt}=e^{dt\,\mathcal{L}_H}$, with $\mathcal{L}_H:=i[H,\cdot\,]=\sum_jc_j\mathcal{L}_j$
(Ref.~\cite{Nakaji2024qSWIFT}, Sec.~III\,B), by the identity $e^{A}Xe^{-A}=e^{[A,\cdot\,]}X$ with $A=i\theta_jP_j$
and $A=i\,dt\,H$ respectively.  With $\delta$ being the error map, expanding both exponentials gives:
\begin{equation}
  \delta \;=\; \sum_jq_j\,\mathcal{V}_j-\mathcal{U}_{dt}
  \;=\; \sum_{p\ge1}\frac{dt^p}{p!}\Bigl(\sum_j\frac{c_j^p}{q_j^{p-1}}\,\mathcal{L}_j^p-\mathcal{L}_H^p\Bigr) .
  \label{eq:s2-defect-series}
\end{equation}
The $p=1$ term vanishes for every $\bq$, since $\sum_jc_j\mathcal{L}_j=\mathcal{L}_H$.  Then, up to fourth order in
$dt$ ($dt^4$), we have the terms:
\begin{equation}
\begin{aligned}
  \mathcal{L}_j^2X &= i\bigl[P_j,i[P_j,X]\bigr] = -\bigl(P_j^2X-2P_jXP_j+XP_j^2\bigr) = 2\bigl(P_jXP_j-X\bigr) ,\\
  \mathcal{L}_j^3X &= -4\,i[P_j,X] , \qquad \mathcal{L}_j^4X = -8\bigl(P_jXP_j-X\bigr) ,
\end{aligned}
  \label{eq:s2-Lpowers}
\end{equation}
where $P_jXP_j-X=:\Delta_j(X)$, and $\mathcal{L}_H^p(X)=i^p[H,[H,\dots[H,X]\dots]]$ with $p$ commutators, so plugging in $\mathcal{L}_j^p$ and $\mathcal{L}_H^p$ up to $p=4$
gives each term in~\eqref{eq:s2-defect}.

To evaluate these on $\ket{\psi}$ with $X=O$, recall the single-step quantities of~\eqref{eq:s1-gamma-beta},
\begin{equation}
  \braket{\Delta_j(O)} = \bra{\psi}P_jOP_j\ket{\psi} - \braket{O} = \gamma_j, \qquad
  \braket{-i[O,P_j]} = -i\bra{\psi}[O,P_j]\ket{\psi} = \beta_j,
\end{equation}
and~\eqref{eq:s2-btot-kc}, the last two with $O$ replaced by $\mathcal{L}_H^2(O) = -[H,[H,O]]$,
\begin{equation}
  \braket{[H,[H,O]]} = -\varphi_{H,H}, \qquad
  \braket{[H,[H,[H,O]]]} = i\,\beta_H(\mathcal{L}_H^2(O)), \qquad
  \braket{[H,[H,[H,[H,O]]]]} = \varphi_{H,H}(\mathcal{L}_H^2(O)).
\end{equation}
Substituting into $\delta_2$, $\delta_3$ and $\delta_4$,
\begin{align}
  \braket{\delta_2(O)} &= \sum_j w_j\gamma_j - \tfrac12\varphi_{H,H} = b_{\gamma,2} - \tfrac12\varphi_{H,H}, \\
  \braket{\delta_3(O)} &= -\tfrac23\sum_j \frac{c_j^3\beta_j}{q_j^2} + \tfrac{i}{6}\cdot i\,\beta_H(\mathcal{L}_H^2(O)) = b_{\beta,3}, \\
  \braket{\delta_4(O)} &= -\tfrac13\sum_j \frac{c_j^4\gamma_j}{q_j^3} - \tfrac1{24}\varphi_{H,H}(\mathcal{L}_H^2(O)) = b_{\gamma,4},
\end{align}
which is~\eqref{eq:s2-defect-eval}.

For a single-term Hamiltonian, $q_1=1$ and $V_1 = e^{-iHdt}$, so $\mathcal{K} = \mathcal{U}_{dt}$ exactly and
every $\delta_p$ vanishes.
\end{proof}}

Now we need to compute the bias at $N$ steps, $B_N = \bra{\psi}(\mathcal{K}^N - \mathcal{U}_T)(O)\ket{\psi}$.  To do so, the
following lemma expands $\mathcal{K}^N$ in terms of the exact evolution $\mathcal{U}$ and the one-step difference $\delta$
of Lemma~\ref{lem:s2-defect}.

\begin{lemma}[Composition]\label{lem:s2-compose}
At fixed $N$,
\begin{equation}
  \mathcal{K}^N = \mathcal{U}_T
    + \sum_{k=0}^{N-1}\mathcal{U}_{(N-1-k)dt}\,\delta\,\mathcal{U}_{k\,dt}
    + \frac{N(N-1)}{2}\,\delta^2 + \mathcal{O}(dt^5).
  \label{eq:s2-compose}
\end{equation}
\end{lemma}

\begin{proof}
The standard telescoping identity for operators,
\begin{equation}
  A^N-B^N \;=\; \sum_{k=0}^{N-1}B^{N-1-k}\,(A-B)\,A^{k} ,
\end{equation}
applies with $A=\mathcal{K}$ and $B=\mathcal{U}_{dt}$, so that $A-B=\delta$, and we get
\begin{equation}
  \mathcal{K}^N-\mathcal{U}_T \;=\; \sum_{k=0}^{N-1}\mathcal{U}_{(N-1-k)dt}\,\delta\,\mathcal{K}^{k} .
\end{equation}
We now apply the same identity to $\mathcal{K}^k$.  By Lemma~\ref{lem:s2-defect}, $\delta=\mathcal{O}(dt^2)$, while
$\mathcal{U}_{s\,dt}$ and $\mathcal{K}^j=(\mathcal{U}_{dt}+\delta)^j$ are both $\mathds{1}+\mathcal{O}(dt)$, so
\begin{equation}
  \mathcal{K}^k-\mathcal{U}_{k\,dt} \;=\; \sum_{j=0}^{k-1}\underbrace{\mathcal{U}_{(k-1-j)dt}}_{\mathds{1}+\mathcal{O}(dt)}\,\delta\,
  \underbrace{\mathcal{K}^{j}}_{\mathds{1}+\mathcal{O}(dt)}
  \;=\; \sum_{j=0}^{k-1}\bigl(\delta+\mathcal{O}(dt^3)\bigr) \;=\; k\,\delta+\mathcal{O}(dt^3) .
\end{equation}
Substituting this into the sum above,
\begin{equation}
  \mathcal{K}^N \;=\; \mathcal{U}_T+\sum_{k=0}^{N-1}\mathcal{U}_{(N-1-k)dt}\,\delta\,\mathcal{U}_{k\,dt}
  +\sum_{k=0}^{N-1}k\,\delta^2+\mathcal{O}(dt^5) ,
\end{equation}
and $\sum_{k=0}^{N-1}k=\sfrac{N(N-1)}{2}$, which gives~\eqref{eq:s2-compose}.
\end{proof}

Applying~\eqref{eq:s2-compose} to $O$, taking the expectation in $\ket{\psi}$ and subtracting
$\bra{\psi}\mathcal{U}_T(O)\ket{\psi} = \braket{O(T)}$,

\begin{equation}
  B_N = \bra{\psi}(\mathcal{K}^N - \mathcal{U}_T)(O)\ket{\psi}
  = \underbrace{\sum_{k=0}^{N-1}\bra{\psi}\mathcal{U}_{(N-1-k)dt}\,\delta\,\mathcal{U}_{k\,dt}(O)\ket{\psi}}_{\text{one factor of }\delta}
  + \underbrace{\frac{N(N-1)}{2}\bra{\psi}\delta^2(O)\ket{\psi}}_{\text{two factors of }\delta}
  + \mathcal{O}(dt^5).
  \label{eq:s2-bias-split}
\end{equation}%
The terms with one factor of $\delta$ are evaluated with the
single-step expectations $\braket{\delta_p(O)}$ of Lemma~\ref{lem:s2-defect}, after expanding the propagators on
either side of $\delta$.  The next lemma gives the expansion of $\braket{\mathcal{P}^{s_L}\,\mathcal{M}_p\,\mathcal{P}^{s_R}(O)}$ for a
general $\mathcal{P}$, which is $\mathcal{U}_{dt}$ for the bias and $\mathcal{K}$ for the variance.

\begin{lemma}[One factor between propagators]\label{lem:s2-bias-one}
Let $\mathcal{P}=\mathds{1}+\mathcal{L}_H\,dt+\mathcal{A}_2\,dt^2+\mathcal{O}(dt^3)$ and
$\mathcal{M}=\sum_{p\ge1}\mathcal{M}_p\,dt^p$ act linearly on operators, with $\mathcal{L}_H=i[H,\cdot\,]$ as in the
proof of Lemma~\ref{lem:s2-defect}.  Then for integers $s_L,s_R\ge0$ and each $p$,
\begin{equation}
  \braket{\mathcal{P}^{s_L}\,\mathcal{M}_p\,\mathcal{P}^{s_R}(O)} \;=\; \braket{\mathcal{M}_p(O)}+m_{p,1}\,dt+m_{p,2}\,dt^2+\mathcal{O}(dt^3),
  \label{eq:s2-sandwich}
\end{equation}
with
\begin{equation}
\begin{aligned}
  m_{p,1} &= \beta_H\bigl(\mathcal{M}_p(O)\bigr)\,s_L + \braket{\mathcal{M}_p(\mathcal{L}_H(O))}\,s_R,\\
  m_{p,2} &= \braket{\mathcal{A}_2(\mathcal{M}_p(O))}\,s_L + \tfrac12\varphi_{H,H}\bigl(\mathcal{M}_p(O)\bigr)\,s_L(s_L{-}1)
     + \braket{\mathcal{M}_p(\mathcal{A}_2(O))}\,s_R\\
  &\quad + \tfrac12\braket{\mathcal{M}_p(\mathcal{L}_H^2(O))}\,s_R(s_R{-}1) + \beta_H\bigl(\mathcal{M}_p(\mathcal{L}_H(O))\bigr)\,s_Ls_R .
\end{aligned}
  \label{eq:s2-sandwich-1}
\end{equation}
\end{lemma}

\begin{proof}
Writing $\mathcal{P}=\mathds{1}+E$ with $E=\mathcal{L}_H\,dt+\mathcal{A}_2\,dt^2+\mathcal{O}(dt^3)$, the binomial expansion
gives
\begin{equation}
  \mathcal{P}^s \;=\; \mathds{1}+sE+\frac{s(s-1)}{2}E^2+\mathcal{O}(dt^3)
  \;=\; \mathds{1} + s\,\mathcal{L}_H\,dt + \Bigl(s\,\mathcal{A}_2+\frac{s(s-1)}{2}\mathcal{L}_H^2\Bigr)dt^2 + \mathcal{O}(dt^3) .
  \label{eq:s2-Ps}
\end{equation}
Substituting~\eqref{eq:s2-Ps} for $\mathcal{P}^{s_L}$ and $\mathcal{P}^{s_R}$ and collecting powers of $dt$,
\begin{equation}
\begin{aligned}
  &\braket{\mathcal{P}^{s_L}\,\mathcal{M}_p\,\mathcal{P}^{s_R}(O)}
  \;=\; \braket{\mathcal{M}_p(O)}
  + \Bigl(s_L\braket{\mathcal{L}_H(\mathcal{M}_p(O))} + s_R\braket{\mathcal{M}_p(\mathcal{L}_H(O))}\Bigr)dt\\
  &+ \Bigl(s_L\braket{\mathcal{A}_2(\mathcal{M}_p(O))} + \tfrac{s_L(s_L-1)}{2}\braket{\mathcal{L}_H^2(\mathcal{M}_p(O))}
     + s_R\braket{\mathcal{M}_p(\mathcal{A}_2(O))}\\
  &\qquad + \tfrac{s_R(s_R-1)}{2}\braket{\mathcal{M}_p(\mathcal{L}_H^2(O))}
     + s_Ls_R\braket{\mathcal{L}_H(\mathcal{M}_p(\mathcal{L}_H(O)))}\Bigr)dt^2 + \mathcal{O}(dt^3) ,
\end{aligned}
\end{equation}
and $\braket{\mathcal{L}_H(Y)}=\beta_H(Y)$ and
$\braket{\mathcal{L}_H^2(Y)}=\varphi_{H,H}(Y)$~\eqref{eq:s2-btot-kc} give~\eqref{eq:s2-sandwich-1}.
\end{proof}

As mentioned, for the bias we have $\mathcal{P}=\mathcal{U}_{dt}=e^{dt\,\mathcal{L}_H}$ (proof of
Lemma~\ref{lem:s2-defect}), whose $dt^2$ term is $\mathcal{A}_2=\tfrac12\mathcal{L}_H^2$.  Restating $m_{p,2}$
from~\eqref{eq:s2-sandwich-1},
\begin{equation}
\begin{aligned}
  m_{p,2} \;=\;& \braket{\mathcal{A}_2(\mathcal{M}_p(O))}\,s_L + \tfrac12\varphi_{H,H}\bigl(\mathcal{M}_p(O)\bigr)\,s_L(s_L{-}1)
     + \braket{\mathcal{M}_p(\mathcal{A}_2(O))}\,s_R\\
  &+ \tfrac12\braket{\mathcal{M}_p(\mathcal{L}_H^2(O))}\,s_R(s_R{-}1) + \beta_H\bigl(\mathcal{M}_p(\mathcal{L}_H(O))\bigr)\,s_Ls_R ,
\end{aligned}
\end{equation}
the two terms with $\mathcal{A}_2$ are
\begin{equation}
  \braket{\mathcal{A}_2(\mathcal{M}_p(O))} \;=\; \tfrac12\braket{\mathcal{L}_H^2(\mathcal{M}_p(O))} \;=\; \tfrac12\varphi_{H,H}\bigl(\mathcal{M}_p(O)\bigr) ,
  \qquad
  \braket{\mathcal{M}_p(\mathcal{A}_2(O))} \;=\; \tfrac12\braket{\mathcal{M}_p(\mathcal{L}_H^2(O))} ,
\end{equation}
so the terms in $s_L$ and in $s_R$ combine as
\begin{equation}
\begin{aligned}
  \braket{\mathcal{A}_2(\mathcal{M}_p(O))}\,s_L + \tfrac12\varphi_{H,H}\bigl(\mathcal{M}_p(O)\bigr)\,s_L(s_L{-}1)
  &= \tfrac12\varphi_{H,H}\bigl(\mathcal{M}_p(O)\bigr)\bigl(s_L+s_L(s_L{-}1)\bigr)
   = \tfrac12\varphi_{H,H}\bigl(\mathcal{M}_p(O)\bigr)\,s_L^2 ,\\
  \braket{\mathcal{M}_p(\mathcal{A}_2(O))}\,s_R + \tfrac12\braket{\mathcal{M}_p(\mathcal{L}_H^2(O))}\,s_R(s_R{-}1)
  &= \tfrac12\braket{\mathcal{M}_p(\mathcal{L}_H^2(O))}\bigl(s_R+s_R(s_R{-}1)\bigr)
   = \tfrac12\braket{\mathcal{M}_p(\mathcal{L}_H^2(O))}\,s_R^2 ,
\end{aligned}
\end{equation}
and so:
\begin{equation}
  m_{p,2} \;=\; \tfrac12\varphi_{H,H}\bigl(\mathcal{M}_p(O)\bigr)\,s_L^2 + \tfrac12\braket{\mathcal{M}_p(\mathcal{L}_H^2(O))}\,s_R^2
  + \beta_H\bigl(\mathcal{M}_p(\mathcal{L}_H(O))\bigr)\,s_Ls_R .
\end{equation}
Finally, we use $\mathcal{M}=\delta$, with $\mathcal{M}_p=\delta_p$ for $p=2,3,4$.  The $k$-th term of the first sum in~\eqref{eq:s2-bias-split} is
$\bra{\psi}\mathcal{U}_{(N-1-k)dt}\,\delta\,\mathcal{U}_{k\,dt}(O)\ket{\psi}$, which we write as
$\bra{\psi}\mathcal{U}_{s_L dt}\,\delta\,\mathcal{U}_{s_R dt}(O)\ket{\psi}$ with $s_L:=N-1-k$ exact steps composed after $\delta$ and
$s_R:=k$ before it, reading right to left.  Inserting
$\delta=\sum_{p=2}^{4}\delta_p\,dt^p+\mathcal{O}(dt^5)$ and writing $\mathcal{U}_{s\,dt}=\mathcal{U}_{dt}^{\,s}$,
\begin{equation}
\begin{aligned}
  \bra{\psi}\mathcal{U}_{s_L dt}\,\delta\,\mathcal{U}_{s_R dt}(O)\ket{\psi}
  \;&=\; \sum_{p=2}^{4}dt^p\,\braket{\mathcal{U}_{dt}^{\,s_L}\,\delta_p\,\mathcal{U}_{dt}^{\,s_R}(O)}+\mathcal{O}(dt^5)\\
  \;&=\; \sum_{p=2}^{4}dt^p\Bigl(\braket{\delta_p(O)}+m_{p,1}\,dt+m_{p,2}\,dt^2\Bigr)+\mathcal{O}(dt^5),
\end{aligned}
\end{equation}
by~\eqref{eq:s2-sandwich} with $\mathcal{P}=\mathcal{U}_{dt}$ and $\mathcal{M}_p=\delta_p$.  Collecting powers of $dt$,
\begin{equation}
  \bra{\psi}\mathcal{U}_{s_L dt}\,\delta\,\mathcal{U}_{s_R dt}(O)\ket{\psi}
  \;=\; \braket{\delta_2(O)}\,dt^2 + \bigl(\braket{\delta_3(O)}+m_{2,1}\bigr)dt^3
  + \bigl(\braket{\delta_4(O)}+m_{3,1}+m_{2,2}\bigr)dt^4 + \mathcal{O}(dt^5),
  \label{eq:s2-bias-one}
\end{equation}
where, by~\eqref{eq:s2-sandwich-1} and the combined $m_{p,2}$ above with $\mathcal{M}_p=\delta_p$,
\begin{equation}
\begin{aligned}
  m_{2,1} &= \beta_H\bigl(\delta_2(O)\bigr)\,s_L + \braket{\delta_2(\mathcal{L}_H(O))}\,s_R ,\qquad
  m_{3,1} = \beta_H\bigl(\delta_3(O)\bigr)\,s_L + \braket{\delta_3(\mathcal{L}_H(O))}\,s_R ,\\
  m_{2,2} &= \tfrac12\varphi_{H,H}\bigl(\delta_2(O)\bigr)\,s_L^2 + \tfrac12\braket{\delta_2(\mathcal{L}_H^2(O))}\,s_R^2
  + \beta_H\bigl(\delta_2(\mathcal{L}_H(O))\bigr)\,s_Ls_R .
\end{aligned}
  \label{eq:s2-bias-m}
\end{equation}
The second part of~\eqref{eq:s2-bias-split}, $\sfrac{N(N-1)}{2}\,\bra{\psi}\delta^2(O)\ket{\psi}$, needs $\delta$ to be applied twice.  We do this in the below.  Inserting
$\delta = \delta_2\,dt^2 + \delta_3\,dt^3 + \delta_4\,dt^4 + \mathcal{O}(dt^5)$ from Lemma~\ref{lem:s2-defect},

\begin{equation}
  \delta^2 = \bigl(\delta_2\,dt^2 + \delta_3\,dt^3 + \dots\bigr)\bigl(\delta_2\,dt^2 + \delta_3\,dt^3 + \dots\bigr)
    = \delta_2\delta_2\,dt^4 + \bigl(\delta_2\delta_3 + \delta_3\delta_2\bigr)\,dt^5 + \mathcal{O}(dt^6),
  \label{eq:s2-delta-sq-order}
\end{equation}%
so to order $dt^4$ only $\braket{\delta_2(\delta_2(O))}$ contributes.  In Lemma~\ref{lem:s2-defect} $\braket{\delta_2(O)}$ is
computed, so the following lemma computes $\braket{\delta_2(\delta_2(O))}$.

\begin{lemma}[Two factors of $\delta$]\label{lem:s2-pair}
Applying $\delta_2$ of Lemma~\ref{lem:s2-defect} twice to $O$ and taking the expectation in $\ket{\psi}$ gives
\begin{equation}
  \Braket{\delta_2\bigl(\delta_2(O)\bigr)} = \sum_{j,k} w_jw_k\,\gamma_{j,k}
    - \tfrac12\sum_j w_j\Bigl[\gamma_j(\mathcal{L}_H^2(O)) + \varphi_{H,H}\bigl(\Delta_j(O)\bigr)\Bigr]
    + \tfrac14\varphi_{H,H}(\mathcal{L}_H^2(O)),
  \label{eq:s2-delta2sq}
\end{equation}
where
\begin{equation}
  \gamma_{j,k} := \braket{P_jP_kOP_kP_j} - \gamma_j - \gamma_k - \braket{O}
\end{equation}
is the shift in $\braket{O}$ when both $P_j$ and $P_k$ are applied, minus the shifts from each alone.  We define~\eqref{eq:s2-delta2sq} as the density
$b_{\gamma\gamma,2} := \braket{\delta_2(\delta_2(O))}$.
\end{lemma}

\begin{proof}
By Lemma~\ref{lem:s2-defect}, $\delta_2(X) = \sum_k w_k\,\Delta_k(X) - \tfrac12\mathcal{L}_H^2(X)$.  Taking
$X = \delta_2(O)$ and inserting $\delta_2(O)$ from the same expression,
\begin{align}
  \delta_2\bigl(\delta_2(O)\bigr)
  &= \sum_j w_j\,\Delta_j\bigl(\delta_2(O)\bigr) - \tfrac12\mathcal{L}_H^2\bigl(\delta_2(O)\bigr) \nonumber\\
  &= \sum_j w_j\,\Delta_j\Bigl(\sum_k w_k\,\Delta_k(O) - \tfrac12\mathcal{L}_H^2(O)\Bigr)
     - \tfrac12\mathcal{L}_H^2\Bigl(\sum_k w_k\,\Delta_k(O) - \tfrac12\mathcal{L}_H^2(O)\Bigr).
\end{align}
Using linearity of $\Delta_j$ and of the commutator,
\begin{equation}
  \delta_2\bigl(\delta_2(O)\bigr) = \sum_{j,k} w_jw_k\,\underbrace{\Delta_j\Delta_k(O)}_{(1)}
    - \tfrac12\sum_j w_j\,\underbrace{\Delta_j\bigl(\mathcal{L}_H^2(O)\bigr)}_{(2)}
    - \tfrac12\sum_k w_k\,\underbrace{\mathcal{L}_H^2\bigl(\Delta_k(O)\bigr)}_{(3)}
    + \tfrac14\underbrace{\mathcal{L}_H^4(O)}_{(4)}.
\end{equation}
To take the expectation of each term, write, for any operator $X$,
\begin{equation}
  \gamma_j(X) := \braket{\Delta_j(X)}, \qquad \varphi_{H,H}(X) := -\braket{[H,[H,X]]},
\end{equation}
and recall $\mathcal{L}_H^2(O) = -[H,[H,O]]$.  Using linearity of $\gamma_j$ and $\varphi_{H,H}$ in $X$, and
$\braket{P_jOP_j} = \gamma_j + \braket{O}$,
\begin{align}
  (1) &\;\Rightarrow\; \Delta_j\Delta_k(O) = P_jP_kOP_kP_j - P_jOP_j - P_kOP_k + O, \nonumber\\
      &\hphantom{\;\Rightarrow\;}\; \braket{\Delta_j\Delta_k(O)} = \braket{P_jP_kOP_kP_j} - \gamma_j - \gamma_k - \braket{O} = \gamma_{j,k}, \\
  (2) &\;\Rightarrow\; \Braket{\Delta_j\bigl(\mathcal{L}_H^2(O)\bigr)} = \gamma_j\bigl(\mathcal{L}_H^2(O)\bigr), \\
  (3) &\;\Rightarrow\; \Braket{\mathcal{L}_H^2\bigl(\Delta_k(O)\bigr)} = \varphi_{H,H}\bigl(\Delta_k(O)\bigr), \\
  (4) &\;\Rightarrow\; \braket{\mathcal{L}_H^4(O)} = \Braket{\mathcal{L}_H^2\bigl(\mathcal{L}_H^2(O)\bigr)} = \varphi_{H,H}\bigl(\mathcal{L}_H^2(O)\bigr).
\end{align}
Substituting (1)--(4) into the expansion of $\delta_2(\delta_2(O))$ gives~\eqref{eq:s2-delta2sq}.
The order of $P_j$ and $P_k$ in $\gamma_{j,k}$ does not matter:
writing $P_jP_k = s\,P_kP_j$ with $s=\pm1$, $P_kP_jOP_jP_k = s^2\,P_jP_kOP_kP_j = P_jP_kOP_kP_j$, so
$\gamma_{j,k} = \gamma_{k,j}$.
\end{proof}

Now that we have all the ingredients, we are in a position to state the general bias and prove it.

\begin{proposition}[Multi-step bias]\label{prop:multistep_bias}
For $N$ independent samples from $\bq$ at step size $dt = \sfrac{T}{N}$, at fixed $N$, the bias
$B_N = \mathbb{E}_{\bq}[\hat O_N] - \braket{O(T)}$ is
\begin{equation}
\begin{split}
  B_N &= N\bigl(b_{\gamma,2} - \tfrac12\varphi_{H,H}\bigr)dt^2 \\
  &\quad + \Bigl[N\,b_{\beta,3}
      + \tfrac{N(N-1)}{2}\bigl(b_{\beta\Delta,2} + b_{\gamma H,2}\bigr)\Bigr]dt^3 \\
  &\quad + \Bigl[N\,b_{\gamma,4}
      + \tfrac{N(N-1)}{2}\bigl(b_{\varphi_{H,j},3} + b_{\varphi_{j,H},3}\bigr)
      + \tfrac{N(N-1)(2N-1)}{12}\bigl(b_{\varphi\Delta,2} + b_{\gamma HH,2}\bigr) \\
  &\qquad + \tfrac{N(N-1)(N-2)}{6}\,b_{\beta\Delta H,2}
      + \tfrac{N(N-1)}{2}\,b_{\gamma\gamma,2}\Bigr]dt^4 \\
  &\quad + \mathcal{O}(dt^5)
\end{split}
\label{eq:s2-BN-superop}
\end{equation}
with the densities of~\eqref{eq:s2-bias-densities}.
\end{proposition}

\begin{proof}[Proof of Proposition~\ref{prop:multistep_bias}]
By~\eqref{eq:s2-bias-split}, \eqref{eq:s2-bias-one} and~\eqref{eq:s2-bias-m} with $s_L = N{-}1{-}k$ and $s_R = k$,
and~\eqref{eq:s2-delta-sq-order} with Lemma~\ref{lem:s2-pair},
\begin{equation}
\begin{split}
  B_N &= \sum_{k=0}^{N-1}\Bigl[\braket{\delta_2(O)}\,dt^2 + \bigl(\braket{\delta_3(O)}+m_{2,1}\bigr)dt^3
  + \bigl(\braket{\delta_4(O)}+m_{3,1}+m_{2,2}\bigr)dt^4\Bigr] \\
  &\quad + \frac{N(N-1)}{2}\,b_{\gamma\gamma,2}\,dt^4 + \mathcal{O}(dt^5).
\end{split}
\label{eq:s2-bias-start}
\end{equation}
The position $k$ of $\delta$ appears in~\eqref{eq:s2-bias-start} via $s_L$ and $s_R$, which occur only in $m_{2,1}$,
$m_{3,1}$ and $m_{2,2}$, so each sum over $k$ is a sum of powers of $s_L=N{-}1{-}k$ and $s_R=k$.  We evaluate each order one at a time, computing each
expectation with $\delta_2$, $\delta_3$ and $\delta_4$ of Lemma~\ref{lem:s2-defect}.

\begin{itemize}[leftmargin=*,topsep=2pt,itemsep=2pt]
\item \emph{Order $dt^2$.}  For this order, we only have $\braket{\delta_2(O)}$, which is independent of $k$, so
\begin{equation}
  \sum_{k=0}^{N-1}\braket{\delta_2(O)}\,dt^2
  = N\,\braket{\delta_2(O)}\,dt^2
  = N\Bigl[\sum_j w_j\,\gamma_j - \tfrac12\varphi_{H,H}\Bigr]dt^2
  = N\bigl(b_{\gamma,2}-\tfrac12\varphi_{H,H}\bigr)dt^2 ,
\end{equation}
where from Lemma~\ref{lem:s2-defect} we have $\delta_2(O) = \sum_j w_j\Delta_j(O) + \tfrac12[H,[H,O]]$ and $\gamma_j = \braket{\Delta_j(O)}$
and $\braket{\tfrac12[H,[H,O]]} = -\tfrac12\varphi_{H,H}$.

\item \emph{Order $dt^3$.}  $m_{2,1}$ is given by:
\begin{equation}
  m_{2,1} = \beta_H\bigl(\delta_2(O)\bigr)\,s_L + \braket{\delta_2(\mathcal{L}_H(O))}\,s_R ,
\end{equation}
so, with $\sum_k s_L = \sum_k s_R = \sfrac{N(N-1)}{2}$,
\begin{equation}
  \sum_{k=0}^{N-1}\bigl(\braket{\delta_3(O)}+m_{2,1}\bigr)
  = N\braket{\delta_3(O)} + \frac{N(N-1)}{2}\Bigl[\beta_H\bigl(\delta_2(O)\bigr) + \braket{\delta_2(\mathcal{L}_H(O))}\Bigr] .
  \label{eq:s2-dt3-sum}
\end{equation}
Again, using Lemma~\ref{lem:s2-defect} to compute each expectation, with $\beta_j = \braket{-i[O,P_j]}$,
\begin{subequations}\label{eq:s2-dt3-terms}
\begin{alignat}{2}
  \braket{\delta_3(O)} &= b_{\beta,3} &\qquad& \text{by~\eqref{eq:s2-defect-eval}}, \label{eq:s2-t3a}\\*
  \beta_H\bigl(\delta_2(O)\bigr) &= b_{\beta\Delta,2} &\qquad& \text{by $\delta_2$ of Lemma~\ref{lem:s2-defect}}, \label{eq:s2-t3b}\\*
  \braket{\delta_2(\mathcal{L}_H(O))} &= b_{\gamma H,2} &\qquad& \text{by $\delta_2$ of Lemma~\ref{lem:s2-defect} and~\eqref{eq:s2-btot-kc}}, \label{eq:s2-t3c}
\end{alignat}
\end{subequations}
with the right-hand sides defined in~\eqref{eq:s2-bias-densities}, where~\eqref{eq:s2-btot-kc} enters through
$\braket{\mathcal{L}_H(Y)} = \beta_H(Y)$.  The $dt^3$ coefficient is therefore $N\,b_{\beta,3} + \tfrac{N(N-1)}{2}\bigl(b_{\beta\Delta,2}+b_{\gamma H,2}\bigr)$.

\item \emph{Order $dt^4$.}  $m_{3,1}$ and $m_{2,2}$ are given by:
\begin{align}
  m_{3,1} &= \beta_H\bigl(\delta_3(O)\bigr)\,s_L + \braket{\delta_3(\mathcal{L}_H(O))}\,s_R , \nonumber\\*
  m_{2,2} &= \tfrac12\varphi_{H,H}\bigl(\delta_2(O)\bigr)\,s_L^2 + \tfrac12\braket{\delta_2(\mathcal{L}_H^2(O))}\,s_R^2
  + \beta_H\bigl(\delta_2(\mathcal{L}_H(O))\bigr)\,s_Ls_R ,
\end{align}
so, with $\sum_k s_L = \sum_k s_R = \sfrac{N(N-1)}{2}$, $\sum_k s_L^2 = \sum_k s_R^2 = \sfrac{N(N-1)(2N-1)}{6}$ and
$\sum_k s_Ls_R = \sfrac{N(N-1)(N-2)}{6}$,
\begin{equation}
\begin{split}
  \sum_{k=0}^{N-1}\bigl(\braket{\delta_4(O)}+m_{3,1}+m_{2,2}\bigr) &= N\braket{\delta_4(O)} \\
  &\quad + \frac{N(N-1)}{2}\Bigl[\beta_H\bigl(\delta_3(O)\bigr) + \braket{\delta_3(\mathcal{L}_H(O))}\Bigr] \\
  &\quad + \frac{N(N-1)(2N-1)}{12}\Bigl[\varphi_{H,H}\bigl(\delta_2(O)\bigr) + \braket{\delta_2(\mathcal{L}_H^2(O))}\Bigr] \\
  &\quad + \frac{N(N-1)(N-2)}{6}\,\beta_H\bigl(\delta_2(\mathcal{L}_H(O))\bigr) .
\end{split}
\label{eq:s2-dt4-sum}
\end{equation}
Again, using Lemma~\ref{lem:s2-defect} to compute each expectation,
\begin{subequations}\label{eq:s2-dt4-terms}
\begin{alignat}{2}
  \braket{\delta_4(O)} &= b_{\gamma,4} &\qquad& \text{by~\eqref{eq:s2-defect-eval}}, \label{eq:s2-t4a}\\*
  \beta_H\bigl(\delta_3(O)\bigr) &= b_{\varphi_{H,j},3} &\qquad& \text{by $\delta_3$ of Lemma~\ref{lem:s2-defect}, \eqref{eq:s2-chi} and~\eqref{eq:s2-btot-kc}}, \label{eq:s2-t4b}\\*
  \braket{\delta_3(\mathcal{L}_H(O))} &= b_{\varphi_{j,H},3} &\qquad& \text{by $\delta_3$ of Lemma~\ref{lem:s2-defect}, \eqref{eq:s2-J} and~\eqref{eq:s2-btot-kc}}, \label{eq:s2-t4c}\\*
  \varphi_{H,H}\bigl(\delta_2(O)\bigr) &= b_{\varphi\Delta,2} &\qquad& \text{by $\delta_2$ of Lemma~\ref{lem:s2-defect}}, \label{eq:s2-t4d}\\*
  \braket{\delta_2(\mathcal{L}_H^2(O))} &= b_{\gamma HH,2} &\qquad& \text{by $\delta_2$ of Lemma~\ref{lem:s2-defect} and~\eqref{eq:s2-btot-kc}}, \label{eq:s2-t4e}\\*
  \beta_H\bigl(\delta_2(\mathcal{L}_H(O))\bigr) &= b_{\beta\Delta H,2} &\qquad& \text{by $\delta_2$ of Lemma~\ref{lem:s2-defect} and~\eqref{eq:s2-btot-kc}}, \label{eq:s2-t4f}
\end{alignat}
\end{subequations}
with the right-hand sides defined in~\eqref{eq:s2-bias-densities}.  \eqref{eq:s2-chi}--\eqref{eq:s2-btot-kc} appear in~\eqref{eq:s2-t4b},
\eqref{eq:s2-t4c}, \eqref{eq:s2-t4e} and~\eqref{eq:s2-t4f} via $\beta_H\bigl(-i[O,P_j]\bigr) = \varphi_{H,j}$, $\braket{-i[\mathcal{L}_H(O),P_j]} = \varphi_{j,H}$,
and $\braket{\mathcal{L}_H(Y)} = \beta_H(Y)$ and $\braket{\mathcal{L}_H^2(Y)} = \varphi_{H,H}(Y)$.
Adding the last term of~\eqref{eq:s2-bias-start}, the $dt^4$ coefficient is therefore
$N\,b_{\gamma,4} + \tfrac{N(N-1)}{2}\bigl(b_{\varphi_{H,j},3}+b_{\varphi_{j,H},3}\bigr)
+ \tfrac{N(N-1)(2N-1)}{12}\bigl(b_{\varphi\Delta,2}+b_{\gamma HH,2}\bigr) + \tfrac{N(N-1)(N-2)}{6}\,b_{\beta\Delta H,2}
+ \tfrac{N(N-1)}{2}\,b_{\gamma\gamma,2}$, which with the $dt^2$ and $dt^3$ coefficients gives~\eqref{eq:s2-BN-superop}.
\end{itemize}
\end{proof}

Proposition~\ref{prop:multistep_bias} gives the bias in terms of $b_x$ factors.  To get from there to the stated form in the main text:
\begin{equation}
\begin{split}
  B_N &=N\bigl(b_{\gamma,2}-\tfrac12\varphi_{H,H}\bigr)dt^2
  +N\bigl[b_{\beta,3}+(N{-}1)B_{3,2}\bigr]dt^3 \\
  &\quad +N\bigl[b_{\gamma,4}+(N{-}1)B_{4,2}+(N{-}1)(N{-}2)B_{4,3}\bigr]dt^4
  +\mathcal O(dt^5),
\end{split}
\tag{\ref{eq:BN_full}}
\end{equation}
we group terms at order $dt^p$ with $k$ factors of $N$ into each $B_{p,k}$~\eqref{eq:methods-bias-groups}.  We also split the prefactor in~\eqref{eq:s2-BN-superop} as:
\begin{equation}
  \frac{N(N-1)(2N-1)}{12} \;=\; \frac{N(N-1)}{4} + \frac{N(N-1)(N-2)}{6} ,
\end{equation}
and group the terms by $N$, $N(N-1)=\mathcal{O}(N^2)$ and $N(N-1)(N-2)=\mathcal{O}(N^3)$.  Note, at fixed $T$, the
$\mathcal{O}(dt^5)$ remainder decays as $\sfrac{1}{N}$, which is the same rate as a function of $N$ as the terms which are
not truncated.  We bound this remainder in Appendix~\ref{app:S7}.

\subsection*{Variance}

Proposition~\ref{prop:multistep_var} at the end of this subsection writes $V_N$ in the following densities, with $b_{\gamma,2}$ as in~\eqref{eq:s2-bias-densities}, $\mathcal{L}_H=i[H,\cdot\,]$ as above and
$\varphi_{H,j}$, $\varphi_{j,H}$, $\varphi_{a,b}$ from~\eqref{eq:s2-chi}--\eqref{eq:s2-J}.  As in
the bias subsection, $\beta_j(X)$, $\beta_H(X)$, $\gamma_j(X)$ and $\varphi_{H,H}(X)$ denote the quantities
of~\eqref{eq:s1-gamma-beta} and~\eqref{eq:s2-btot-kc} with $O$ replaced by $X$, so that $\beta_j(O) = \beta_j$,
$\beta_H(O)=\beta_H$, $\gamma_j(O)=\gamma_j$ and $\varphi_{H,H}(O) = \varphi_{H,H}$.  Two weighted sums over the terms are also
needed,
\begin{equation}
  \Delta_{\bq}(X) := \sum_j w_j\,\Delta_j(X), \qquad\qquad
  \gamma_{\bq}(X) := \sum_j w_j\,\gamma_j(X) = \Braket{\Delta_{\bq}(X)} ,
  \label{eq:s2-var-argdefs}
\end{equation}
with $w_j = \sfrac{c_j^2}{q_j}$ as in the bias subsection.  The densities are
{\small\setlength{\jot}{8pt}%
\begin{equation}
\begin{gathered}
\begin{aligned}
  v_{\beta^2,2} &:= \sum\nolimits_j w_j\beta_j^2 - \beta_H^2, &
  v_{\beta\gamma,3} &:= 2\Bigl(\sum\nolimits_j \frac{c_j^3\beta_j\gamma_j}{q_j^2} - \beta_H b_{\gamma,2}\Bigr), \\
  v_\gamma &:= \sum\nolimits_j \frac{c_j^4\gamma_j^2}{q_j^3} - b_{\gamma,2}^2, &
  v_{\beta^2,4} &:= \frac43\Bigl(\sum\nolimits_j \frac{c_j^4\beta_j^2}{q_j^3} - \beta_H \sum\nolimits_j \frac{c_j^3\beta_j}{q_j^2}\Bigr), \\
  v_{\beta\cdot\varphi_{H,j},2} &:= \sum\nolimits_j w_j\beta_j\varphi_{H,j} - \beta_H\varphi_{H,H}, &
  v_{\beta\cdot\varphi_{j,H},2} &:= \sum\nolimits_j w_j\beta_j\varphi_{j,H} - \beta_H\varphi_{H,H}, \\
  v_{\gamma\cdot\varphi_{H,j},3} &:= \sum\nolimits_j \frac{c_j^3\gamma_j\varphi_{H,j}}{q_j^2}, &
  v_{\gamma\cdot\varphi_{j,H},3} &:= \sum\nolimits_j \frac{c_j^3\gamma_j\varphi_{j,H}}{q_j^2}, \\
  v_{\varphi_{H,j}^2,2} &:= \sum\nolimits_j w_j\varphi_{H,j}^2, &
  v_{\varphi_{j,H}^2,2} &:= \sum\nolimits_j w_j\varphi_{j,H}^2, \\
  v_{\varphi_{\ssub{H,j}}\varphi_{\ssub{j,H}},2} &:= \sum\nolimits_j w_j\varphi_{H,j}\varphi_{j,H}, &
  v_\varphi &:= \sum\nolimits_{a,b} w_aw_b\,\varphi_{a,b}^2,
\end{aligned}\\
\begin{aligned}
  v_{\beta\cdot\varphi} &:= \sum_j \frac{c_j^3\beta_j}{q_j^2}\Bigl(\gamma_j(\mathcal{L}_H(O)) + \beta_H\bigl(\Delta_j(O)\bigr)\Bigr)
    + \sum_j w_j\beta_j\Bigl(\beta_j\bigl(\Delta_{\bq}(O)\bigr) + \gamma_{\bq}\bigl(i[P_j,O]\bigr)\Bigr)\\
    &\qquad - 2\beta_H\Bigl(\beta_H\bigl(\Delta_{\bq}(O)\bigr) + \gamma_{\bq}(\mathcal{L}_H(O))\Bigr),\\
  v_{\beta\cdot\psi} &:= \tfrac13\Bigl[\sum_j w_j\beta_j\Bigl(\beta_j(\mathcal{L}_H^2(O))
    + \beta_H\bigl(i[P_j,\mathcal{L}_H(O)]\bigr)
     + \varphi_{H,H}\bigl(i[P_j,O]\bigr)\Bigr) - 3\beta_H\,\beta_H(\mathcal{L}_H^2(O))\Bigr].
\end{aligned}
\end{gathered}
  \label{eq:s2-var-densities}
\end{equation}}
As with the bias, where present $v_{x,p}$ denotes the power of $c_j$ ($c_j^p$) in its first term.

We compute the variance in the same way as the bias, returning to
the rotated observable $\mathcal{V}_j$, the mean step $\mathcal{K}$ of~\eqref{eq:s2-meanchannel} and the trajectory
estimator~\eqref{eq:s2-Vmap} of the bias subsection.  In the case of the bias, the mean of the observable collapses to
$\mathcal{K}^N$ since each step could be analysed independently.  For the variance this is not possible, as it is quadratic
in $\hat O_N$ and couples different \qdrift{} steps.

We compute $V_N$ to order $dt^4$ in three steps.  The first writes $V_N$ on two copies of the state, Lemma~\ref{lem:s2-var-gexp} evaluates the terms with one factor of the single-step covariance $\mathcal{C}$ of~\eqref{eq:s2-var-C}, and Lemma~\ref{lem:s2-var-pair} evaluates the term with two.

For maps $\mathcal{A}$,
$\mathcal{B}$ and operators $X$, $Y$, we have $(\mathcal{A}\otimes\mathcal{B})(X\otimes Y) := \mathcal{A}(X)\otimes\mathcal{B}(Y)$
and $\braket{X\otimes Y} := \braket{X}\braket{Y}$, where
$\braket{X\otimes Y} = \bigl(\bra{\psi}\otimes\bra{\psi}\bigr)\,(X\otimes Y)\,\bigl(\ket{\psi}\otimes\ket{\psi}\bigr)$.  We define the following:
\begin{equation}
\begin{aligned}
  \mathcal{K}^{(2)} &:= \sum_j q_j\,\mathcal{V}_j\otimes\mathcal{V}_j , \\
  \mathcal{D}_j &:= \mathcal{V}_j - \mathcal{K} = (\mathds{1}+r_j) - \Bigl(\mathds{1}+\sum\nolimits_i q_i\,r_i\Bigr) = r_j - \sum\nolimits_i q_i\,r_i ,
    \qquad \sum\nolimits_j q_j\,\mathcal{D}_j = \mathcal{K} - \mathcal{K} = 0 , \\
  \mathcal{C} &:= \mathcal{K}^{(2)} - \mathcal{K}\otimes\mathcal{K}
    = \sum_j q_j\,\mathcal{D}_j\otimes\mathcal{D}_j + \mathcal{K}\otimes\underbrace{\sum\nolimits_j q_j\,\mathcal{D}_j}_{0}
      + \underbrace{\sum\nolimits_j q_j\,\mathcal{D}_j}_{0}\otimes\mathcal{K}
    = \sum_j q_j\,\mathcal{D}_j\otimes\mathcal{D}_j ,
\end{aligned}
  \label{eq:s2-var-C}
\end{equation}
with $\mathcal{V}_j = \mathds{1}+r_j$ and $\mathcal{K} = \mathds{1}+\sum_iq_ir_i$ from~\eqref{eq:s2-rj} and~\eqref{eq:s2-rj-mean}.

The first step is to write the variance on two copies of the state.  Recall from~\eqref{eq:s2-Vmap} that
$\hat O_N := \Braket{\mathcal{V}_{j_1}\cdots\mathcal{V}_{j_N}(O)}$ for a trajectory $(j_1,\dots,j_N)$.  Its square is
\begin{equation}
\begin{alignedat}{2}
  \hat O_N^2 &= \Braket{\mathcal{V}_{j_1}\cdots\mathcal{V}_{j_N}(O)}\Braket{\mathcal{V}_{j_1}\cdots\mathcal{V}_{j_N}(O)} && \\
  &= \Braket{\mathcal{V}_{j_1}\cdots\mathcal{V}_{j_N}(O)\otimes\mathcal{V}_{j_1}\cdots\mathcal{V}_{j_N}(O)}
    &\qquad& \text{by $\braket{X\otimes Y} := \braket{X}\braket{Y}$}, \\
  &= \Braket{(\mathcal{V}_{j_1}\otimes\mathcal{V}_{j_1})\bigl(\mathcal{V}_{j_2}\cdots\mathcal{V}_{j_N}(O)\otimes\mathcal{V}_{j_2}\cdots\mathcal{V}_{j_N}(O)\bigr)}
    && \text{by $(\mathcal{A}\otimes\mathcal{B})(X\otimes Y) := \mathcal{A}(X)\otimes\mathcal{B}(Y)$}, \\
  &= \Braket{(\mathcal{V}_{j_1}\otimes\mathcal{V}_{j_1})\cdots(\mathcal{V}_{j_N}\otimes\mathcal{V}_{j_N})(O\otimes O)}
    && \text{repeating for $j_2,\dots,j_N$}.
\end{alignedat}
  \label{eq:s2-var-square}
\end{equation}
The same $j_k$ acts on both copies, since both come from one trajectory.  The variance is then
\begin{equation}
\begin{alignedat}{2}
  V_N &= \mathbb{E}_{\bq}\bigl[\hat O_N^2\bigr] - \mathbb{E}_{\bq}\bigl[\hat O_N\bigr]^2 && \\
      &= \mathbb{E}_{\bq}\Braket{(\mathcal{V}_{j_1}\otimes\mathcal{V}_{j_1})\cdots(\mathcal{V}_{j_N}\otimes\mathcal{V}_{j_N})(O\otimes O)}
         - \Braket{\mathcal{K}^N(O)}^2
      &\quad& \text{by~\eqref{eq:s2-var-square}}, \\
      &= \Braket{\bigl(\mathcal{K}^{(2)}\bigr)^N(O\otimes O)} - \Braket{(\mathcal{K}\otimes\mathcal{K})^N(O\otimes O)}
      && \text{the $j_k$ independent, $\mathbb{E}_{\bq}[\mathcal{V}_j\otimes\mathcal{V}_j]=\mathcal{K}^{(2)}$ of~\eqref{eq:s2-var-C}}, \\
      &= \Braket{\bigl((\mathcal{K}\otimes\mathcal{K} + \mathcal{C})^N - (\mathcal{K}\otimes\mathcal{K})^N\bigr)(O\otimes O)}
      && \text{by~\eqref{eq:s2-var-C}}.
\end{alignedat}
  \label{eq:s2-var-copies}
\end{equation}

Expanding the $N$-th power in~\eqref{eq:s2-var-copies}, each of the $N$ factors contributes either $\mathcal{K}\otimes\mathcal{K}$
or $\mathcal{C}$, and the term with no $\mathcal{C}$ cancels $(\mathcal{K}\otimes\mathcal{K})^N$, so
\begin{equation}
\begin{aligned}
  V_N &= \sum_{k=0}^{N-1}\Braket{(\mathcal{K}\otimes\mathcal{K})^{N-1-k}\,\mathcal{C}\,(\mathcal{K}\otimes\mathcal{K})^{k}(O\otimes O)} \\
  &\quad + \sum_{0\le k<l\le N-1}\Braket{(\mathcal{K}\otimes\mathcal{K})^{N-1-l}\,\mathcal{C}\,(\mathcal{K}\otimes\mathcal{K})^{l-k-1}\,\mathcal{C}\,(\mathcal{K}\otimes\mathcal{K})^{k}(O\otimes O)} + \dots
\end{aligned}
  \label{eq:s2-var-expand}
\end{equation}
Each $r_j$ is of order $dt$ by~\eqref{eq:s2-rj}, so $\mathcal{D}_j = r_j - \sum_iq_i\,r_i$ is of order $dt$, and each factor
$\mathcal{C} = \sum_jq_j\,\mathcal{D}_j\otimes\mathcal{D}_j$ is of order $dt^2$.  We therefore drop the terms with three or more
factors of $\mathcal{C}$, which are of order $dt^6$.  The second sum of~\eqref{eq:s2-var-expand} has $\sfrac{N(N-1)}{2}$ terms, one for each pair $k<l$, each with two factors of $\mathcal{C}$.  We have $\mathcal{K} = \mathds{1} + \mathcal{O}(dt)$, and $\mathcal{K}\otimes\mathcal{K} = \mathds{1} + \mathcal{O}(dt)$.  So, each of these terms is:
\begin{equation}
  \Braket{(\mathcal{K}\otimes\mathcal{K})^{N-1-l}\,\mathcal{C}\,(\mathcal{K}\otimes\mathcal{K})^{l-k-1}\,\mathcal{C}\,(\mathcal{K}\otimes\mathcal{K})^{k}(O\otimes O)}
  = \Braket{\mathcal{C}^2(O\otimes O)} + \mathcal{O}(dt^5) \qquad \forall\,k<l .
\end{equation}
Then, to order $dt^4$,~\eqref{eq:s2-var-expand} becomes:
\begin{equation}
  V_N = \sum_{k=0}^{N-1}\Braket{(\mathcal{K}\otimes\mathcal{K})^{N-1-k}\,\mathcal{C}\,(\mathcal{K}\otimes\mathcal{K})^{k}(O\otimes O)}
   + \frac{N(N-1)}{2}\Braket{\mathcal{C}^2(O\otimes O)} + \mathcal{O}(dt^5) .
  \label{eq:s2-var-split}
\end{equation}
In the computation of~\eqref{eq:s2-var-split}, we will evaluate several terms of the form:
\begin{equation}
  \mathbb{E}_{\bq}\Bigl[\frac{c_j^p}{q_j^p}\,x_j\Bigr] = \sum_j \frac{c_j^p\,x_j}{q_j^{p-1}} , \qquad p = 1,\dots,4.
  \label{eq:s2-var-moments}
\end{equation}
In particular
\begin{equation}
  \mathbb{E}_{\bq}\Bigl[\frac{c_j}{q_j}\beta_j\Bigr] = \beta_H, \qquad
  \mathbb{E}_{\bq}\Bigl[\frac{c_j}{q_j}\varphi_{H,j}\Bigr] = \mathbb{E}_{\bq}\Bigl[\frac{c_j}{q_j}\varphi_{j,H}\Bigr] = \varphi_{H,H}, \qquad
  \mathbb{E}_{\bq}\Bigl[\frac{c_j^2}{q_j^2}\gamma_j\Bigr] = b_{\gamma,2}.
  \label{eq:s2-var-moments2}
\end{equation}

In the $k$-th term of~\eqref{eq:s2-var-split}, $s_R := k$ mean steps act on $O$ before the sampled
step and $s_L := N-1-k$ after it, as in the bias subsection.  The term is
\begin{equation}
\begin{alignedat}{2}
  &\Braket{(\mathcal{K}\otimes\mathcal{K})^{s_L}\,\mathcal{C}\,(\mathcal{K}\otimes\mathcal{K})^{s_R}(O\otimes O)} && \\
  &\quad= \sum_j q_j\Braket{(\mathcal{K}^{s_L}\otimes\mathcal{K}^{s_L})(\mathcal{D}_j\otimes\mathcal{D}_j)(\mathcal{K}^{s_R}\otimes\mathcal{K}^{s_R})(O\otimes O)}
    &\qquad& \text{by $\mathcal{C} = \sum_jq_j\,\mathcal{D}_j\otimes\mathcal{D}_j$ of~\eqref{eq:s2-var-C}}, \\
  &\quad= \sum_j q_j\Braket{\mathcal{K}^{s_L}\mathcal{D}_j\mathcal{K}^{s_R}(O)\otimes\mathcal{K}^{s_L}\mathcal{D}_j\mathcal{K}^{s_R}(O)}
    && \text{by $(\mathcal{A}\otimes\mathcal{B})(X\otimes Y) := \mathcal{A}(X)\otimes\mathcal{B}(Y)$}, \\
  &\quad= \sum_j q_j\Braket{\mathcal{K}^{s_L}\mathcal{D}_j\mathcal{K}^{s_R}(O)}^2
    && \text{by $\braket{X\otimes Y} := \braket{X}\braket{Y}$}.
\end{alignedat}
  \label{eq:s2-var-gstep}
\end{equation}
Substituting $\mathcal{D}_j = r_j - \sum_iq_i\,r_i$ from~\eqref{eq:s2-var-C} into each expectation gives
\begin{equation}
  \Braket{\mathcal{K}^{s_L}\mathcal{D}_j\mathcal{K}^{s_R}(O)} = g(j) - \sum_iq_i\,g(i), \qquad
  g(j) := \Braket{\mathcal{K}^{s_L}r_j\mathcal{K}^{s_R}(O)} ,
  \label{eq:s2-var-gdef}
\end{equation}
so $g(j)$ applies $s_R$ mean steps to $O$, then $r_j = \mathcal{V}_j - \mathds{1}$, then $s_L$ mean steps, and takes the expectation in $\ket{\psi}$.
The term is therefore
\begin{equation}
  \Braket{(\mathcal{K}\otimes\mathcal{K})^{s_L}\,\mathcal{C}\,(\mathcal{K}\otimes\mathcal{K})^{s_R}(O\otimes O)}
  = \sum_j q_j\Bigl(g(j) - \sum_iq_i\,g(i)\Bigr)^{\!2} = \Var_{\bq}[g] .
  \label{eq:s2-var-g}
\end{equation}
Hence the $k$-th term of~\eqref{eq:s2-var-split} can be written as the variance of the random variable $g(j)$, with $j$ sampled from $\bq$, which is the difference
$g(j) = \braket{\mathcal{K}^{s_L}\mathcal{V}_j\mathcal{K}^{s_R}(O)} - \braket{\mathcal{K}^{s_L}\mathcal{K}^{s_R}(O)}$ (using $r_j = \mathcal{V}_j - \mathds{1}$) in the expectation of $O$ when 1) step $k$ applies term $j$, and 2) every other step is replaced by the mean step, $\mathcal{K}$, relative to the same expectation with step $k$ removed.  In the following, we compute $g$ up to $\mathcal{O}(dt^4)$.

\begin{lemma}[One factor of $\mathcal{C}$]\label{lem:s2-var-gexp}
Writing $g(j) = g_1(j)\,dt + g_2(j)\,dt^2 + g_3(j)\,dt^3 + \mathcal{O}(dt^4)$,
\begin{align}
  g_1(j) &= \frac{c_j}{q_j}\,\beta_j, \label{eq:s2-var-g1}\\
  g_2(j) &= \frac{c_j^2}{q_j^2}\,\gamma_j + \frac{c_j}{q_j}\bigl(\varphi_{H,j}\,s_L + \varphi_{j,H}\,s_R\bigr), \label{eq:s2-var-g2}\\
  g_3(j) &= \frac{c_j}{q_j}\Bigl(\gamma_{\bq}\bigl(i[P_j,O]\bigr)\,s_L + \beta_j\bigl(\Delta_{\bq}(O)\bigr)\,s_R
     + \tfrac12\varphi_{H,H}\bigl(i[P_j,O]\bigr)\,s_L(s_L{-}1) + \tfrac12\beta_j(\mathcal{L}_H^2(O))\,s_R(s_R{-}1) \nonumber\\
   &\quad + \beta_H\bigl(i[P_j,\mathcal{L}_H(O)]\bigr)\,s_Ls_R\Bigr)
     + \frac{c_j^2}{q_j^2}\Bigl(\beta_H\bigl(\Delta_j(O)\bigr)\,s_L + \gamma_j(\mathcal{L}_H(O))\,s_R\Bigr)
     - \frac23\,\frac{c_j^3}{q_j^3}\,\beta_j . \label{eq:s2-var-g3}
\end{align}
\end{lemma}

\begin{proof}
By~\eqref{eq:s2-var-gdef}, $g(j) = \braket{\mathcal{K}^{s_L}\,r_j\,\mathcal{K}^{s_R}(O)}$ is a single map sandwiched between two
powers of a propagator, i.e.\ $\braket{\mathcal{P}^{s_L}\mathcal{M}\,\mathcal{P}^{s_R}(O)}$ of
Lemma~\ref{lem:s2-bias-one}, with
\begin{itemize}[leftmargin=*,topsep=2pt,itemsep=2pt]
  \item $\mathcal{P} = \mathcal{K} = \mathds{1} + \mathcal{L}_H\,dt + \Delta_{\bq}\,dt^2 + \mathcal{O}(dt^3)$ by
    Lemma~\ref{lem:s2-meanstep}, so $\mathcal{A}_2 = \Delta_{\bq}$;
  \item $\mathcal{M} = r_j$, whose first three orders are, by~\eqref{eq:s2-rj},
    $\mathcal{M}_1=\sfrac{c_j}{q_j}\,\mathcal{L}_j$, $\mathcal{M}_2=\sfrac{c_j^2}{q_j^2}\,\Delta_j$ and
    $\mathcal{M}_3=-\tfrac23\sfrac{c_j^3}{q_j^3}\,\mathcal{L}_j$.
\end{itemize}
Summing~\eqref{eq:s2-sandwich} over $p$ and collecting powers of $dt$, the coefficient of $dt^n$ in $g$ is
$g_n(j) = \braket{\mathcal{M}_n(O)} + \sum_{p=1}^{n-1} m_{p,n-p}$, so
\begin{equation}
  g_1 = \braket{\mathcal{M}_1(O)}, \qquad
  g_2 = \braket{\mathcal{M}_2(O)} + m_{1,1}, \qquad
  g_3 = \braket{\mathcal{M}_3(O)} + m_{2,1} + m_{1,2} .
  \label{eq:s2-var-gm}
\end{equation}
With these $\mathcal{M}_n$ and $\mathcal{A}_2 = \Delta_{\bq}$, the expectations $\braket{\mathcal{M}_n(O)}$ and the terms
$m_{p,1}$, $m_{p,2}$ of~\eqref{eq:s2-sandwich-1} are
\begin{equation}
\begin{aligned}
  \braket{\mathcal{M}_1(O)} &= \frac{c_j}{q_j}\braket{i[P_j,O]}, \qquad
  \braket{\mathcal{M}_2(O)} = \frac{c_j^2}{q_j^2}\braket{\Delta_j(O)}, \qquad
  \braket{\mathcal{M}_3(O)} = -\frac23\,\frac{c_j^3}{q_j^3}\braket{i[P_j,O]}, \\
  m_{1,1} &= \frac{c_j}{q_j}\Bigl(\beta_H\bigl(i[P_j,O]\bigr)\,s_L + \braket{i[P_j,\mathcal{L}_H(O)]}\,s_R\Bigr), \\
  m_{2,1} &= \frac{c_j^2}{q_j^2}\Bigl(\beta_H\bigl(\Delta_j(O)\bigr)\,s_L + \braket{\Delta_j(\mathcal{L}_H(O))}\,s_R\Bigr), \\
  m_{1,2} &= \frac{c_j}{q_j}\Bigl(\braket{\Delta_{\bq}(i[P_j,O])}\,s_L + \tfrac12\varphi_{H,H}\bigl(i[P_j,O]\bigr)\,s_L(s_L{-}1)
     + \braket{i[P_j,\Delta_{\bq}(O)]}\,s_R \\
  &\qquad + \tfrac12\braket{i[P_j,\mathcal{L}_H^2(O)]}\,s_R(s_R{-}1) + \beta_H\bigl(i[P_j,\mathcal{L}_H(O)]\bigr)\,s_Ls_R\Bigr) .
\end{aligned}
  \label{eq:s2-var-mexplicit}
\end{equation}
Plugging these into~\eqref{eq:s2-var-gm},
\begin{equation}
\begin{aligned}
  g_1 &= \frac{c_j}{q_j}\braket{i[P_j,O]}, \\
  g_2 &= \frac{c_j^2}{q_j^2}\braket{\Delta_j(O)}
     + \frac{c_j}{q_j}\Bigl(\beta_H\bigl(i[P_j,O]\bigr)\,s_L + \braket{i[P_j,\mathcal{L}_H(O)]}\,s_R\Bigr), \\
  g_3 &= \frac{c_j}{q_j}\Bigl(\braket{\Delta_{\bq}(i[P_j,O])}\,s_L + \tfrac12\varphi_{H,H}\bigl(i[P_j,O]\bigr)\,s_L(s_L{-}1)
     + \braket{i[P_j,\Delta_{\bq}(O)]}\,s_R + \tfrac12\braket{i[P_j,\mathcal{L}_H^2(O)]}\,s_R(s_R{-}1) \\
  &\quad + \beta_H\bigl(i[P_j,\mathcal{L}_H(O)]\bigr)\,s_Ls_R\Bigr)
     + \frac{c_j^2}{q_j^2}\Bigl(\beta_H\bigl(\Delta_j(O)\bigr)\,s_L + \braket{\Delta_j(\mathcal{L}_H(O))}\,s_R\Bigr)
     - \frac23\,\frac{c_j^3}{q_j^3}\braket{i[P_j,O]} .
\end{aligned}
  \label{eq:s2-var-gn-explicit}
\end{equation}
The expectations here are evaluated with
\begin{equation}
\begin{gathered}
  \braket{i[P_j,Y]} = \beta_j(Y), \qquad
  \braket{\Delta_j(Y)} = \gamma_j(Y), \qquad
  \braket{\Delta_{\bq}(Y)} = \gamma_{\bq}(Y), \\
  \beta_j(O) = \beta_j, \qquad \gamma_j(O) = \gamma_j, \qquad
  \beta_H\bigl(i[P_j,O]\bigr) = \varphi_{H,j}, \qquad
  \beta_j(\mathcal{L}_H(O)) = \varphi_{j,H} ,
\end{gathered}
  \label{eq:s2-var-gid}
\end{equation}
and plugging these into~\eqref{eq:s2-var-gn-explicit} gives Lemma~\ref{lem:s2-var-gexp}, \eqref{eq:s2-var-g1}--\eqref{eq:s2-var-g3}.

\end{proof}

The next lemma evaluates the second term of~\eqref{eq:s2-var-split}.

\begin{lemma}[Two factors of $\mathcal{C}$]\label{lem:s2-var-pair}
\begin{equation}
  \Braket{\mathcal{C}^2(O\otimes O)}
  = \bigl(v_\varphi - v_{\varphi_{H,j}^2,2} - v_{\varphi_{j,H}^2,2} + \varphi_{H,H}^2\bigr)\,dt^4 + \mathcal{O}(dt^5).
  \label{eq:s2-var-pair}
\end{equation}
\end{lemma}

\begin{proof}
With $a$ and $b$ the terms sampled at the two steps, the same steps as in~\eqref{eq:s2-var-gstep} give
\begin{equation}
  \Braket{\mathcal{C}^2(O\otimes O)} = \sum_{a,b}q_aq_b\,\Braket{\mathcal{D}_a\mathcal{D}_b(O)}^2 .
  \label{eq:s2-var-pair-1}
\end{equation}
To leading order in $dt$, \eqref{eq:s2-rj} and $\sum_jc_jP_j = H$ give
\begin{equation}
\begin{aligned}
  r_a &= \frac{c_a}{q_a}\,i[P_a,\cdot\,]\,dt + \mathcal{O}(dt^2), \qquad
  \sum_iq_i\,r_i = \sum_ic_i\,i[P_i,\cdot\,]\,dt + \mathcal{O}(dt^2) = i[H,\cdot\,]\,dt + \mathcal{O}(dt^2), \\
  \mathcal{D}_a &= r_a - \sum_iq_i\,r_i = i\bigl[A_a,\cdot\,\bigr]\,dt + \mathcal{O}(dt^2), \qquad
  A_a := \frac{c_a}{q_a}P_a - H ,
\end{aligned}
\end{equation}
the last by~\eqref{eq:s2-var-C}.  Applying $\mathcal{D}_b$ and then $\mathcal{D}_a$ to $O$, and expanding the double commutator
term by term with $\varphi_{a,b} = \braket{[P_a,[O,P_b]]}$ and its contractions~\eqref{eq:s2-chi}--\eqref{eq:s2-btot-kc},
\begin{equation}
\begin{aligned}
  \Braket{\mathcal{D}_a\mathcal{D}_b(O)}
  &= i^2\braket{[A_a,[A_b,O]]}\,dt^2 + \mathcal{O}(dt^3)
   = \braket{[A_a,[O,A_b]]}\,dt^2 + \mathcal{O}(dt^3) \\
  &= \Bigl(\frac{c_ac_b}{q_aq_b}\braket{[P_a,[O,P_b]]} - \frac{c_b}{q_b}\braket{[H,[O,P_b]]}
     - \frac{c_a}{q_a}\braket{[P_a,[O,H]]} + \braket{[H,[O,H]]}\Bigr)dt^2 + \mathcal{O}(dt^3) \\
  &= \Bigl(\Phi - \frac{c_b}{q_b}\,\varphi_{H,b} - \frac{c_a}{q_a}\,\varphi_{a,H} + \varphi_{H,H}\Bigr)dt^2 + \mathcal{O}(dt^3),
\end{aligned}
  \label{eq:s2-var-pair-2}
\end{equation}
with
\begin{equation}
  \Phi := \frac{c_ac_b}{q_aq_b}\,\varphi_{a,b} .
\end{equation}
Substituting~\eqref{eq:s2-var-pair-2} into~\eqref{eq:s2-var-pair-1},
\begin{equation}
\begin{aligned}
  \Braket{\mathcal{C}^2(O\otimes O)} &= \sum_{a,b}q_aq_b\,\Braket{\mathcal{D}_a\mathcal{D}_b(O)}^2
  = \mathbb{E}_{a,b}\bigl[\Braket{\mathcal{D}_a\mathcal{D}_b(O)}^2\bigr] \\
  &= \mathbb{E}_{a,b}\Bigl[\Bigl(\Phi - \frac{c_b}{q_b}\varphi_{H,b} - \frac{c_a}{q_a}\varphi_{a,H} + \varphi_{H,H}\Bigr)^{\!2}\Bigr]dt^4 + \mathcal{O}(dt^5) .
\end{aligned}
  \label{eq:s2-var-pair-sq}
\end{equation}
We can write the other terms in~\eqref{eq:s2-var-pair-sq} as averages over $\Phi$ as follows, with
$\varphi_{a,b} = \braket{\psi|[P_a,[O,P_b]]|\psi}$, $\sum_a c_aP_a = H$ and~\eqref{eq:s2-chi}--\eqref{eq:s2-btot-kc}:
\begin{equation}
\begin{aligned}
  \frac{c_b}{q_b}\varphi_{H,b} &= \frac{c_b}{q_b}\braket{\psi|[H,[O,P_b]]|\psi}
    = \sum_a q_a\frac{c_ac_b}{q_aq_b}\,\varphi_{a,b} = \mathbb{E}_a[\Phi] =: X, \\
  \frac{c_a}{q_a}\varphi_{a,H} &= \frac{c_a}{q_a}\braket{\psi|[P_a,[O,H]]|\psi}
    = \sum_b q_b\frac{c_ac_b}{q_aq_b}\,\varphi_{a,b} = \mathbb{E}_b[\Phi] =: Y, \\
  \varphi_{H,H} &= \braket{\psi|[H,[O,H]]|\psi}
    = \sum_{a,b} c_ac_b\,\varphi_{a,b} = \mathbb{E}_{a,b}[\Phi] =: Z .
\end{aligned}
  \label{eq:s2-var-pair-3}
\end{equation}
Plugging~\eqref{eq:s2-var-pair-3} into~\eqref{eq:s2-var-pair-sq} and expanding the square,
\begin{equation}
\begin{split}
  \mathbb{E}_{a,b}\Bigl[\bigl(\Phi - X - Y + Z\bigr)^2\Bigr] &= \mathbb{E}_{a,b}\bigl[\Phi^2\bigr] + \mathbb{E}_b\bigl[X^2\bigr] + \mathbb{E}_a\bigl[Y^2\bigr] + Z^2 - 2\mathbb{E}_{a,b}\bigl[\Phi\,X\bigr] \\
  &\quad - 2\mathbb{E}_{a,b}\bigl[\Phi\,Y\bigr] + 2\mathbb{E}_{a,b}\bigl[\Phi\,Z\bigr] + 2\mathbb{E}_{a,b}\bigl[X\,Y\bigr] - 2\mathbb{E}_{a,b}\bigl[X\,Z\bigr] - 2\mathbb{E}_{a,b}\bigl[Y\,Z\bigr] .
\end{split}
\label{eq:s2-var-pair-expand}
\end{equation}
The cross terms simplify because $X$ does not depend on $a$, $Y$ does not depend on $b$, $Z$ is a constant, and $a$
and $b$ are sampled independently.  With $\mathbb{E}_b[X] = \mathbb{E}_a[Y] = \mathbb{E}_{a,b}[\Phi] = Z$,
\begin{equation}
\begin{gathered}
  \mathbb{E}_{a,b}\bigl[\Phi\,X\bigr] = \mathbb{E}_b\bigl[\mathbb{E}_a[\Phi]\,X\bigr] = \mathbb{E}_b\bigl[X^2\bigr], \qquad
  \mathbb{E}_{a,b}\bigl[\Phi\,Y\bigr] = \mathbb{E}_a\bigl[Y^2\bigr], \\
  \mathbb{E}_{a,b}\bigl[\Phi\,Z\bigr] = \mathbb{E}_{a,b}\bigl[X\,Y\bigr] = \mathbb{E}_{a,b}\bigl[X\,Z\bigr] = \mathbb{E}_{a,b}\bigl[Y\,Z\bigr] = Z^2 .
\end{gathered}
\end{equation}  Substituting these into~\eqref{eq:s2-var-pair-expand},
\begin{equation}
\begin{aligned}
  \mathbb{E}_{a,b}\Bigl[\bigl(\Phi - X - Y + Z\bigr)^2\Bigr]
  &= \mathbb{E}_{a,b}\bigl[\Phi^2\bigr] + \mathbb{E}_b\bigl[X^2\bigr] + \mathbb{E}_a\bigl[Y^2\bigr] + Z^2 \\
  &\quad - 2\,\mathbb{E}_b\bigl[X^2\bigr] - 2\,\mathbb{E}_a\bigl[Y^2\bigr] + 2Z^2 + 2Z^2 - 2Z^2 - 2Z^2 \\
  &= \mathbb{E}_{a,b}\bigl[\Phi^2\bigr] - \mathbb{E}_b\bigl[X^2\bigr] - \mathbb{E}_a\bigl[Y^2\bigr] + Z^2 .
\end{aligned}
  \label{eq:s2-var-pair-4}
\end{equation}
By the moments~\eqref{eq:s2-var-moments} and~\eqref{eq:s2-var-pair-3}, these four terms are
\begin{equation}
\begin{gathered}
  \mathbb{E}_{a,b}\bigl[\Phi^2\bigr] = \sum_{a,b}w_aw_b\,\varphi_{a,b}^2 = v_\varphi, \qquad
  \mathbb{E}_b\bigl[X^2\bigr] = \sum_b w_b\,\varphi_{H,b}^2 = v_{\varphi_{H,j}^2,2}, \\
  \mathbb{E}_a\bigl[Y^2\bigr] = \sum_a w_a\,\varphi_{a,H}^2 = v_{\varphi_{j,H}^2,2}, \qquad
  Z^2 = \varphi_{H,H}^2 .
\end{gathered}
\end{equation}
Plugging these back into~\eqref{eq:s2-var-pair-4},
\begin{equation}
  \mathbb{E}_{a,b}\Bigl[\bigl(\Phi - X - Y + Z\bigr)^2\Bigr] = v_\varphi - v_{\varphi_{H,j}^2,2} - v_{\varphi_{j,H}^2,2} + \varphi_{H,H}^2 ,
\end{equation}
which is~\eqref{eq:s2-var-pair}.
\end{proof}

Now that we have all the ingredients, we are in a position to state the general variance and prove it.

\begin{proposition}[Multi-step variance]\label{prop:multistep_var}
For $N$ independent samples from $\bq$ at step size $dt = \sfrac{T}{N}$, at fixed $N$, the variance
$V_N := \Var_{\bq}[\hat O_N]$ is
\begin{equation}
\begin{split}
  V_N &= N\,v_{\beta^2,2}\,dt^2
    + \Bigl[N\,v_{\beta\gamma,3} + N(N{-}1)\bigl(v_{\beta\cdot\varphi_{H,j},2} + v_{\beta\cdot\varphi_{j,H},2}\bigr)\Bigr]dt^3 \\
  &\quad + \Bigl[N\bigl(v_\gamma - v_{\beta^2,4}\bigr)
      + N(N{-}1)\bigl(v_{\gamma\cdot\varphi_{H,j},3} + v_{\gamma\cdot\varphi_{j,H},3} + \tfrac12 v_\varphi + v_{\beta\cdot\varphi}
        - \tfrac12\varphi_{H,H}^2 - 2b_{\gamma,2}\varphi_{H,H}\bigr) \\
  &\qquad + N(N{-}1)(N{-}2)\bigl(\tfrac13\bigl(v_{\varphi_{H,j}^2,2} + v_{\varphi_{j,H}^2,2} + v_{\varphi_{\ssub{H,j}}\varphi_{\ssub{j,H}},2}\bigr)
        + v_{\beta\cdot\psi} - \varphi_{H,H}^2\bigr)\Bigr]dt^4 \\
  &\quad + \mathcal{O}(dt^5)
\end{split}
\label{eq:s2-VN}
\end{equation}
with the densities of~\eqref{eq:s2-var-densities}.
\end{proposition}

\begin{proof}[Proof of Proposition~\ref{prop:multistep_var}]
By~\eqref{eq:s2-var-split}, \eqref{eq:s2-var-g} and Lemma~\ref{lem:s2-var-pair},
\begin{equation}
  V_N = \sum_{k=0}^{N-1}\Var_{\bq}[g]
   + \frac{N(N-1)}{2}\bigl(v_\varphi - v_{\varphi_{H,j}^2,2} - v_{\varphi_{j,H}^2,2} + \varphi_{H,H}^2\bigr)dt^4
   + \mathcal{O}(dt^5),
  \label{eq:s2-var-start}
\end{equation}
and with $g = g_1\,dt + g_2\,dt^2 + g_3\,dt^3 + \mathcal{O}(dt^4)$ of Lemma~\ref{lem:s2-var-gexp},
\begin{equation}
  \Var_{\bq}[g] = \Var_{\bq}[g_1]\,dt^2 + 2\Cov_{\bq}[g_1,g_2]\,dt^3
    + \bigl(\Var_{\bq}[g_2] + 2\Cov_{\bq}[g_1,g_3]\bigr)dt^4 + \mathcal{O}(dt^5).
  \label{eq:s2-var-orders}
\end{equation}
The position $k$ of $\mathcal{C}$ appears in~\eqref{eq:s2-var-start} via $s_L$ and $s_R$, which occur only in $g_2$ and $g_3$, so each sum over $k$ is a sum of powers of $s_L=N{-}1{-}k$ and $s_R=k$.  We evaluate each order one at a time, computing each variance and covariance with the moments~\eqref{eq:s2-var-moments} and~\eqref{eq:s2-var-moments2}.

\begin{itemize}[leftmargin=*,topsep=2pt,itemsep=2pt]
\item \emph{Order $dt^2$.}  For this order, we only have $\Var_{\bq}[g_1]$, which is independent of $k$, so by~\eqref{eq:s2-var-g1}
\begin{equation}
  \sum_{k=0}^{N-1}\Var_{\bq}[g_1]
  = N\,\Var_{\bq}\Bigl[\frac{c_j}{q_j}\,\beta_j\Bigr]
  = N\Bigl[\sum_jw_j\beta_j^2 - \beta_H^2\Bigr]
  = N\,v_{\beta^2,2} .
\end{equation}

\item \emph{Order $dt^3$.}  $2\Cov_{\bq}[g_1,g_2]$ is given by~\eqref{eq:s2-var-g1} and~\eqref{eq:s2-var-g2}:
\begin{equation}
  2\Cov_{\bq}[g_1,g_2]
  = 2\Cov_{\bq}\Bigl[\frac{c_j}{q_j}\,\beta_j,\,\frac{c_j^2}{q_j^2}\,\gamma_j\Bigr]
  + 2\Cov_{\bq}\Bigl[\frac{c_j}{q_j}\,\beta_j,\,\frac{c_j}{q_j}\,\varphi_{H,j}\Bigr]\,s_L
  + 2\Cov_{\bq}\Bigl[\frac{c_j}{q_j}\,\beta_j,\,\frac{c_j}{q_j}\,\varphi_{j,H}\Bigr]\,s_R ,
\end{equation}
so, with $\sum_k s_L = \sum_k s_R = \sfrac{N(N-1)}{2}$,
\begin{equation}
  \sum_{k=0}^{N-1}2\Cov_{\bq}[g_1,g_2]
  = 2N\Cov_{\bq}\Bigl[\frac{c_j}{q_j}\,\beta_j,\,\frac{c_j^2}{q_j^2}\,\gamma_j\Bigr]
  + N(N{-}1)\Bigl(\Cov_{\bq}\Bigl[\frac{c_j}{q_j}\,\beta_j,\,\frac{c_j}{q_j}\,\varphi_{H,j}\Bigr]
  + \Cov_{\bq}\Bigl[\frac{c_j}{q_j}\,\beta_j,\,\frac{c_j}{q_j}\,\varphi_{j,H}\Bigr]\Bigr) .
\end{equation}
Again, using $\Cov_{\bq}[x_j,y_j] := \mathbb{E}_{\bq}[x_jy_j] - \mathbb{E}_{\bq}[x_j]\,\mathbb{E}_{\bq}[y_j]$ and the moments~\eqref{eq:s2-var-moments} and~\eqref{eq:s2-var-moments2} to compute each covariance,
\begin{subequations}\label{eq:s2-var-t3}
\begin{align}
  2\Cov_{\bq}\Bigl[\frac{c_j}{q_j}\,\beta_j,\,\frac{c_j^2}{q_j^2}\,\gamma_j\Bigr] &= 2\Bigl(\sum_j\frac{c_j^3\beta_j\gamma_j}{q_j^2} - \beta_H\,b_{\gamma,2}\Bigr) = v_{\beta\gamma,3}, \\*
  \Cov_{\bq}\Bigl[\frac{c_j}{q_j}\,\beta_j,\,\frac{c_j}{q_j}\,\varphi_{H,j}\Bigr] &= \sum_jw_j\beta_j\varphi_{H,j} - \beta_H\varphi_{H,H} = v_{\beta\cdot\varphi_{H,j},2}, \\*
  \Cov_{\bq}\Bigl[\frac{c_j}{q_j}\,\beta_j,\,\frac{c_j}{q_j}\,\varphi_{j,H}\Bigr] &= \sum_jw_j\beta_j\varphi_{j,H} - \beta_H\varphi_{H,H} = v_{\beta\cdot\varphi_{j,H},2},
\end{align}
\end{subequations}
with the right-hand sides defined in~\eqref{eq:s2-var-densities}.  The $dt^3$ coefficient is therefore $N\,v_{\beta\gamma,3} + N(N{-}1)\bigl(v_{\beta\cdot\varphi_{H,j},2} + v_{\beta\cdot\varphi_{j,H},2}\bigr)$.

\item \emph{Order $dt^4$.}  Recall from~\eqref{eq:s2-var-g2} that
\begin{equation}
  g_2(j) = \frac{c_j^2}{q_j^2}\,\gamma_j + \frac{c_j}{q_j}\bigl(\varphi_{H,j}\,s_L + \varphi_{j,H}\,s_R\bigr) .
\end{equation}
With $\Var_{\bq}[x_j] := \mathbb{E}_{\bq}[x_j^2] - \mathbb{E}_{\bq}[x_j]^2$, the covariance above and the
moments~\eqref{eq:s2-var-moments} and~\eqref{eq:s2-var-moments2}, its variance is
\begin{equation}
\begin{split}
  \Var_{\bq}[g_2] &= \Var_{\bq}\Bigl[\frac{c_j^2}{q_j^2}\gamma_j\Bigr] + 2\Cov_{\bq}\Bigl[\frac{c_j^2}{q_j^2}\gamma_j,\,\frac{c_j}{q_j}\varphi_{H,j}\Bigr]\,s_L + 2\Cov_{\bq}\Bigl[\frac{c_j^2}{q_j^2}\gamma_j,\,\frac{c_j}{q_j}\varphi_{j,H}\Bigr]\,s_R \\
  &\quad + \Var_{\bq}\Bigl[\frac{c_j}{q_j}\varphi_{H,j}\Bigr]\,s_L^2 + \Var_{\bq}\Bigl[\frac{c_j}{q_j}\varphi_{j,H}\Bigr]\,s_R^2 + 2\Cov_{\bq}\Bigl[\frac{c_j}{q_j}\varphi_{H,j},\,\frac{c_j}{q_j}\varphi_{j,H}\Bigr]\,s_Ls_R \\
  &= v_\gamma
   + 2\bigl(v_{\gamma\cdot\varphi_{H,j},3} - b_{\gamma,2}\varphi_{H,H}\bigr)s_L
   + 2\bigl(v_{\gamma\cdot\varphi_{j,H},3} - b_{\gamma,2}\varphi_{H,H}\bigr)s_R \\
  &\quad + \bigl(v_{\varphi_{H,j}^2,2} - \varphi_{H,H}^2\bigr)s_L^2
   + \bigl(v_{\varphi_{j,H}^2,2} - \varphi_{H,H}^2\bigr)s_R^2
   + 2\bigl(v_{\varphi_{\ssub{H,j}}\varphi_{\ssub{j,H}},2} - \varphi_{H,H}^2\bigr)s_Ls_R .
\end{split}
\end{equation}
For $2\Cov_{\bq}[g_1,g_3]$, write $g_3 = -\tfrac23(\sfrac{c_j}{q_j})^3\beta_j + g_{3,1} + g_{3,2}$, with $g_{3,1}$ the
terms of~\eqref{eq:s2-var-g3} in $s_L$ and $s_R$, and $g_{3,2}$ those in $s_L(s_L{-}1)$, $s_R(s_R{-}1)$ and $s_Ls_R$.  The
first term gives $-\tfrac43\Cov_{\bq}\bigl[\sfrac{c_j}{q_j}\,\beta_j,\sfrac{c_j^3}{q_j^3}\,\beta_j\bigr] = -v_{\beta^2,4}$.
Again, using the moments~\eqref{eq:s2-var-moments} to compute each mean that enters the remaining terms,
\begin{subequations}\label{eq:s2-var-t4}
\begin{alignat}{2}
  \mathbb{E}_{\bq}\Bigl[\frac{c_j}{q_j}\,\gamma_{\bq}(i[P_j,O])\Bigr] &= \gamma_{\bq}(\mathcal{L}_H(O)) &\qquad& \text{at $p=1$}, \label{eq:s2-var-t4a}\\*
  \mathbb{E}_{\bq}\Bigl[\frac{c_j}{q_j}\,\beta_j(\Delta_{\bq}(O))\Bigr] &= \beta_H(\Delta_{\bq}(O)) && \text{at $p=1$}, \label{eq:s2-var-t4b}\\*
  \mathbb{E}_{\bq}\Bigl[\frac{c_j^2}{q_j^2}\,\beta_H(\Delta_j(O))\Bigr] &= \beta_H(\Delta_{\bq}(O)) && \text{at $p=2$, with~\eqref{eq:s2-var-argdefs}}, \label{eq:s2-var-t4c}\\*
  \mathbb{E}_{\bq}\Bigl[\frac{c_j^2}{q_j^2}\,\gamma_j(\mathcal{L}_H(O))\Bigr] &= \gamma_{\bq}(\mathcal{L}_H(O)) && \text{at $p=2$, with~\eqref{eq:s2-var-argdefs}}, \label{eq:s2-var-t4d}\\*
  \mathbb{E}_{\bq}\Bigl[\frac{c_j}{q_j}\,\varphi_{H,H}(i[P_j,O])\Bigr] &= \varphi_{H,H}(\mathcal{L}_H(O)) = \beta_H(\mathcal{L}_H^2(O)) && \text{at $p=1$, with~\eqref{eq:s2-btot-kc}}, \label{eq:s2-var-t4e}\\*
  \mathbb{E}_{\bq}\Bigl[\frac{c_j}{q_j}\,\beta_j(\mathcal{L}_H^2(O))\Bigr] &= \beta_H(\mathcal{L}_H^2(O)) && \text{at $p=1$}, \label{eq:s2-var-t4f}\\*
  \mathbb{E}_{\bq}\Bigl[\frac{c_j}{q_j}\,\beta_H(i[P_j,\mathcal{L}_H(O)])\Bigr] &= \beta_H(\mathcal{L}_H^2(O)) && \text{at $p=1$}, \label{eq:s2-var-t4g}
\end{alignat}
\end{subequations}
Each term of $g_{3,1}$ is $\sfrac{c_j}{q_j}\,x_j$ or $\sfrac{c_j^2}{q_j^2}\,x_j$, whose covariance with $g_1$ is, by the
moments, $\sum_jw_j\beta_jx_j - \beta_H\,\mathbb{E}_{\bq}[\sfrac{c_j}{q_j}\,x_j]$ or
$\sum_j\sfrac{c_j^3\beta_j}{q_j^2}\,x_j - \beta_H\,\mathbb{E}_{\bq}[\sfrac{c_j^2}{q_j^2}\,x_j]$.  With
$\sum_ks_L = \sum_ks_R = \tfrac{N(N-1)}{2}$,
\begin{align}
  \sum_k 2\Cov_{\bq}[g_1,g_{3,1}]
  &= N(N{-}1)\Bigl[\sum_jw_j\beta_j\,\bigl[\gamma_{\bq}(i[P_j,O])\bigr] - \beta_H\bigl[\gamma_{\bq}(\mathcal{L}_H(O))\bigr] \nonumber\\
  &\qquad + \sum_jw_j\beta_j\,\bigl[\beta_j(\Delta_{\bq}(O))\bigr] - \beta_H\bigl[\beta_H(\Delta_{\bq}(O))\bigr] \nonumber\\
  &\qquad + \sum_j\frac{c_j^3\beta_j}{q_j^2}\,\bigl[\beta_H(\Delta_j(O))\bigr] - \beta_H\bigl[\beta_H(\Delta_{\bq}(O))\bigr] \nonumber\\
  &\qquad + \sum_j\frac{c_j^3\beta_j}{q_j^2}\,\bigl[\gamma_j(\mathcal{L}_H(O))\bigr] - \beta_H\bigl[\gamma_{\bq}(\mathcal{L}_H(O))\bigr]\Bigr] \nonumber\\
  &= N(N{-}1)\,v_{\beta\cdot\varphi} ,
  \label{eq:s2-var-g31}
\end{align}
the four terms being those of $g_{3,1}$ in the order of~\eqref{eq:s2-var-g3}, with the means~\eqref{eq:s2-var-t4a}--\eqref{eq:s2-var-t4d}, and the last line by~\eqref{eq:s2-var-densities}.  In $g_{3,2}$ every term is $\sfrac{c_j}{q_j}\,x_j$.  With
$\sum_ks(s{-}1) = \tfrac{N(N-1)(N-2)}{3}$ for $s=s_L$ or $s_R$ and $\sum_ks_Ls_R = \tfrac{N(N-1)(N-2)}{6}$, the $\tfrac12$ of the
$s(s{-}1)$ terms and the $1$ of the $s_Ls_R$ term each combine with the factor $2$ of $2\Cov_{\bq}$ to a prefactor
$\tfrac{N(N-1)(N-2)}{3}$,
\begin{align}
  \sum_k 2\Cov_{\bq}[g_1,g_{3,2}]
  &= \tfrac{N(N-1)(N-2)}{3}\Bigl[\sum_jw_j\beta_j\,\bigl[\varphi_{H,H}(i[P_j,O])\bigr] + \sum_jw_j\beta_j\,\bigl[\beta_j(\mathcal{L}_H^2(O))\bigr] \nonumber\\
  &\qquad + \sum_jw_j\beta_j\,\bigl[\beta_H(i[P_j,\mathcal{L}_H(O)])\bigr] - 3\beta_H\bigl[\beta_H(\mathcal{L}_H^2(O))\bigr]\Bigr] \nonumber\\
  &= N(N{-}1)(N{-}2)\,v_{\beta\cdot\psi} ,
  \label{eq:s2-var-g32}
\end{align}
where each term of $g_{3,2}$ has the form $\frac{c_j}{q_j}\,x_j$, with $x_j = \varphi_{H,H}(i[P_j,O])$,
$\beta_j(\mathcal{L}_H^2(O))$ or $\beta_H(i[P_j,\mathcal{L}_H(O)])$.  By~\eqref{eq:s2-var-t4e}--\eqref{eq:s2-var-t4g} all three
have the same mean,
\begin{equation}
  \mathbb{E}_{\bq}\Bigl[\frac{c_j}{q_j}\,\varphi_{H,H}(i[P_j,O])\Bigr]
  = \mathbb{E}_{\bq}\Bigl[\frac{c_j}{q_j}\,\beta_j(\mathcal{L}_H^2(O))\Bigr]
  = \mathbb{E}_{\bq}\Bigl[\frac{c_j}{q_j}\,\beta_H(i[P_j,\mathcal{L}_H(O)])\Bigr]
  = \beta_H(\mathcal{L}_H^2(O)) ,
\end{equation}
so each covariance with $g_1$ is
$\Cov_{\bq}\bigl[\sfrac{c_j}{q_j}\,\beta_j,\,\sfrac{c_j}{q_j}\,x_j\bigr] = \sum_jw_j\beta_j\,x_j - \beta_H\,\beta_H(\mathcal{L}_H^2(O))$,
and the three together give the $-3\beta_H\,\beta_H(\mathcal{L}_H^2(O))$ in~\eqref{eq:s2-var-g32}.  The last line
of~\eqref{eq:s2-var-g32} is the definition of $v_{\beta\cdot\psi}$ in~\eqref{eq:s2-var-densities},
\begin{equation}
  v_{\beta\cdot\psi} := \tfrac13\Bigl[\sum_j w_j\beta_j\Bigl(\beta_j(\mathcal{L}_H^2(O))
    + \beta_H\bigl(i[P_j,\mathcal{L}_H(O)]\bigr) + \varphi_{H,H}\bigl(i[P_j,O]\bigr)\Bigr) - 3\beta_H\,\beta_H(\mathcal{L}_H^2(O))\Bigr] .
\end{equation}
Summing $\Var_{\bq}[g_2]$ over $k$ with $\sum_ks = \tfrac{N(N-1)}{2}$ and $\sum_ks^2 = \tfrac{N(N-1)}{2} + \tfrac{N(N-1)(N-2)}{3}$ for $s=s_L$ or $s_R$, and $\sum_ks_Ls_R = \tfrac{N(N-1)(N-2)}{6}$, and adding the pair term of~\eqref{eq:s2-var-start}, the $dt^4$ coefficient is therefore
\begin{equation}
\begin{aligned}
  &N\bigl(v_\gamma - v_{\beta^2,4}\bigr)
   + N(N{-}1)\Bigl[v_{\gamma\cdot\varphi_{H,j},3} + v_{\gamma\cdot\varphi_{j,H},3} - 2b_{\gamma,2}\varphi_{H,H} + v_{\beta\cdot\varphi}
   + \tfrac12\bigl(v_{\varphi_{H,j}^2,2} + v_{\varphi_{j,H}^2,2} - 2\varphi_{H,H}^2\bigr) \\
  &\qquad + \tfrac12\bigl(v_\varphi - v_{\varphi_{H,j}^2,2} - v_{\varphi_{j,H}^2,2} + \varphi_{H,H}^2\bigr)\Bigr]
   + N(N{-}1)(N{-}2)\Bigl[\tfrac13\bigl(v_{\varphi_{H,j}^2,2} + v_{\varphi_{j,H}^2,2} + v_{\varphi_{\ssub{H,j}}\varphi_{\ssub{j,H}},2}\bigr) + v_{\beta\cdot\psi} - \varphi_{H,H}^2\Bigr] \\
  &= N\bigl(v_\gamma - v_{\beta^2,4}\bigr)
   + N(N{-}1)\bigl(v_{\gamma\cdot\varphi_{H,j},3} + v_{\gamma\cdot\varphi_{j,H},3} + \tfrac12 v_\varphi + v_{\beta\cdot\varphi} - \tfrac12\varphi_{H,H}^2 - 2b_{\gamma,2}\varphi_{H,H}\bigr) \\
  &\quad + N(N{-}1)(N{-}2)\bigl(\tfrac13\bigl(v_{\varphi_{H,j}^2,2} + v_{\varphi_{j,H}^2,2} + v_{\varphi_{\ssub{H,j}}\varphi_{\ssub{j,H}},2}\bigr) + v_{\beta\cdot\psi} - \varphi_{H,H}^2\bigr) .
\end{aligned}
\end{equation}
In the $N(N{-}1)$ term the second $\tfrac12(\cdots)$ is the pair term of~\eqref{eq:s2-var-start}, and the rest comes from the
terms with one factor of $s$ and the $\tfrac{N(N-1)}{2}$ part of $\sum_ks^2$.  The $N(N{-}1)(N{-}2)$ term comes from the terms
in $s(s{-}1)$ and $s_Ls_R$ and the $\tfrac{N(N-1)(N-2)}{3}$ part of $\sum_ks^2$.  This with the $dt^2$ and $dt^3$ coefficients gives~\eqref{eq:s2-VN}.

\end{itemize}
\end{proof}

Proposition~\ref{prop:multistep_var} gives the variance in terms of $v_x$ factors.  To get from there to the stated form in the main text:
\begin{equation}
\begin{split}
  V_N &=N\,v_{\beta^2,2}\,dt^2+N\bigl[v_{\beta\gamma,3}+(N{-}1)V_{3,2}\bigr]dt^3 \\
  &\quad +N\bigl[v_\gamma-v_{\beta^2,4}+(N{-}1)V_{4,2}+(N{-}1)(N{-}2)V_{4,3}\bigr]dt^4+\mathcal O(dt^5),
\end{split}
\tag{\ref{eq:VN_full}}
\end{equation}
we group terms at order $dt^p$ with $k$ factors of $N$ into each $V_{p,k}$~\eqref{eq:methods-variance-groups}.  As for the bias, at fixed $T$ the $\mathcal{O}(dt^5)$ remainder decays as $\sfrac{1}{N}$.

\section{Special cases of the general formulas}\label{app:S3}

Propositions~\ref{prop:multistep_bias} and~\ref{prop:multistep_var} hold for any fixed depth, any reference state and any
observable.  We can make a number of simplifications to remove terms in each.  Specifically, using $N=1$, a
real-amplitude reference, taking the observable to be the Hamiltonian ($O=H$), and $\ket{\psi}$ being an energy
eigenstate each zero some terms, and this section describes the simplified formulae in each case.

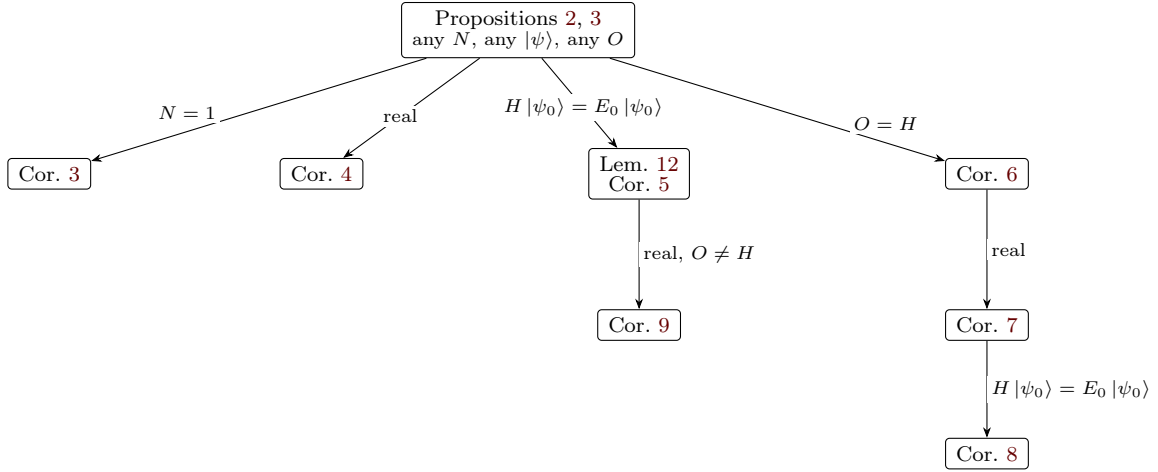
\begin{figure}[H]
\centering
\footnotesize
\begin{tikzpicture}[
  >={Stealth[length=4pt]},
  box/.style={draw,rounded corners=1.5pt,inner xsep=4pt,inner ysep=3pt,align=center},
  lab/.style={font=\scriptsize,inner sep=1.5pt,fill=white},
]
  \node[box] (gen) at (0,0) {Propositions~\ref{prop:multistep_bias},~\ref{prop:multistep_var}\\[-2pt]
     {\scriptsize any $N$, any $\ket{\psi}$, any $O$}};
  \node[box] (n1)  at (-6.2,-1.9) {Cor.~\ref{prop:single-step-var}};
  \node[box] (re)  at (-2.6,-1.9) {Cor.~\ref{cor:beta0}};
  \node[box] (eig) at ( 1.6,-1.9) {Lem.~\ref{lem:collapse}\\[-2pt]Cor.~\ref{cor:eig-general}};
  \node[box] (oh)  at ( 6.2,-1.9) {Cor.~\ref{cor:energy}};
  \node[box] (obs) at ( 1.6,-3.9) {Cor.~\ref{cor:obs-eig}};
  \node[box] (ohr) at ( 6.2,-3.9) {Cor.~\ref{cor:energy-real}};
  \node[box] (ohe) at ( 6.2,-5.6) {Cor.~\ref{cor:energy-eig}};
  \draw[->] (gen) -- (n1)  node[lab,pos=0.62,above left=-1pt]{$N=1$};
  \draw[->] (gen) -- (re)  node[lab,pos=0.58]{real};
  \draw[->] (gen) -- (eig) node[lab,pos=0.55]{$H\ket{\psi_0}=E_0\ket{\psi_0}$};
  \draw[->] (gen) -- (oh)  node[lab,pos=0.72,above right=-1pt]{$O=H$};
  \draw[->] (eig) -- (obs) node[lab,pos=0.5,right]{real, $O\neq H$};
  \draw[->] (oh)  -- (ohr) node[lab,pos=0.5,right]{real};
  \draw[->] (ohr) -- (ohe) node[lab,pos=0.5,right]{$H\ket{\psi_0}=E_0\ket{\psi_0}$};
\end{tikzpicture}
\caption{\textbf{Special cases of the general bias and variance.}  Each arrow imposes one further condition on
Propositions~\ref{prop:multistep_bias} and~\ref{prop:multistep_var}: a single step, a real-amplitude reference
(Lemma~\ref{lem:beta}, which gives $\beta_j=0$), an energy eigenstate, or the observable equal to the Hamiltonian.  The two
lower chains compose conditions, and each box names the corollary that records what survives.}
\label{fig:s2-special-cases}
\end{figure}

\newpage
\subsection*{A single step ($N=1$)}

\begin{corollary}[$N=1$ bias and variance]\label{prop:single-step-var}
For a single \qdrift{} step,
\begin{align}
  B_1 &= \bigl(b_{\gamma,2} - \tfrac12\varphi_{H,H}\bigr)dt^2 + b_{\beta,3}\,dt^3
    + b_{\gamma,4}\,dt^4 + \mathcal{O}(dt^5),
  \label{eq:D12b}\\
  V_1 &= v_{\beta^2,2}\,dt^2 + v_{\beta\gamma,3}\,dt^3
    + \bigl(v_\gamma - v_{\beta^2,4}\bigr)dt^4 + \mathcal{O}(dt^5) .
  \label{eq:D12}
\end{align}
\end{corollary}

\begin{proof}
At $N=1$ every term of~\eqref{eq:BN_full} and~\eqref{eq:s2-VN} carrying a factor $N-1$ drops out.
\end{proof}

\subsection*{Real amplitudes ($\beta_j = 0$)}

\begin{corollary}[Real-amplitude regime]\label{cor:beta0}
If $\ket{\psi}$ has real amplitudes in the computational basis and every $P_j$ and every Pauli string of $O$ contain an
even number of $Y$ operators, then by Lemma~\ref{lem:beta} $\beta_j = 0$ for all $j$ (hence
$\beta_H = 0$), and
\begin{align}
  B_1 &= \bigl(b_{\gamma,2} - \tfrac12\varphi_{H,H}\bigr)dt^2 + b_{\gamma,4}\,dt^4
    + \mathcal{O}(dt^6),
  \label{eq:D13b}\\
  V_1 &= v_\gamma\,dt^4 + \mathcal{O}(dt^6) .
  \label{eq:D13}
\end{align}
\end{corollary}

\begin{proof}
Each $Y$ contributes a factor of $i$, so an even count leaves the Pauli string real and Lemma~\ref{lem:beta}
applies, giving $\beta_j = 0$ for every $j$ and hence $\beta_H = \sum_j c_j\beta_j = 0$.
Corollary~\ref{prop:single-step-var} gives
\begin{align}
  B_1 &= \bigl(b_{\gamma,2} - \tfrac12\varphi_{H,H}\bigr)dt^2 + b_{\beta,3}\,dt^3
    + b_{\gamma,4}\,dt^4 + \mathcal{O}(dt^5), \nonumber\\
  V_1 &= v_{\beta^2,2}\,dt^2 + v_{\beta\gamma,3}\,dt^3
    + \bigl(v_\gamma - v_{\beta^2,4}\bigr)dt^4 + \mathcal{O}(dt^5) ,
\end{align}
in which $v_{\beta^2,2}$, $v_{\beta\gamma,3}$ and $v_{\beta^2,4}$ each carry a factor $\beta_j$ or
$\beta_H$ in every term and so vanish.  The density $b_{\beta,3}$ of~\eqref{eq:s2-bias-densities} also vanishes,
\begin{equation}
  b_{\beta,3} = -\frac23\sum_j\frac{c_j^3}{q_j^2}\,\underbrace{\beta_j}_{0}
    - \frac16\,\underbrace{\beta_H\bigl(\mathcal{L}_H^2(O)\bigr)}_{-i\braket{[H,[H,[H,O]]]}\,=\,0} = 0 ,
\end{equation}
the second part because $[H,[H,[H,O]]]$ is real and antisymmetric for real symmetric $H$ and $O$, and so has zero
expectation in a real-amplitude state, as in the proof of Corollary~\ref{cor:energy-real}.  This leaves~\eqref{eq:D13b}--\eqref{eq:D13}.
\end{proof}

The condition on $\ket{\psi}$ holds for a UCCSD ansatz with real amplitudes on a real Hartree--Fock reference, and the condition on $P_j$ and $O$ holds whenever $H$ and $O$ are real matrices.

\paragraph{Evenness in $dt$.}  Corollary~\ref{cor:beta0} is stated at $N=1$.  However, this removal of the odd powers of $dt$ also applies at every $N$, given the same assumptions.  This is because, when $H$, $O$ and $\ket{\psi}$ are real, running a trajectory backwards in time is the same as taking its complex conjugate, and a real expectation value is unchanged by complex conjugation.  Specifically, we have that $e^{-i\theta P_j}=\bigl(e^{i\theta P_j}\bigr)^{*}$ in the real basis.  As such, the product of all $N$ rotations along a given trajectory, $U(dt)=e^{-i\theta_{j_N}P_{j_N}}\cdots e^{-i\theta_{j_1}P_{j_1}}$ with $\theta_j=\sfrac{c_j\,dt}{q_j}$, obeys $U(-dt)=U(dt)^{*}$, since every angle changes sign with $dt$.  Its expectation $f(dt)=\bra{\psi}U(dt)^\dagger O\,U(dt)\ket{\psi}$ satisfies
\begin{equation}
  f(-dt) \;=\; \bra{\psi}U(dt)^{\mathsf T}\,O\,U(dt)^{*}\ket{\psi} \;=\; f(dt)^{*} \;=\; f(dt) .
  \label{eq:s2-even-dt}
\end{equation}
The exact expectation is even in $dt$ in the same way, with $e^{-iH\,dt}$ in place of $U(dt)$.  The bias and the variance are built from the trajectory averages of $f$ and $f^2$ and from the exact expectation, so both are even in $dt$.

\subsection*{Energy eigenstates}

At an energy eigenstate several quantities in Propositions~\ref{prop:multistep_bias} and~\ref{prop:multistep_var}
vanish, which simplifies the allocations of the main text.

\begin{lemma}[Energy eigenstate]\label{lem:collapse}
If the reference $\ket{\psi_0}$ is an energy eigenstate, $H\ket{\psi_0} = E_0\ket{\psi_0}$, then $\beta_H$ and
$\varphi_{H,H}$ vanish on every argument,
\begin{equation}
  \beta_H(X) = i\bra{\psi_0}[H,X]\ket{\psi_0} = 0,
  \qquad
  \varphi_{H,H}(X) = -\bra{\psi_0}[H,[H,X]]\ket{\psi_0} = 0,
  \qquad\text{for every operator } X ,
  \label{eq:s2-eig-all}
\end{equation}
and in particular $\beta_H = 0$,
$\varphi_{H,a} = 0$ for every $a$ and $\varphi_{H,H} = 0$, while
$\varphi_{a,H} = \bra{\psi_0}[P_a,[O,H]]\ket{\psi_0}$ need not vanish.
\end{lemma}

\begin{proof}
For any operator $X$, using $H\ket{\psi_0} = E_0\ket{\psi_0}$ and $\bra{\psi_0}H = E_0\bra{\psi_0}$,
\begin{equation}
  \bra{\psi_0}[H,X]\ket{\psi_0} = \bra{\psi_0}HX\ket{\psi_0} - \bra{\psi_0}XH\ket{\psi_0}
  = E_0\bra{\psi_0}X\ket{\psi_0} - E_0\bra{\psi_0}X\ket{\psi_0} = 0 .
  \label{eq:s2-eig-comm}
\end{equation}
Both identities of~\eqref{eq:s2-eig-all} follow: the first is~\eqref{eq:s2-eig-comm} itself, and the second
is~\eqref{eq:s2-eig-comm} with $[H,X]$ in place of $X$.
Every quantity whose outer commutator is with $H$ therefore vanishes.  From~\eqref{eq:s2-chi}
and~\eqref{eq:s2-btot-kc},
\begin{alignat}{2}
  \beta_H &= i\bra{\psi_0}[H,O]\ket{\psi_0} = 0 &\qquad& (X = O), \\
  \varphi_{H,a} &= \bra{\psi_0}[H,[O,P_a]]\ket{\psi_0} = 0 && (X = [O,P_a]), \\
  \varphi_{H,H} &= \bra{\psi_0}[H,[O,H]]\ket{\psi_0} = 0 && (X = [O,H]).
\end{alignat}
The quantity $\varphi_{a,H} = \bra{\psi_0}[P_a,[O,H]]\ket{\psi_0}$ has its outer commutator with $P_a$, and
$\ket{\psi_0}$ is in general not an eigenstate of $P_a$, so~\eqref{eq:s2-eig-comm} does not apply.
Computing $\varphi_{a,H}$ by direct expansion of the commutator gives
\begin{align}
  \varphi_{a,H}
  &= \bra{\psi_0}P_aOH\ket{\psi_0} - \bra{\psi_0}P_aHO\ket{\psi_0}
    - \bra{\psi_0}OHP_a\ket{\psi_0} + \bra{\psi_0}HOP_a\ket{\psi_0} \notag\\
  &= E_0\bra{\psi_0}P_aO\ket{\psi_0} - \bra{\psi_0}P_a\,H\,O\ket{\psi_0}
    - \bra{\psi_0}O\,H\,P_a\ket{\psi_0} + E_0\bra{\psi_0}OP_a\ket{\psi_0} \notag\\
  &= -\bra{\psi_0}\bigl[P_a(H - E_0)O + O(H - E_0)P_a\bigr]\ket{\psi_0}
   = -2\,\mathrm{Re}\,\bra{\psi_0}P_a(H - E_0)O\ket{\psi_0} ,
  \label{eq:s2-eig-phiaH}
\end{align}
Therefore, $\varphi_{a,H}$ vanishes exactly when $\mathrm{Re}\,\bra{\psi_0}P_a(H-E_0)O\ket{\psi_0}=0$, for instance if
$P_a\ket{\psi_0}$ is orthogonal to $(H-E_0)O\ket{\psi_0}$.  Nothing forces that for a given $P_a$, so
$\varphi_{a,H}\neq0$ in general.  However, its weighted sum does vanish:
\begin{equation}
  \sum_a c_a\,\varphi_{a,H}
   = \sum_a c_a\bra{\psi_0}[P_a,[O,H]]\ket{\psi_0}
   = \bra{\psi_0}\Bigl[\sum_a c_aP_a,\,[O,H]\Bigr]\ket{\psi_0}
   = \bra{\psi_0}[H,[O,H]]\ket{\psi_0}
   = \varphi_{H,H} = 0,
\end{equation}
using $H = \sum_a c_aP_a$ and then~\eqref{eq:s2-eig-comm} with $X = [O,H]$.
So the $\varphi_{a,H}$ cancel in this sum without each being zero.
\end{proof}

\begin{corollary}[Energy eigenstate, general observable]\label{cor:eig-general}
Let the reference be an energy eigenstate, $H\ket{\psi_0} = E_0\ket{\psi_0}$.  Then
\begin{equation}
  b_{\beta\Delta,2} = b_{\varphi_{H,j},3} = b_{\varphi\Delta,2} = b_{\beta\Delta H,2} = 0,
  \qquad
  v_{\beta\cdot\varphi_{H,j},2} = v_{\gamma\cdot\varphi_{H,j},3} = v_{\varphi_{H,j}^2,2}
    = v_{\varphi_{\ssub{H,j}}\varphi_{\ssub{j,H}},2} = 0 ,
  \label{eq:s2-eig-vanish}
\end{equation}
so that Propositions~\ref{prop:multistep_bias} and~\ref{prop:multistep_var} reduce to
\begin{align}
  B_N &= N\,b_{\gamma,2}\,dt^2
    + \Bigl[N\,b_{\beta,3} + \tfrac{N(N-1)}{2}\,b_{\gamma H,2}\Bigr]dt^3 \notag\\
  &\quad + \Bigl[N\,b_{\gamma,4} + \tfrac{N(N-1)}{2}\,b_{\varphi_{j,H},3}
    + \tfrac{N(N-1)(2N-1)}{12}\,b_{\gamma HH,2}
    + \tfrac{N(N-1)}{2}\,b_{\gamma\gamma,2}\Bigr]dt^4 + \mathcal{O}(dt^5),
  \label{eq:s2-eig-BN}\\
  V_N &= N\,v_{\beta^2,2}\,dt^2
    + \Bigl[N\,v_{\beta\gamma,3} + N(N{-}1)\,v_{\beta\cdot\varphi_{j,H},2}\Bigr]dt^3 \notag\\
  &\quad + \Bigl[N\bigl(v_\gamma - v_{\beta^2,4}\bigr)
    + N(N{-}1)\bigl(v_{\gamma\cdot\varphi_{j,H},3} + \tfrac12 v_\varphi + v_{\beta\cdot\varphi}\bigr)
    + N(N{-}1)(N{-}2)\bigl(\tfrac13 v_{\varphi_{j,H}^2,2} + v_{\beta\cdot\psi}\bigr)\Bigr]dt^4 \notag\\
  &\quad + \mathcal{O}(dt^5),
  \label{eq:s2-eig-VN}
\end{align}
\end{corollary}

\begin{proof}
By Lemma~\ref{lem:collapse}, $\beta_H(X) = \varphi_{H,H}(X) = 0$ for every operator $X$, and
$\varphi_{H,j} = 0$ for every $j$, on an energy eigenstate.  For example, with $b_{\beta\Delta,2}$, we have:
\begin{equation}
  b_{\beta\Delta,2}
   = \sum_j w_j\,\beta_H\bigl(\Delta_j(O)\bigr) - \tfrac12\,\beta_H\bigl(-[H,[H,O]]\bigr)
   = \sum_j w_j\underbrace{\beta_H\bigl(\Delta_j(O)\bigr)}_{0\ \text{by}~\eqref{eq:s2-eig-all}}
     - \tfrac12\underbrace{\beta_H\bigl(-[H,[H,O]]\bigr)}_{0\ \text{by}~\eqref{eq:s2-eig-all}}
   = 0 .
  \label{eq:s2-eig-example}
\end{equation}
The densities $b_{\varphi_{H,j},3}$, $b_{\varphi\Delta,2}$ and $b_{\beta\Delta H,2}$ in the bias, and
$v_{\beta\cdot\varphi_{H,j},2}$, $v_{\gamma\cdot\varphi_{H,j},3}$, $v_{\varphi_{H,j}^2,2}$ and
$v_{\varphi_{\ssub{H,j}}\varphi_{\ssub{j,H}},2}$ in the variance, vanish in the same way, every term carrying
$\beta_H(\cdot)$, $\varphi_{H,H}(\cdot)$ or $\varphi_{H,j}$, which proves~\eqref{eq:s2-eig-vanish}.

For the rest of the terms, $\beta_H(\cdot)$ and $\varphi_{H,H}(\cdot)$ appear only as correction terms.
For example:
\begin{equation}
  b_{\beta,3}
   = -\frac23\sum_j \frac{c_j^3\beta_j}{q_j^2}
     - \frac16\underbrace{\beta_H(\mathcal{L}_H^2(O))}_{0\ \text{by}~\eqref{eq:s2-eig-all}}
   = -\frac23\sum_j \frac{c_j^3\beta_j}{q_j^2} .
  \label{eq:s2-eig-example2}
\end{equation}
The densities $b_{\gamma H,2}$, $b_{\gamma,4}$, $b_{\varphi_{j,H},3}$, $b_{\gamma HH,2}$, $b_{\gamma\gamma,2}$,
$v_{\beta^2,2}$, $v_{\beta\gamma,3}$, $v_{\beta^2,4}$, $v_{\beta\cdot\varphi}$ and $v_{\beta\cdot\psi}$
simplify in the same way.  In the variance, $\varphi_{H,H} = 0$ also removes the corrections $-\tfrac12\varphi_{H,H}^2$, $-2b_{\gamma,2}\varphi_{H,H}$ and
$-\varphi_{H,H}^2$ of~\eqref{eq:s2-VN}.

The terms $\beta_j(\Delta_{\bq}(O))$, $\gamma_{\bq}(i[P_j,O])$ and $\gamma_j(\mathcal{L}_H(O))$ of $v_{\beta\cdot\varphi}$
in~\eqref{eq:s2-var-densities} survive.  Writing out $\Delta_k(X) = P_kXP_k - X$ and $\Delta_{\bq} = \sum_kw_k\Delta_k$,
\begin{equation}
\begin{aligned}
  \beta_j(\Delta_{\bq}(O)) &= \sum_k w_k\Bigl(\bra{\psi_0}i\bigl[P_j,P_kOP_k\bigr]\ket{\psi_0} - \bra{\psi_0}i[P_j,O]\ket{\psi_0}\Bigr), \\
  \gamma_{\bq}(i[P_j,O]) &= \sum_k w_k\Bigl(\bra{\psi_0}P_k\,i[P_j,O]\,P_k\ket{\psi_0} - \bra{\psi_0}i[P_j,O]\ket{\psi_0}\Bigr), \\
  \gamma_j(\mathcal{L}_H(O)) &= \bra{\psi_0}P_j\,i[H,O]\,P_j\ket{\psi_0} - \underbrace{\bra{\psi_0}i[H,O]\ket{\psi_0}}_{\beta_H\,=\,0} .
\end{aligned}
  \label{eq:s2-eig-vbphi}
\end{equation}
Lemma~\ref{lem:collapse} removes only expectations of the form $\bra{\psi_0}[H,X]\ket{\psi_0}$, and none of the remaining terms in~\eqref{eq:s2-eig-vbphi} are of this form.  Two of the surviving densities of~\eqref{eq:s2-eig-BN}, namely
$b_{\gamma H,2} = \sum_j w_j\gamma_j(\mathcal{L}_H(O))$ at order $dt^3$ and $b_{\gamma HH,2} = \sum_j w_j\gamma_j(\mathcal{L}_H^2(O))$
at order $dt^4$, are sums of $\gamma_j(\mathcal{L}_H(O))$ and $\gamma_j(\mathcal{L}_H^2(O))$, which simplify in turn.
Applying~\eqref{eq:s2-eig-comm} with $X = O$ and with $X = [H,O]$ gives
$\bra{\psi_0}\mathcal{L}_H(O)\ket{\psi_0} = \bra{\psi_0}\mathcal{L}_H^2(O)\ket{\psi_0} = 0$, so the subtracted term of
$\gamma_j(X) = \bra{\psi_0}P_jXP_j\ket{\psi_0} - \bra{\psi_0}X\ket{\psi_0}$ drops for both,
\begin{equation}
  \gamma_j(\mathcal{L}_H(O)) = \bra{\psi_0}P_j\,\mathcal{L}_H(O)\,P_j\ket{\psi_0} ,
  \qquad
  \gamma_j(\mathcal{L}_H^2(O)) = \bra{\psi_0}P_j\,\mathcal{L}_H^2(O)\,P_j\ket{\psi_0} .
  \label{eq:s2-eig-gammadot}
\end{equation}
Substituting into Propositions~\ref{prop:multistep_bias} and~\ref{prop:multistep_var}
gives~\eqref{eq:s2-eig-BN}--\eqref{eq:s2-eig-VN}.
\end{proof}

\subsection*{Energy estimation ($O = H$)}

\begin{corollary}[$O = H$, general reference: energy estimation]\label{cor:energy}
Let $O = H$ and let $\ket{\psi}$ be any reference state.  Then
\begin{equation}
\begin{gathered}
  \mathcal{L}_H(O) = i[H,H] = 0, \quad \mathcal{L}_H^2(O) = -[H,[H,H]] = 0, \\
  \beta_H = i\braket{[H,H]} = 0, \quad \varphi_{H,H} = 0, \quad \varphi_{j,H} = \braket{[P_j,[H,H]]} = 0 ,
\end{gathered}
  \label{eq:s2-energy-zero}
\end{equation}
so
\begin{equation}
  b_{\gamma H,2} = b_{\varphi_{j,H},3} = b_{\gamma HH,2} = b_{\beta\Delta H,2} = 0,
  \qquad
  v_{\beta\cdot\varphi_{j,H},2} = v_{\gamma\cdot\varphi_{j,H},3} = v_{\varphi_{j,H}^2,2}
    = v_{\varphi_{\ssub{H,j}}\varphi_{\ssub{j,H}},2} = 0 ,
  \label{eq:s2-energy-vanish}
\end{equation}
and Propositions~\ref{prop:multistep_bias} and~\ref{prop:multistep_var} reduce to
\begin{align}
  B_N &= N\,b_{\gamma,2}\,dt^2
    + \Bigl[N\,b_{\beta,3} + \tfrac{N(N-1)}{2}\,b_{\beta\Delta,2}\Bigr]dt^3 \notag\\
  &\quad + \Bigl[N\,b_{\gamma,4} + \tfrac{N(N-1)}{2}\,b_{\varphi_{H,j},3}
    + \tfrac{N(N-1)(2N-1)}{12}\,b_{\varphi\Delta,2}
    + \tfrac{N(N-1)}{2}\,b_{\gamma\gamma,2}\Bigr]dt^4 + \mathcal{O}(dt^5),
  \label{eq:s2-energy-BN}\\
  V_N &= N\,v_{\beta^2,2}\,dt^2
    + \Bigl[N\,v_{\beta\gamma,3} + N(N{-}1)\,v_{\beta\cdot\varphi_{H,j},2}\Bigr]dt^3 \notag\\
  &\quad + \Bigl[N\bigl(v_\gamma - v_{\beta^2,4}\bigr)
    + N(N{-}1)\bigl(v_{\gamma\cdot\varphi_{H,j},3} + \tfrac12 v_\varphi + v_{\beta\cdot\varphi}\bigr)
    + N(N{-}1)(N{-}2)\bigl(\tfrac13 v_{\varphi_{H,j}^2,2} + v_{\beta\cdot\psi}\bigr)\Bigr]dt^4 \notag\\
  &\quad + \mathcal{O}(dt^5),
  \label{eq:s2-energy-VN}
\end{align}
with $\gamma_j = \braket{P_jHP_j} - \braket{H}$ and $\varphi_{H,j} = \braket{[H,[H,P_j]]}$.
\end{corollary}

\begin{proof}
The three scalars in~\eqref{eq:s2-energy-zero} are the definitions~\eqref{eq:s2-chi}--\eqref{eq:s2-btot-kc} at $O=H$,
and $\mathcal{L}_H(O)$, $\mathcal{L}_H^2(O)$ are the zero operator because $H$ commutes with itself.  Every density evaluated at the zero
operator vanishes, so
\begin{equation}
  \gamma_j(\mathcal{L}_H(O)) = \gamma_j(\mathcal{L}_H^2(O)) = \beta_j(\mathcal{L}_H^2(O)) = \beta_H(\mathcal{L}_H^2(O)) = \varphi_{H,H}(\mathcal{L}_H^2(O))
    = \beta_H(\Delta_j\mathcal{L}_H(O)) = \gamma_{\bq}(\mathcal{L}_H(O)) = 0 ,
  \label{eq:s2-energy-p1}
\end{equation}
which removes $b_{\gamma H,2}$, $b_{\gamma HH,2}$ and $b_{\beta\Delta H,2}$.  With $\varphi_{j,H} = 0$,
\begin{equation}
  b_{\varphi_{j,H},3} = -\tfrac23\sum_j \frac{c_j^3}{q_j^2}\underbrace{\varphi_{j,H}}_{0}
    - \tfrac16\underbrace{\varphi_{H,H}(\mathcal{L}_H^2(O))}_{0} = 0 ,
  \label{eq:s2-energy-p2}
\end{equation}
and $\varphi_{j,H} = 0$ removes $v_{\beta\cdot\varphi_{j,H},2}$, $v_{\gamma\cdot\varphi_{j,H},3}$,
$v_{\varphi_{j,H}^2,2}$ and $v_{\varphi_{\ssub{H,j}}\varphi_{\ssub{j,H}},2}$, each of which carries it as a factor, which
proves~\eqref{eq:s2-energy-vanish}.

Simplifying the remaining densities of Propositions~\ref{prop:multistep_bias} and~\ref{prop:multistep_var} as in the
proof of Corollary~\ref{cor:eig-general} gives~\eqref{eq:s2-energy-BN}--\eqref{eq:s2-energy-VN}.
\end{proof}

\begin{corollary}[Energy at a real-amplitude reference]\label{cor:energy-real}
Let $O = H$ and let a basis exist in which $\ket{\psi}$ has real amplitudes and every $P_j$ has real matrix entries,
as in Lemma~\ref{lem:beta} and Corollary~\ref{cor:beta-pauli}.  Then every odd order of
Corollary~\ref{cor:energy} vanishes and
\begin{align}
  B_N &= N\,b_{\gamma,2}\,dt^2
    + \Bigl[N\,b_{\gamma,4} + \tfrac{N(N-1)}{2}\,b_{\varphi_{H,j},3}
    + \tfrac{N(N-1)(2N-1)}{12}\,b_{\varphi\Delta,2}
    + \tfrac{N(N-1)}{2}\,b_{\gamma\gamma,2}\Bigr]dt^4 + \mathcal{O}(dt^6),
  \label{eq:s2-energy-real-BN}\\
  V_N &= \Bigl[N\,v_\gamma + N(N{-}1)\bigl(v_{\gamma\cdot\varphi_{H,j},3} + \tfrac12 v_\varphi\bigr)
    + \tfrac13N(N{-}1)(N{-}2)\,v_{\varphi_{H,j}^2,2}\Bigr]dt^4 + \mathcal{O}(dt^6),
  \label{eq:s2-energy-real-VN}
\end{align}
with the densities of Corollary~\ref{cor:energy}.  These are the energy bias and variance~\eqref{eq:V1B1-energy} of
the main text.
\end{corollary}

\begin{proof}
Lemma~\ref{lem:beta} gives $\beta_j = 0$ for every $j$, hence $\beta_H = 0$, so every density of
Corollary~\ref{cor:energy} carrying an explicit factor $\beta_j$ vanishes: every density at order $dt^3$, and
$v_{\beta^2,4}$, $v_{\beta\cdot\varphi}$ and $v_{\beta\cdot\psi}$ at order $dt^4$.  The exception is
$b_{\beta\Delta,2} = \sum_j w_j\beta_H(\Delta_jH)$, which carries no such factor.

For it, apply the argument of Lemma~\ref{lem:beta} to the pair $(H,\Delta_jH)$ in place of $(O,P_j)$.  In the given
basis $H$ and $P_j$ are real and symmetric, so
\begin{equation}
  \Delta_jH = P_jHP_j - H \quad\text{is real and symmetric}, \qquad
  [H,\Delta_jH] \quad\text{is real and antisymmetric},
  \label{eq:s2-energy-real-p0}
\end{equation}
and a real antisymmetric operator has vanishing expectation in a state with real amplitudes.  For $A$ real with
$A^{\mathsf T} = -A$ and $\ket{\psi}$ real in that basis, $\bra{\psi}A\ket{\psi}$ is a real scalar and so equals
its own transpose,
\begin{equation}
  \bra{\psi}A\ket{\psi}
   = \bigl(\bra{\psi}A\ket{\psi}\bigr)^{\mathsf T}
   = \bra{\psi}A^{\mathsf T}\ket{\psi}
   = -\bra{\psi}A\ket{\psi}
  \qquad\Rightarrow\qquad \bra{\psi}A\ket{\psi} = 0 .
  \label{eq:s2-energy-real-p0b}
\end{equation}
With $A = [H,\Delta_jH]$,
\begin{equation}
  \beta_H(\Delta_jH) = i\braket{[H,\Delta_jH]} = 0
  \qquad\Rightarrow\qquad
  b_{\beta\Delta,2} = \sum_j w_j\underbrace{\beta_H(\Delta_jH)}_{0} = 0 .
  \label{eq:s2-energy-real-p1}
\end{equation}
A second commutator with $H$ returns a real \emph{symmetric} operator, since $H$ is symmetric
and $[H,\Delta_jH]$ is antisymmetric by~\eqref{eq:s2-energy-real-p0},
\begin{equation}
  \bigl[H,[H,\Delta_jH]\bigr]^{\mathsf T}
   = [H,\Delta_jH]^{\mathsf T}H^{\mathsf T} - H^{\mathsf T}[H,\Delta_jH]^{\mathsf T}
   = -[H,\Delta_jH]\,H + H\,[H,\Delta_jH]
   = \bigl[H,[H,\Delta_jH]\bigr] ,
  \label{eq:s2-energy-real-p2}
\end{equation}
whose expectation in a real-amplitude state is nonzero in general, so
$\varphi_{H,H}(\Delta_jH) = -\braket{[H,[H,\Delta_jH]]} \neq 0$ and as such
$b_{\varphi\Delta,2} = \sum_j w_j\varphi_{H,H}(\Delta_jH) \neq 0$.  Substituting
\begin{equation}
  b_{\beta,3} = b_{\beta\Delta,2} = 0, \qquad
  v_{\beta^2,2} = v_{\beta\gamma,3} = v_{\beta\cdot\varphi_{H,j},2}
   = v_{\beta^2,4} = v_{\beta\cdot\varphi} = v_{\beta\cdot\psi} = 0
  \label{eq:s2-energy-real-p3}
\end{equation}
into~\eqref{eq:s2-energy-BN}--\eqref{eq:s2-energy-VN}
gives~\eqref{eq:s2-energy-real-BN}--\eqref{eq:s2-energy-real-VN}.
\end{proof}

\begin{corollary}[Energy at a real-amplitude energy eigenstate]\label{cor:energy-eig}
Let $O = H$ under the hypotheses of both Corollary~\ref{cor:energy-real} and Lemma~\ref{lem:collapse}.  Then
\begin{align}
  B_N &= N\,b_{\gamma,2}\,dt^2
    + \Bigl[-\tfrac{N}{3}\sum_j \frac{c_j^4\gamma_j}{q_j^3}
      + \tfrac{N(N-1)}{2}\sum_{j,k} w_jw_k\,\gamma_{j,k}\Bigr]dt^4 + \mathcal{O}(dt^6),
  \label{eq:s2-energy-eig-BN}\\
  V_N &= \Bigl[N\,v_\gamma + \tfrac{N(N-1)}{2}\,v_\varphi\Bigr]dt^4 + \mathcal{O}(dt^6) .
  \label{eq:s2-energy-eig-VN}
\end{align}
The variance is the one used in the mean-squared error~\eqref{eq:mse-beta0}, and the bias at order $dt^2$ is $Nb_{\gamma,2}dt^2$.
\end{corollary}

\begin{proof}
Start from Corollary~\ref{cor:energy-real} and apply Lemma~\ref{lem:collapse}.  It gives $\varphi_{H,j} = 0$ for
every $j$, so
\begin{equation}
  b_{\varphi_{H,j},3} = 0, \qquad
  v_{\gamma\cdot\varphi_{H,j},3} = 0, \qquad
  v_{\varphi_{H,j}^2,2} = 0 ,
  \label{eq:s2-energy-eig-p1}
\end{equation}
and~\eqref{eq:s2-eig-all} with $X = \Delta_jH$ gives $\varphi_{H,H}(\Delta_jH) = 0$, so
\begin{equation}
  b_{\varphi\Delta,2} = \sum_j w_j\underbrace{\varphi_{H,H}(\Delta_jH)}_{0} = 0,
  \qquad
  b_{\gamma\gamma,2} = \sum_{j,k} w_jw_k\gamma_{j,k}
    - \tfrac12\sum_j w_j\underbrace{\varphi_{H,H}(\Delta_jH)}_{0}
  = \sum_{j,k} w_jw_k\gamma_{j,k} .
  \label{eq:s2-energy-eig-p2}
\end{equation}
With $b_{\gamma,4} = -\tfrac13\sum_j \sfrac{c_j^4\gamma_j}{q_j^3}$ from Corollary~\ref{cor:energy},
substituting~\eqref{eq:s2-energy-eig-p1}--\eqref{eq:s2-energy-eig-p2}
into~\eqref{eq:s2-energy-real-BN}--\eqref{eq:s2-energy-real-VN}
gives~\eqref{eq:s2-energy-eig-BN}--\eqref{eq:s2-energy-eig-VN}.
\end{proof}

\begin{corollary}[General observable at a real-amplitude energy eigenstate]\label{cor:obs-eig}
Let $O \neq H$, let $\ket{\psi_0}$ be an energy eigenstate, and let a basis exist in which $\ket{\psi_0}$ has real
amplitudes and $O$ and every $P_j$ have real matrix entries.  Then
\begin{align}
  B_N &= N\,b_{\gamma,2}\,dt^2
    + \Bigl[N\,b_{\gamma,4} + \tfrac{N(N-1)}{2}\,b_{\varphi_{j,H},3}
    + \tfrac{N(N-1)(2N-1)}{12}\,b_{\gamma HH,2}
    + \tfrac{N(N-1)}{2}\,b_{\gamma\gamma,2}\Bigr]dt^4 + \mathcal{O}(dt^6),
  \label{eq:s2-obs-eig-BN}\\
  V_N &= \Bigl[N\,v_\gamma + N(N{-}1)\bigl(v_{\gamma\cdot\varphi_{j,H},3} + \tfrac12 v_\varphi\bigr)
    + \tfrac13N(N{-}1)(N{-}2)\,v_{\varphi_{j,H}^2,2}\Bigr]dt^4 + \mathcal{O}(dt^6),
  \label{eq:s2-obs-eig-VN}
\end{align}
with $b_{\gamma,4}$, $b_{\varphi_{j,H},3}$, $b_{\gamma HH,2}$ and $b_{\gamma\gamma,2}$ as in
Corollary~\ref{cor:eig-general}.
\end{corollary}

\begin{proof}
Corollary~\ref{cor:eig-general} applies, and Lemma~\ref{lem:beta} adds $\beta_j = 0$ for every $j$.  Every reduced
density of Corollary~\ref{cor:eig-general} that carries an explicit factor $\beta_j$ therefore vanishes, namely
$b_{\beta,3}$, $v_{\beta^2,2}$, $v_{\beta\gamma,3}$, $v_{\beta^2,4}$, $v_{\beta\cdot\varphi_{j,H},2}$,
$v_{\beta\cdot\varphi}$ and $v_{\beta\cdot\psi}$.  At order $dt^3$ that leaves only
$b_{\gamma H,2} = \sum_j w_j\gamma_j(\mathcal{L}_H(O))$, which has no such factor.  In the given basis $H$ and $O$ are real and symmetric, so $[H,O]$ is
real and antisymmetric, and so is $P_j[H,O]P_j$; both have vanishing expectation in a state with real amplitudes,
whence
\begin{equation}
  \gamma_j(\mathcal{L}_H(O)) = i\Bigl(\underbrace{\bra{\psi_0}P_j[H,O]P_j\ket{\psi_0}}_{0}
    - \underbrace{\bra{\psi_0}[H,O]\ket{\psi_0}}_{0}\Bigr) = 0
  \qquad\Rightarrow\qquad
  b_{\gamma H,2} = 0 ,
  \label{eq:s2-obs-eig-p1}
\end{equation}
and the whole $dt^3$ order of~\eqref{eq:s2-eig-BN} vanishes.  The density $b_{\gamma HH,2} = \sum_j w_j\gamma_j(\mathcal{L}_H^2(O))$
is not removed: $\mathcal{L}_H^2(O) = -[H,[H,O]]$ is real and symmetric, so $\bra{\psi_0}P_j\,\mathcal{L}_H^2(O)\,P_j\ket{\psi_0}$ is nonzero in
general.  Substituting into~\eqref{eq:s2-eig-BN}--\eqref{eq:s2-eig-VN}
gives~\eqref{eq:s2-obs-eig-BN}--\eqref{eq:s2-obs-eig-VN}.
\end{proof}

\section{Allocation theory}\label{app:S4}

In this section we derive the bias- and variance-optimal allocations where analytic derivations are possible, and
where they are not, we describe the methodology for heuristically optimised allocations.  We treat the
$\gamma$-driven and $\beta$-driven regimes, show that the rule of Ref.~\cite{Wu2025fluctuation} is a special case
of the first, and combine the two into the one-parameter family of Lemma~\ref{lem:monotone} and
Corollary~\ref{prop:joint-main} in the main text.

\subsection*{Variance-optimal distributions}

We minimise the estimator variance over the simplex $\bigl\{\bq : \sum_j q_j = 1,\; q_j > 0\bigr\}$.  The
minimiser depends on whether $\beta_j = 0$ at the reference state, which by Corollary~\ref{cor:beta-pauli} holds for
a real-amplitude $\ket{\psi}$ with $O$ and every $P_j$ even in $Y$, as in electronic structure over real orbitals;
complex references and odd-$Y$ terms give the $\beta$-driven regime.

Before proving the optimal allocations, we first prove the following lemma, which is useful across several of the
proofs below.

\begin{lemma}\label{lem:combined}
Let $c_k \neq 0$ and $f_k > 0$ for every $k$.  Then
\begin{equation}
  \sum_j \frac{c_j^2 f_j}{q_j}
  \label{eq:s3-recip}
\end{equation}
is minimised over $\bigl\{\sum_j q_j = 1,\, q_j > 0\bigr\}$ at $q_k \propto |c_k|\sqrt{f_k}$, with minimum value
$\bigl(\sum_j |c_j|\sqrt{f_j}\bigr)^2$.
\end{lemma}

\begin{proof}
Insert $\sum_j q_j = 1$ and apply Cauchy--Schwarz, splitting each term as
$\sqrt{q_j}\cdot\sfrac{|c_j|\sqrt{f_j}}{\sqrt{q_j}}$,
\begin{equation}
  \sum_j \frac{c_j^2f_j}{q_j}
   = \underbrace{\left(\sum_j q_j\right)}_{1}\left(\sum_j \frac{c_j^2f_j}{q_j}\right)
   \;\ge\; \left(\sum_j \sqrt{q_j}\,\frac{|c_j|\sqrt{f_j}}{\sqrt{q_j}}\right)^{\!2}
   = \left(\sum_j |c_j|\sqrt{f_j}\right)^{\!2} ,
  \label{eq:s3-recip-cs}
\end{equation}
a bound on~\eqref{eq:s3-recip} that holds for every $\bq$.  Cauchy--Schwarz is an equality if and only if the two factors are
proportional,
\begin{equation}
  \sqrt{q_j} \propto \frac{|c_j|\sqrt{f_j}}{\sqrt{q_j}}
  \qquad\Longleftrightarrow\qquad
  q_k \propto |c_k|\sqrt{f_k} ,
  \label{eq:s3-recip-eq}
\end{equation}
which lies on the simplex after normalisation.  The bound is therefore attained there, and that allocation is the
minimiser with minimum value $\bigl(\sum_j|c_j|\sqrt{f_j}\bigr)^2$.
\end{proof}

Next, we begin deriving the lemmata relating to certain simplifications of the bias and variance formulae.  We
begin with a single \qdrift{} step ($N=1$) below.

\subsubsection*{Single-step variance ($N = 1$)}

Corollary~\ref{prop:single-step-var} reduced the general variance to a single step,
\begin{equation}
  V_1(\bq,\psi) \;=\; v_{\beta^2,2}\,dt^2 + v_{\beta\gamma,3}\,dt^3 + \bigl(v_\gamma - v_{\beta^2,4}\bigr)dt^4 + \mathcal{O}(dt^5) .
  \label{eq:S4}
\end{equation}
The allocation enters $V_1$ only through these four densities, and $v_{\beta\gamma,3}$ and $v_{\beta^2,4}$ vanish
at $\beta_j = 0$.

Each closed form below minimises the leading density of $V_1$ in its regime, so for $N>1$ it minimises only the
part of $V_N$ proportional to $N$.

\subsubsection*{Real-amplitude regime ($\beta_j = 0$): $v_\gamma$ dominant}

By Corollary~\ref{cor:beta0}, at $\beta_j = 0$ the densities $v_{\beta^2,2}$, $v_{\beta\gamma,3}$ and
$v_{\beta^2,4}$ vanish and $V_1 = v_\gamma\,dt^4$, which can be made zero when the $\gamma_j$ share a sign.

\begin{proposition}[$\gamma$-driven optimum]\label{prop:real-neyman}
Let $\beta_j = 0$, let $\gamma_j \neq 0$ for every $j$ and let $\mathrm{sign}(\gamma_i) = \mathrm{sign}(\gamma_j)$ for all
$i,j$.  Then $v_\gamma$ is minimised on the simplex at
\begin{equation}
  q_k^\star \;\propto\; |c_k|\sqrt{|\gamma_k|}
  \label{eq:S11}
\end{equation}
where $v_\gamma = 0$ and $b_{\gamma,2}(\bq^\star) = \mathrm{sign}(\gamma)\bigl(\sum_j |c_j|\sqrt{|\gamma_j|}\bigr)^2$.
\end{proposition}

\begin{proof}
Let $j$ be sampled from $\bq$ and
\begin{equation}
  X_j := \Bigl(\frac{c_j}{q_j}\Bigr)^2\gamma_j .
\end{equation}
Its first two moments are
\begin{equation}
  \mathbb{E}_{\bq}[X] = \sum_j q_j\,\frac{c_j^2\gamma_j}{q_j^2} = b_{\gamma,2}, \qquad
  \mathbb{E}_{\bq}[X^2] = \sum_j q_j\,\frac{c_j^4\gamma_j^2}{q_j^4} = \sum_j \frac{c_j^4\gamma_j^2}{q_j^3},
\end{equation}
so, as in~\eqref{eq:Vgamma-def},
\begin{equation}
  v_\gamma = \mathbb{E}_{\bq}[X^2] - \mathbb{E}_{\bq}[X]^2 = \Var_{\bq}[X] \;\ge\; 0,
\end{equation}
with equality exactly when $X_j$ takes one common value $x$ for every $j$.  Solving for $q_j$,
\begin{equation}
  \frac{c_j^2\gamma_j}{q_j^2} = x
  \;\;\Rightarrow\;\; q_j^2 = \frac{c_j^2\gamma_j}{x}
  \;\;\Rightarrow\;\; q_j = |c_j|\sqrt{\frac{|\gamma_j|}{|x|}},
\end{equation}
which has a solution only if every $\gamma_j$ has the sign of $x$.  The normalisation $\sum_j q_j = 1$ fixes
\begin{equation}
  \sqrt{|x|} = \sum_j |c_j|\sqrt{|\gamma_j|},
\end{equation}
which gives~\eqref{eq:S11}.  Since $X_j = x$ for every $j$, $b_{\gamma,2}(\bq^\star) = \mathbb{E}_{\bq^\star}[X] = x
= \mathrm{sign}(\gamma)\bigl(\sum_j|c_j|\sqrt{|\gamma_j|}\bigr)^2$.
\end{proof}

Minimising $v_\gamma$ is not the same as minimising the bias.  At $\beta_j = 0$ and $N=1$,
Corollary~\ref{cor:beta0} gives
\begin{equation}
  B_1 = \bigl(b_{\gamma,2} - \tfrac12\varphi_{H,H}\bigr)dt^2 + b_{\gamma,4}\,dt^4 + \mathcal{O}(dt^5),
  \qquad
  V_1 = v_\gamma\,dt^4 + \mathcal{O}(dt^5),
  \label{eq:s3-B1V1-beta0}
\end{equation}
so the leading bias is removed by the cancellation $b_{\gamma,2} = \tfrac12\varphi_{H,H}$, not by making
$b_{\gamma,2}$ small, while $\bq^\star_\gamma$ sends $b_{\gamma,2}$ to $b^\star_\gamma$ of~\eqref{eq:S11}.

The two objectives coincide if $\tfrac12\varphi_{H,H}\le b^\star_\gamma$ when every $\gamma_j>0$ or $\tfrac12\varphi_{H,H}\ge b^\star_\gamma$ when every $\gamma_j<0$, as for the energy $O=H$, where $\varphi_{H,H}=0$ and so $|b_{\gamma,2}-\tfrac12\varphi_{H,H}|=|b_{\gamma,2}|$.  There $\bq^\star_\gamma$
minimises $|b_{\gamma,2}-\tfrac12\varphi_{H,H}|$ as well, by Lemma~\ref{lem:combined} at $f_j = |\gamma_j|$, so it minimises
\begin{equation}
  \mathrm{MSE}_1 = \bigl(b_{\gamma,2}-\tfrac12\varphi_{H,H}\bigr)^2\,dt^4 + \frac{v_\gamma\,dt^4}{M}
  \label{eq:s3-mse1}
\end{equation}
for every shot budget $M$.  Away from these cases, there is a trade-off between optimising the two, which is why we
generally optimise the mean-squared error to find a joint optimum of both quantities.

\paragraph{Relation to the fluctuation rule.}
The state-aware rule of Ref.~\cite{Wu2025fluctuation} weights each term by its fluctuation in the state,
$\Delta H_j = |c_j|\sqrt{1-\braket{P_j}^2}$.  It is the case $O = \ket{\psi}\bra{\psi}$ of
Proposition~\ref{prop:real-neyman}, where $\beta_j = 0$ and $\gamma_j = \braket{P_j}^2 - 1$, so
\begin{equation}
  q^\star_j \;\propto\; |c_j|\sqrt{|\gamma_j|} \;=\; |c_j|\sqrt{1-\braket{P_j}^2} \;=\; \Delta H_j .
  \label{eq:fwz-limit}
\end{equation}

Here every $\gamma_j\le0$, so every allocation has $b_{\gamma,2}\le b^\star_\gamma$, while $\tfrac12\varphi_{H,H}\ge b^\star_\gamma$.  The leading bias can then vanish only if $b_{\gamma,2}=b^\star_\gamma=\tfrac12\varphi_{H,H}$, which needs $\bq=\bq^\star_\gamma$ and $\sqrt{\Var_\psi(H)}=\sum_j\Delta H_j$ (Corollary~\ref{cor:fwz-nocancel}).  To see when the last equality holds, write $\Var_\psi(H)=\sum_{j,k}c_jc_k\Cov_\psi(P_j,P_k)$ with $\Cov_\psi(P_j,P_k)=\tfrac12\braket{\{P_j,P_k\}}-\braket{P_j}\braket{P_k}$.  Since $P_j^2=\mathds{1}$, $\Cov_\psi(P_j,P_j)=1-\braket{P_j}^2$, so $\Delta H_j^2=c_j^2\Cov_\psi(P_j,P_j)$ by~\eqref{eq:fwz-limit}, and
\begin{equation}
  \Bigl(\sum_j\Delta H_j\Bigr)^{\!2}-\Var_\psi(H)
  \;=\; \sum_{j,k}\Bigl(\Delta H_j\Delta H_k-c_jc_k\Cov_\psi(P_j,P_k)\Bigr) .
\end{equation}
Each term is non-negative, since $|c_jc_k\Cov_\psi(P_j,P_k)|\le\Delta H_j\Delta H_k$ by Cauchy--Schwarz.  The last equality therefore holds only when every term vanishes,
\begin{equation}
  c_jc_k\Cov_\psi(P_j,P_k) \;=\; \Delta H_j\Delta H_k \qquad \text{for all } j,k ,
\end{equation}
that is, when every pair of terms $c_jP_j$ and $c_kP_k$ is perfectly correlated in $\ket{\psi}$.  This does not hold in general, so the bias cannot be cancelled.

\begin{corollary}[No cancellation at $O = \ket{\psi}\bra{\psi}$]\label{cor:fwz-nocancel}
At $O = \ket{\psi}\bra{\psi}$, every allocation on the simplex obeys
\begin{equation}
  \bigl|b_{\gamma,2} - \tfrac12\varphi_{H,H}\bigr|
  \;\ge\; \Bigl(\sum_j \Delta H_j\Bigr)^{\!2} - \Var_\psi(H) \;\ge\; 0 ,
  \label{eq:fwz-nocancel}
\end{equation}
with equality in the second only when $\sqrt{\Var_\psi(H)} = \sum_j \Delta H_j$.
\end{corollary}

\begin{proof}
Since $[H,[O,H]] = 2HOH - H^2O - OH^2$ and $O = \ket{\psi}\bra{\psi}$,
\begin{equation}
  \tfrac12\varphi_{H,H} \;=\; \braket{H}^2 - \braket{H^2} \;=\; -\Var_\psi(H) .
\end{equation}
Every $\gamma_j = \braket{P_j}^2 - 1 \le 0$, so $b_{\gamma,2} \le 0$, and, using $|c_j|\sqrt{|\gamma_j|} = \Delta H_j$ from~\eqref{eq:fwz-limit} and Cauchy--Schwarz with $\sum_jq_j=1$,
\begin{equation}
  \Bigl(\sum_j \Delta H_j\Bigr)^{\!2}
  \;=\; \Bigl(\sum_j \sqrt{q_j}\,\frac{\Delta H_j}{\sqrt{q_j}}\Bigr)^{\!2}
  \;\le\; \sum_j q_j\,\sum_j \frac{\Delta H_j^2}{q_j}
  \;=\; \sum_j \frac{c_j^2|\gamma_j|}{q_j}
  \;=\; |b_{\gamma,2}| .
\end{equation}
The standard deviation is subadditive, so $\sqrt{\Var_\psi(H)} \le \sum_j \Delta H_j$ for $H = \sum_j c_jP_j$.
Combining the two bounds,
\begin{equation}
  |b_{\gamma,2}| \;\ge\; \Bigl(\sum_j\Delta H_j\Bigr)^{\!2} \;\ge\; \Var_\psi(H) .
  \label{eq:fwz-two-bounds}
\end{equation}
Since both $b_{\gamma,2}$ and $\tfrac12\varphi_{H,H}$ are non-positive, with $\tfrac12\varphi_{H,H}=-\Var_\psi(H)$,
\begin{equation}
  b_{\gamma,2}-\tfrac12\varphi_{H,H} \;=\; -|b_{\gamma,2}|+\Var_\psi(H) \;\le\; 0
\end{equation}
by~\eqref{eq:fwz-two-bounds}, so
\begin{equation}
\begin{split}
  \bigl|b_{\gamma,2} - \tfrac12\varphi_{H,H}\bigr|
  &\;=\; \bigl|-\bigl(|b_{\gamma,2}|-\Var_\psi(H)\bigr)\bigr|
  \;=\; \bigl|\underbrace{|b_{\gamma,2}|-\Var_\psi(H)}_{\ge0}\bigr|
  \;=\; |b_{\gamma,2}| - \Var_\psi(H) \\
  &\;\ge\; \Bigl(\sum_j\Delta H_j\Bigr)^{\!2}-\Var_\psi(H) \;\ge\; 0 ,
\end{split}
\end{equation}
where the first inequality is the left inequality of~\eqref{eq:fwz-two-bounds}, $|b_{\gamma,2}|\ge(\sum_j\Delta H_j)^2$, and the second is its right inequality, $(\sum_j\Delta H_j)^2\ge\Var_\psi(H)$.  This is~\eqref{eq:fwz-nocancel}.
\end{proof}

The left-hand side of~\eqref{eq:fwz-nocancel} is the objective of Ref.~\cite{Wu2025fluctuation} itself.  Since
$c_j^2|\gamma_j| = (\Delta H_j)^2$ and $\tfrac12\varphi_{H,H} = -(\Delta H)^2$, the fidelity expansion given there,
\begin{equation}
  F \;=\; 1 - \frac{T^2}{N^2}\Bigl[\sum_j \frac{(\Delta H_j)^2}{q_j} - (\Delta H)^2\Bigr] ,
\end{equation}
has leading infidelity $\sfrac{T^2}{N^2}\bigl|b_{\gamma,2}-\tfrac12\varphi_{H,H}\bigr|$, and~\eqref{eq:fwz-limit}
attains the bound.  The gap in~\eqref{eq:fwz-nocancel} does not depend on $\bq$, so it cannot be closed by a more optimal choice of allocation.  The cancellation that produces the savings of Figure~\ref{fig:chem} is therefore available only for other
observables.

Ref.~\cite{Fan2025adaptive} measures the same second-order error in a different norm.  Its cost is the $\bq$-dependent part of a bound on the
Hilbert--Schmidt norm of the channel defect, $\sum_j \sfrac{\|D_{jj}(\rho)\|}{q_j}$ with
$D_{jj}(\rho) = -[H_j,[H_j,\rho]]$, and at $H_j = c_jP_j$ on a pure state
\begin{equation}
  \bigl\|D_{jj}(\rho)\bigr\| \;=\; 2\sqrt{2}\,c_j^2\sqrt{|\gamma_j|} ,
\end{equation}
so that cost is~\eqref{eq:s3-recip} at $f_j = \sqrt{|\gamma_j|}$, and Lemma~\ref{lem:combined} minimises it at
\begin{equation}
  q_j \;\propto\; |c_j|\,|\gamma_j|^{\sfrac14} .
  \label{eq:adarc-limit}
\end{equation}
Both rules are therefore Lemma~\ref{lem:combined} at $O = \ket{\psi}\bra{\psi}$, differing only in $f_j$:
$|\gamma_j|$ for the fidelity of Ref.~\cite{Wu2025fluctuation} and $\sqrt{|\gamma_j|}$ for the Hilbert--Schmidt
norm of Ref.~\cite{Fan2025adaptive}.  Both target the $\mathcal{O}(dt^2)$ channel error; the fourth moments of
Ref.~\cite{Fan2025adaptive} come from that norm being quadratic in $D_{jj}$, not from a higher order in $dt$.
Corollary~\ref{cor:fwz-nocancel} bounds the leading bias at every allocation on the simplex, so it applies to
both.

\subsubsection*{$\beta$-driven regime ($\beta_j \neq 0$): $v_{\beta^2,2}$ dominant}

With $\beta_j \neq 0$ the leading term of $V_1$ in~\eqref{eq:S4} is $v_{\beta^2,2}\,dt^2$, the remaining terms
being smaller for small enough $dt$.

\begin{proposition}[$\beta$-driven optimum]\label{prop:beta-neyman}
If $c_k\beta_k\neq0$ for every $k$, the leading variance density $v_{\beta^2,2}$ of $V_1$ is minimised on the
simplex at
\begin{equation}
  q_k^\star \;\propto\; |c_k|\,|\beta_k|, \qquad
  v_{\beta^2,2}^{\min} \;=\; \left(\sum_j|c_j||\beta_j|\right)^{\!2}
  {}-\beta_H^2 \;\ge\; 0 .
  \label{eq:S15}
\end{equation}
\end{proposition}

\begin{proof}
Of the two terms in
\begin{equation}
  v_{\beta^2,2} = \sum_j w_j\beta_j^2 - \beta_H^2 ,
  \label{eq:s3-vbeta2}
\end{equation}
only the first depends on the allocation, since $\beta_H^2$ of~\eqref{eq:per-step-defs} is $\bq$-independent.
Since it is $\bq$-independent, it shifts $v_{\beta^2,2}$ without moving its minimiser, and for $O = H$ it vanishes
outright,
$\beta_H = i\braket{[H,H]} = 0$.  That first term is~\eqref{eq:s3-recip} at $f_j = \beta_j^2$, so
it is minimised at $q_k\propto|c_k||\beta_k|$ with minimum $\bigl(\sum_j|c_j||\beta_j|\bigr)^2$.  Subtracting
$\beta_H^2$ gives~\eqref{eq:S15}.
\end{proof}

\subsection*{Combined family and joint improvement}

The previous two subsections each gave a closed form for one regime: $\bq^\star_\gamma$ of~\eqref{eq:S11} at
$\beta_j = 0$, and $\bq^\star_\beta$ of~\eqref{eq:S15} when $v_{\beta^2,2}$ dominates.  Neither covers the general
case, where $\bgamma$ and $\bbeta$ are both non-zero.

There the conflict is between two densities that enter at the same order in $dt$, one in the bias and one in the
variance.  By Corollary~\ref{prop:single-step-var},
\begin{equation}
  B_1 = \Bigl(\underbrace{b_{\gamma,2}}_{\text{min.\ at }\bq^\star_\gamma} - \tfrac12\varphi_{H,H}\Bigr)dt^2
    + \mathcal{O}(dt^3) ,
  \qquad
  V_1 = \underbrace{v_{\beta^2,2}}_{\text{min.\ at }\bq^\star_\beta}\,dt^2 + \mathcal{O}(dt^3) ,
  \label{eq:s3-conflict}
\end{equation}
where $b_{\gamma,2} = \sum_j w_j\gamma_j$ is minimised by $\bq^\star_\gamma$, by Lemma~\ref{lem:combined} at
$f_j = \gamma_j$ when every $\gamma_j>0$, and $v_{\beta^2,2} = \sum_j w_j\beta_j^2 - \beta_H^2$ by $\bq^\star_\beta$, by
Proposition~\ref{prop:beta-neyman}.  The two allocations agree only when $\sqrt{|\gamma_k|} \propto |\beta_k|$.
Otherwise the optimal allocation for one term can have detrimental effects on the other.  For example, using
$\bq^\star_\gamma$ to minimise $b_{\gamma,2}$ can make $v_{\beta^2,2}$ larger than at the vanilla \qdrift{}
allocation $\bq^c$, and vice versa for $\bq^\star_\beta$ with $b_{\gamma,2}$.

In this section we derive the theoretical background for the one-parameter family
\begin{equation}
  \bq^{(\alpha)} \;:\; q_j \;\propto\; |c_j|\sqrt{|\gamma_j| + \alpha\beta_j^2} , \qquad \alpha \ge 0 ,
  \label{eq:s3-family}
\end{equation}
of the main text~\eqref{eq:combined-neyman}, shown in Figure~\ref{fig:joint-crossover}, aiming to jointly optimise
both bias and variance simultaneously.

We work with $S_\beta := \sum_j \sfrac{c_j^2\beta_j^2}{q_j}$, the $\bq$-dependent part of $v_{\beta^2,2}$, so that
$V_1 = \bigl(S_\beta - \beta_H^2\bigr)dt^2 + \mathcal{O}(dt^3)$ by~\eqref{eq:s3-conflict}.  It shares the minimisers of $v_{\beta^2,2}$.  Both $b_{\gamma,2}$ and $S_\beta$ have the form
$\sum_j \sfrac{c_j^2f_j}{q_j}$, so for $\alpha \ge 0$ their weighted sum is also of the same form,
\begin{equation}
  J_\alpha(\bq) \;:=\; b_{\gamma,2} + \alpha\,S_\beta
  \;=\; \sum_j \frac{c_j^2(\gamma_j + \alpha\beta_j^2)}{q_j} ,
  \label{eq:S58}
\end{equation}
whose minimiser, when every $\gamma_j>0$, is the allocation $\bq^{(\alpha)}$.

\begin{lemma}\label{lem:family}
Let $\alpha \ge 0$, $c_k\neq0$ and $\gamma_k + \alpha\beta_k^2 > 0$ for every $k$.  Then $J_\alpha$ of~\eqref{eq:S58} is
minimised on the simplex, uniquely, at
\begin{equation}
  q_k^{(\alpha)} \;\propto\; |c_k|\sqrt{\gamma_k + \alpha\beta_k^2} , \qquad
  J_\alpha\bigl(\bq^{(\alpha)}\bigr) = \Bigl(\sum_j|c_j|\sqrt{\gamma_j+\alpha\beta_j^2}\Bigr)^{\!2} .
  \label{eq:S61}
\end{equation}
\end{lemma}

\begin{proof}
By~\eqref{eq:S58}, $J_\alpha$ is the sum~\eqref{eq:s3-recip} of Lemma~\ref{lem:combined} at
$f_j = \gamma_j + \alpha\beta_j^2$, which is positive by hypothesis.  That lemma gives the minimiser, the minimum value, and uniqueness through its equality
condition $q_k \propto |c_k|\sqrt{f_k}$.
\end{proof}

The ends of the family are the two optima above.
\begin{itemize}[topsep=2pt,itemsep=2pt,leftmargin=*]
  \item $\alpha = 0$: the objective is $b_{\gamma,2}$ alone and
    $q_k^{(0)} \propto |c_k|\sqrt{\gamma_k}$, the $\gamma$-driven optimum $\bq^\star_\gamma$ of~\eqref{eq:S11}.
  \item $\alpha \to \infty$: since
    $\sqrt{\gamma_k+\alpha\beta_k^2} = \sqrt{\alpha}\,|\beta_k| + \mathcal{O}(\alpha^{-\sfrac12})$ for
    $\beta_k \neq 0$, the $\sqrt{\alpha}$ cancels against the normalisation and
    \begin{equation}
      q_k^{(\alpha)} \;\longrightarrow\; \frac{|c_k||\beta_k|}{\sum_j|c_j||\beta_j|} ,
      \label{eq:s3-family-ends}
    \end{equation}
    the $\beta$-driven optimum $\bq^\star_\beta$ of~\eqref{eq:S15}.
\end{itemize}

Lemma~\ref{lem:family} assumes $\gamma_k + \alpha\beta_k^2 > 0$ for every $k$.  If $\gamma_k+\alpha\beta_k^2<0$ for some $k$, the term $\sfrac{c_k^2(\gamma_k+\alpha\beta_k^2)}{q_k}$ of $J_\alpha$ tends to $-\infty$ as $q_k\to0$.  If $\gamma_k+\alpha\beta_k^2=0$, that term is zero for every $q_k$, and moving its probability onto the other terms lowers $J_\alpha$.  In the first case $J_\alpha$ is unbounded below, and in the second it decreases towards $\bigl(\sum_{j\neq k}|c_j|\sqrt{\gamma_j+\alpha\beta_j^2}\bigr)^2$, reached only at $q_k=0$.  Both cases together imply $J_\alpha$ cannot be minimised when both $q_k>0$ and $\gamma_k+\alpha\beta_k^2\le0$ hold.  The main text therefore uses $|\gamma_j|$
in~\eqref{eq:combined_obj}--\eqref{eq:combined-neyman}, which makes every $|\gamma_j|+\alpha\beta_j^2\ge0$ and is $J_\alpha$ itself when the $\gamma_j$ are
positive, the case treated below.  For mixed-sign $\bgamma$ nothing below applies: the signed $b_{\gamma,2}(\alpha)$
need not be monotone along~\eqref{eq:combined-neyman}, the conditions~\eqref{eq:interval-boundaries} need not
define an interval, and $\alpha$ is instead chosen by minimising the mean-squared error
(Fig.~\ref{fig:joint-crossover}c).

The family replaces the $L-1$ free parameters of $\bq$ with $\alpha$ alone.  Densities that are $\bq$-independent,
such as the shift $\tfrac12\varphi_{H,H}$, or of mixed sign, such as $v_{\beta\gamma,3}$, cannot be built into the
allocation at all, and enter the allocation implicitly only via the optimisation of the mean-squared error.  In the next result we prove
a regime in which the bias density $b_{\gamma,2}$ and the variance part $S_\beta$ can be optimised jointly.

\subsubsection*{The $\gamma_j>0$ assumption}
We begin with Lemma~\ref{lem:sign-reduction}, which allows us to focus on the case where $\gamma_j>0$.

\begin{lemma}[Sign reduction]\label{lem:sign-reduction}
Let $\gamma_j<0\ \forall\,j$, and write $\tilde\gamma_j:=|\gamma_j|=-\gamma_j>0$ and $\tilde\varphi:=-\varphi_{H,H}$.  Then $\bgamma$ and $\tilde\bgamma$ give the same allocations $\bq^{(\alpha)}$, the same $\bq^\star_\gamma$ and the same $S_\beta(\alpha)$, and at every $\bq$, we have:
\begin{equation}
\begin{aligned}
  b_{\gamma,2}(\bq) &= \sum_j\frac{c_j^2\gamma_j}{q_j} = -\sum_j\frac{c_j^2\tilde\gamma_j}{q_j} = -\,b_{\tilde\gamma,2}(\bq), \\
  b_{\gamma,2}(\bq)-\tfrac12\varphi_{H,H} &= -\bigl(b_{\tilde\gamma,2}(\bq)-\tfrac12\tilde\varphi\bigr) .
\end{aligned}
  \label{eq:s3-negative-gamma}
\end{equation}
As a result, $|b_{\gamma,2}(\bq)|=b_{\tilde\gamma,2}(\bq)$ and $|b_{\gamma,2}(\bq)-\tfrac12\varphi_{H,H}|=|b_{\tilde\gamma,2}(\bq)-\tfrac12\tilde\varphi|$.
\end{lemma}

\begin{proof}
The family and the $\gamma$-driven optimum are built from $|\gamma_j|=\tilde\gamma_j$,
\begin{equation}
  q^{(\alpha)}_j \;\propto\; |c_j|\sqrt{|\gamma_j|+\alpha\beta_j^2} \;=\; |c_j|\sqrt{\tilde\gamma_j+\alpha\beta_j^2} ,
  \qquad
  q^\star_{\gamma,j} \;\propto\; |c_j|\sqrt{|\gamma_j|} \;=\; |c_j|\sqrt{\tilde\gamma_j} ,
\end{equation}
and $S_\beta(\bq)=\sum_j\sfrac{c_j^2\beta_j^2}{q_j}$ does not involve $\bgamma$.  Substituting $\gamma_j=-\tilde\gamma_j$ into $b_{\gamma,2}(\bq)=\sum_j\sfrac{c_j^2\gamma_j}{q_j}$, and $\varphi_{H,H}=-\tilde\varphi$ into $b_{\gamma,2}(\bq)-\tfrac12\varphi_{H,H}$, gives the two lines of~\eqref{eq:s3-negative-gamma}.
\end{proof}

\subsubsection*{Monotonicity and joint improvement for $\gamma_j>0$}
From Lemma~\ref{lem:sign-reduction}, every result below for positive $\gamma_j$ applies to $\tilde\bgamma$, with $\tilde\varphi$ in place of $\varphi_{H,H}$, and by the first line of~\eqref{eq:s3-negative-gamma} its statements about $b_{\tilde\gamma,2}$ are statements about $|b_{\gamma,2}|$.

In the below, we use the following notation for clarity:
\begin{equation}
\begin{gathered}
  f_j(\alpha) := \gamma_j + \alpha\beta_j^2, \qquad \frac{df_j}{d\alpha} = \beta_j^2, \qquad
  Z(\alpha) := \sum_j|c_j|\sqrt{f_j(\alpha)}, \qquad
  q^{(\alpha)}_j = \frac{|c_j|\sqrt{f_j(\alpha)}}{Z(\alpha)}, \\
  b_{\gamma,2}(\alpha) := b_{\gamma,2}\bigl(\bq^{(\alpha)}\bigr), \qquad S_\beta(\alpha) := S_\beta\bigl(\bq^{(\alpha)}\bigr), \qquad
  q^c_j = \frac{|c_j|}{\lambda}, \qquad \lambda = \sum_j|c_j| .
\end{gathered}
  \label{eq:proof-joint-notation}
\end{equation}
At $\alpha=0$ the family gives $\bq^{(0)}=\bq^\star_\gamma$, which minimises the bias density over all allocations.  By Lemma~\ref{lem:combined} at $f_j=\gamma_j$,
\begin{equation}
  b_{\gamma,2}(\bq) \;\ge\; b_{\gamma,2}(0) \quad\text{for every } \bq ,
  \qquad\text{with equality if and only if } q_k \propto |c_k|\sqrt{\gamma_k} .
  \label{eq:proof-joint-eqcond}
\end{equation}

\begin{lemma}[Monotone distribution allocation densities, Lemma~\ref{lem:monotone} restated]\label{lem:monotone-si}
Let $\gamma_j \neq 0$ and
$\mathrm{sign}(\gamma_i) = \mathrm{sign}(\gamma_j)\;\;\forall\,i,j$.  Then along the
family~\eqref{eq:combined-neyman}, $|b_{\gamma,2}(\alpha)|$ is non-decreasing and
$v_{\beta^2,2}(\alpha)$ is non-increasing in $\alpha$.
\end{lemma}

\begin{proof}
\medskip\noindent\textbf{$S_\beta(\alpha)$ is non-increasing.}\\
As discussed above, we assume $\gamma_j>0\ \forall\,j$.  Inserting $q^{(\alpha)}_j$ into $S_\beta$,
\begin{equation}
  S_\beta(\alpha) = \sum_j\frac{c_j^2\beta_j^2\,Z}{|c_j|\sqrt{f_j}}
   = Z\sum_j\frac{|c_j|\beta_j^2}{\sqrt{f_j}} .
\end{equation}
Differentiating $Z$ twice,
\begin{equation}
  Z' = \frac12\sum_j\frac{|c_j|\beta_j^2}{\sqrt{f_j}}, \qquad
  Z'' = -\frac14\sum_j\frac{|c_j|\beta_j^4}{f_j^{\sfrac32}} \;\le\; 0,
\end{equation}
so
\begin{equation}
  S_\beta = 2ZZ', \qquad
  S_\beta' = \underbrace{2(Z')^2}_{(1)} + \underbrace{2ZZ''}_{(2)} .
  \label{eq:proof-joint-sbeta}
\end{equation}
Term (1) is non-negative and term (2) is non-positive, so $S_\beta' \le 0$ is the statement
$(Z')^2 \le -ZZ''$.  We return to Cauchy--Schwarz,
\begin{equation}
  \Bigl(\sum_j x_jy_j\Bigr)^{\!2} \;\le\; \Bigl(\sum_j x_j^2\Bigr)\Bigl(\sum_j y_j^2\Bigr) .
  \label{eq:s3-cs-generic}
\end{equation}
We set $x_j := \sqrt{|c_j|}\,f_j^{\sfrac14}$ and $y_j := \sqrt{|c_j|}\,\beta_j^2f_j^{-\sfrac34}$, for which
\begin{equation}
  \sum_jx_jy_j = \sum_j\frac{|c_j|\beta_j^2}{\sqrt{f_j}} = 2Z', \qquad
  \sum_jx_j^2 = \sum_j|c_j|\sqrt{f_j} = Z, \qquad
  \sum_jy_j^2 = \sum_j\frac{|c_j|\beta_j^4}{f_j^{\sfrac32}} = -4Z'' ,
\end{equation}
and filling these into~\eqref{eq:s3-cs-generic} gives
\begin{equation}
  (2Z')^2 \le Z\,(-4Z'')
  \;\;\Rightarrow\;\; (Z')^2 + ZZ'' \le 0
  \;\;\Rightarrow\;\; S_\beta' = 2\bigl[(Z')^2 + ZZ''\bigr] \le 0 .
  \label{eq:proof-joint-cs2}
\end{equation}
\medskip\noindent\textbf{$b_{\gamma,2}(\alpha)$ is non-decreasing.}\\
By Lemma~\ref{lem:family}, $J_\alpha(\bq^{(\alpha)}) = Z(\alpha)^2$, that is
\begin{equation}
  b_{\gamma,2}(\alpha) + \alpha\,S_\beta(\alpha) = Z(\alpha)^2 .
\end{equation}
Differentiating both sides in $\alpha$ and using $S_\beta = 2ZZ'$ from~\eqref{eq:proof-joint-sbeta},
\begin{align}
  b_{\gamma,2}'(\alpha) + S_\beta(\alpha) + \alpha\,S_\beta'(\alpha)
    &= 2Z(\alpha)Z'(\alpha) = S_\beta(\alpha), \\
  b_{\gamma,2}'(\alpha) &= \underbrace{-\alpha}_{\le 0}\;\underbrace{S_\beta'(\alpha)}_{\le 0\ \text{by~\eqref{eq:proof-joint-cs2}}}
    \;\ge\; 0 .
  \label{eq:proof-joint-bmono}
\end{align}
Since $v_{\beta^2,2}=S_\beta-\beta_H^2$,~\eqref{eq:proof-joint-cs2} and~\eqref{eq:proof-joint-bmono} prove Lemma~\ref{lem:monotone} for every $\gamma_j>0$.  For every $\gamma_j<0$, Lemma~\ref{lem:sign-reduction} leaves $\bq^{(\alpha)}$ and $S_\beta(\alpha)$ unchanged and gives $|b_{\gamma,2}|=b_{\tilde\gamma,2}$, so the same two results applied to $\tilde\bgamma$ prove it.
\end{proof}

Proposition~\ref{prop:beta-neyman} assumes every $c_k\beta_k\neq0$.  However, if some $\beta_j=0$, the proofs below still hold.  They need only two facts from that proposition, which the next lemma proves regardless of whether some $\beta_j=0$ or not.

\begin{lemma}[The $\beta$-driven bound when some $\beta_j=0$]\label{lem:beta-zero}
For every allocation $\bq$ with every $q_j>0$, and along the family~\eqref{eq:S61},
\begin{equation}
  S_\beta(\bq) \;\ge\; \Bigl(\sum_j|c_j||\beta_j|\Bigr)^{\!2} \ \text{ for every } \bq ,
  \qquad
  \lim_{\alpha\to\infty}S_\beta(\alpha) \;=\; \Bigl(\sum_j|c_j||\beta_j|\Bigr)^{\!2} .
\end{equation}
The first holds with equality only at $\bq=\bq^\star_\beta$, which requires every $\beta_j\neq0$.
\end{lemma}

\begin{proof}
Terms with $\beta_j=0$ contribute nothing to $S_\beta$, and the remaining $q_j$ sum to at most $1$, so Cauchy--Schwarz gives the first,
\begin{equation}
  \Bigl(\sum_j|c_j||\beta_j|\Bigr)^{\!2}
  \;=\; \Bigl(\sum_{\beta_j\neq0}\sqrt{q_j}\,\frac{|c_j||\beta_j|}{\sqrt{q_j}}\Bigr)^{\!2}
  \;\le\; \sum_{\beta_j\neq0}q_j\,\sum_{\beta_j\neq0}\frac{c_j^2\beta_j^2}{q_j}
  \;\le\; S_\beta(\bq) .
  \label{eq:beta-zero-cs}
\end{equation}
For the second, dividing $q^{(\alpha)}_j\propto|c_j|\sqrt{\gamma_j+\alpha\beta_j^2}$ by $\sqrt{\alpha}$ gives $q^{(\alpha)}_j\to\sfrac{|c_j||\beta_j|}{\sum_k|c_k||\beta_k|}$ as $\alpha\to\infty$, which is zero when $\beta_j=0$, so
\begin{equation}
  S_\beta(\alpha) \;=\; \sum_{\beta_j\neq0}\frac{c_j^2\beta_j^2}{q^{(\alpha)}_j}
  \;\longrightarrow\; \sum_{\beta_j\neq0}c_j^2\beta_j^2\,\frac{\sum_k|c_k||\beta_k|}{|c_j||\beta_j|}
  \;=\; \Bigl(\sum_j|c_j||\beta_j|\Bigr)^{\!2} .
\end{equation}
Equality in~\eqref{eq:beta-zero-cs} requires:
\begin{itemize}[topsep=2pt,itemsep=2pt,leftmargin=*]
  \item $q_j\propto|c_j||\beta_j|$ on the terms with $\beta_j\neq0$, for equality in the first inequality (Cauchy--Schwarz),
  \item no probability on the terms with $\beta_j=0$, for equality in the second inequality.
\end{itemize}
These conditions, when taken together, are equivalent to $\bq=\bq^\star_\beta$.  When some $\beta_j=0$, the second condition cannot hold for an allocation with every $q_j>0$, so $S_\beta(\bq)>(\sum_j|c_j||\beta_j|)^2$ strictly.
\end{proof}

\begin{lemma}[From densities to the single-step error]\label{lem:densities-to-error}
Let every $\gamma_j>0$ and $\tfrac12\varphi_{H,H}\le b_{\gamma,2}(\bq^\star_\gamma)$, let the leading coefficients of vanilla \qdrift{} be non-zero, $b_{\gamma,2}(\bq^c)\neq\tfrac12\varphi_{H,H}$ and $S_\beta(\bq^c)\neq\beta_H^2$, and let $\alpha\ge0$ have
\begin{equation}
  b_{\gamma,2}(\alpha)\;\le\; b_{\gamma,2}(\bq^c) ,
  \qquad
  S_\beta(\alpha)\;\le\; S_\beta(\bq^c) .
  \label{eq:dens-hyp}
\end{equation}
Then
\begin{equation}
  \lim_{dt\to0}\frac{\bigl|B_1(\bq^{(\alpha)})\bigr|}{\bigl|B_1(\bq^c)\bigr|} \le 1
  \quad\text{and}\quad
  \lim_{dt\to0}\frac{V_1\bigl(\bq^{(\alpha)}\bigr)}{V_1(\bq^c)} \le 1 .
  \label{eq:dens-ratios}
\end{equation}
\end{lemma}

\begin{proof}
By~\eqref{eq:proof-joint-eqcond} and~\eqref{eq:dens-hyp},
$b_{\gamma,2}(\bq^\star_\gamma) = b_{\gamma,2}(0) \le b_{\gamma,2}(\alpha) \le b_{\gamma,2}(\bq^c)$.
Subtracting $\tfrac12\varphi_{H,H}$ and using $\tfrac12\varphi_{H,H}\le b_{\gamma,2}(\bq^\star_\gamma)$, every term of
\begin{equation}
  0 \;\le\; b_{\gamma,2}(\alpha) - \tfrac12\varphi_{H,H} \;\le\; b_{\gamma,2}(\bq^c) - \tfrac12\varphi_{H,H}
  \label{eq:dens-order}
\end{equation}
is non-negative, so the absolute values obey the same ordering,
\begin{equation}
  \bigl|b_{\gamma,2}(\alpha)-\tfrac12\varphi_{H,H}\bigr| \;\le\; \bigl|b_{\gamma,2}(\bq^c)-\tfrac12\varphi_{H,H}\bigr| .
  \label{eq:sm-shifted-bias}
\end{equation}
By~\eqref{eq:V1B1-general},
\begin{equation}
  B_1 = \bigl(b_{\gamma,2}-\tfrac12\varphi_{H,H}\bigr)dt^2 + \mathcal{O}(dt^3), \qquad
  V_1 = \bigl(S_\beta - \beta_H^2\bigr)dt^2 + \mathcal{O}(dt^3) ,
\end{equation}
so in each ratio the $dt^2$ cancels and the $\mathcal{O}(dt^3)$ terms vanish as $dt\to0$, leaving
\begin{equation}
  \lim_{dt\to0}\frac{|B_1(\bq^{(\alpha)})|}{|B_1(\bq^c)|}
    = \frac{\bigl|b_{\gamma,2}(\alpha)-\tfrac12\varphi_{H,H}\bigr|}{\bigl|b_{\gamma,2}(\bq^c)-\tfrac12\varphi_{H,H}\bigr|} \le 1,
  \qquad
  \lim_{dt\to0}\frac{V_1(\bq^{(\alpha)})}{V_1(\bq^c)} = \frac{S_\beta(\alpha)-\beta_H^2}{S_\beta(\bq^c)-\beta_H^2} \le 1 .
  \label{eq:dens-limits}
\end{equation}
The first inequality follows from~\eqref{eq:sm-shifted-bias}, and the second
from $S_\beta(\alpha)\le S_\beta(\bq^c)$, since $\beta_H$ does not depend on $\bq$.  Both denominators are non-zero, since the lemma assumes $b_{\gamma,2}(\bq^c)\neq\tfrac12\varphi_{H,H}$ and $S_\beta(\bq^c)\neq\beta_H^2$.  The variance denominator is also positive, since $V_1\ge0$ makes its leading coefficient $S_\beta(\bq^c)-\beta_H^2$ non-negative.
\end{proof}

\noindent\textbf{Weak form.}  In the main text, we stated Corollary~\ref{prop:joint-main}.  This was a weaker version with $\le$, for better-or-equal performance relative to vanilla \qdrift{}.  In the below, we prove this, and then show how it can be made strict in Proposition~\ref{prop:joint} with further assumptions beyond those for the weak version.

\begin{corollary}[Joint improvement, Corollary~\ref{prop:joint-main} restated]\label{cor:joint-main-si}
Let the hypotheses of Lemma~\ref{lem:monotone} hold, and let
$\tfrac12\varphi_{H,H}\le b_{\gamma,2}(\bq^\star_\gamma)$ when every $\gamma_j>0$, or
$b_{\gamma,2}(\bq^\star_\gamma)\le\tfrac12\varphi_{H,H}$ when every $\gamma_j<0$, and let the $|\beta_j|$ not all be equal.
If the leading bias and variance coefficients of vanilla \qdrift{}, $q_j^c\propto|c_j|$, are non-zero, then some
$\alpha\ge0$ has
\begin{equation}
  \lim_{dt\to0}\frac{\bigl|B_1(\bq^{(\alpha)})\bigr|}{\bigl|B_1(\bq^c)\bigr|} \le 1
  \quad\text{and}\quad
  \lim_{dt\to0}\frac{V_1\bigl(\bq^{(\alpha)}\bigr)}{V_1(\bq^c)} \le 1 ,
  \label{eq:joint-weak-main-si}
\end{equation}
with $B_1$ and $V_1$ the full single-step
bias and variance, whose expansions are given in~\eqref{eq:V1B1-general}.
\end{corollary}

To do this, we first need the two propositions with the weak and strict inequalities respectively.

\begin{proposition}[Weak joint improvement]\label{prop:joint-weak}
Let every $\gamma_j>0$.  For every $\bq\neq\bq^\star_\beta$ (including vanilla $\bq^c$ when the $|\beta_j|$ are not all equal), some $\alpha\ge0$ has
\begin{equation}
  b_{\gamma,2}\bigl(\bq^{(\alpha)}\bigr)\le b_{\gamma,2}(\bq)
  \quad\text{and}\quad
  S_\beta\bigl(\bq^{(\alpha)}\bigr)\le S_\beta(\bq) .
  \label{eq:joint-weak}
\end{equation}
\end{proposition}

\begin{proof}
We need to show that, for every possible allocation $\bq$, there exists a value for $\alpha$ that is no worse than $\bq$ in both $b_{\gamma,2}$ and $S_\beta$.  For the parameterised allocation we have that $S_\beta(\alpha)$ starts at $S_\beta(0)$ and only decreases as $\alpha$ increases.  We then need to identify the relationship of all other allocations, through $S_\beta(\bq)$, relative to $S_\beta(0)$.  The two cases below are the possibilities.

\medskip\noindent\textbf{Case 1: $S_\beta(\bq)\ge S_\beta(0)$.}\\
First, if $S_\beta(\bq)\ge S_\beta(0)$ take $\alpha=0$, where $b_{\gamma,2}(0)\le b_{\gamma,2}(\bq)$
by~\eqref{eq:proof-joint-eqcond}.  Together with $S_\beta(0)\le S_\beta(\bq)$, both inequalities of~\eqref{eq:joint-weak} hold.

\medskip\noindent\textbf{Case 2: $S_\beta(\bq)<S_\beta(0)$.}\\
By Lemma~\ref{lem:beta-zero}, $S_\beta(\bq)>S_\beta(\bq^\star_\beta)=\lim_{\alpha\to\infty}S_\beta(\alpha)$.  The inequality is strict since we assume $\bq\neq\bq^\star_\beta$.  Then we have:
\begin{equation}
  S_\beta(0) \;>\; S_\beta(\bq) \;>\; \lim_{\alpha\to\infty}S_\beta(\alpha) .
\end{equation}
The function $S_\beta(\alpha)$ is continuous and non-increasing (\eqref{eq:proof-joint-cs2}, which needs only Cauchy--Schwarz), so by the intermediate value theorem some finite $\alpha>0$ has
$S_\beta(\alpha)=S_\beta(\bq)$.  The allocation $\bq^{(\alpha)}$ minimises $J_\alpha$ (Lemma~\ref{lem:family}), so $J_\alpha(\bq^{(\alpha)})\le J_\alpha(\bq)$, that is
\begin{equation}
  b_{\gamma,2}(\alpha)+\alpha S_\beta(\alpha)\;\le\; b_{\gamma,2}(\bq)+\alpha S_\beta(\bq) .
  \label{eq:weak-J}
\end{equation}
With $S_\beta(\alpha)=S_\beta(\bq)$, the terms $\alpha S_\beta(\alpha)$ and $\alpha S_\beta(\bq)$ on the two sides are equal and cancel, which gives
$b_{\gamma,2}(\alpha)\le b_{\gamma,2}(\bq)$, and both inequalities of~\eqref{eq:joint-weak} hold.
\end{proof}

\paragraph{Strict form.}  With the further conditions stated next, the inequalities of~\eqref{eq:joint-weak} at $\bq=\bq^c$ become strict for every $\alpha$ in an interval.  This is not needed for Corollary~\ref{prop:joint-main}, but will be for the strict version of it in Corollary~\ref{cor:joint-sm}.

\begin{proposition}[Existence of joint improvement]\label{prop:joint}
Let
\begin{equation}
  S_\beta(\bq^\star_\gamma) \;>\; S_\beta(\bq^c)
  \label{eq:nontrivial}
\end{equation}
and
\begin{enumerate}[label=(\alph*),topsep=2pt,itemsep=1pt]
  \item $\gamma_j > 0$ for all $j$,
  \item $\bigl(\sum_j|c_j||\beta_j|\bigr)^2 < S_\beta(\bq^c)$, which by Cauchy--Schwarz holds exactly when
    $|\beta_i|\neq|\beta_j|$ for some pair $i,j$,
  \item no $\alpha\ge0$ makes $\gamma_j+\alpha\beta_j^2$ the same for every term, so that $\bq^c$ is not a member of the
    family $\{\bq^{(\alpha)}\}$.
\end{enumerate}
Then there exists $\alpha > 0$ such that
\begin{equation}
  b_{\gamma,2}\!\bigl(\bq^{(\alpha)}\bigr) \;<\; b_{\gamma,2}(\bq^c)
  \quad\text{and}\quad
  S_\beta\!\bigl(\bq^{(\alpha)}\bigr) \;<\; S_\beta(\bq^c) .
  \label{eq:joint-improvement}
\end{equation}
\end{proposition}

The proof shows that $S_\beta(\alpha)$ is strictly decreasing and $b_{\gamma,2}(\alpha)$ non-decreasing.  Hence
there is a unique $\alpha_0 > 0$ with
\begin{equation}
  S_\beta(\alpha) < S_\beta(\bq^c) \iff \alpha > \alpha_0 ,
  \qquad
  b_{\gamma,2}(\alpha) < b_{\gamma,2}(\bq^c) \iff \alpha \in [0,\alpha_1) ,
  \label{eq:s3-crossings}
\end{equation}
for some $\alpha_1 \in (0,\infty]$.  The two thresholds are ordered as $\alpha_0 < \alpha_1$, which comes from
$\bq^{(\alpha_0)}$ being the minimiser of $J_{\alpha_0}$, so the interval $(\alpha_0,\alpha_1)$ is non-empty and
every $\alpha$ in it satisfies~\eqref{eq:joint-improvement}.

\begin{proof}
We use the notation and the results~\eqref{eq:proof-joint-notation}--\eqref{eq:proof-joint-bmono} above.

\medskip\noindent\textbf{Step~1: $b_{\gamma,2}(\alpha) < b_{\gamma,2}(\bq^c)$ near $\alpha = 0$.}\\
Take $\bq = \bq^c$ in~\eqref{eq:proof-joint-eqcond}, so that $q_k \propto |c_k|$.  Equality in~\eqref{eq:proof-joint-eqcond} would then require
$|c_k| \propto |c_k|\sqrt{\gamma_k}$, that is $\gamma_k$ the same for every $k$, which condition~(c) at $\alpha=0$ excludes.  The
inequality is therefore strict,
\begin{equation}
  b_{\gamma,2}(0) \;<\; b_{\gamma,2}(\bq^c) .
  \label{eq:proof-joint-cs1}
\end{equation}
Since $b_{\gamma,2}(\alpha)$ is continuous, the gap persists for small $\alpha$.  Define
\begin{equation}
  \alpha_1 := \sup\bigl\{\alpha > 0 : b_{\gamma,2}(\alpha') < b_{\gamma,2}(\bq^c)\ \text{for all}\ \alpha'\in[0,\alpha]\bigr\} \;\in\; (0,\infty],
\end{equation}
so $b_{\gamma,2}(\alpha) < b_{\gamma,2}(\bq^c)$ for all $\alpha \in [0, \alpha_1)$, possibly with $\alpha_1 = \infty$.

\medskip\noindent\textbf{Step~2: $S_\beta(\alpha)$ is strictly decreasing.}\\
By~\eqref{eq:proof-joint-cs2}, $S_\beta'\le0$.  Equality in~\eqref{eq:s3-cs-generic} needs $x_j \propto y_j$, which after cancelling $\sqrt{|c_j|}$ and
multiplying by $f_j^{\sfrac34}$ reads
\begin{equation}
  f_j^{\sfrac14} \propto \beta_j^2 f_j^{-\sfrac34}
  \qquad\Longleftrightarrow\qquad
  f_j \propto \beta_j^2 .
\end{equation}
Since $f_j = \gamma_j + \alpha\beta_j^2$, this forces $\gamma_j \propto \beta_j^2$, and then $f_j \propto \beta_j^2$
at every $\alpha$, so
\begin{equation}
  q^{(\alpha)}_j \;\propto\; |c_j|\sqrt{f_j} \;\propto\; |c_j||\beta_j|
  \qquad\text{for every } \alpha .
\end{equation}
In particular $\bq^\star_\gamma = \bq^{(0)} = \bq^\star_\beta$, so by Proposition~\ref{prop:beta-neyman} and
condition~(b),
\begin{equation}
  S_\beta(\bq^\star_\gamma) = \Bigl(\sum_j|c_j||\beta_j|\Bigr)^{\!2} \;<\; S_\beta(\bq^c) ,
\end{equation}
which contradicts~\eqref{eq:nontrivial}.  Hence $S_\beta'(\alpha) < 0$.

\medskip\noindent\textbf{Step~3: $S_\beta(\alpha)$ crosses $S_\beta(\bq^c)$ once.}\\
By~\eqref{eq:nontrivial}, $S_\beta(\bq^c)<S_\beta(0)$, and $\bq^c\neq\bq^\star_\beta$ by condition~(b), so $\bq=\bq^c$ falls in Case~2 of the proof of Proposition~\ref{prop:joint-weak}.  That case gives an $\alpha_0>0$ with $S_\beta(\alpha_0)=S_\beta(\bq^c)$.  By Step~2 this $\alpha_0$ is unique, and $S_\beta(\alpha)<S_\beta(\bq^c)$ for every $\alpha>\alpha_0$.

\medskip\noindent\textbf{Step~4: $\alpha_0 < \alpha_1$, so the interval is non-empty.}\\
At $\alpha_0$,~\eqref{eq:weak-J} with $\bq=\bq^c$ gives
\begin{equation}
  b_{\gamma,2}(\alpha_0)+\alpha_0 S_\beta(\alpha_0) \;\le\; b_{\gamma,2}(\bq^c)+\alpha_0 S_\beta(\bq^c) .
\end{equation}
This inequality is strict, because $\bq^{(\alpha_0)}$ is the unique minimiser of $J_{\alpha_0}$ (Lemma~\ref{lem:family}) and $\bq^{(\alpha_0)}\neq\bq^c$ by condition~(c).  The terms $\alpha_0S_\beta(\alpha_0)$ and $\alpha_0S_\beta(\bq^c)$ are equal, since $S_\beta(\alpha_0)=S_\beta(\bq^c)$ by Step~3, so
\begin{equation}
  b_{\gamma,2}(\alpha_0) \;<\; b_{\gamma,2}(\bq^c) .
\end{equation}
For $\alpha\le\alpha_0$, $b_{\gamma,2}$ is non-decreasing by~\eqref{eq:proof-joint-bmono}, so
\begin{equation}
  b_{\gamma,2}(\alpha) \;\le\; b_{\gamma,2}(\alpha_0) \;<\; b_{\gamma,2}(\bq^c)
  \qquad\text{for every } \alpha\in[0,\alpha_0] .
\end{equation}
For $\alpha$ just above $\alpha_0$, $b_{\gamma,2}$ is continuous, so the gap $b_{\gamma,2}(\bq^c)-b_{\gamma,2}(\alpha_0)>0$ persists, and some $\varepsilon>0$ has $b_{\gamma,2}(\alpha)<b_{\gamma,2}(\bq^c)$ for every $\alpha\in[\alpha_0,\alpha_0+\varepsilon]$.  The strict inequality therefore holds on all of $[0,\alpha_0+\varepsilon]$, and since Step~1 defines $\alpha_1$ as the supremum of the $\alpha$ for which it holds on all of $[0,\alpha]$, we get $\alpha_1\ge\alpha_0+\varepsilon>\alpha_0$.  Finally, by Steps~1 and~3,
\begin{equation}
  b_{\gamma,2}(\alpha) \;<\; b_{\gamma,2}(\bq^c) \quad\text{for } \alpha\in[0,\alpha_1) ,
  \qquad
  S_\beta(\alpha) \;<\; S_\beta(\bq^c) \quad\text{for } \alpha\in(\alpha_0,\infty) ,
\end{equation}
so every $\alpha$ in the non-empty interval $(\alpha_0,\alpha_1)$ satisfies~\eqref{eq:joint-improvement}.
\end{proof}

With Propositions~\ref{prop:joint-weak} and~\ref{prop:joint}, we are in a position to prove Corollary~\ref{cor:joint-main-si}, which is Corollary~\ref{prop:joint-main} of the main text, and Corollary~\ref{cor:joint-sm} below.

\begin{proof}[Proof of Corollary~\ref{cor:joint-main-si}]
\medskip\noindent\textbf{Densities.}\\
Let every $\gamma_j>0$.  The first part of the proof is to show that some $\alpha\ge0$ has
\begin{equation}
  b_{\gamma,2}(\alpha) \;\le\; b_{\gamma,2}(\bq^c) ,
  \qquad
  S_\beta(\alpha) \;\le\; S_\beta(\bq^c) .
  \label{eq:cor-dens}
\end{equation}
This follows from Proposition~\ref{prop:joint-weak} at $\bq=\bq^c$, which applies because $\bq^c\neq\bq^\star_\beta$.  Vanilla has $q^c_j\propto|c_j|$ and $\bq^\star_\beta$ has $q^\star_{\beta,j}\propto|c_j||\beta_j|$, and the $|\beta_j|$ are not all equal.

\medskip\noindent\textbf{From densities to $B_1$ and $V_1$.}\\
The second part is to turn~\eqref{eq:cor-dens}, a statement about the densities, into~\eqref{eq:joint-weak-main-si}, a statement about the single-step bias $B_1$ and variance $V_1$.  Lemma~\ref{lem:densities-to-error} does this, and we assume the necessary hypotheses of that lemma here, namely $\tfrac12\varphi_{H,H}\le b_{\gamma,2}(\bq^\star_\gamma)$ and non-zero leading coefficients of vanilla \qdrift{}, $b_{\gamma,2}(\bq^c)\neq\tfrac12\varphi_{H,H}$ and $S_\beta(\bq^c)\neq\beta_H^2$.  With~\eqref{eq:cor-dens} as~\eqref{eq:dens-hyp}, the lemma gives~\eqref{eq:joint-weak-main-si}, that is,
\begin{equation}
  \lim_{dt\to0}\frac{\bigl|B_1(\bq^{(\alpha)})\bigr|}{\bigl|B_1(\bq^c)\bigr|} \;\le\; 1
  \quad\text{and}\quad
  \lim_{dt\to0}\frac{V_1\bigl(\bq^{(\alpha)}\bigr)}{V_1(\bq^c)} \;\le\; 1 .
\end{equation}

\medskip\noindent\textbf{The case $\gamma_j<0\ \forall\,j$.}\\
The two parts above assume every $\gamma_j>0$.  Corollary~\ref{cor:joint-main-si} also covers every $\gamma_j<0$, and for this we apply the case above to $\tilde\bgamma$ of Lemma~\ref{lem:sign-reduction}.  To do so, we need to first show that the assumptions of the case $\gamma_j>0$ also hold for $\tilde\bgamma$.  We find the assumptions hold, as follows.  For every $\gamma_j<0$ the corollary assumes $b_{\gamma,2}(\bq^\star_\gamma)\le\tfrac12\varphi_{H,H}$.  Using $b_{\gamma,2}=-b_{\tilde\gamma,2}$ from~\eqref{eq:s3-negative-gamma} and $\tilde\varphi=-\varphi_{H,H}$, this is equivalent to $\tfrac12\tilde\varphi\le b_{\tilde\gamma,2}(\bq^\star_\gamma)$.  And so we have the following if and only if statements:
\begin{equation}
  b_{\gamma,2}(\bq^\star_\gamma)\le\tfrac12\varphi_{H,H}
  \;\;\Longleftrightarrow\;\;
  -b_{\tilde\gamma,2}(\bq^\star_\gamma)\le-\tfrac12\tilde\varphi
  \;\;\Longleftrightarrow\;\;
  \tfrac12\tilde\varphi\le b_{\tilde\gamma,2}(\bq^\star_\gamma) .
\end{equation}
The second line of~\eqref{eq:s3-negative-gamma}, $b_{\gamma,2}-\tfrac12\varphi_{H,H}=-\bigl(b_{\tilde\gamma,2}-\tfrac12\tilde\varphi\bigr)$, implies that the leading bias coefficient of vanilla is non-zero for $\tilde\bgamma$ exactly when it is for $\bgamma$, and $S_\beta$, which sets the leading variance coefficient, is unchanged.  Therefore,~\eqref{eq:joint-weak-main-si} also holds for $\tilde\bgamma$.  To get back from $\tilde\bgamma$ to $\bgamma$ (and show that~\eqref{eq:joint-weak-main-si} holds not only when $\tilde\bgamma$ is used but also for $\bgamma$ when $\gamma_j<0$), we again use~\eqref{eq:s3-negative-gamma}, which gives $|b_{\gamma,2}(\bq)-\tfrac12\varphi_{H,H}|=|b_{\tilde\gamma,2}(\bq)-\tfrac12\tilde\varphi|$ at every $\bq$.  The limiting bias ratio is therefore the same for $\bgamma$ and $\tilde\bgamma$, and we have:
\begin{equation}
  \frac{\bigl|b_{\gamma,2}(\alpha)-\tfrac12\varphi_{H,H}\bigr|}{\bigl|b_{\gamma,2}(\bq^c)-\tfrac12\varphi_{H,H}\bigr|}
  \;=\;
  \frac{\bigl|b_{\tilde\gamma,2}(\alpha)-\tfrac12\tilde\varphi\bigr|}{\bigl|b_{\tilde\gamma,2}(\bq^c)-\tfrac12\tilde\varphi\bigr|}
  \;\le\; 1 .
\end{equation}
The limiting variance ratio, $\sfrac{(S_\beta(\alpha)-\beta_H^2)}{(S_\beta(\bq^c)-\beta_H^2)}$, does not involve $\bgamma$ at all.  Hence~\eqref{eq:joint-weak-main-si} holds for $\bgamma$.

\end{proof}

Finally, we prove the stricter version of Corollary~\ref{cor:joint-main-si}, which uses strict inequalities.  For this we use Proposition~\ref{prop:joint} instead of Proposition~\ref{prop:joint-weak}.

\begin{corollary}[Joint improvement, strict form of Corollary~\ref{prop:joint-main}]\label{cor:joint-sm}
Assume the hypotheses of Proposition~\ref{prop:joint}, with $\alpha_0$ and $\alpha_1$
as in~\eqref{eq:s3-crossings}, and let the shift $\tfrac12\varphi_{H,H}$ lie below the whole family,
\begin{equation}
  \tfrac12\varphi_{H,H} \;\le\; b_{\gamma,2}(\bq^\star_\gamma) .
  \label{eq:sm-joint-side}
\end{equation}
Then for every $\alpha\in(\alpha_0,\alpha_1)$,
\begin{equation}
  \lim_{dt\to0}\frac{\bigl|B_1(\bq^{(\alpha)})\bigr|}{\bigl|B_1(\bq^c)\bigr|} < 1
  \quad\text{and}\quad
  \lim_{dt\to0}\frac{V_1\bigl(\bq^{(\alpha)}\bigr)}{V_1(\bq^c)} < 1 ,
  \label{eq:sm-joint-improvement}
\end{equation}
with $B_1$ and $V_1$ the full single-step bias and variance of~\eqref{eq:V1B1-general}.
\end{corollary}

\begin{proof}
The proof follows that of Corollary~\ref{cor:joint-main-si}, with Proposition~\ref{prop:joint} in place of Proposition~\ref{prop:joint-weak}.

\medskip\noindent\textbf{Densities.}\\
For every $\alpha\in(\alpha_0,\alpha_1)$, Proposition~\ref{prop:joint} gives~\eqref{eq:cor-dens} with strict inequalities, which is~\eqref{eq:joint-improvement}.

\medskip\noindent\textbf{From densities to $B_1$ and $V_1$.}\\
The strict inequalities make the leading coefficients of vanilla \qdrift{} non-zero,
\begin{equation}
  b_{\gamma,2}(\bq^c) \;>\; b_{\gamma,2}(\alpha) \;\ge\; \tfrac12\varphi_{H,H} ,
  \qquad
  S_\beta(\bq^c) \;>\; S_\beta(\alpha) \;\ge\; \Bigl(\sum_j|c_j||\beta_j|\Bigr)^{\!2} \;\ge\; \beta_H^2 ,
\end{equation}
the first by~\eqref{eq:proof-joint-eqcond} and~\eqref{eq:sm-joint-side}, and the second by Lemma~\ref{lem:beta-zero} and $|\beta_H|\le\sum_j|c_j||\beta_j|$.  Lemma~\ref{lem:densities-to-error} therefore applies and gives~\eqref{eq:sm-joint-improvement} with $\le$ in place of $<$.

It remains to show that both limits are strictly below $1$.  For the bias, $b_{\gamma,2}(\alpha)<b_{\gamma,2}(\bq^c)$ makes the second inequality of~\eqref{eq:dens-order} strict,
\begin{equation}
  0 \;\le\; b_{\gamma,2}(\alpha) - \tfrac12\varphi_{H,H} \;<\; b_{\gamma,2}(\bq^c) - \tfrac12\varphi_{H,H} ,
\end{equation}
so
\begin{equation}
  \lim_{dt\to0}\frac{\bigl|B_1(\bq^{(\alpha)})\bigr|}{\bigl|B_1(\bq^c)\bigr|}
  \;=\; \frac{\bigl|b_{\gamma,2}(\alpha)-\tfrac12\varphi_{H,H}\bigr|}{\bigl|b_{\gamma,2}(\bq^c)-\tfrac12\varphi_{H,H}\bigr|} \;<\; 1 .
\end{equation}
For the variance, $S_\beta(\alpha)<S_\beta(\bq^c)$ and the denominator $S_\beta(\bq^c)-\beta_H^2$ is positive, so
\begin{equation}
  \lim_{dt\to0}\frac{V_1\bigl(\bq^{(\alpha)}\bigr)}{V_1(\bq^c)}
  \;=\; \frac{S_\beta(\alpha)-\beta_H^2}{S_\beta(\bq^c)-\beta_H^2} \;<\; 1 .
\end{equation}
\end{proof}

\subsubsection*{Mixing with vanilla \qdrift{}}
The above assumes we are using pure \statedrift{}-optimised allocations.  However, in practice we use the version with \qdrift{}-weighted mixing~\eqref{eq:tau-mixing},
\begin{equation}
  \bq=\tau\bq^c+(1-\tau)\bq^{\mathsf{SD}},
  \qquad \tau\in(0,1] .
\end{equation}
Both Corollaries~\ref{cor:joint-main-si} and~\ref{cor:joint-sm} also apply in this case.  For simplicity, assume $\gamma_j>0\ \forall\,j$.  To show this, we just need to show that the first part of the proof of Corollary~\ref{cor:joint-main-si},~\eqref{eq:cor-dens}, applies for the mixture $\bq_\tau=\tau\bq^c+(1-\tau)\bq^{(\alpha)}$ in place of $\bq^{(\alpha)}$.  The second part follows directly, and Corollary~\ref{cor:joint-sm} can be shown analogously.  Both densities, $b_{\gamma,2}$ and $S_\beta$, are convex in $\bq$, due to the fact that each of their terms, of the form $\sfrac{c_j^2f_j}{q_j}$ with $f_j\ge0$, is convex in $\bq$.  As such, mixing $\bq^{(\alpha)}$ with vanilla \qdrift{} keeps the densities $\le$ their vanilla versions:
\begin{equation}
  b_{\gamma,2}(\bq_\tau)\;\le\;\tau\,b_{\gamma,2}(\bq^c)+(1-\tau)\,b_{\gamma,2}(\alpha)\;\le\;b_{\gamma,2}(\bq^c),
  \qquad
  S_\beta(\bq_\tau)\;\le\;\tau\,S_\beta(\bq^c)+(1-\tau)\,S_\beta(\alpha)\;\le\;S_\beta(\bq^c) .
\end{equation}

\subsection*{Bounded reweighting and \qdrift{} guarantees}

The allocations of this section are chosen from $\ket{\psi}$ and $O$, and each is a rule $\bq^{\mathsf{SD}}$ that may
place little or no weight on a term: the $\gamma$-driven optimum~\eqref{eq:S11} gives $q^{\mathsf{SD}}_k = 0$ wherever
$\gamma_k = 0$, and the combined family~\eqref{eq:combined-neyman} wherever $\gamma_k = \beta_k = 0$.  We therefore
deploy none of them raw, but mixed with vanilla in a fixed proportion,
\begin{equation}
  q_j \;=\; \tau\,q^c_j + (1-\tau)\,q^{\mathsf{SD}}_j ,
  \qquad \tau \in (0,1] ,
  \label{eq:s3-tau-mixing}
\end{equation}
which is~\eqref{eq:tau-mixing} of the main text.  The mixing weight $\tau$ is the one parameter it introduces, and
two properties fix its role.  Every term keeps at least a fraction $\tau$ of its vanilla weight,
$q_j \ge \tau q^c_j$, whatever the rule proposes; and at $\tau = 1$ the mixture returns $\bq^c$ exactly, so vanilla
lies inside the family.  In practice $\tau$ limits how far \statedrift{} can move from vanilla \qdrift{},
and it also lets the allocation inherit the \qdrift{} guarantees.

The \qdrift{} bounds hold for an arbitrary fixed $\bq$, and $\bq$ enters them only through the probability ratio $\omega_j = \sfrac{q^c_j}{q_j}$ of Lemma~\ref{lem:minweight}, as in Ref.~\cite{KGR2023}.  The rotation angle is
our $\theta_j = \sfrac{c_j\,dt}{q_j}$ and the average channel is our $\mathcal{K}$ of~\eqref{eq:s2-meanchannel}, so
a depth-$N$ run is $\mathcal{K}^N$ at $dt = \sfrac{T}{N}$.

On states, write $\mathcal{A}_j:=-i[P_j,\cdot\,]$ and $\mathcal{A}:=\sum_jc_j\mathcal{A}_j$, so that $\|\mathcal{A}_j\|_\diamond\le2$ and $\|\mathcal{A}\|_\diamond\le2\lambda$.  One exact step is $e^{dt\,\mathcal{A}}$ and one \qdrift{} step is $\sum_jq_j\,e^{\theta_j\mathcal{A}_j}$.  Since $\sum_jq_j\theta_j\mathcal{A}_j=dt\,\mathcal{A}$, the first-order terms cancel, and the Taylor remainder $\|e^{\mathcal{B}}-1-\mathcal{B}\|_\diamond\le\tfrac12\|\mathcal{B}\|_\diamond^2e^{\|\mathcal{B}\|_\diamond}$, applied to each exponential, gives
\begin{equation}
  \Bigl\|e^{dt\,\mathcal{A}}-\sum_jq_j\,e^{\theta_j\mathcal{A}_j}\Bigr\|_\diamond
  \;\le\; 2\lambda^2dt^2e^{2\lambda\,dt}+2\,dt^2\sum_j\frac{c_j^2}{q_j}\,e^{2|\theta_j|}
  \;\le\; 2\lambda^2dt^2\bigl(1+\Expect_{\bq^c}[\omega]\bigr)\,e^{2\lambda\,dt\max_j\omega_j} ,
  \label{eq:s4-step-bound}
\end{equation}
using $\sum_jc_j^2/q_j=\lambda^2\Expect_{\bq^c}[\omega]$, $|\theta_j|=\lambda\,dt\,\omega_j$ and $\max_j\omega_j\ge1$.  Both channels are contractive in the diamond norm, so $N$ steps err by at most $N$ times one step.  In our notation, the allocation enters the bias of the mean channel and the depth and
realisations $NM$ sufficient for a target accuracy~\cite{KGR2023} through one factor each:
\begin{align}
  \bigl\|e^{-iHT}(\cdot)\,e^{iHT} - \mathcal{E}^N\bigr\|_\diamond
    &\;\le\; \frac{2\lambda^2T^2}{N}\,\bigl(1+\Expect_{\bq^c}[\omega]\bigr)\,e^{2\lambda T\max_j\omega_j/N},
    \label{eq:kgr-bias}\\
  NM \;&\propto\; \bigl(1+\max_j\omega_j\bigr)^2 ,
    \label{eq:kgr-conc}
\end{align}
The omitted terms in the proportionality do not depend on $\bq$.  Vanilla \qdrift{} has $\omega_j = 1$, so there
$1+\Expect_{\bq^c}[\omega] = 2$ and $\bigl(1+\max_j\omega_j\bigr)^2 = 4$.  Ref.~\cite{KGR2023} takes $\omega$ as
whatever the chosen $\bq$ produces, evaluating $\Expect_{\bq^c}[\omega]$ for the particular cost-based rule the
authors propose.  The $\tau$-mixing bounds it for any rule.

\begin{lemma}[Bounded reweighting, Lemma~\ref{lem:minweight} restated]\label{lem:minweight-si}
Let $\tau\in(0,1]$, let $\bq^{\mathsf{SD}}$ be any probability vector with $q^{\mathsf{SD}}_j \ge 0$, and let
\begin{equation}
  q_j \;=\; \tau\,q^c_j + (1-\tau)\,q^{\mathsf{SD}}_j
  \label{eq:s3-mix-lemma}
\end{equation}
be its mixture with vanilla \qdrift{}.  With $\omega_j := \sfrac{q^c_j}{q_j}$ the probability ratio relative to
vanilla,
\begin{equation}
  \max_j \omega_j \le \frac{1}{\tau},
  \qquad
  \Expect_{\bq^c}[\omega]=\sum_j\frac{(q_j^c)^2}{q_j}\le\frac{1}{\tau}.
  \label{eq:omega-bound-si}
\end{equation}
\end{lemma}

\begin{proof}
Since $q^{\mathsf{SD}}_j \ge 0$, every term of~\eqref{eq:s3-mix-lemma} obeys $q_j \ge \tau q^c_j$, so for every $j$
carrying $c_j\neq0$,
\begin{equation}
  \omega_j \;=\; \frac{q^c_j}{q_j} \;\le\; \frac{q^c_j}{\tau q^c_j} \;=\; \frac{1}{\tau},
\end{equation}
so $\max_j\omega_j \le \sfrac1\tau$.  Averaging this over $\bq^c$, and using $\sum_j q^c_j = 1$,
\begin{equation}
  \Expect_{\bq^c}[\omega] \;=\; \sum_j q^c_j\,\omega_j \;\le\; \max_j\omega_j \;\le\; \frac{1}{\tau} .
\end{equation}

Neither bound in~\eqref{eq:omega-bound-si} can be improved: $\max_j\omega_j = \sfrac1\tau$ whenever some $q^{\mathsf{SD}}_k = 0$, and allocation rules and Hamiltonians can be found where $\Expect_{\bq^c}[\omega]$ comes arbitrarily close to $\sfrac1\tau$.  The bound on $\max_j\omega_j$ is
saturated by any rule allocating $q^{\mathsf{SD}}_k = 0$ for some $k$,
\begin{equation}
  q^{\mathsf{SD}}_k = 0
  \;\;\Longrightarrow\;\;
  q_k = \tau q^c_k
  \;\;\Longrightarrow\;\;
  \omega_k = \frac{1}{\tau} ,
  \label{eq:s3-omega-tight}
\end{equation}
The bound on $\Expect_{\bq^c}[\omega]$ is never saturated for $\tau<1$, since $\sum_j q^{\mathsf{SD}}_j = 1$ forces
$q^{\mathsf{SD}}_m > 0$ for at least one $m$, and then
\begin{equation}
  \omega_m \;=\; \frac{q^c_m}{\tau q^c_m + (1-\tau)q^{\mathsf{SD}}_m} \;<\; \frac{1}{\tau}
  \qquad\Longrightarrow\qquad
  \Expect_{\bq^c}[\omega] \;=\; \sum_j q^c_j\,\omega_j \;<\; \frac{1}{\tau} .
  \label{eq:s3-omega-strict}
\end{equation}
It is approached nonetheless: concentrating $\bq^{\mathsf{SD}}$ on a single term $m$ leaves $\omega_j = \sfrac1\tau$
on every other term, so $\Expect_{\bq^c}[\omega]\to\sfrac1\tau$ as $q^c_m\to0$.  For a particular $\bq$ both
quantities in~\eqref{eq:omega-bound-si} are computable from $\bq$ alone, and are typically well below
$\sfrac1\tau$.
\end{proof}

The bounds~\eqref{eq:kgr-bias}--\eqref{eq:kgr-conc} apply to $\bq$ under two conditions, and both
hold here.
\begin{enumerate}[label=(\roman*),leftmargin=*,topsep=3pt,itemsep=6pt]
  \item \emph{The allocation is fixed rather than adaptive.}  The bounds ask for one channel composed
    $N$ times, not for $\bq$ to be independent of the input state.  Ours is fixed at $t=0$ and held for all $N$
    steps and all $M$ realisations, so it is state-\emph{chosen} but not state-\emph{adaptive}.  A rule that
    re-derived $\bq$ from the evolving state would make the run a product of different channels, and in general
    not even a fixed sequence of them, since the distribution at step $k$ would depend on which terms were
    sampled before it.  The bounds above then do not apply as stated.  This is the case for the adaptive rules
    of Refs.~\cite{Fan2025adaptive,Wu2025fluctuation}, which recompute the distribution from the current state
    at every step.
  \item \emph{Every weight is strictly positive.}  The mixture~\eqref{eq:s3-mix-lemma} gives
    $q_j\ge\tau q^c_j>0$ on every term carrying $c_j\neq0$, so no weight vanishes and
    $\Expect_{\bq^c}[\omega]$ is finite, bounded by $\sfrac1\tau$ in~\eqref{eq:omega-bound-si}.  At
    $q_k=0$ the term $c_kP_k$ is never sampled, so the mixture reproduces $H-c_kP_k$ to first order rather
    than $H$, and no bound for $H$ applies to it at all.
\end{enumerate}

By Cauchy--Schwarz $\Expect_{\bq^c}[\omega]\ge1$, with equality only at $\bq=\bq^c$, so vanilla
\qdrift{} minimises the right-hand side of~\eqref{eq:kgr-bias}~\cite{KGR2023}.  No state-aware allocation can
tighten this worst-case bound.  The question is how much reweighting loosens it.

Equation~\eqref{eq:kgr-bias} follows from~\eqref{eq:s4-step-bound}, and Ref.~\cite{KGR2023} gives~\eqref{eq:kgr-conc}.  With $\rho=\ket{\psi}\bra{\psi}$, the bias of the estimator of $\braket{O(T)}$ inherits the channel bound,
\begin{equation}
\begin{split}
  |B_N|
  &\;=\; \Bigl|\mathrm{tr}\Bigl[O\,\bigl(e^{-iHT}\rho\,e^{iHT}-\mathcal{E}^N(\rho)\bigr)\Bigr]\Bigr|
  \;\le\; \|O\|\,\bigl\|e^{-iHT}(\cdot)\,e^{iHT} - \mathcal{E}^N\bigr\|_\diamond \\
  &\;\le\; \|O\|\,\frac{2\lambda^2T^2}{N}\,\bigl(1+\Expect_{\bq^c}[\omega]\bigr)\,e^{2\lambda T\max_j\omega_j/N} .
\end{split}
\end{equation}
Combining this and~\eqref{eq:kgr-conc} with our bounds~\eqref{eq:omega-bound-si}, $\Expect_{\bq^c}[\omega]\le\sfrac1\tau$ and $\max_j\omega_j\le\sfrac1\tau$, gives
\begin{equation}
  |B_N| \;\le\; \|O\|\,\frac{2\lambda^2T^2}{N}\Bigl(1+\frac{1}{\tau}\Bigr)\,e^{2\lambda T/(\tau N)} ,
  \qquad
  NM \;\propto\; \bigl(1+\max_j\omega_j\bigr)^2 \;\le\; \Bigl(1+\frac{1}{\tau}\Bigr)^{\!2} .
\end{equation}
As such, write $\bar B_N(\bq):=\|O\|\,\frac{2\lambda^2T^2}{N}\bigl(1+\Expect_{\bq^c}[\omega]\bigr)$ for the bias bound of an allocation $\bq$ without its exponential factor, which tends to one as $N$ grows, and $(NM)(\bq)$ for its cost in~\eqref{eq:kgr-conc}.  Vanilla \qdrift{} has $\omega_j=1$ for every $j$, so $\bar B_N(\bq^c)=2\|O\|\,\frac{\lambda^2T^2}{N}$ and $(NM)(\bq^c)\propto4$, and we have
\begin{align}
  \frac{\bar B_N(\bq)}{\bar B_N(\bq^c)} &\;=\; \frac{1+\Expect_{\bq^c}[\omega]}{2} \;\le\; \frac{1}{2}\Bigl(1+\frac{1}{\tau}\Bigr) \;=\; 3 \quad\text{at } \tau=0.2 , \\
  \frac{(NM)(\bq)}{(NM)(\bq^c)} &\;=\; \frac{\bigl(1+\max_j\omega_j\bigr)^2}{4} \;\le\; \frac{1}{4}\Bigl(1+\frac{1}{\tau}\Bigr)^{\!2} \;=\; 9 \quad\text{at } \tau=0.2 .
\end{align}
As such, this reweighting and mixing with vanilla \qdrift{} allows us to retain the bounds on vanilla \qdrift{}, and never worsen them by more than constant factors which depend only on $\tau$, i.e.\ are problem-independent.  Reducing $\tau$ gives our allocation more freedom at the expense of worse worst-case bounds.  Note also that for any allocation, $\Expect_{\bq^c}[\omega]$ and $\max_j\omega_j$ can be computed from $\bq$ before any run, and we find them typically to be much smaller than $\sfrac1\tau$.

\section{Deployed allocation and its optimisation}\label{app:S5}

In this section, we describe in greater detail the general allocations for $N>1$, the optimisation objective
function and constraints, along with relevant gradients for gradient-based optimisation.

\subsection*{Depth-resolved allocation}

At $N=1$ the family~\eqref{eq:combined-neyman} suffices, since the cross-step densities vanish.  For $N>1$ the variance
expansion~\eqref{eq:VN_full},
\begin{equation}
\begin{split}
  V_N &= Nv_{\beta^2,2}\,dt^2 + N\bigl[v_{\beta\gamma,3}+(N{-}1)V_{3,2}\bigr]dt^3 \\
  &\quad + N\bigl[v_\gamma-v_{\beta^2,4}+(N{-}1)V_{4,2}+(N{-}1)(N{-}2)V_{4,3}\bigr]dt^4 + \mathcal O(dt^5) ,
\end{split}
\label{eq:s4-VN}
\end{equation}
adds the cross-step groups $V_{3,2}$, $V_{4,2}$ and $V_{4,3}$ defined in Appendix~\ref{app:densities},~\eqref{eq:methods-variance-groups}.

Which densities can be targeted by the parameterised allocation is fixed by Lemma~\ref{lem:combined}: a density
$\sum_j\frac{c_j^2f_j}{q_j}$ has the minimiser $q_j\propto|c_j|\sqrt{f_j}$ only for $f_j\ge0$, and
Lemma~\ref{lem:family} extends this to a positive combination, $q_j\propto|c_j|\sqrt{\sum_ka_kf^{(k)}_j}$.  To
$dt^4$, the parameterisation~(\ref{eq:joint-N}, \ref{eq:s4-joint-N}) has five parameter vectors,
$\boldsymbol\rho$, $\boldsymbol\alpha$, $\boldsymbol\mu$, $\boldsymbol\nu$ and $\boldsymbol\nu'$, to target
particular densities, marked below:
\begin{equation}
  b_{\gamma,2} = \sum_j\underbrace{\frac{c_j^2\gamma_j}{q_j}}_{\rho_j},\;\;
  v_\gamma = \sum_j\underbrace{\frac{c_j^4\gamma_j^2}{q_j^3}}_{\rho_j} - \underbrace{b_{\gamma,2}^2}_{\rho_j},\;\;
  v_{\beta^2,2} = \sum_j\underbrace{\frac{c_j^2\beta_j^2}{q_j}}_{\alpha_j} - \beta_H^2,\;\;
  v_\varphi = \sum_{a,b}\underbrace{\frac{c_a^2c_b^2\varphi_{a,b}^2}{q_aq_b}}_{\mu_a,\mu_b},
  \label{eq:s4-weight-targets}
\end{equation}
\begin{equation}
  v_{\varphi_{H,j}^2,2} = \sum_j\underbrace{\frac{c_j^2\varphi_{H,j}^2}{q_j}}_{\nu_j},\;\;
  v_{\varphi_{j,H}^2,2} = \sum_j\underbrace{\frac{c_j^2\varphi_{j,H}^2}{q_j}}_{\nu'_j} ,
  \label{eq:s4-weight-targets2}
\end{equation}
the last two being the two orderings of the nested commutator, $\varphi_{H,j} = \varphi_{H,P_j}(O)$ and
$\varphi_{j,H} = \varphi_{P_j,H}(O)$ of~\eqref{eq:H-sums}, which are different quantities and enter
$V_{4,3}$ with the same prefactor.  Note that the term $v_\varphi$ is not of the form $\sum_j\sfrac{c_j^2f_j}{q_j}$ of
Lemma~\ref{lem:combined}, as it couples two probabilities from two different terms, $a$, $b$.  However, its
derivative still has the same shape as the terms which are in the required form, e.g.
\begin{equation}
  \frac{\partial b_{\gamma,2}}{\partial q_k} \;=\; -\frac{c_k^2}{q_k^2}\,\gamma_k ,
  \label{eq:s4-db-gamma}
\end{equation}
and we have
\begin{equation}
  \frac{\partial v_\varphi}{\partial q_k} \;=\; -\frac{c_k^2}{q_k^2}\sum_b w_b\bigl(\varphi_{k,b}^2+\varphi_{b,k}^2\bigr),
  \qquad w_b=\frac{c_b^2}{q_b} ,
  \label{eq:s4-dv-phi}
\end{equation}
with both orderings because $q_k$ appears in $v_\varphi$ in either index.  The optimality condition on the
simplex is therefore the same as for a density in $\sfrac{1}{q_k}$, and the minimiser of $v_\varphi$ satisfies
$q_k\propto|c_k|\bigl(\sum_bw_b(\varphi_{k,b}^2+\varphi_{b,k}^2)\bigr)^{1/2}$, the square-root rule of
Lemma~\ref{lem:combined}.  The factors $w_b$ depend on $\bq$, so this is a self-consistent condition rather than a
closed form.  At uniform $\bq$, $q_b=\sfrac1L$, so $w_b=Lc_b^2\propto c_b^2$, and replacing $w_b$ by $c_b^2$ gives the
$\bq$-independent density
\begin{equation}
  \Phi_j \;:=\; \tfrac12\sum_b c_b^2\bigl(\varphi_{j,b}^2+\varphi_{b,j}^2\bigr) ,
  \label{eq:s4-phibar}
\end{equation}
Therefore, we use $\Phi_j$ rather than an individual $\varphi_{j,b}$ in the parameterised family, weighted by $\mu_j$.  In $v_\gamma$, we analogously have $q_j^{-3}$, so we apply the same justification for the single targeting parameter, $\rho_j$.  The complete family is then
\begin{equation}
  q^{(\bxi)}_j \;\propto\;
  |c_j|\sqrt{\,\rho_j|\gamma_j| + \alpha_j\beta_j^2 + \mu_j\Phi_j + \nu_j\varphi_{H,j}^2
    + \nu'_j\varphi_{j,H}^2\,},
  \qquad
  \rho_j>0,\quad \alpha_j,\mu_j,\nu_j,\nu'_j \ge 0 .
  \label{eq:s4-joint-N}
\end{equation}
The remaining variance densities to $dt^4$ either have no fixed sign or, like $v_{\beta^2,4}$, appear in $V_N$ with a minus sign,
\begin{equation}
  v_{\beta\gamma,3},\;\; v_{\beta^2,4},\;\; v_{\beta\cdot\varphi_{H,j},2},\;\; v_{\beta\cdot\varphi_{j,H},2},\;\;
  v_{\gamma\cdot\varphi_{H,j},3},\;\; v_{\gamma\cdot\varphi_{j,H},3},\;\; v_{\varphi_{\ssub{H,j}}\varphi_{\ssub{j,H}},2},\;\;
  v_{\beta\cdot\varphi},\;\; v_{\beta\cdot\psi},\;\; -2b_{\gamma,2}\varphi_{H,H} ,
\end{equation}
and a per-term contribution that can be negative has no Cauchy--Schwarz minimiser, since it drives the objective to
$-\infty$ as its $q_j\to0$.  These enter the objective~\eqref{eq:mseN} but not the family, which is why the number of weights
we optimise per term stops at five.

As a reminder, the weights $\bxi$ follow from the constrained minimisation~\eqref{eq:deploy-opt}.  We optimise
the mean-squared error, the bias or the variance,
\begin{equation}
  \bxi^\star \;=\; \arg\min_{\bxi}\ \mathcal{F}_N\bigl(\bq^{(\bxi)}\bigr),
  \qquad \mathcal{F}_N \in \Bigl\{\underbrace{B_N^2 + \frac{V_N}{M}}_{\mathrm{MSE}_N},\;\; |B_N|,\;\; V_N\Bigr\},
  \label{eq:s4-deploy}
\end{equation}
subject to
\begin{equation}
  |B_N| \;\le\; (1+\eta)\,\bigl|B_N^{\mathrm{van}}\bigr|,
  \qquad
  V_N \;\le\; (1+\eta)\,V_N^{\mathrm{van}} ,
  \qquad \frac{|c_j|\,dt}{q_j} \;\le\; \theta_{\max}=0.5 ,
  \label{eq:s4-deploy-constraints}
\end{equation}
both evaluated at the same $N$ and $dt$ as the objective.  The bound on the angle keeps all rotations within the range of validity for the expansion.  The margin $\eta$ lets whichever quantity is not
being minimised rise a little above its vanilla value, so that a large gain in one can be bought with a small loss
in the other.  We use $\eta=5\%$ throughout.  Which $\mathcal{F}_N$ is used follows from what the figure
reports: an error at a fixed depth is minimised directly, whereas a gate count or a resolving depth is set by the
bias alone, since $B_N$ falls as $\sfrac{1}{N}$ at fixed $T$ while the variance is reduced by averaging over
realisations, so there we minimise $|B_N|$ under the variance cap and re-solve at every depth.  The constraint is
written on $|B_N|$ rather than $B_N$ because the bias is signed and changes sign as the allocation moves, so a
constraint on $B_N$ would be satisfied by an allocation that overshoots the cancellation and lands at a large bias
of the opposite sign.  The variance needs no modulus.

\paragraph{Single step.}
At $N=1$ the cross-step quantities $\Phi_j$, $\varphi_{H,j}$ and $\varphi_{j,H}$ no longer enter the objective, so
$\mu_j=\nu_j=\nu'_j=0$ and
\begin{equation}
  q_j \;\propto\; |c_j|\sqrt{\rho_j|\gamma_j| + \alpha_j\beta_j^2}.
\end{equation}
Sharing the parameters, $\rho_j=1$ and $\alpha_j=\alpha$, gives the family~\eqref{eq:combined-neyman} with the
optimum $\alpha^\star(M)$ of Figure~\ref{fig:joint-crossover}.

\paragraph{Energy eigenstate at $\beta_j = 0$.}
Here $\ket{\psi} = \ket{\psi_0}$ with $H\ket{\psi_0} = E\ket{\psi_0}$, so $\varphi_{H,j}=0$ and
$v_{\beta^2,2}=0$, and only $\rho_j$, $\mu_j$ and $\nu'_j$ are active,
\begin{equation}
  q_j \;\propto\; |c_j|\sqrt{\rho_j|\gamma_j| + \mu_j\Phi_j + \nu'_j\varphi_{j,H}^2},
\end{equation}
which is~\eqref{eq:joint-beta0}.  At $N=1$ with $O = H$, $\varphi_{H,H} = 0$ and sign-definite $\bgamma$, the choice $\rho_j=1$ gives
$\bq^\star_\gamma$ for $\bq^{\mathsf{SD}}$.  It minimises both $|b_{\gamma,2}|$ (Lemma~\ref{lem:combined} at $f_j = |\gamma_j|$) and $v_\gamma$
(Proposition~\ref{prop:real-neyman}), so it minimises $\mathrm{MSE}_1 = b_{\gamma,2}^2\,dt^4 + \sfrac{v_\gamma\,dt^4}{M}$ for every $M$, so the
$\gamma$-driven allocation does not depend on $M$ there.

\paragraph{Per-term weights.}
Since the weights are free and positive the bracket takes any positive value term by term, so with the
$\tau$-mixing the family reaches every allocation with $q_j > \tau q^c_j$, provided $|\gamma_j|+\beta_j^2+\Phi_j+\varphi_{H,j}^2+\varphi_{j,H}^2>0$ for every $j$.  For the $\gamma$-driven molecular
quenches $\beta_j=0$, so $\alpha_j$ drops and the deployed weights are
\begin{equation}
  q_j \;\propto\; |c_j|\sqrt{\rho_j|\gamma_j| + \mu_j\Phi_j + \nu_j\varphi_{H,j}^2 + \nu'_j\varphi_{j,H}^2}\,,
  \label{eq:joint-perterm}
\end{equation}
with $\mathcal{O}(L)$ parameters.

\paragraph{Shared scalars.}
A further simplification is to share the parameters over all terms $j$.  Setting $\rho_j=1$ and
$\alpha_j=\alpha$, $\mu_j=\mu$, $\nu_j=\nu$, $\nu'_j=\nu'$ leaves four numbers,
\begin{equation}
  q_j \;\propto\; |c_j|\sqrt{|\gamma_j| + \alpha\beta_j^2 + \mu\Phi_j + \nu\varphi_{H,j}^2
    + \nu'\varphi_{j,H}^2}\,,
  \label{eq:joint-shared}
\end{equation}
which is a genuine restriction of~\eqref{eq:s4-joint-N}.

\subsection*{Gradients of the deployed family}

Each objective of~\eqref{eq:s4-deploy} is minimised by gradient descent over the $5L$ weights $\bxi$
of~\eqref{eq:s4-joint-N}.  Since the mean-squared error is $N$-dependent, given the number of steps $N$ we wish
to simulate, we re-optimise the allocation.  With the minimum-weight mixing written out, the \statedrift{}
allocation is
\begin{equation}
  q_j = \tau\,q^c_j+(1-\tau)\,q^{\mathsf{SD}}_j,\qquad
  q^{\mathsf{SD}}_j = \frac{|c_j|\sqrt{f_j}}{\sum_k|c_k|\sqrt{f_k}},\qquad
  f_j := \rho_j|\gamma_j|+\alpha_j\beta_j^2+\mu_j\Phi_j+\nu_j\varphi_{H,j}^2+\nu'_j\varphi_{j,H}^2,
  \label{eq:family-map}
\end{equation}
We assume $f_k>0$ for every $k$.  Differentiating $f_k$ with respect to each weight
$\xi_k\in\{\rho_k,\alpha_k,\mu_k,\nu_k,\nu'_k\}$ gives
\begin{equation}
  \frac{\partial f_k}{\partial\rho_k} = |\gamma_k|, \quad
  \frac{\partial f_k}{\partial\alpha_k} = \beta_k^2, \quad
  \frac{\partial f_k}{\partial\mu_k} = \Phi_k, \quad
  \frac{\partial f_k}{\partial\nu_k} = \varphi_{H,k}^2, \quad
  \frac{\partial f_k}{\partial\nu'_k} = \varphi_{k,H}^2 .
\end{equation}
Differentiating the quotient in~\eqref{eq:family-map},
\begin{equation}
  \frac{\partial q_j}{\partial \xi_k}
   = (1-\tau)\,\frac{q^{\mathsf{SD}}_k}{2f_k}\,\frac{\partial f_k}{\partial \xi_k}
     \Bigl(\delta_{jk} - q^{\mathsf{SD}}_j\Bigr).
  \label{eq:dq-dp}
\end{equation}
The term $-q^{\mathsf{SD}}_j$ comes from the normalisation, so raising one weight changes every $q_j$.  The chain
rule then gives, for any of the objectives
$\mathcal{F}_N\in\bigl\{B_N^2+\sfrac{V_N}{M},\,|B_N|,\,V_N\bigr\}$ of~\eqref{eq:s4-deploy},
\begin{equation}
  \frac{\partial\,\mathcal{F}_N}{\partial \xi_k}
   = \sum_j \frac{\partial\,\mathcal{F}_N}{\partial q_j}\,\frac{\partial q_j}{\partial \xi_k}
   = (1-\tau)\,\frac{q^{\mathsf{SD}}_k}{2f_k}\,\frac{\partial f_k}{\partial \xi_k}
     \Bigl(\frac{\partial\,\mathcal{F}_N}{\partial q_k}
       - \sum_j q^{\mathsf{SD}}_j\,\frac{\partial\,\mathcal{F}_N}{\partial q_j}\Bigr),
  \label{eq:dmseN-dp}
\end{equation}
so a single evaluation of $\sfrac{\partial\mathcal{F}_N}{\partial\bq}$ gives the gradient in all $5L$ weights.  In practice the weights are optimised through their logarithms, $\xi_k=e^{\theta_k}$.  This keeps each parameter (and hence the term in the square root) positive without further constraints.  Other options are possible, for example free optimisation followed by positive projection, which could be explored in future work.

The three objectives differentiate to
\begin{equation}
  \frac{\partial |B_N|}{\partial q_j} = \mathrm{sgn}(B_N)\,\frac{\partial B_N}{\partial q_j},
  \qquad
  \frac{\partial V_N}{\partial q_j},
  \qquad
  \frac{\partial\,\mathrm{MSE}_N}{\partial q_j}
   = \frac{\partial B_N^2}{\partial q_j} + \frac{1}{M}\,\frac{\partial V_N}{\partial q_j} ,
  \label{eq:s4-obj-derivs}
\end{equation}
the first discontinuous where $B_N=0$.  All three need $\sfrac{\partial B_N}{\partial q_j}$,
$\sfrac{\partial B_N^2}{\partial q_j}$ and $\sfrac{\partial V_N}{\partial q_j}$, which we now compute.  Writing
$b_2 := b_{\gamma,2} - \tfrac12\varphi_{H,H}$ for the leading bias density, \eqref{eq:BN_full}
and~\eqref{eq:VN_full} read
\begin{align}
  B_N &= N b_2\,dt^2
    + N\bigl[b_{\beta,3} + (N{-}1)B_{3,2}\bigr]dt^3
    + N\bigl[b_{\gamma,4} + (N{-}1)B_{4,2} + (N{-}1)(N{-}2)B_{4,3}\bigr]dt^4 ,
  \label{eq:s4-BN-repeat}\\
  B_N^2 &= N^2 b_2^2\,dt^4 + \mathcal{O}(dt^5) ,
  \label{eq:s4-BNsq}\\
  V_N &= N v_{\beta^2,2}\,dt^2
    + N\bigl[v_{\beta\gamma,3} + (N{-}1)V_{3,2}\bigr]dt^3
    + N\bigl[v_\gamma - v_{\beta^2,4} + (N{-}1)V_{4,2} + (N{-}1)(N{-}2)V_{4,3}\bigr]dt^4 .
  \label{eq:s4-VN-repeat}
\end{align}
Since $\tfrac12\varphi_{H,H}$, $\beta_H$ and the constant subtracted from each density do not depend on $\bq$,
every density derivative comes from
\begin{equation}
  \frac{\partial}{\partial q_j}\sum_k \frac{c_k^{\,p}a_k}{q_k^{\,p-1}}
   = -(p-1)\,\frac{c_j^{\,p}a_j}{q_j^{\,p}}
  \label{eq:s4-generic-partial}
\end{equation}
and the product rule.  For the bias densities of~\eqref{eq:s4-BN-repeat}, that is $b_2$, $b_{\beta,3}$,
$b_{\gamma,4}$ and the densities making up $B_{3,2}$, $B_{4,2}$ and $B_{4,3}$, we have
\begin{alignat}{3}
  \frac{\partial b_2}{\partial q_j} = \frac{\partial b_{\gamma,2}}{\partial q_j}
    &= -\frac{c_j^2\gamma_j}{q_j^2}, &\quad
  \frac{\partial b_{\beta,3}}{\partial q_j} &= \frac{4}{3}\,\frac{c_j^3\beta_j}{q_j^3}, &\quad
  \frac{\partial b_{\gamma,4}}{\partial q_j} &= \frac{c_j^4\gamma_j}{q_j^4}, \notag\\
  \frac{\partial b_{\beta\Delta,2}}{\partial q_j}
    &= -\frac{c_j^2}{q_j^2}\,\beta_H\bigl(\Delta_j(O)\bigr), &\quad
  \frac{\partial b_{\gamma H,2}}{\partial q_j} &= -\frac{c_j^2}{q_j^2}\,\gamma_j(\mathcal{L}_H(O)), &\quad
  \frac{\partial b_{\gamma HH,2}}{\partial q_j} &= -\frac{c_j^2}{q_j^2}\,\gamma_j(\mathcal{L}_H^2(O)), \notag\\
  \frac{\partial b_{\varphi_{H,j},3}}{\partial q_j}
    &= \frac{4}{3}\,\frac{c_j^3\varphi_{H,j}}{q_j^3}, &\quad
  \frac{\partial b_{\varphi_{j,H},3}}{\partial q_j} &= \frac{4}{3}\,\frac{c_j^3\varphi_{j,H}}{q_j^3}, &\quad
  \frac{\partial b_{\varphi\Delta,2}}{\partial q_j}
    &= -\frac{c_j^2}{q_j^2}\,\varphi_{H,H}\bigl(\Delta_j(O)\bigr), \notag\\
  \frac{\partial b_{\beta\Delta H,2}}{\partial q_j}
    &= -\frac{c_j^2}{q_j^2}\,\beta_H\bigl(\Delta_j(\mathcal{L}_H(O))\bigr), &&&&
  \label{eq:mse-partials}
\end{alignat}
and, for $b_{\gamma\gamma,2}$, which sums over two sampled terms at once, using
$\gamma_{j,k} = \gamma_{k,j}$ (Lemma~\ref{lem:s2-pair}) and $w_k = \sfrac{c_k^2}{q_k}$,
\begin{equation}
  \frac{\partial b_{\gamma\gamma,2}}{\partial q_j}
   = -\frac{2c_j^2}{q_j^2}\sum_k w_k\,\gamma_{j,k}
     - \frac12\Bigl(\frac{\partial b_{\varphi\Delta,2}}{\partial q_j}
       + \frac{\partial b_{\gamma HH,2}}{\partial q_j}\Bigr) .
  \label{eq:s4-dbgg}
\end{equation}
For the variance densities of~\eqref{eq:s4-VN-repeat}, that is $v_{\beta^2,2}$, $v_{\beta\gamma,3}$,
$v_\gamma$, $v_{\beta^2,4}$ and the densities making up $V_{3,2}$, $V_{4,2}$ and $V_{4,3}$, we have
\begin{alignat}{2}
  \frac{\partial v_{\beta^2,2}}{\partial q_j} &= -\frac{c_j^2\beta_j^2}{q_j^2}, &\quad
  \frac{\partial v_{\beta\gamma,3}}{\partial q_j} &= -\frac{4c_j^3\beta_j\gamma_j}{q_j^3}
    + \frac{2\beta_H c_j^2\gamma_j}{q_j^2}, \notag\\
  \frac{\partial v_\gamma}{\partial q_j} &= -\frac{3c_j^4\gamma_j^2}{q_j^4}
    + \frac{2b_{\gamma,2}\,c_j^2\gamma_j}{q_j^2}, &\quad
  \frac{\partial v_{\beta^2,4}}{\partial q_j} &= -\frac{4c_j^4\beta_j^2}{q_j^4}
    + \frac{8\beta_H c_j^3\beta_j}{3q_j^3}, \notag\\
  \frac{\partial v_{\beta\cdot\varphi_{H,j},2}}{\partial q_j}
    &= -\frac{c_j^2\beta_j\varphi_{H,j}}{q_j^2}, &\quad
  \frac{\partial v_{\beta\cdot\varphi_{j,H},2}}{\partial q_j}
    &= -\frac{c_j^2\beta_j\varphi_{j,H}}{q_j^2}, \notag\\
  \frac{\partial v_{\gamma\cdot\varphi_{H,j},3}}{\partial q_j}
    &= -\frac{2c_j^3\gamma_j\varphi_{H,j}}{q_j^3}, &\quad
  \frac{\partial v_{\gamma\cdot\varphi_{j,H},3}}{\partial q_j}
    &= -\frac{2c_j^3\gamma_j\varphi_{j,H}}{q_j^3}, \notag\\
  \frac{\partial v_{\varphi_{H,j}^2,2}}{\partial q_j} &= -\frac{c_j^2\varphi_{H,j}^2}{q_j^2}, &\quad
  \frac{\partial v_{\varphi_{j,H}^2,2}}{\partial q_j} &= -\frac{c_j^2\varphi_{j,H}^2}{q_j^2}, \notag\\
  \frac{\partial v_{\varphi_{\ssub{H,j}}\varphi_{\ssub{j,H}},2}}{\partial q_j}
    &= -\frac{c_j^2\varphi_{H,j}\varphi_{j,H}}{q_j^2}, &\quad
  \frac{\partial v_\varphi}{\partial q_j}
    &= -\frac{c_j^2}{q_j^2}\sum_b w_b\bigl(\varphi_{j,b}^2+\varphi_{b,j}^2\bigr),
  \label{eq:VN-partials}
\end{alignat}
where, for $v_\varphi=\sum_{a,b}w_aw_b\varphi_{a,b}^2$, we get $\sfrac{\partial v_\varphi}{\partial q_j}$ as follows:
\begin{equation}
  \frac{\partial v_\varphi}{\partial q_j}
  = \underbrace{\frac{\partial w_j}{\partial q_j}\sum_b w_b\,\varphi_{j,b}^2}_{a=j}
  + \underbrace{\frac{\partial w_j}{\partial q_j}\sum_a w_a\,\varphi_{a,j}^2}_{b=j} ,
\end{equation}  The two
remaining fourth-order densities $v_{\beta\cdot\varphi}$ of~\eqref{eq:s2-var-densities} and $v_{\beta\cdot\psi}$
of~\eqref{eq:s2-var-densities} are sums and products of the same form and are differentiated the same way.  The
$\bq$-dependent correction carried by $V_{4,2}$ gives
\begin{equation}
  \frac{\partial}{\partial q_j}\bigl[-2b_{\gamma,2}\,\varphi_{H,H}\bigr]
   = -2\varphi_{H,H}\,\frac{\partial b_{\gamma,2}}{\partial q_j}
   = 2\varphi_{H,H}\,\frac{c_j^2\gamma_j}{q_j^2} ,
\end{equation}
while $-\tfrac12\varphi_{H,H}^2$ in $V_{4,2}$ and $-\varphi_{H,H}^2$ in $V_{4,3}$ are $\bq$-independent and
differentiate to zero.  Combining these as in~\eqref{eq:methods-bias-groups}
and~\eqref{eq:methods-variance-groups} gives the following:
\begin{align}
  \frac{\partial B_{3,2}}{\partial q_j} &= -\frac{c_j^2}{2q_j^2}
    \Bigl[\beta_H\bigl(\Delta_j(O)\bigr) + \gamma_j(\mathcal{L}_H(O))\Bigr], \notag\\
  \frac{\partial B_{4,2}}{\partial q_j} &= \frac{2c_j^3}{3q_j^3}\bigl(\varphi_{H,j}+\varphi_{j,H}\bigr)
    - \frac{c_j^2}{q_j^2}\sum_k w_k\,\gamma_{j,k}, \notag\\
  \frac{\partial B_{4,3}}{\partial q_j} &= -\frac{c_j^2}{6q_j^2}
    \Bigl[\varphi_{H,H}\bigl(\Delta_j(O)\bigr) + \gamma_j(\mathcal{L}_H^2(O)) + \beta_H\bigl(\Delta_j(\mathcal{L}_H(O))\bigr)\Bigr], \notag\\
  \frac{\partial V_{3,2}}{\partial q_j} &= -\frac{c_j^2\beta_j}{q_j^2}\bigl(\varphi_{H,j}+\varphi_{j,H}\bigr), \notag\\
  \frac{\partial V_{4,2}}{\partial q_j} &= -\frac{2c_j^3\gamma_j}{q_j^3}\bigl(\varphi_{H,j}+\varphi_{j,H}\bigr)
    - \frac{c_j^2}{2q_j^2}\sum_b w_b\bigl(\varphi_{j,b}^2+\varphi_{b,j}^2\bigr)
    + \frac{2\varphi_{H,H}\,c_j^2\gamma_j}{q_j^2}
    + \frac{\partial v_{\beta\cdot\varphi}}{\partial q_j}, \notag\\
  \frac{\partial V_{4,3}}{\partial q_j} &= -\frac{c_j^2}{3q_j^2}
    \bigl(\varphi_{H,j}^2 + \varphi_{j,H}^2 + \varphi_{H,j}\varphi_{j,H}\bigr)
    + \frac{\partial v_{\beta\cdot\psi}}{\partial q_j} ,
  \label{eq:s4-dgroups}
\end{align}
where the $\tfrac14\bigl(b_{\varphi\Delta,2}+b_{\gamma HH,2}\bigr)$ of $B_{4,2}$ cancels against the same
combination inside $\tfrac12 b_{\gamma\gamma,2}$.  Adding these with the prefactors of~\eqref{eq:s4-BN-repeat} and~\eqref{eq:s4-VN-repeat} gives
$\sfrac{\partial B_N}{\partial q_j}$ and $\sfrac{\partial V_N}{\partial q_j}$, and by the chain rule
on~\eqref{eq:s4-BNsq},
\begin{equation}
  \frac{\partial B_N^2}{\partial q_j} = 2N^2 b_2\,\frac{\partial b_2}{\partial q_j}\,dt^4
   = -2N^2 b_2\,\frac{c_j^2\gamma_j}{q_j^2}\,dt^4 .
  \label{eq:s4-dBNsq-full}
\end{equation}

\section{Gate cost of bias reduction}\label{app:S6}

In the main text we count the reduction in \qdrift{} steps along with the reduction in two-qubit gate
counts required to compile each Pauli term.  In this section we analyse this further.  Here we compare to
cost-aware \qdrift{}~\cite{KGR2023}, and investigate whether adding a cost-aware metric to our optimisation is
helpful.

\subsection*{Gate cost, and cost-aware \qdrift{}}

Each term $P_j$, exponentiated and compiled by a CNOT ladder over its support, costs $C_j = 2(w_j-1)$ two-qubit
gates for a weight-$w_j$ Pauli string~\cite{Whitfield2011Simulation,Seeley2012BravyiKitaev,CAS2026}.  A \qdrift{} step samples $j$ from $\bq$, so its expected cost is
\begin{equation}
  \kappa(\bq) \;=\; \sum_j q_j\,C_j ,
  \label{eq:s5-kappa}
\end{equation}
which is~\eqref{eq:kappa}, and a depth-$N$ trajectory costs $N\kappa(\bq)$ gates in expectation.  Two allocations
at the same $N$ therefore differ in gates, and a choice of \qdrift{} sampling probabilities saves overall gate
depth only if the reduction in the steps $N$ needed to reach a given accuracy is not accompanied by a cancelling
rise in $\kappa$.

Ref.~\cite{KGR2023} weights the sampling itself by this cost, using
\begin{equation}
  q^{\rm cost}_j \;=\; \frac{|c_j|/C_j}{\sum_\ell |c_\ell|/C_\ell}
  \label{eq:s5-kgr}
\end{equation}
with the rotation angles rescaled to $\sfrac{c_j\,dt}{q^{\rm cost}_j}$, and assigning a single-qubit term the cost
$0.1$ and a multi-qubit term its CNOT count.

Since this increases the relative weight of cheap terms, they are sampled more often, and the expected gate
count reduces.  To reach the same accuracy, however, the number of \qdrift{} steps $N$ must rise.  To leading order in $\sfrac1N$, the channel bound~\eqref{eq:kgr-bias} depends on
$\bq$ only through $\Expect_{\bq^c}[\omega] = \sum_j \sfrac{(q^c_j)^2}{q_j} \ge 1$, which is minimised with the $\bq^c$ allocation from vanilla \qdrift{},
and falls as $\sfrac1N$, so setting it equal to a target $\epsilon_\diamond$ and solving for $N$,
\begin{equation}
  N_q \;=\; \frac{2\lambda^2T^2}{\epsilon_\diamond}\bigl(1+\Expect_{\bq^c}[\omega]\bigr),
  \qquad
  \frac{N_q}{N_{\rm van}} \;=\; \frac{1+\Expect_{\bq^c}[\omega]}{2} ,
  \label{eq:s5-kgr-depth}
\end{equation}
the second using $\omega_j = 1$ at vanilla.  A trajectory of $N$ \qdrift{} steps costs $N\kappa$ gates, so the two allocations
compare as
\begin{equation}
  \frac{N_q\,\kappa(\bq^{\rm cost})}{N_{\rm van}\,\kappa(\bq^c)}
   \;=\; \underbrace{\frac{1+\Expect_{\bq^c}[\omega]}{2}}_{\text{steps paid}}\;
     \underbrace{\frac{\kappa(\bq^{\rm cost})}{\kappa(\bq^c)}}_{\text{cost saved}} ,
  \label{eq:s5-kgr-gates}
\end{equation}
and the rule is worth using when the second factor falls by more than the first rises.  Measuring $\kappa$ with
the same $C_j$ that set the allocation~\eqref{eq:s5-kgr}, $\Expect_{\bq^c}[\omega] =
\sfrac{\kappa(\bq^c)}{\kappa(\bq^{\rm cost})}$, so~\eqref{eq:s5-kgr-gates} becomes
\begin{equation}
  \frac{N_q\,\kappa(\bq^{\rm cost})}{N_{\rm van}\,\kappa(\bq^c)}
   \;=\; \frac12\biggl(1+\frac{\kappa(\bq^{\rm cost})}{\kappa(\bq^c)}\biggr) \;\le\; 1 ,
  \label{eq:s5-kgr-result}
\end{equation}
since $\kappa(\bq^{\rm cost})\le\kappa(\bq^c)$ by Cauchy--Schwarz.  This is the result of Ref.~\cite{KGR2023}:
for the same worst-case channel error the rule needs fewer gates than vanilla \qdrift{}.

\subsection*{\statedrift{} versus cost-aware \qdrift{}}

\begin{figure}[H]
\centering
\includegraphics[width=\textwidth]{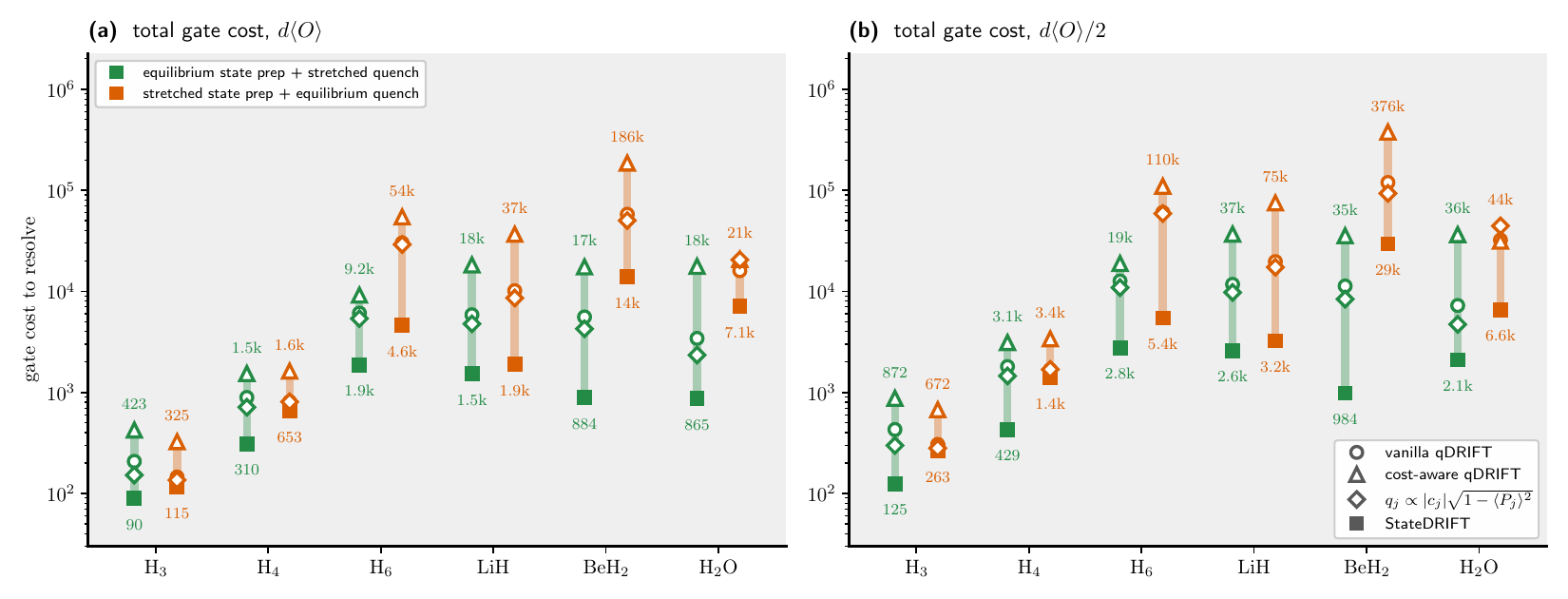}
\caption{\textbf{\statedrift{} versus cost-aware~\cite{KGR2023} and fluctuation-guided~\cite{Wu2025fluctuation} \qdrift{}.}  References, where applicable, of vanilla \qdrift{} (open circles) and \statedrift{} (filled squares) are those of
Figure~\ref{fig:chem} in the main text.  Open triangles are the cost-aware allocation~\eqref{eq:s5-kgr}.  Open diamonds are the allocation $q_j\propto|c_j|\sqrt{1-\braket{P_j}^2}$ of Ref.~\cite{Wu2025fluctuation}, evaluated on the reference state and held fixed at every step.  Total gate cost to
resolve, with $2(w_j-1)$ for a weight-$w_j$ term~\cite{Whitfield2011Simulation,Seeley2012BravyiKitaev} and $0.1$ for a single-qubit term as in Ref.~\cite{KGR2023}, at \textbf{(a)}~$\varepsilon=d\braket{O}$ and
\textbf{(b)}~$\varepsilon=\sfrac{d\braket{O}}{2}$.  The numbers are the costs at the two ends of each
bar.}
\label{fig:cost-compare}
\end{figure}

In Figure~\ref{fig:cost-compare} we compare the cost-aware \qdrift{} allocation~\eqref{eq:s5-kgr} of Ref.~\cite{KGR2023} to \statedrift{}.  Its
resolving depths reach $1.2\times10^6$, far beyond those of Figure~\ref{fig:chem}.  We sample it as in Figure~\ref{fig:chem}, with $M=5000$ trajectories, and take its resolving depth $N^\star_{\rm cost}$ from~\eqref{eq:s4-nstar}, so its gate count is $N^\star_{\rm cost}\,\kappa(\bq^{\rm cost})$.

In general, for our problem setup, the cost-aware allocation rule lowers the expected gate cost per step relative to
vanilla \qdrift{}, but it raises the number of steps $N^\star$ needed to resolve the bias.  For example, for the equilibrium ground state of water evolved under the stretched-bond Hamiltonian, at $\varepsilon=d\braket{O}$ it lowers the cost per step by $15.5\times$ but raises $N^\star$ by
$80\times$, from $1256$ to $100\,738$, so its total gate cost is $5.2\times$ that of vanilla \qdrift{}.  It is built for the worst-case bound, which weights every term by $|c_j|$ alone, whereas the
bias is set by the terms that move the observable, and these are the expensive terms it samples least.
We report the results in a cost model that assigns $2(w_j-1)$ two-qubit gates to a weight-$w_j$ Pauli term~\cite{Whitfield2011Simulation,Seeley2012BravyiKitaev} and, as in Ref.~\cite{KGR2023}, $0.1$ to each single-qubit term, and \statedrift{} requires $2.4$--$36\times$ fewer gates than the cost-weighted allocation on the six molecules compared.  \statedrift{} also needs $1.1$--$11\times$ fewer gates than the fluctuation-guided distribution of Ref.~\cite{Wu2025fluctuation}, evaluated on the reference state and held fixed at every step (Figure~\ref{fig:cost-compare}).

\paragraph{Reweighting the outcome instead.}
Ref.~\cite{CAS2026} makes the sampling cost-aware differently.  It keeps the vanilla channel, samples each trajectory
from a distribution $q$ in place of $\bq^c$, and multiplies the measured outcome by $\omega=\sfrac{q^c}{q}$.  The
bias, and with it $N^\star$, stays at its vanilla value.  It minimises
\begin{equation}
  \mathbb{E}_q\bigl[C_{\rm traj}\bigr]\,\mathbb{E}_q\bigl[\omega^2\bigr],
  \qquad
  C_{\rm traj} \;=\; \sum_{k=1}^N C_{j_k},
  \label{eq:s5-netcost}
\end{equation}
and at the optimum the cost falls by the factor
\begin{equation}
  \frac{\mathbb{E}_{\bq^c}\bigl[\sqrt{C_{\rm traj}}\bigr]^2}{\mathbb{E}_{\bq^c}[C_{\rm traj}]}
  \;=\; 1-\frac{\sigma_C^2}{4N\kappa(\bq^c)^2}+o\bigl(N^{-1}\bigr),
  \label{eq:s5-cas-asym}
\end{equation}
with $\sigma_C^2$ the variance of the cost of one step.  The saving vanishes with depth, as Ref.~\cite{CAS2026} also finds for independent
steps.  Because the bias is unchanged, the reweighted run needs the same number of steps $N^\star$ as vanilla
\qdrift{}, and every trajectory is still a vanilla circuit.  The reweighting only changes how often each circuit is
run, lowering the expected gate count per trajectory while raising the number of trajectories needed to bound the
worst-case variance, with net effect~\eqref{eq:s5-cas-asym} at the vanilla $N^\star$.  On the references of
Figure~\ref{fig:chem}, $\sfrac{\sigma_C^2}{\kappa(\bq^c)^2}$ is at most $3$ and $N^\star$ at least $60$, so this
net effect is below $1.3\%$.

The reweighting is orthogonal to \statedrift{}, which chooses the channel and so the bias and $N^\star$, and the two
compose: keep the \statedrift{} allocation $\bq$, sample trajectories from
$q^\star\propto\sfrac{p}{\sqrt{C_{\rm traj}}}$ with $p=\prod_k q_{j_k}$, and multiply each outcome by
$\sfrac{p}{q^\star}$.  On the benchmarks of Figure~\ref{fig:cost-compare} the saving~\eqref{eq:s5-cas-asym} is then
below $1\%$, for example $0.05\%$ for the equilibrium ground state of LiH evolved under the stretched-bond Hamiltonian, because \statedrift{} concentrates on the expensive terms
and the step cost varies little.

\subsection*{A cost-aware \statedrift{} allocation}

The cost weighting of Ref.~\cite{KGR2023} applied to \statedrift{} would weight the family~\eqref{eq:joint-N} against $C_j$.  To find the right
weighting, take a density $b_{x,2} = \sum_j \sfrac{c_j^2x_j}{q_j}$ with $x_j>0$, as in Lemma~\ref{lem:combined},
and minimise the product $b_{x,2}\,\kappa$ with $\kappa$ of~\eqref{eq:s5-kappa} and $C_j>0$.  By Cauchy--Schwarz,
\begin{equation}
  b_{x,2}\,\kappa
   \;=\; \Bigl(\sum_j \frac{c_j^2x_j}{q_j}\Bigr)\Bigl(\sum_j C_jq_j\Bigr)
   \;\ge\; \Bigl(\sum_j |c_j|\sqrt{x_jC_j}\Bigr)^{\!2} ,
\end{equation}
with equality when $\sfrac{c_j^2x_j}{q_j}\propto C_jq_j$, that is at
\begin{equation}
  q_j \;\propto\; |c_j|\sqrt{\frac{x_j}{C_j}} ,
\end{equation}
so the cost appears as $C_j^{-\sfrac12}$.

Taking that weighting gives the family~\eqref{eq:joint-N} divided by $\sqrt{C_j}$,
\begin{equation}
  q_j \;\propto\; |c_j|\sqrt{\frac{\rho_j|\gamma_j| + \alpha_j\beta_j^2 + \mu_j\Phi_j
  + \nu_j\varphi_{H,j}^2 + \nu'_j\varphi_{j,H}^2}{C_j}}\,.
  \label{eq:s5-cost-sqrt}
\end{equation}

However, this uniform weighting of the cost over each term is problematic.  Each weight in~\eqref{eq:s5-cost-sqrt} targets one per-term quantity,
and each of these enters the $dt^4$ bias~\eqref{eq:BN_full} and variance~\eqref{eq:VN_full} through several
densities, with different powers of $q$ in the denominator:
\begin{center}
\renewcommand{\arraystretch}{1.3}
\begin{tabular}{llllll}
\toprule
weight & quantity & $\sfrac{1}{q_j}$ & $\sfrac{1}{q_j^2}$ & $\sfrac{1}{q_j^3}$ & $\sfrac{1}{q_jq_k}$ \\
\midrule
$\rho_j$ & $\gamma_j$ & $b_{\gamma,2}$
  & $v_{\beta\gamma,3}$, $v_{\gamma\cdot\varphi_{H,j},3}$, $v_{\gamma\cdot\varphi_{j,H},3}$
  & $v_\gamma$, $b_{\gamma,4}$ & $b_{\gamma\gamma,2}$ \\
$\alpha_j$ & $\beta_j$ & $v_{\beta^2,2}$, $v_{\beta\cdot\varphi_{H,j},2}$, $v_{\beta\cdot\varphi_{j,H},2}$
  & $b_{\beta,3}$, $v_{\beta\gamma,3}$ & $v_{\beta^2,4}$ & \\
$\mu_j$ & $\Phi_j$ & & & & $v_\varphi$ \\
$\nu_j$ & $\varphi_{H,j}$
  & $v_{\varphi_{H,j}^2,2}$, $v_{\beta\cdot\varphi_{H,j},2}$, $v_{\varphi_{\ssub{H,j}}\varphi_{\ssub{j,H}},2}$
  & $b_{\varphi_{H,j},3}$, $v_{\gamma\cdot\varphi_{H,j},3}$ & & \\
$\nu'_j$ & $\varphi_{j,H}$
  & $v_{\varphi_{j,H}^2,2}$, $v_{\beta\cdot\varphi_{j,H},2}$, $v_{\varphi_{\ssub{H,j}}\varphi_{\ssub{j,H}},2}$
  & $b_{\varphi_{j,H},3}$, $v_{\gamma\cdot\varphi_{j,H},3}$ & & \\
\bottomrule
\end{tabular}
\end{center}
The Cauchy--Schwarz argument above gives $C_j^{-\sfrac12}$ only for a density in $\sfrac{1}{q_j}$.  Every weight
except $\mu_j$ also acts on densities in $\sfrac{1}{q_j^2}$ or $\sfrac{1}{q_j^3}$, and $\mu_j$ acts only on
$v_\varphi$, in $\sfrac{1}{q_jq_k}$, so no single power of $C_j$ is the right cost weighting for any of them.  As in
Ref.~\cite{KGR2023}, a single-qubit term is assigned the cost $C_j=0.1$.

Nevertheless, optimising~\eqref{eq:s5-cost-sqrt} to minimise $|B_N|$ makes no major difference over the cost-blind version.  On LiH, the cost-blind version resolves with a $12.1\times$ step advantage over vanilla for the equilibrium ground state evolved under the stretched-bond Hamiltonian and $15.9\times$ for the stretched ground state evolved under the equilibrium Hamiltonian, while the cost-aware version achieves $12.0\times$ and $15.7\times$.

\section{Truncation error of the fourth-order bias}\label{app:S7}

In this section, we analyse the error of truncating the bias at fourth order.  The allocations of
Figure~\ref{fig:chem} are chosen by minimising $B_N^{(4)}$ of~\eqref{eq:BN_full}, but the bias we report is the
exact $B_N$, estimated by sampling \qdrift{} trajectories~\eqref{eq:s4-nstar}.  The gap between the two is
the error in the formula from truncation.  We bound the terms $B_N^{(4)}$ drops and find how each scales with
$N$ at fixed $T$.  For vanilla \qdrift{}, the exact expansion of
$\mathcal{K}^N$ in $\delta=\mathcal{K}-\mathcal{U}_{dt}$~\cite{Nakaji2024qSWIFT} and the expansion of the bias in
$\sfrac{1}{N}$ at fixed $T$~\cite{Watson2024RandomlyCompiled} are known, and we use both in our notation.  The
bounds are worst-case and serve to identify which parameters control the truncation.

\subsection*{Decomposition of the error}

Proposition~\ref{prop:multistep_bias} gives the bias to order $dt^4$.  To bound what it drops we start from the
exact bias,
\begin{equation}
  B_N \;=\; \bra{\psi}\bigl(\mathcal{K}^N-\mathcal{U}_T\bigr)(O)\ket{\psi} .
  \label{eq:s7-BN-exact}
\end{equation}
Write the mean step as the exact step plus the difference $\delta$ of Lemma~\ref{lem:s2-defect},
$\mathcal{K}=\mathcal{U}_{dt}+\delta$, and expand the $N$-fold product as Ref.~\cite{Nakaji2024qSWIFT} (Eq.~(29)) expands
the exact evolution, with the roles of the exact and \qdrift{} steps exchanged,
\begin{equation}
  \mathcal{K}^N \;=\; \bigl(\mathcal{U}_{dt}+\delta\bigr)^N
  \;=\; \underbrace{\mathcal{U}_{dt}^N}_{\mathcal{U}_T} + \sum_{m=1}^{N}\mathcal{S}_m ,
  \label{eq:s7-KN-expand}
\end{equation}
where $\mathcal{S}_m=M_{m,N-m}(\delta,\mathcal{U}_{dt})$, the mixture function of that reference, is the sum of the
$\binom{N}{m}$ products with $\delta$ at $m$ of the $N$ steps,
\begin{equation}
  \mathcal{S}_m := \sum_{0\le k_1<\dots<k_m\le N-1}
  \mathcal{U}_{(N-1-k_m)dt}\,\delta\,\mathcal{U}_{(k_m-k_{m-1}-1)dt}\,\delta\cdots\delta\,
  \mathcal{U}_{(k_2-k_1-1)dt}\,\delta\,\mathcal{U}_{k_1dt},
  \label{eq:s7-Sm}
\end{equation}
with $k_i$ the step at which the $i$-th factor of $\delta$ acts, counted from the first step applied.  Substituting~\eqref{eq:s7-KN-expand} into~\eqref{eq:s7-BN-exact},
the exact evolution cancels,
\begin{equation}
  B_N \;=\; \Braket{\Bigl(\mathcal{U}_T + \sum_{m=1}^{N}\mathcal{S}_m - \mathcal{U}_T\Bigr)(O)}
  \;=\; \sum_{m=1}^{N}\braket{\mathcal{S}_m(O)} .
  \label{eq:s8-words}
\end{equation}
This holds exactly, for any fixed $\bq$.

The fourth-order bias of Proposition~\ref{prop:multistep_bias} is~\eqref{eq:s8-words} truncated at order $dt^4$.
From $\mathcal{S}_1$ it keeps the terms to $dt^4$, which we write $B^{(4)}_{N,1}$, and by
Lemma~\ref{lem:s2-compose} it keeps from $\mathcal{S}_2$ only $\sfrac{N(N-1)}{2}\,\delta_2^2\,dt^4$ and nothing from
$\mathcal{S}_{m\ge3}$, so
\begin{equation}
  B_N^{(4)} \;=\; B^{(4)}_{N,1} + \frac{N(N-1)}{2}\Braket{\delta_2\bigl(\delta_2(O)\bigr)}\,dt^4 .
\end{equation}  Subtracting this from~\eqref{eq:s8-words}, the error $R_N:=B_N-B_N^{(4)}$ is
\begin{equation}
  R_N = \underbrace{\braket{\mathcal{S}_1(O)} - B^{(4)}_{N,1}}_{(1)}
      + \underbrace{\braket{\mathcal{S}_2(O)} - \frac{N(N-1)}{2}\braket{\delta_2(\delta_2(O))}\,dt^4}_{(2)}
      + \underbrace{\sum_{m\ge3}\braket{\mathcal{S}_m(O)}}_{(3)} .
  \label{eq:s7-RN-parts}
\end{equation}
We collect (2) and (3), which carry two or more factors of $\delta$, as $R_\delta$.  Term (1) comes from
$\mathcal{S}_1$, which by~\eqref{eq:s7-Sm} at $m=1$, writing $s_L:=N{-}1{-}k$ and $s_R:=k$ for the exact steps after
and before $\delta$ in the order of composition, is
\begin{equation}
  \braket{\mathcal{S}_1(O)} \;=\; \sum_{k=0}^{N-1}\braket{\mathcal{U}_{s_Ldt}\,\delta\,\mathcal{U}_{s_Rdt}(O)} .
\end{equation}
The fourth-order bias truncates this in two places: it keeps $\delta$ only to $dt^4$, as
$\delta^{(4)}:=\delta_2\,dt^2+\delta_3\,dt^3+\delta_4\,dt^4$, and it keeps only the first few commutators from the
exact steps on either side.  Adding and subtracting $\delta^{(4)}$ separates the two,
\begin{equation}
  (1) \;=\; \underbrace{\sum_{k=0}^{N-1}\Braket{\mathcal{U}_{s_Ldt}\bigl(\delta-\delta^{(4)}\bigr)\mathcal{U}_{s_Rdt}(O)}}_{R_{\rm step}}
  \;+\; \underbrace{\sum_{k=0}^{N-1}\braket{\mathcal{U}_{s_Ldt}\,\delta^{(4)}\,\mathcal{U}_{s_Rdt}(O)} - B^{(4)}_{N,1}}_{R_H} ,
  \label{eq:s7-split1}
\end{equation}
so that
\begin{equation}
  R_N=R_{\rm step}+R_H+R_\delta .
  \label{eq:s8-split}
\end{equation}
Throughout, for any operator $A$, maps $\mathcal{A}$ and $\mathcal{B}$, and time $t$, we use
\begin{equation}
  |\bra{\psi}A\ket{\psi}|\le\|A\| , \qquad \|\mathcal{A}\,\mathcal{B}\|\le\|\mathcal{A}\|\,\|\mathcal{B}\| , \qquad
  \|\mathcal{K}\|=\|\mathcal{U}_t\|=1 , \qquad \|[H,X]\|\le2\lambda\|X\| ,
  \label{eq:s8-norm-facts}
\end{equation}
the third because $\mathcal{K}$ and $\mathcal{U}_t$ are unital and completely positive, and the last with
$\lambda=\sum_j|c_j|$.

We now bound the three parts for any reference state, observable and fixed $\bq$, as functions of $N$ at fixed
$T=N\,dt$, to compare with the leading bias $\sfrac{T^2}{N}\bigl(b_{\gamma,2}-\tfrac12\varphi_{H,H}\bigr)$
of~\eqref{eq:BN_full}.  The bounds on $R_H$ and $R_\delta$ use
\begin{equation}
  D_{\bq} \;:=\; 2\sum_j\frac{c_j^2}{q_j}+2\lambda^2 \;\ge\; \|\delta_2\| ,
  \label{eq:s7-D}
\end{equation}
which follows from $\delta_2$ of Lemma~\ref{lem:s2-defect} with $\|\Delta_j\|\le2$ and $\|[H,[H,X]]\|\le4\lambda^2\|X\|$.  In the same way, $\|\delta_3\|\le\sfrac43\bigl(\sum_j\sfrac{|c_j|^3}{q_j^2}+\lambda^3\bigr)$ and $\|\delta_4\|\le\sfrac23\bigl(\sum_j\sfrac{c_j^4}{q_j^3}+\lambda^4\bigr)$.

\paragraph{Bounding $R_{\rm step}$.}

\begin{proposition}[Single-step remainder]\label{prop:s7-step}
\begin{equation}
  |R_{\rm step}| \;\le\; \Bigl[\frac{4}{15}\Bigl(\sum_j\frac{|c_j|^5}{q_j^4}+\lambda^5\Bigr)\frac{T^5}{N^4}
    + \frac{4}{45}\sum_j\frac{c_j^6}{q_j^5}\,\frac{T^6}{N^5}\Bigr]\|O\| ,
  \label{eq:s7-Rstep}
\end{equation}
which falls as $N^{-4}$, while the leading $\mathcal{O}(dt^2)$ bias falls as $N^{-1}$.
\end{proposition}

\begin{proof}
We reduce $R_{\rm step}$ to the norm of $\delta-\delta^{(4)}$, write both maps exactly, and bound their difference
term by term.

\emph{Step 1: one step.}  Each term of $R_{\rm step}$ is the expectation of $\delta-\delta^{(4)}$ between two exact
steps, so by~\eqref{eq:s8-norm-facts}
\begin{equation}
  |R_{\rm step}|
  \;\le\; \sum_{k=0}^{N-1}\|\mathcal{U}_{s_Ldt}\|\,\bigl\|\delta-\delta^{(4)}\bigr\|\,\|\mathcal{U}_{s_Rdt}\|\,\|O\|
  \;=\; N\,\bigl\|\delta-\delta^{(4)}\bigr\|\,\|O\| ,
  \label{eq:s7-Rstep-norm}
\end{equation}
and it remains to bound $\|\delta-\delta^{(4)}\|$.

\emph{Step 2: the tail.}  By the exact one-step change~\eqref{eq:s2-rj-exact} of
Lemma~\ref{lem:s2-onestep}, $\mathcal{K}=\mathds{1}+\sum_jq_j\bigl(\sin^2\theta_j\,\Delta_j+\tfrac12\sin2\theta_j\,\mathcal{L}_j\bigr)$
with $\theta_j=\sfrac{c_j\,dt}{q_j}$, and $\delta^{(4)}$ keeps the Taylor polynomials of $\sin^2\theta_j$ and
$\tfrac12\sin2\theta_j$ to $\theta_j^4$ and of $\mathcal{U}_{dt}=e^{dt\,\mathcal{L}_H}$ to $dt^4$, so
\begin{equation}
  \delta-\delta^{(4)} = \sum_jq_j\Bigl[\Bigl(\sin^2\theta_j-\theta_j^2+\frac{\theta_j^4}{3}\Bigr)\Delta_j
  + \Bigl(\frac12\sin2\theta_j-\theta_j+\frac{2\theta_j^3}{3}\Bigr)\mathcal{L}_j\Bigr]
  - \sum_{p\ge5}\frac{dt^p}{p!}\,\mathcal{L}_H^p .
  \label{eq:s7-delta-rem}
\end{equation}
The Lagrange remainders of $\cos2\theta$ and $\sin2\theta$ bound the two brackets by $\sfrac{2\theta_j^6}{45}$
and $\sfrac{2|\theta_j|^5}{15}$, and Taylor's theorem with $\|e^{s\mathcal{L}_H}\|=1$ and $\|\mathcal{L}_H\|\le2\lambda$
bounds the last sum by $\sfrac{(2\lambda\,dt)^5}{5!}$.  With $\|\Delta_j\|,\|\mathcal{L}_j\|\le2$,
\begin{equation}
  \bigl\|\delta-\delta^{(4)}\bigr\|
  \;\le\; \frac{4}{45}\sum_j\frac{c_j^6}{q_j^5}\,dt^6 + \frac{4}{15}\sum_j\frac{|c_j|^5}{q_j^4}\,dt^5
    + \frac{4}{15}\,\lambda^5dt^5 ,
  \label{eq:s7-delta-rem-norm}
\end{equation}
and substituting into~\eqref{eq:s7-Rstep-norm} with $N\,dt^5=\sfrac{T^5}{N^4}$ and $N\,dt^6=\sfrac{T^6}{N^5}$
gives~\eqref{eq:s7-Rstep}.
\end{proof}

\paragraph{Bounding $R_H$.}

\begin{proposition}[Propagator remainder]\label{prop:s7-RH}
\begin{equation}
  |R_H|\;\le\;\Bigl[\|\delta_2\|\,\frac{(2\lambda T)^3}{3!}\,\frac{T^2}{N}
  +\|\delta_3\|\,\frac{(2\lambda T)^2}{2!}\,\frac{T^3}{N^2}
  +\|\delta_4\|\,2\lambda T\,\frac{T^4}{N^3}\Bigr]\|O\| ,
  \label{eq:s8-RH}
\end{equation}
with $\|\delta_2\|\le D_{\bq}$ from~\eqref{eq:s7-D}.
\end{proposition}

\begin{proof}
We write $R_H$ as a sum of Taylor remainders, bound each with Taylor's theorem, and sum over the step
$k=0,\dots,N-1$ at which the factor $\delta$ in $\mathcal{S}_1$~\eqref{eq:s7-Sm} acts.

\emph{Step 1: $R_H$ as a sum of Taylor remainders.}  By~\eqref{eq:s7-split1}, $R_H$ is the sum over $k$ of
$\braket{\mathcal{U}_{s_Ldt}\,\delta^{(4)}\,\mathcal{U}_{s_Rdt}(O)}$, minus $B^{(4)}_{N,1}$.  For the step $k$ at which $\delta$
acts, add an interpolating parameter $t$ to the propagators such that
\begin{equation}
  \mathcal{U}_{s_Ldt}\;\to\;\mathcal{U}_{t\,s_Ldt},\qquad \mathcal{U}_{s_Rdt}\;\to\;\mathcal{U}_{t\,s_Rdt},\qquad t\in[0,1],
  \label{eq:s7-interp}
\end{equation}
which defines
\begin{equation}
  f_p(t) \;:=\; \braket{\mathcal{U}_{t\,s_Ldt}\,\delta_p\,\mathcal{U}_{t\,s_Rdt}(O)} ,
  \label{eq:s7-fp}
\end{equation}
which is real since each $\delta_p$ maps Hermitian operators to Hermitian operators.  Inserting
$\delta^{(4)}=\sum_{p=2}^4\delta_p\,dt^p$, using that $\mathcal{U}_{s_Ldt}$ and the expectation are linear, and
reading off~\eqref{eq:s7-fp} at $t=1$, where the propagators are unscaled,
\begin{equation}
  \braket{\mathcal{U}_{s_Ldt}\,\delta^{(4)}\,\mathcal{U}_{s_Rdt}(O)}
  \;=\; \Braket{\mathcal{U}_{s_Ldt}\Bigl(\sum_{p=2}^4dt^p\,\delta_p\bigl(\mathcal{U}_{s_Rdt}(O)\bigr)\Bigr)}
  \;=\; \sum_{p=2}^4dt^p\,\braket{\mathcal{U}_{s_Ldt}\,\delta_p\,\mathcal{U}_{s_Rdt}(O)}
  \;=\; \sum_{p=2}^4dt^p\,f_p(1) .
\end{equation}
By Lemma~\ref{lem:s2-bias-one}, $dt^p f_p^{(r)}(0)$ is of order $dt^{p+r}$ and the fourth-order bias keeps the
terms with $p+r\le4$, so it keeps only the Taylor polynomial
$\sum_{r=0}^{4-p}\frac{1}{r!}f_p^{(r)}(0)$ of each $f_p$ at $t=1$, and sums it over $k=0,\dots,N-1$ to give
$B^{(4)}_{N,1}$.  Subtracting,
\begin{equation}
  R_H \;=\; \sum_{k=0}^{N-1}\sum_{p=2}^4dt^p\,\Bigl[f_p(1)-\sum_{r=0}^{4-p}\frac{f_p^{(r)}(0)}{r!}\Bigr] .
  \label{eq:s7-RH-exact}
\end{equation}

\emph{Step 2: Taylor's theorem.}  For a function $f$ that is $m+1$ times continuously differentiable on $[0,1]$, Taylor's
theorem with the integral remainder gives
\begin{equation}
  f(1) \;=\; \sum_{r=0}^{m}\frac{f^{(r)}(0)}{r!} + \int_0^1\frac{(1-u)^m}{m!}\,f^{(m+1)}(u)\,du ,
\end{equation}
and since $\int_0^1\frac{(1-u)^m}{m!}\,du=\frac{1}{(m+1)!}$,
\begin{equation}
  \Bigl|f(1)-\sum_{r=0}^{m}\frac{f^{(r)}(0)}{r!}\Bigr| \;\le\; \frac{1}{(m+1)!}\,\sup_{t\in[0,1]}\bigl|f^{(m+1)}(t)\bigr| .
  \label{eq:s7-taylor-rem}
\end{equation}
With $f=f_p$ and $m=4-p$, each bracket in~\eqref{eq:s7-RH-exact} is therefore bounded by the $(5-p)$-th derivative of
$f_p$.

\emph{Step 3: the derivatives of $f_p$.}  Since $\delta_p$ is linear, $f_p(t)$ is linear in each of its two
$t$-dependent factors, the left propagator $\mathcal{U}_{t\,s_Ldt}$ and $\mathcal{U}_{t\,s_Rdt}(O)$, so its derivatives follow the
general Leibniz rule for a product, $(uv)^{(r)}=\sum_{a+b=r}\binom{r}{a}u^{(a)}v^{(b)}$.  The derivatives of each factor
follow from $\mathcal{U}_t=e^{t\,\mathcal{L}_H}$ (proof of Lemma~\ref{lem:s2-defect}) at time $t\,s\,dt$, applied $a$ times,
\begin{equation}
  \frac{d^a}{dt^a}\,\mathcal{U}_{t\,s\,dt}(Y) \;=\; (i\,s\,dt)^a\,\mathcal{U}_{t\,s\,dt}\bigl(\underbrace{[H,\dots[H}_{a},Y]\dots]\bigr) ,
\end{equation}
so $a$ derivatives on the left propagator and $b$ on the right give
\begin{equation}
  f_p^{(r)}(t) = \sum_{a+b=r}\binom{r}{a}\,(i\,s_L\,dt)^a(i\,s_R\,dt)^b\;
  \Braket{\mathcal{U}_{t\,s_Ldt}\Bigl(\underbrace{[H,\dots[H}_{a},\,\delta_p\bigl(\mathcal{U}_{t\,s_Rdt}
  (\underbrace{[H,\dots[H}_{b},O]\dots])\bigr)]\dots]\Bigr)} .
\end{equation}
The propagators have norm one and $\|[H,Y]\|\le2\lambda\|Y\|$, so with $x_L:=2\lambda\,s_L\,dt$ and
$x_R:=2\lambda\,s_R\,dt$ each term is at most $x_L^ax_R^b\,\|\delta_p\|\,\|O\|$ in absolute value.  The exact steps
after and before $\delta$ add to $s_L+s_R=(N-1-k)+k=N-1$ for every $k$, so
$x_L+x_R=2\lambda(N-1)\,dt\le2\lambda N\,dt=2\lambda T$, and
\begin{equation}
  \bigl|f_p^{(r)}(t)\bigr| \;\le\; \sum_{a+b=r}\binom{r}{a}\,x_L^a x_R^b\,\|\delta_p\|\,\|O\|
  \;=\; (x_L+x_R)^r\,\|\delta_p\|\,\|O\| \;\le\; (2\lambda T)^r\,\|\delta_p\|\,\|O\| .
\end{equation}

\emph{Step 4: summing.}  By~\eqref{eq:s7-taylor-rem} with $m=4-p$ and Step~3 at $r=5-p$, each bracket
in~\eqref{eq:s7-RH-exact} is at most $\frac{(2\lambda T)^{5-p}}{(5-p)!}\,\|\delta_p\|\,\|O\|$, for every $k$.  Summing over
the $N$ values of $k$,
\begin{equation}
  |R_H| \;\le\; N\sum_{p=2}^4 dt^p\,\frac{(2\lambda T)^{5-p}}{(5-p)!}\,\|\delta_p\|\,\|O\| ,
\end{equation}
and $N\,dt^p=\sfrac{T^p}{N^{p-1}}$ gives~\eqref{eq:s8-RH}.
\end{proof}

At fixed $T$ the first term of~\eqref{eq:s8-RH} falls as $N^{-1}$, at the same rate
as the leading bias $\frac{T^2}{N}\bigl(b_{\gamma,2}-\tfrac12\varphi_{H,H}\bigr)$, and the other two as $N^{-2}$ and
$N^{-3}$.  At fixed $T$ the bias is a series $B_N=\sum_{j\ge1}\sfrac{A_j}{N^j}$ in $\sfrac1N$
(Ref.~\cite{Watson2024RandomlyCompiled}, Eq.~(12)), and~\eqref{eq:BN_full} keeps each coefficient $A_j$ only to order
$T^4$, so the part of $A_1$ it drops falls as $\sfrac1N$ like the leading bias.  The first term of~\eqref{eq:s8-RH} bounds
these dropped terms of $A_1$: it is the $\delta_2$ contribution to $R_H$, the Taylor remainder of the propagators around $\delta_2$
beyond second order.

\paragraph{Bounding $R_\delta$.}

\begin{proposition}[More than one factor of $\delta$]\label{prop:s7-Rdelta}
\begin{equation}
  |R_\delta| \;\le\; \Bigl[\frac{(N\|\delta\|)^2}{2!}\,e^{N\|\delta\|} + \|\delta_2\|^2\,\frac{T^4}{2N^2}\Bigr]\|O\| ,
  \label{eq:s8-Rdelta}
\end{equation}
with $\|\delta_2\|\le D_{\bq}$ from~\eqref{eq:s7-D}.
\end{proposition}

\begin{proof}
By~\eqref{eq:s7-RN-parts}, $R_\delta=(2)+(3)$, where (2) is what the fourth-order bias drops from the terms with two
factors of $\delta$, and (3) collects the terms with three or more.  Adding them,
\begin{equation}
  R_\delta \;=\; \sum_{m\ge2}\braket{\mathcal{S}_m(O)} - \frac{N(N-1)}{2}\Braket{\delta_2\bigl(\delta_2(O)\bigr)}\,dt^4 .
  \label{eq:s8-Rdelta-parts}
\end{equation}
We bound the two terms separately with~\eqref{eq:s8-norm-facts}.

\emph{Step 1: the norm of $\delta$.}  Since $\delta^{(4)}=\delta_2\,dt^2+\delta_3\,dt^3+\delta_4\,dt^4$,
\begin{equation}
  \delta \;=\; \delta_2\,dt^2+\delta_3\,dt^3+\delta_4\,dt^4+\bigl(\delta-\delta^{(4)}\bigr) ,
\end{equation}
and the triangle inequality gives
\begin{equation}
  \|\delta\| \;\le\; \|\delta_2\|\,dt^2+\|\delta_3\|\,dt^3+\|\delta_4\|\,dt^4+\bigl\|\delta-\delta^{(4)}\bigr\| ,
  \label{eq:s8-eta}
\end{equation}
with $\|\delta-\delta^{(4)}\|$ bounded in~\eqref{eq:s7-delta-rem-norm}.

\emph{Step 2: the sum over $m\ge2$.}  Ref.~\cite{Nakaji2024qSWIFT} (Appendix~A) bounds the norm of its
mixture function by $\binom{N}{m}$ times the product of the norms of its factors.  For
$\mathcal{S}_m=M_{m,N-m}(\delta,\mathcal{U}_{dt})$, with $\|\mathcal{U}_{dt}\|=1$ from~\eqref{eq:s8-norm-facts} and
$\binom{N}{m}\le\sfrac{N^m}{m!}$, this gives
\begin{equation}
  \Bigl|\sum_{m\ge2}\braket{\mathcal{S}_m(O)}\Bigr| \;\le\; \sum_{m\ge2}\frac{(N\|\delta\|)^m}{m!}\,\|O\|
  \;\le\; \frac{(N\|\delta\|)^2}{2!}\,e^{N\|\delta\|}\,\|O\| ,
  \label{eq:s8-three}
\end{equation}
the second inequality term by term, since $m!\ge2!\,(m-2)!$ for $m\ge2$.
To see how~\eqref{eq:s8-three} scales with $N$ at fixed $T$, multiply~\eqref{eq:s8-eta} by $N$ and use
$N\,dt^p=\sfrac{T^p}{N^{p-1}}$,
\begin{equation}
  N\|\delta\| \;\le\; \|\delta_2\|\,\frac{T^2}{N}+\|\delta_3\|\,\frac{T^3}{N^2}+\|\delta_4\|\,\frac{T^4}{N^3}
  + N\bigl\|\delta-\delta^{(4)}\bigr\| ,
  \label{eq:s8-Ndelta}
\end{equation}
where the last term is $\mathcal{O}(N^{-4})$ by~\eqref{eq:s7-delta-rem-norm}.  So
$N\|\delta\|\le\|\delta_2\|\,\sfrac{T^2}{N}+\mathcal{O}(N^{-2})$, hence $e^{N\|\delta\|}=1+\mathcal{O}(N^{-1})$ and
\begin{equation}
  \frac{(N\|\delta\|)^2}{2!}\,e^{N\|\delta\|} \;\le\; \|\delta_2\|^2\,\frac{T^4}{2N^2}+\mathcal{O}\bigl(N^{-3}\bigr) \;=\; \mathcal{O}\bigl(N^{-2}\bigr) .
  \label{eq:s8-three-leading}
\end{equation}

\emph{Step 3: the retained term.}  By~\eqref{eq:s8-norm-facts}, $|\braket{\delta_2(\delta_2(O))}|\le\|\delta_2\|^2\|O\|$,
and with $\sfrac{N(N-1)}{2}\le\sfrac{N^2}{2}$ and $dt=\sfrac{T}{N}$,
\begin{equation}
  \frac{N(N-1)}{2}\bigl|\Braket{\delta_2\bigl(\delta_2(O)\bigr)}\bigr|\,dt^4 \;\le\; \frac{N^2}{2}\,\|\delta_2\|^2\,\frac{T^4}{N^4}\,\|O\|
  \;=\; \|\delta_2\|^2\,\frac{T^4}{2N^2}\,\|O\| .
  \label{eq:s8-2b}
\end{equation}

\emph{Step 4: summing.}  The triangle inequality applied to~\eqref{eq:s8-Rdelta-parts}, with~\eqref{eq:s8-three}
and~\eqref{eq:s8-2b}, gives~\eqref{eq:s8-Rdelta}.
\end{proof}

At fixed $T$, both terms of~\eqref{eq:s8-Rdelta} scale as $N^{-2}$, the first by~\eqref{eq:s8-three-leading}, so
$R_\delta$ is $\mathcal{O}(N^{-2})$, relative to $\mathcal{O}(N^{-1})$ for the leading bias.

\paragraph{Total truncation error.}

\begin{proposition}[Total truncation error]\label{prop:s8-total}
The error $R_N=B_N-B_N^{(4)}$ of the fourth-order bias obeys
\begin{align}
  |R_N| \;\le\;& \biggl[\|\delta_2\|\,\frac{(2\lambda T)^3}{3!}\,\frac{T^2}{N} \notag\\
  &+ \Bigl[\|\delta_3\|\,\frac{(2\lambda T)^2}{2!}\,T^3 + \frac{\|\delta_2\|^2}{2}\,T^4\Bigr]\frac{1}{N^2}
    + \frac{(N\|\delta\|)^2}{2!}\,e^{N\|\delta\|} \notag\\
  &+ \|\delta_4\|\,2\lambda T\,\frac{T^4}{N^3} \notag\\
  &+ \frac{4}{15}\Bigl(\sum_j\frac{|c_j|^5}{q_j^4}+\lambda^5\Bigr)\frac{T^5}{N^4}
    + \frac{4}{45}\sum_j\frac{c_j^6}{q_j^5}\,\frac{T^6}{N^5}\biggr]\|O\| ,
  \label{eq:s8-RN-total}
\end{align}
where, at fixed $T$, the terms in each of the four lines scale as $N^{-1}$, $N^{-2}$, $N^{-3}$ and $N^{-4}$
or faster respectively.  To leading order in $N$,
this error is
\begin{equation}
  |R_N| \;\le\; \|\delta_2\|\,\frac{(2\lambda T)^3}{3!}\,\frac{T^2}{N}\,\|O\| + \mathcal{O}\bigl(N^{-2}\bigr) .
  \label{eq:s8-RN-leading}
\end{equation}
\end{proposition}

\begin{proof}
By~\eqref{eq:s8-split} and the triangle inequality, $|R_N|\le|R_{\rm step}|+|R_H|+|R_\delta|$.  Adding the bounds of
Propositions~\ref{prop:s7-step}, \ref{prop:s7-RH} and~\ref{prop:s7-Rdelta} and ordering the terms by power of
$\sfrac1N$ gives~\eqref{eq:s8-RN-total}, where the term $\frac{(N\|\delta\|)^2}{2!}e^{N\|\delta\|}$ begins at $N^{-2}$
by~\eqref{eq:s8-three-leading}.
\end{proof}

At fixed $T$, the only term of~\eqref{eq:s8-RN-total} that scales as $N^{-1}$, like the leading bias, is the first,
from $R_H$.  Every other term scales as $N^{-2}$ or faster, so at large $N$ the error bound on the truncation is
dominated by $\|\delta_2\|\,\frac{(2\lambda T)^3}{3!}\,\frac{T^2}{N}\,\|O\|$, and therefore by $\lambda T$.
Against the leading bias $\sfrac{T^2}{N}\bigl|b_{\gamma,2}-\tfrac12\varphi_{H,H}\bigr|$ it tends to
\begin{equation}
  \frac{\|\delta_2\|\,(2\lambda T)^3\,\|O\|}{3!\,\bigl|b_{\gamma,2}-\frac12\varphi_{H,H}\bigr|} .
  \label{eq:s8-RN-ratio}
\end{equation}

From Proposition~\ref{prop:s8-total}, the StateDRIFT allocation leaves a remainder outside the $dt^4$ truncation.
The bound on this remainder vanishes with $N$, but only as $N^{-1}$, at the same rate as the leading bias, so the ratio~\eqref{eq:s8-RN-ratio} is set by $\lambda T$ and does not reduce with depth.  In Appendix~\ref{app:S8} we discuss in detail that despite this, the truncated allocation is useful in
determining StateDRIFT effectiveness.  We also discuss how a practitioner could use it in practice, and future
extensions to potentially close this gap.

\section{Validity metric for \statedrift{}}\label{app:S8}

In this work, we optimise \qdrift{} probability allocations by minimising the expansion of the fourth-order bias
(Eq.~\eqref{eq:BN_full} in the main text).  However, in Figure~\ref{fig:chem} we report the exact bias, estimated by Monte Carlo sampling.  In the previous section, we derived a bound on the truncation error $R_N$.  If the truncated bias is not accurate enough relative to its true value, the allocation will give unreliable results, and may lose to vanilla \qdrift{}.
To alleviate this, here we elaborate on the validity metric from the main text, and its failure modes, to help decide if we can use \statedrift{} in practice for the problem of interest.

By definition, we want the deployed allocation $\bq^\star$ to have a smaller bias than vanilla \qdrift{},
$|B_N(\bq^\star)|<|B_N(\bq^c)|$.  To see when this holds, first, since the optimiser minimises the truncated bias, by construction we have:
\begin{equation}
  \bigl|B_N^{(4)}(\bq^\star)\bigr| \;\le\; \bigl|B_N^{(4)}(\bq^c)\bigr| .
\end{equation}  The
exact bias of each allocation is its fourth-order bias plus the error $R_N$ of
Proposition~\ref{prop:s8-total},
\begin{equation}
  B_N(\bq) \;=\; B_N^{(4)}(\bq) + R_N(\bq) ,
\end{equation}
so the triangle inequality $|a+b|\le|a|+|b|$, applied to $\bq^\star$, and its reverse form $|a+b|\ge|a|-|b|$, applied to
$\bq^c$, give
\begin{align}
  |B_N(\bq^\star)| &\;\le\; |B_N^{(4)}(\bq^\star)| + |R_N(\bq^\star)| , \label{eq:s8-upper}\\
  |B_N(\bq^c)| &\;\ge\; |B_N^{(4)}(\bq^c)| - |R_N(\bq^c)| . \label{eq:s8-lower}
\end{align}
Now using~\eqref{eq:s8-upper} and~\eqref{eq:s8-lower} we can compare the true bias as follows:
\begin{align}
  |B_N(\bq^\star)| \;&\le\; |B_N^{(4)}(\bq^\star)| + |R_N(\bq^\star)| \label{eq:s8-chain}\\
  &\overset{?}{<}\; |B_N^{(4)}(\bq^c)| - |R_N(\bq^c)| \label{eq:s8-chain-mid}\\
  &\le\; |B_N(\bq^c)| , \label{eq:s8-chain-end}
\end{align}
where the question mark marks the inequality to be checked, so $|B_N(\bq^\star)|<|B_N(\bq^c)|$
whenever~\eqref{eq:s8-chain-mid} holds.  Rearranging, it suffices that
\begin{equation}
  \underbrace{|B_N^{(4)}(\bq^c)|-|B_N^{(4)}(\bq^\star)|}_{\text{margin}\;\ge\;0}
  \;>\; \underbrace{|R_N(\bq^\star)|+|R_N(\bq^c)|}_{\text{sum of the remainders}} .
  \label{eq:s8-rank}
\end{equation}

Obviously, we cannot evaluate the right-hand side of~\eqref{eq:s8-rank}, because it would involve exact simulation
of the full \qdrift{} channel.  If the bound of Proposition~\ref{prop:s8-total} was tighter, we could use it to estimate the right-hand side of~\eqref{eq:s8-rank}, which could be compared directly to the left-hand side.  Tightening it so this can be done is a primary direction for future work.  Given this, we instead proposed the heuristic test in the main text, which we elaborate on below.

\subsection*{\statedrift{} on H$_2$O}

\begin{table}[!htb]
\centering

\caption{\textbf{H$_2$O at $T=1$ and $T=\sfrac12$.}  For each preparation, $\lambda T$, losses out of depths, and
medians over depths of the fourth-order bias of vanilla \qdrift{} and its last kept group, of $R_{\rm val}^{(4)}$, of
the remainder of the deployed allocation, $R_N(\bq^\star)$, estimated as its sampled bias minus
$B_N^{(4)}(\bq^\star)$, relative to the fourth-order bias of vanilla \qdrift{}, and of the bias ratio of \statedrift{}
to vanilla \qdrift{}.}
\label{tab:s8-h2o}
\small
\begin{tabular}{lrrrrrrrr}
\toprule
preparation & $T$ & $\lambda T$ & losses & $|B_N^{(4)}(\bq^c)|$ & $|B_N^{(4)}-B_N^{(3)}|(\bq^c)$ & $R_{\rm val}^{(4)}$
  & $\sfrac{|R_N(\bq^\star)|}{|B_N^{(4)}(\bq^c)|}$ & $\sfrac{|B_N(\bq^\star)|}{|B_N(\bq^c)|}$ \\
\midrule
equilibrium & $1$          & $68.1$ & $7/31$  & $0.496$  & $3.12$   & $0.49$ & $1.01$ & $0.65$ \\
            & $\sfrac12$   & $34.1$ & $0/42$  & $0.515$  & $0.402$  & $2.09$ & $0.09$ & $0.08$ \\
stretched   & $1$          & $72.0$ & $27/37$ & $0.0493$ & $0.325$  & $0.15$ & $3.20$ & $1.32$ \\
            & $\sfrac12$   & $36.0$ & $0/41$  & $0.0826$ & $0.0395$ & $2.09$ & $0.10$ & $0.14$ \\
\bottomrule
\end{tabular}
\end{table}

In the main text, we ran \statedrift{} on H$_2$O up to $T=\sfrac12$.  This is because H$_2$O at $T=1$ exhibits a
failure (in the sense of not achieving a lower-than-vanilla-\qdrift{} bias) under \statedrift{}, which is related to the value of $\lambda T$ and the cancellation effect.  Note, it is not a feature of
$\lambda T$ alone, as NH$_3$ has a comparable $\lambda T$ and yet \statedrift{} still finds a lower bias at $T=1$.  In
H$_2$O, $\gamma_j$ are almost equally divided between $\mathrm{sign}(\gamma_j)=+1$ and $\mathrm{sign}(\gamma_j)=-1$.  As a result, the optimiser can be driven to cancel the leading bias altogether:
\begin{equation}
  B_N \;=\; \underbrace{\frac{T^2}{N}\bigl(b_{\gamma,2}-\tfrac12\varphi_{H,H}\bigr)}_{=\,0}
  \;+\; \mathcal{O}\bigl(T^3\bigr) \;+\; R_N .
  \label{eq:s8-cancel}
\end{equation}
$R_N$ is then compared with the remaining $\mathcal{O}(T^3)$ terms and above.  For water at $T=1$ the deployed allocation cancels the kept groups, and the terms it drops are as large as
the whole fourth-order bias of vanilla \qdrift{} (Table~\ref{tab:s8-h2o}), so~\eqref{eq:s8-rank} fails.  Based on this, in the next section we define the heuristic which we use as a
one-sided validity metric---to rule out cases where \statedrift{} should not be used over vanilla \qdrift{} to reduce
the bias.

\subsection*{Validity metric}

For an asymptotic series the error is of the order of the first term dropped as the expansion parameter tends to zero~\cite{DLMF}.  We cannot use this directly in our case because it holds only in the limit $dt\to0$, with an unknown constant, and at fixed $T$ the $dt^2$, $dt^3$ and $dt^4$ groups all scale as $\sfrac1N$, so they are not ordered by size.  However, we still find it to be a useful estimate to approximate the first groups dropped, $dt^5$ and $dt^6$, by the last ones kept of the same parity. In our case, these are the $dt^3$ and $dt^4$ terms which are the
odd and even parity respectively, so we take
\begin{equation}
  \bigl|R_N(\bq)\bigr| \;\approx\; \bigl|B_N^{(3)}(\bq)-B_N^{(2)}(\bq)\bigr|+\bigl|B_N^{(4)}(\bq)-B_N^{(3)}(\bq)\bigr| ,
  \label{eq:s8-R-estimate}
\end{equation}
with $B_N^{(p)}$ the bias~\eqref{eq:BN_full} truncated at order $dt^p$.  Under the reality condition of Lemma~\ref{lem:beta} the odd orders vanish and the $dt^3$ group is zero.  When
the deployed allocation cancels the terms its fourth-order bias keeps, $B_N^{(4)}(\bq^\star)=0$, the limit the optimiser aims for,
and~\eqref{eq:s8-rank} becomes
\begin{equation}
  \bigl|B_N^{(4)}(\bq^c)\bigr| \;>\; \bigl|R_N(\bq^\star)\bigr| + \bigl|R_N(\bq^c)\bigr| .
  \label{eq:s8-rank-cancel}
\end{equation}
We further assume that $|R_N(\bq^\star)|\approx|R_N(\bq^c)|$ and subbing~\eqref{eq:s8-R-estimate}
into~\eqref{eq:s8-rank-cancel} gives:
\begin{equation}
  \begin{aligned}
  \bigl|B_N^{(4)}(\bq^c)\bigr| &\;>\; \bigl|R_N(\bq^\star)\bigr| + \bigl|R_N(\bq^c)\bigr|\\
  &\;\approx\; 2\,\bigl|R_N(\bq^c)\bigr|\\
  &\;\approx\; 2\Bigl(\bigl|B_N^{(3)}(\bq^c)-B_N^{(2)}(\bq^c)\bigr|+\bigl|B_N^{(4)}(\bq^c)-B_N^{(3)}(\bq^c)\bigr|\Bigr) .
  \end{aligned}
  \label{eq:s8-rank-vanilla}
\end{equation}
Rearranging:
\begin{equation}
  R_{\rm val}^{(4)} \;:=\; \frac{\bigl|B_N^{(4)}(\bq^c)\bigr|}{\bigl|B_N^{(3)}(\bq^c) - B_N^{(2)}(\bq^c)\bigr|+\bigl|B_N^{(4)}(\bq^c) - B_N^{(3)}(\bq^c)\bigr|} \;\gtrsim\; 2 ,
  \label{eq:s8-rvan}
\end{equation}
Intuitively, this means that the truncated bias has terms added at subsequent orders which are less and less
impactful (i.e.\ the $dt^3$ and $dt^4$ groups are small relative to the $dt$ and $dt^2$ terms).  If this was not the
case, then the truncated terms are as large as those kept, so the $dt^5$ and beyond orders are also likely to be large
relative to $B_N^{(4)}$.

Computing the metric requires only terms which are already required by the allocation
and optimisation.  Also, we find in practice that across all the instances we test in this work (shown in Table~\ref{tab:s8-rval}) it is able to
separate those problem instances where \statedrift{} has an advantage in bias reduction over vanilla \qdrift{}.

\begin{table}[H]
\centering
\caption{\textbf{Heuristic validity metric.}  Across all problem instances and parameter regimes tested in this work,
every depth with $R_{\rm val}^{(4)}>2$ shows a \statedrift{} advantage in bias.  For $R_{\rm val}^{(4)}\le2$, we may
have a win or a loss (note that most of these cases are still \statedrift{} wins) due to the conservative one-way nature
of the metric.}
\label{tab:s8-rval}

\begin{tabular}{lrrrrrr}
\toprule
 & \multicolumn{2}{c}{molecules} & \multicolumn{2}{c}{lattice, $\beta_j=0$} & \multicolumn{2}{c}{lattice, $\beta_j\neq0$} \\
\cmidrule(lr){2-3}\cmidrule(lr){4-5}\cmidrule(lr){6-7}
$R_{\rm val}^{(4)}$ & wins or ties & loses & wins or ties & loses & wins or ties & loses \\
\midrule
$>2$    & $424$ & $0$  & $54$ & $0$  & $24$ & $0$ \\
$\le2$  & $109$  & $36$ & $55$ & $11$ & $89$ & $7$ \\
\bottomrule
\end{tabular}
\end{table}

Evaluating the test at the fixed vanilla
allocation removes the optimiser and tests the model alone.  Since~\eqref{eq:s8-R-estimate} is a heuristic, the test
follows from~\eqref{eq:s8-rank} only heuristically, and the table is its validation.
With complex amplitudes the odd orders no longer vanish, and the $dt^3$ group in~\eqref{eq:s8-R-estimate}
is then needed.  For the transverse-field Ising chain at $\lambda T=7.5$, the $dt^4$ group alone would give
$R_{\rm val}^{(4)}=47.5$, yet the $dt^4$ bias of vanilla \qdrift{} is about $50$ times its exact value and
\statedrift{} loses by a factor of $14$.  There the $dt^3$ group is about $90$ times the $dt^4$ one, and with it
$R_{\rm val}^{(4)}=0.50$.

\section{Numerical benchmarks}\label{app:S9}

In this section, we give the details, ablations and sensitivity studies behind the
benchmarks of the main text.

\subsection*{Molecular references}

Table~\ref{tab:s7-mol} lists the molecular references of Figure~\ref{fig:chem}.  Every Hamiltonian is the full
STO-3G Hamiltonian at its own geometry, written in its Hartree--Fock orbitals (restricted open-shell for H$_3$) with
PySCF~\cite{Sun2020} and mapped to $n$ qubits by the Jordan--Wigner transformation~\cite{JordanWigner1928}.  The quench
evolves the ground state of one qubit-encoded Hamiltonian under the other. No orbital rotation connects the two geometries.  Each Hamiltonian uses its own converged Hartree--Fock orbitals, with the signs and degenerate orientations that PySCF returns, fixed once and reused for every run.  As such, the quench is a benchmark of the method rather than the physical response of the molecule to a bond stretch.

\begin{table}[h]
\centering
\caption{\textbf{Details of molecular configuration setups for Figure~\ref{fig:chem}.}  The equilibrium (eq) and
$2\times$ stretched (str) bond lengths are given by $R$, and $H_A\to H_B$, with $A,B\in\{\mathrm{eq},\mathrm{str}\}$,
denotes the ground state of $H_A$ evolved under $H_B$ for a time $T$, in atomic units ($\hbar/E_{\rm h}\approx24.2$\,as).  $L$ counts the non-identity Pauli terms,
and $\lambda T$ refers to $H_B$.  H$_k$ are linear chains, BeH$_2$ is linear, and in H$_2$O, NH$_3$ and CH$_4$ every bond is stretched to $2\times$ its equilibrium length with the bond angles held at their equilibrium values.  The prepared state is a UCCSD (SD) or UCCGSD
(GSD) reference (UCC), and $\Delta E$ (mHa) and $F$ are its energy above, and fidelity with, the exact ground state.  All references are singlets except H$_3^{\dagger}$, a doublet with a two-fold
degenerate ground state, for which $F$ is the fidelity with the ground-state subspace, summed over the two
degenerate states.}
\label{tab:s7-mol}
\small
\setlength{\tabcolsep}{3pt}
\begin{tabular*}{\textwidth}{@{\extracolsep{\fill}}lrrcrr rrcrr rrcrr@{}}
\toprule
 & & & & \multicolumn{2}{c}{$R$ (\AA)} & \multicolumn{5}{c}{$H_{\rm eq}\to H_{\rm str}$} & \multicolumn{5}{c}{$H_{\rm str}\to H_{\rm eq}$} \\
\cmidrule(lr){5-6}\cmidrule(lr){7-11}\cmidrule(l){12-16}
Molecule & $n$ & $L$ & $T$ & eq & str & $\lambda T$ & $d\braket{O}$ & UCC & $\Delta E$ & $F$ & $\lambda T$ & $d\braket{O}$ & UCC & $\Delta E$ & $F$ \\
\midrule
H$_3^{\dagger}$ & 6 & 61 & 1 & 0.735 & 1.470 & 3.0 & 0.253 & SD & ${<}10^{-6}$ & 1.0000 & 4.5 & 0.443 & SD & ${<}10^{-6}$ & 1.0000 \\
H$_4$ & 8 & 184 & 1 & 0.735 & 1.470 & 5.7 & 0.540 & SD & 0.009 & 1.0000 & 8.8 & 0.829 & SD & 1.2 & 0.9984 \\
H$_6$ & 12 & 918 & 1 & 0.735 & 1.470 & 14.4 & 1.43 & SD & 0.21 & 0.9999 & 21.5 & 0.376 & SD & 4.4 & 0.9944 \\
LiH & 12 & 630 & 1 & 1.596 & 3.192 & 12.3 & 0.282 & SD & 0.011 & 1.0000 & 12.3 & 0.0993 & SD & 0.10 & 1.0000 \\
BeH$_2$ & 14 & 665 & 1 & 1.330 & 2.660 & 19.1 & 0.621 & SD & 0.38 & 0.9997 & 21.5 & 0.105 & GSD & 0.030 & 1.0000 \\
H$_2$O & 14 & 1085 & $\sfrac{1}{2}$ & 0.958 & 1.916 & 34.1 & 0.362 & SD & 0.10 & 1.0000 & 36.0 & 0.0414 & GSD & 0.006 & 1.0000 \\
NH$_3$ & 16 & 3608 & 1 & 1.012 & 2.024 & 63.1 & 2.56 & SD & 0.18 & 0.9999 & 66.9 & 0.183 & GSD & 0.29 & 0.9998 \\
CH$_4$ & 18 & 6891 & 1 & 1.087 & 2.174 & 59.1 & 2.90 & SD & 0.19 & 0.9999 & 65.1 & 0.400 & GSD & 6.4 & 0.9875 \\
\bottomrule
\end{tabular*}
\end{table}

\subsection*{Experimental parameters}

In this section we detail some experimental parameters for reproducibility of the results in the main text.

\subsubsection*{Optimisation of allocation}

To optimise the parameters, $\bxi=\{\rho_j,\alpha_j,\mu_j,\nu_j,\nu'_j\}_{j\in[1,L]}$ of~\eqref{eq:family-map} and
various subsets, we use the gradient descent (GD) based
Adam optimiser~\cite{Kingma2014Adam} with analytic gradients as computed in Appendix~\ref{app:S5} (Gradients of the
deployed family), applied to the log-weights $\theta_k=\log\xi_k$.
\begin{itemize}[leftmargin=*,topsep=2pt,itemsep=1pt]
  \item As mentioned above we can simplify the allocations by sharing parameters across the Hamiltonian terms.  We
    warm start the optimisation by first optimising the simpler allocation $\rho_j=1$, $\alpha_j=\alpha$,
    $\mu_j=\mu$, $\nu_j=\nu$, $\nu'_j=\nu'$ of~\eqref{eq:joint-shared} over $4$ random seeds.  We use $400$ steps of GD with learning rate $0.1$ and
    standard Adam $\beta_1$, $\beta_2$ parameters.
  \item From the best shared allocation, $(\rho_j,\alpha_j,\mu_j,\nu_j,\nu'_j)=(1,\alpha^*,\mu^*,\nu^*,\nu'^*)$ for every $j$, we
    then optimise the per-term allocations for a
    further $600$ GD iterations ($1000$ total) with a lower learning rate of $0.05$.
  \item For each allocation, we also optimise over $\tau$ in the vanilla \qdrift{} mixing,
    \begin{equation}
      \bq(\tau) \;=\; \tau\,\bq^c+(1-\tau)\,\bq^{\mathsf{SD}} ,
    \end{equation}
    of~\eqref{eq:tau-mixing}, by a search over $\tau\in[0.2,1]$ that minimises the fourth-order bias $B_N^{(4)}$ subject to
    $V_N\le(1+\eta)\,V_N^{\mathrm{van}}$~\eqref{eq:s4-deploy-constraints}.
\end{itemize}

\subsubsection*{Resolving depth}

The accuracy target is set by the movement of the observable over the simulated time,
$d\braket{O} := \bigl|\braket{O(T)} - \braket{O(0)}\bigr|$.  This is a minimal criterion for
`resolution', as if the bias is larger than this movement, the algorithm will not even be able to certify that any
evolution happened at all.  We define
$\varepsilon_O := f\bigl(d\braket{O}\bigr)$ for the threshold, with $f(x)=c\,x$, and $c=1$ and $c=\sfrac12$ in
Figure~\ref{fig:chem}.  We select a grid of depths, $N\in\{N_1,\dots,N_K\}$, and run Monte Carlo sampling over $M$
\qdrift{} trajectories over the grid to approximate the true bias (not the estimated bias from our truncation, which
may differ by the error in Appendix~\ref{app:S7}).  We define the `resolving' depth as $N^\star$ with $N^\star$
given by:
\begin{equation}
  N^\star := \min\Bigl\{ N \;:\;
    \bigl|\widehat B_{N'}\bigr| + 2\,\hat\sigma_{N'} < \varepsilon_O
    \ \text{ at } N' = N \text{ and at every deeper sampled } N' \Bigr\} ,
  \label{eq:s4-nstar}
\end{equation}
where $\widehat B_N$ is the estimated bias and $\hat\sigma_N=\sqrt{\sfrac{\widehat V_N}{M}}$ its standard error, with
$\widehat V_N$ the sample variance of the $M$ trajectory estimates $\hat O_N^{(m)}$.  The allocation is re-optimised at every depth of the grid.  Once bounds on the crossing are found by two values of $N$, these are refined to find the true crossing within a certain confidence.
We include the $2\hat\sigma_N$ buffer to account for the statistical error from $M$ runs and distinguish between values of $N$ close together.  We do not report $N^\star$ as the \emph{first} depth which resolves under $\varepsilon_O$ via the sampled runs, but
we also ensure that the bias resolves for all subsequent depths.  This is because the bias can be non-monotonic in $N$, so
selecting just the first resolution may not be reliable.  Every $N$ we select for the grid is such that $N\ge2\lambda T$, such that every vanilla rotation angle
$|\theta_j|=\lambda\,dt\le\sfrac12$.

\subsection*{\statedrift{} probability reallocations}

\begin{figure}[!htbp]
\centering
\includegraphics[width=0.84\textwidth]{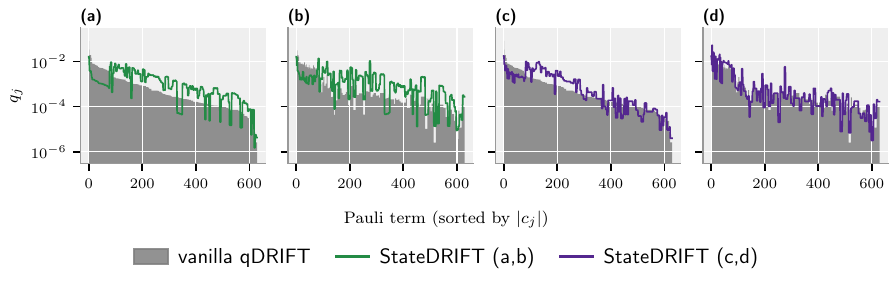}
\caption{\textbf{Sampling weights across reference states and observables, LiH STO-3G
($L=630$, $N=256$).}  In each panel, the vanilla \qdrift{} weights $q_j=\sfrac{|c_j|}{\lambda}$
(grey area) and the deployed \statedrift{} weights~\eqref{eq:joint-N} (line), on one shared log
axis, with the $630$ Pauli terms ordered by $|c_j|$.  The gap between the two is the
redistribution, and the line colour denotes the observable.
\textbf{(a,b)} Green: the weighted density $O=-\sfrac{1}{2}\sum_p (p{+}1) Z_p$, on the equilibrium
and the stretched ground state.
\textbf{(c,d)} Purple: the $z$-component of the electronic dipole
$O=-\sum_{pq} z_{pq}\,a^\dagger_p a_q$, whose off-diagonal $p\neq q$ block is retained so that it
carries non-commuting Pauli terms, on the same two reference states.}
\label{fig:alloc-dist}
\end{figure}

Figure~\ref{fig:alloc-dist} illustrates the reallocation of probabilities over Pauli terms enforced by \statedrift{}, on LiH across the two reference states and two observables.

\subsection*{Numerical check of the expansion}

\begin{figure}[H]
\centering
\includegraphics[width=0.84\textwidth]{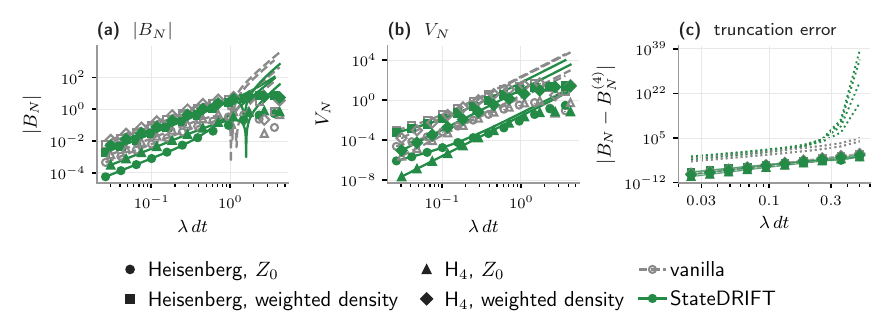}
\caption{\textbf{The fourth-order formulas against exact values, at $N=4$.}  \textbf{(a)} $|B_N|$ and
\textbf{(b)} $V_N$ against $\lambda\,dt$ for vanilla \qdrift{} (grey) and \statedrift{} (green,
$q_j\propto|c_j|\sqrt{|\gamma_j|+\beta_j^2}$ at $\tau=0.2$).  Marker shapes are the references, the Heisenberg chain
(Heis.) and H$_4$, each with $O=Z_0$ and the weighted density (wtd.).  Solid lines are the truncated $dt^4$ formulae, with the markers being
either \textbf{(a)} the exact bias computed exactly, or \textbf{(b)} the true variance (approximated by sampling over
$M=10^5$ trajectories).  The dips in \textbf{(a)} are values of $\lambda\,dt$ where $B_N^{(4)}=0$.  \textbf{(c)} The truncation error
$|B_N-B_N^{(4)}|$ (markers) and the bound of Proposition~\ref{prop:s8-total} (dotted), up to
$\lambda\,dt=\sfrac12$.}
\label{fig:supp-dt-check}
\end{figure}

In this section, we provide a numerical example for the truncation and error formulae we derive in this work.  To
compute and compare against the exact channel bias, we use examples small enough to completely simulate.  However,
we still sample the true variance because of the expense of computing it exactly.
We use the $\beta_j\neq0$ Heisenberg chain on $8$ qubits and H$_4$ as the demonstration, shown in
Figure~\ref{fig:supp-dt-check}.  For $N=4$, the truncation formulae are accurate up to $\lambda\,dt\approx1$.

\subsection*{Dependence on the accuracy target}

\begin{figure}[!htbp]
\centering
\includegraphics[width=\linewidth]{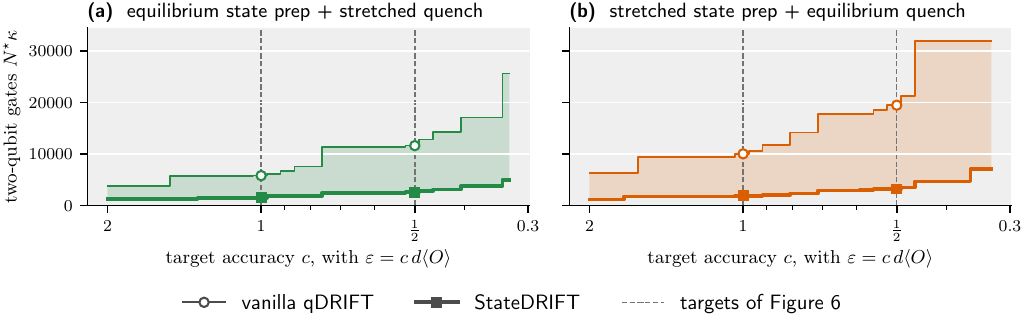}
\caption{\textbf{Two-qubit gates versus target accuracy.}  Two-qubit gates $N^\star\kappa$ that vanilla
\qdrift{} (thin line) and \statedrift{} (thick line) need to resolve $\varepsilon=c\,d\braket{O}$, with the gap shaded, for LiH with the weighted-density observable, prepared
at \textbf{(a)}~equilibrium and \textbf{(b)}~the stretched geometry.}
\label{fig:target-sweep}
\end{figure}

Figure~\ref{fig:chem} reports the saving at two targets, $\varepsilon=d\braket{O}$ and $\sfrac{d\braket{O}}{2}$.
In Figure~\ref{fig:target-sweep} we extend this for multiple values of $f(d\braket{O})=c\,d\braket{O}$ for LiH with
$O=-\sfrac{1}{2}\sum_p (p{+}1)Z_p$ to show that the gap in gate count between \statedrift{} and vanilla \qdrift{} grows as we request greater precision.

\subsection*{Simplified allocations}

\begin{figure}[H]
\centering
\includegraphics[width=\linewidth]{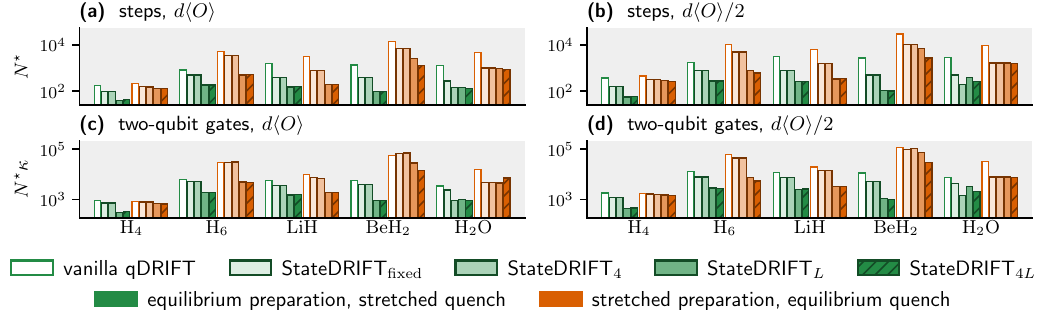}
\caption{\textbf{Simplified allocations.}  Steps $N^\star$ \textbf{(a,b)} and two-qubit gates
$N^\star\kappa$ \textbf{(c,d)} to resolve $\varepsilon=d\braket{O}$ \textbf{(a,c)} and
$\varepsilon=\sfrac{d\braket{O}}{2}$ \textbf{(b,d)}, as in Figure~\ref{fig:chem}.}
\label{fig:simplified-alloc}
\end{figure}

The allocation we propose in the main text is the more general form, but it is overparameterised and may not be
strictly necessary in all cases.  To test this, in Figure~\ref{fig:simplified-alloc} we evaluate the following simpler configurations on a subset of the
molecules from Figure~\ref{fig:chem} in the main text.
\begin{enumerate}[leftmargin=*,nosep]
  \item vanilla \qdrift{}, $q_j\propto|c_j|$, $N_{\rm params}=0$
  \item \statedrift{}$_{\rm fixed}$, $q_j\propto|c_j|\sqrt{|\gamma_j|}$, $N_{\rm params}=0$
  \item \statedrift{}$_{4}$, $q_j\propto|c_j|\sqrt{\rho|\gamma_j|+\mu\Phi_j+\nu\varphi_{H,j}^2+\nu'\varphi_{j,H}^2}$,
    $N_{\rm params}=4$
  \item \statedrift{}$_{L}$, $q_j\propto|c_j|\sqrt{\rho_j|\gamma_j|}$, $N_{\rm params}=L$
  \item \statedrift{}$_{4L}$, $q_j\propto|c_j|\sqrt{\rho_j|\gamma_j|+\mu_j\Phi_j+\nu_j\varphi_{H,j}^2+\nu'_j\varphi_{j,H}^2}$
    of~\eqref{eq:joint-N}, $N_{\rm params}=4L$
\end{enumerate}
In the above, only $N_{\rm params}-1$ are free as the last one is fixed by normalisation, and most of the saving needs
one weight per term: \statedrift{}$_{L}$ recovers most of the saving of the full rule, while allocations which are not
term-dependent reduce the gate count far less.

\begin{thebibliography}{76}%
\makeatletter
\providecommand \@ifxundefined [1]{%
 \@ifx{#1\undefined}
}%
\providecommand \@ifnum [1]{%
 \ifnum #1\expandafter \@firstoftwo
 \else \expandafter \@secondoftwo
 \fi
}%
\providecommand \@ifx [1]{%
 \ifx #1\expandafter \@firstoftwo
 \else \expandafter \@secondoftwo
 \fi
}%
\providecommand \natexlab [1]{#1}%
\providecommand \enquote  [1]{``#1''}%
\providecommand \bibnamefont  [1]{#1}%
\providecommand \bibfnamefont [1]{#1}%
\providecommand \citenamefont [1]{#1}%
\providecommand \href@noop [0]{\@secondoftwo}%
\providecommand \href [0]{\begingroup \@sanitize@url \@href}%
\providecommand \@href[1]{\@@startlink{#1}\@@href}%
\providecommand \@@href[1]{\endgroup#1\@@endlink}%
\providecommand \@sanitize@url [0]{\catcode `\\12\catcode `\$12\catcode
  `\&12\catcode `\#12\catcode `\^12\catcode `\_12\catcode `\%12\relax}%
\providecommand \@@startlink[1]{}%
\providecommand \@@endlink[0]{}%
\providecommand \url  [0]{\begingroup\@sanitize@url \@url }%
\providecommand \@url [1]{\endgroup\@href {#1}{\urlprefix }}%
\providecommand \urlprefix  [0]{URL }%
\providecommand \Eprint [0]{\href }%
\providecommand \doibase [0]{https://doi.org/}%
\providecommand \selectlanguage [0]{\@gobble}%
\providecommand \bibinfo  [0]{\@secondoftwo}%
\providecommand \bibfield  [0]{\@secondoftwo}%
\providecommand \translation [1]{[#1]}%
\providecommand \BibitemOpen [0]{}%
\providecommand \bibitemStop [0]{}%
\providecommand \bibitemNoStop [0]{.\EOS\space}%
\providecommand \EOS [0]{\spacefactor3000\relax}%
\providecommand \BibitemShut  [1]{\csname bibitem#1\endcsname}%
\let\auto@bib@innerbib\@empty
\bibitem [{\citenamefont {Feynman}(1982)}]{Feynman1982}%
  \BibitemOpen
  \bibfield  {author} {\bibinfo {author} {\bibfnamefont {R.~P.}\ \bibnamefont
  {Feynman}},\ }\href {https://doi.org/10.1007/BF02650179} {\bibfield
  {journal} {\bibinfo  {journal} {Int. J. Theor. Phys.}\ }\textbf {\bibinfo
  {volume} {21}},\ \bibinfo {pages} {467} (\bibinfo {year} {1982})}\BibitemShut
  {NoStop}%
\bibitem [{\citenamefont {Lloyd}(1996)}]{Lloyd1996}%
  \BibitemOpen
  \bibfield  {author} {\bibinfo {author} {\bibfnamefont {S.}~\bibnamefont
  {Lloyd}},\ }\href {https://doi.org/10.1126/science.273.5278.1073} {\bibfield
  {journal} {\bibinfo  {journal} {Science}\ }\textbf {\bibinfo {volume}
  {273}},\ \bibinfo {pages} {1073} (\bibinfo {year} {1996})}\BibitemShut
  {NoStop}%
\bibitem [{\citenamefont {Bauer}\ \emph {et~al.}(2020)\citenamefont {Bauer},
  \citenamefont {Bravyi}, \citenamefont {Motta},\ and\ \citenamefont
  {Chan}}]{Bauer2020}%
  \BibitemOpen
  \bibfield  {author} {\bibinfo {author} {\bibfnamefont {B.}~\bibnamefont
  {Bauer}}, \bibinfo {author} {\bibfnamefont {S.}~\bibnamefont {Bravyi}},
  \bibinfo {author} {\bibfnamefont {M.}~\bibnamefont {Motta}},\ and\ \bibinfo
  {author} {\bibfnamefont {G.~K.-L.}\ \bibnamefont {Chan}},\ }\href
  {https://doi.org/10.1021/acs.chemrev.9b00829} {\bibfield  {journal} {\bibinfo
   {journal} {Chem. Rev.}\ }\textbf {\bibinfo {volume} {120}},\ \bibinfo
  {pages} {12685} (\bibinfo {year} {2020})}\BibitemShut {NoStop}%
\bibitem [{\citenamefont {Suzuki}(1991)}]{Suzuki1991}%
  \BibitemOpen
  \bibfield  {author} {\bibinfo {author} {\bibfnamefont {M.}~\bibnamefont
  {Suzuki}},\ }\href {https://doi.org/10.1063/1.529425} {\bibfield  {journal}
  {\bibinfo  {journal} {J. Math. Phys.}\ }\textbf {\bibinfo {volume} {32}},\
  \bibinfo {pages} {400} (\bibinfo {year} {1991})}\BibitemShut {NoStop}%
\bibitem [{\citenamefont {Childs}\ \emph {et~al.}(2021)\citenamefont {Childs},
  \citenamefont {Su}, \citenamefont {Tran}, \citenamefont {Wiebe},\ and\
  \citenamefont {Zhu}}]{Childs2021}%
  \BibitemOpen
  \bibfield  {author} {\bibinfo {author} {\bibfnamefont {A.~M.}\ \bibnamefont
  {Childs}}, \bibinfo {author} {\bibfnamefont {Y.}~\bibnamefont {Su}}, \bibinfo
  {author} {\bibfnamefont {M.~C.}\ \bibnamefont {Tran}}, \bibinfo {author}
  {\bibfnamefont {N.}~\bibnamefont {Wiebe}},\ and\ \bibinfo {author}
  {\bibfnamefont {S.}~\bibnamefont {Zhu}},\ }\href
  {https://doi.org/10.1103/PhysRevX.11.011020} {\bibfield  {journal} {\bibinfo
  {journal} {Phys. Rev. X}\ }\textbf {\bibinfo {volume} {11}},\ \bibinfo
  {pages} {011020} (\bibinfo {year} {2021})}\BibitemShut {NoStop}%
\bibitem [{\citenamefont {Childs}\ and\ \citenamefont
  {Wiebe}(2012)}]{ChildsWiebe2012}%
  \BibitemOpen
  \bibfield  {author} {\bibinfo {author} {\bibfnamefont {A.~M.}\ \bibnamefont
  {Childs}}\ and\ \bibinfo {author} {\bibfnamefont {N.}~\bibnamefont {Wiebe}},\
  }\href {https://doi.org/10.26421/QIC12.11-12-1} {\bibfield  {journal}
  {\bibinfo  {journal} {Quantum Inf. Comput.}\ }\textbf {\bibinfo {volume}
  {12}},\ \bibinfo {pages} {901} (\bibinfo {year} {2012})}\BibitemShut
  {NoStop}%
\bibitem [{\citenamefont {Berry}\ \emph {et~al.}(2015)\citenamefont {Berry},
  \citenamefont {Childs}, \citenamefont {Cleve}, \citenamefont {Kothari},\ and\
  \citenamefont {Somma}}]{Berry2015Taylor}%
  \BibitemOpen
  \bibfield  {author} {\bibinfo {author} {\bibfnamefont {D.~W.}\ \bibnamefont
  {Berry}}, \bibinfo {author} {\bibfnamefont {A.~M.}\ \bibnamefont {Childs}},
  \bibinfo {author} {\bibfnamefont {R.}~\bibnamefont {Cleve}}, \bibinfo
  {author} {\bibfnamefont {R.}~\bibnamefont {Kothari}},\ and\ \bibinfo {author}
  {\bibfnamefont {R.~D.}\ \bibnamefont {Somma}},\ }\href
  {https://doi.org/10.1103/PhysRevLett.114.090502} {\bibfield  {journal}
  {\bibinfo  {journal} {Phys. Rev. Lett.}\ }\textbf {\bibinfo {volume} {114}},\
  \bibinfo {pages} {090502} (\bibinfo {year} {2015})}\BibitemShut {NoStop}%
\bibitem [{\citenamefont {Low}\ and\ \citenamefont
  {Chuang}(2017)}]{LowChuang2017}%
  \BibitemOpen
  \bibfield  {author} {\bibinfo {author} {\bibfnamefont {G.~H.}\ \bibnamefont
  {Low}}\ and\ \bibinfo {author} {\bibfnamefont {I.~L.}\ \bibnamefont
  {Chuang}},\ }\href {https://doi.org/10.1103/PhysRevLett.118.010501}
  {\bibfield  {journal} {\bibinfo  {journal} {Phys. Rev. Lett.}\ }\textbf
  {\bibinfo {volume} {118}},\ \bibinfo {pages} {010501} (\bibinfo {year}
  {2017})}\BibitemShut {NoStop}%
\bibitem [{\citenamefont {Low}\ and\ \citenamefont
  {Chuang}(2019)}]{LowChuang2019}%
  \BibitemOpen
  \bibfield  {author} {\bibinfo {author} {\bibfnamefont {G.~H.}\ \bibnamefont
  {Low}}\ and\ \bibinfo {author} {\bibfnamefont {I.~L.}\ \bibnamefont
  {Chuang}},\ }\href {https://doi.org/10.22331/q-2019-07-12-163} {\bibfield
  {journal} {\bibinfo  {journal} {Quantum}\ }\textbf {\bibinfo {volume} {3}},\
  \bibinfo {pages} {163} (\bibinfo {year} {2019})}\BibitemShut {NoStop}%
\bibitem [{\citenamefont {Childs}\ \emph {et~al.}(2019)\citenamefont {Childs},
  \citenamefont {Ostrander},\ and\ \citenamefont
  {Su}}]{Childs2019Randomization}%
  \BibitemOpen
  \bibfield  {author} {\bibinfo {author} {\bibfnamefont {A.~M.}\ \bibnamefont
  {Childs}}, \bibinfo {author} {\bibfnamefont {A.}~\bibnamefont {Ostrander}},\
  and\ \bibinfo {author} {\bibfnamefont {Y.}~\bibnamefont {Su}},\ }\href
  {https://doi.org/10.22331/q-2019-09-02-182} {\bibfield  {journal} {\bibinfo
  {journal} {Quantum}\ }\textbf {\bibinfo {volume} {3}},\ \bibinfo {pages}
  {182} (\bibinfo {year} {2019})}\BibitemShut {NoStop}%
\bibitem [{\citenamefont {Campbell}(2019)}]{Campbell2019}%
  \BibitemOpen
  \bibfield  {author} {\bibinfo {author} {\bibfnamefont {E.~T.}\ \bibnamefont
  {Campbell}},\ }\href {https://doi.org/10.1103/PhysRevLett.123.070503}
  {\bibfield  {journal} {\bibinfo  {journal} {Phys. Rev. Lett.}\ }\textbf
  {\bibinfo {volume} {123}},\ \bibinfo {pages} {070503} (\bibinfo {year}
  {2019})}\BibitemShut {NoStop}%
\bibitem [{\citenamefont {Nakaji}\ \emph {et~al.}(2024)\citenamefont {Nakaji},
  \citenamefont {Bagherimehrab},\ and\ \citenamefont
  {Aspuru-Guzik}}]{Nakaji2024qSWIFT}%
  \BibitemOpen
  \bibfield  {author} {\bibinfo {author} {\bibfnamefont {K.}~\bibnamefont
  {Nakaji}}, \bibinfo {author} {\bibfnamefont {M.}~\bibnamefont
  {Bagherimehrab}},\ and\ \bibinfo {author} {\bibfnamefont {A.}~\bibnamefont
  {Aspuru-Guzik}},\ }\href {https://doi.org/10.1103/PRXQuantum.5.020330}
  {\bibfield  {journal} {\bibinfo  {journal} {PRX Quantum}\ }\textbf {\bibinfo
  {volume} {5}},\ \bibinfo {pages} {020330} (\bibinfo {year}
  {2024})}\BibitemShut {NoStop}%
\bibitem [{\citenamefont {Kitaev}(1995)}]{Kitaev1995}%
  \BibitemOpen
  \bibfield  {author} {\bibinfo {author} {\bibfnamefont {A.~Y.}\ \bibnamefont
  {Kitaev}},\ }\href {https://arxiv.org/abs/quant-ph/9511026} {\bibinfo {title}
  {Quantum measurements and the {A}belian {S}tabilizer {P}roblem}} (\bibinfo
  {year} {1995}),\ \Eprint {https://arxiv.org/abs/quant-ph/9511026}
  {arXiv:quant-ph/9511026} \BibitemShut {NoStop}%
\bibitem [{\citenamefont {Abrams}\ and\ \citenamefont
  {Lloyd}(1999)}]{AbramsLloyd1999}%
  \BibitemOpen
  \bibfield  {author} {\bibinfo {author} {\bibfnamefont {D.~S.}\ \bibnamefont
  {Abrams}}\ and\ \bibinfo {author} {\bibfnamefont {S.}~\bibnamefont {Lloyd}},\
  }\href {https://doi.org/10.1103/PhysRevLett.83.5162} {\bibfield  {journal}
  {\bibinfo  {journal} {Phys. Rev. Lett.}\ }\textbf {\bibinfo {volume} {83}},\
  \bibinfo {pages} {5162} (\bibinfo {year} {1999})}\BibitemShut {NoStop}%
\bibitem [{\citenamefont {Ding}\ and\ \citenamefont
  {Lin}(2023)}]{DingLin2023QCELS}%
  \BibitemOpen
  \bibfield  {author} {\bibinfo {author} {\bibfnamefont {Z.}~\bibnamefont
  {Ding}}\ and\ \bibinfo {author} {\bibfnamefont {L.}~\bibnamefont {Lin}},\
  }\href {https://doi.org/10.1103/PRXQuantum.4.020331} {\bibfield  {journal}
  {\bibinfo  {journal} {PRX Quantum}\ }\textbf {\bibinfo {volume} {4}},\
  \bibinfo {pages} {020331} (\bibinfo {year} {2023})}\BibitemShut {NoStop}%
\bibitem [{\citenamefont {Ding}\ \emph {et~al.}(2024)\citenamefont {Ding},
  \citenamefont {Li}, \citenamefont {Lin}, \citenamefont {Ni}, \citenamefont
  {Ying},\ and\ \citenamefont {Zhang}}]{Ding2024QMEGS}%
  \BibitemOpen
  \bibfield  {author} {\bibinfo {author} {\bibfnamefont {Z.}~\bibnamefont
  {Ding}}, \bibinfo {author} {\bibfnamefont {H.}~\bibnamefont {Li}}, \bibinfo
  {author} {\bibfnamefont {L.}~\bibnamefont {Lin}}, \bibinfo {author}
  {\bibfnamefont {H.}~\bibnamefont {Ni}}, \bibinfo {author} {\bibfnamefont
  {L.}~\bibnamefont {Ying}},\ and\ \bibinfo {author} {\bibfnamefont
  {R.}~\bibnamefont {Zhang}},\ }\href
  {https://doi.org/10.22331/q-2024-10-02-1487} {\bibfield  {journal} {\bibinfo
  {journal} {Quantum}\ }\textbf {\bibinfo {volume} {8}},\ \bibinfo {pages}
  {1487} (\bibinfo {year} {2024})}\BibitemShut {NoStop}%
\bibitem [{\citenamefont {McArdle}\ \emph {et~al.}(2019)\citenamefont
  {McArdle}, \citenamefont {Jones}, \citenamefont {Endo}, \citenamefont {Li},
  \citenamefont {Benjamin},\ and\ \citenamefont {Yuan}}]{McArdle2019VarQITE}%
  \BibitemOpen
  \bibfield  {author} {\bibinfo {author} {\bibfnamefont {S.}~\bibnamefont
  {McArdle}}, \bibinfo {author} {\bibfnamefont {T.}~\bibnamefont {Jones}},
  \bibinfo {author} {\bibfnamefont {S.}~\bibnamefont {Endo}}, \bibinfo {author}
  {\bibfnamefont {Y.}~\bibnamefont {Li}}, \bibinfo {author} {\bibfnamefont
  {S.~C.}\ \bibnamefont {Benjamin}},\ and\ \bibinfo {author} {\bibfnamefont
  {X.}~\bibnamefont {Yuan}},\ }\href
  {https://doi.org/10.1038/s41534-019-0187-2} {\bibfield  {journal} {\bibinfo
  {journal} {npj Quantum Inf.}\ }\textbf {\bibinfo {volume} {5}},\ \bibinfo
  {pages} {75} (\bibinfo {year} {2019})}\BibitemShut {NoStop}%
\bibitem [{\citenamefont {Motta}\ \emph {et~al.}(2020)\citenamefont {Motta},
  \citenamefont {Sun}, \citenamefont {Tan}, \citenamefont {O'Rourke},
  \citenamefont {Ye}, \citenamefont {Minnich}, \citenamefont {Brand{\~a}o},\
  and\ \citenamefont {Chan}}]{Motta2020QITE}%
  \BibitemOpen
  \bibfield  {author} {\bibinfo {author} {\bibfnamefont {M.}~\bibnamefont
  {Motta}}, \bibinfo {author} {\bibfnamefont {C.}~\bibnamefont {Sun}}, \bibinfo
  {author} {\bibfnamefont {A.~T.~K.}\ \bibnamefont {Tan}}, \bibinfo {author}
  {\bibfnamefont {M.~J.}\ \bibnamefont {O'Rourke}}, \bibinfo {author}
  {\bibfnamefont {E.}~\bibnamefont {Ye}}, \bibinfo {author} {\bibfnamefont
  {A.~J.}\ \bibnamefont {Minnich}}, \bibinfo {author} {\bibfnamefont {F.~G.
  S.~L.}\ \bibnamefont {Brand{\~a}o}},\ and\ \bibinfo {author} {\bibfnamefont
  {G.~K.-L.}\ \bibnamefont {Chan}},\ }\href
  {https://doi.org/10.1038/s41567-019-0704-4} {\bibfield  {journal} {\bibinfo
  {journal} {Nat. Phys.}\ }\textbf {\bibinfo {volume} {16}},\ \bibinfo {pages}
  {205} (\bibinfo {year} {2020})}\BibitemShut {NoStop}%
\bibitem [{\citenamefont {Huggins}\ \emph {et~al.}(2022)\citenamefont
  {Huggins}, \citenamefont {O'Gorman}, \citenamefont {Rubin}, \citenamefont
  {Reichman}, \citenamefont {Babbush},\ and\ \citenamefont
  {Lee}}]{Huggins2022QCQMC}%
  \BibitemOpen
  \bibfield  {author} {\bibinfo {author} {\bibfnamefont {W.~J.}\ \bibnamefont
  {Huggins}}, \bibinfo {author} {\bibfnamefont {B.~A.}\ \bibnamefont
  {O'Gorman}}, \bibinfo {author} {\bibfnamefont {N.~C.}\ \bibnamefont {Rubin}},
  \bibinfo {author} {\bibfnamefont {D.~R.}\ \bibnamefont {Reichman}}, \bibinfo
  {author} {\bibfnamefont {R.}~\bibnamefont {Babbush}},\ and\ \bibinfo {author}
  {\bibfnamefont {J.}~\bibnamefont {Lee}},\ }\href
  {https://doi.org/10.1038/s41586-021-04351-z} {\bibfield  {journal} {\bibinfo
  {journal} {Nature}\ }\textbf {\bibinfo {volume} {603}},\ \bibinfo {pages}
  {416} (\bibinfo {year} {2022})}\BibitemShut {NoStop}%
\bibitem [{\citenamefont {Zhang}\ \emph {et~al.}(2025)\citenamefont {Zhang},
  \citenamefont {Huang}, \citenamefont {Sun}, \citenamefont {Lv},\ and\
  \citenamefont {Yuan}}]{Zhang2025QCQMC}%
  \BibitemOpen
  \bibfield  {author} {\bibinfo {author} {\bibfnamefont {Y.}~\bibnamefont
  {Zhang}}, \bibinfo {author} {\bibfnamefont {Y.}~\bibnamefont {Huang}},
  \bibinfo {author} {\bibfnamefont {J.}~\bibnamefont {Sun}}, \bibinfo {author}
  {\bibfnamefont {D.}~\bibnamefont {Lv}},\ and\ \bibinfo {author}
  {\bibfnamefont {X.}~\bibnamefont {Yuan}},\ }\href
  {https://doi.org/10.1103/jt8s-hzhd} {\bibfield  {journal} {\bibinfo
  {journal} {Phys. Rev. A}\ }\textbf {\bibinfo {volume} {112}},\ \bibinfo
  {pages} {022428} (\bibinfo {year} {2025})}\BibitemShut {NoStop}%
\bibitem [{\citenamefont {Kanno}\ \emph {et~al.}(2024)\citenamefont {Kanno},
  \citenamefont {Nakamura}, \citenamefont {Kobayashi}, \citenamefont {Gocho},
  \citenamefont {Hatanaka}, \citenamefont {Yamamoto},\ and\ \citenamefont
  {Gao}}]{Kanno2024QCQMC}%
  \BibitemOpen
  \bibfield  {author} {\bibinfo {author} {\bibfnamefont {S.}~\bibnamefont
  {Kanno}}, \bibinfo {author} {\bibfnamefont {H.}~\bibnamefont {Nakamura}},
  \bibinfo {author} {\bibfnamefont {T.}~\bibnamefont {Kobayashi}}, \bibinfo
  {author} {\bibfnamefont {S.}~\bibnamefont {Gocho}}, \bibinfo {author}
  {\bibfnamefont {M.}~\bibnamefont {Hatanaka}}, \bibinfo {author}
  {\bibfnamefont {N.}~\bibnamefont {Yamamoto}},\ and\ \bibinfo {author}
  {\bibfnamefont {Q.}~\bibnamefont {Gao}},\ }\href
  {https://doi.org/10.1038/s41534-024-00851-8} {\bibfield  {journal} {\bibinfo
  {journal} {npj Quantum Inf.}\ }\textbf {\bibinfo {volume} {10}},\ \bibinfo
  {pages} {56} (\bibinfo {year} {2024})}\BibitemShut {NoStop}%
\bibitem [{\citenamefont {Buonaiuto}\ \emph {et~al.}(2026)\citenamefont
  {Buonaiuto}, \citenamefont {M{\'a}rquez~Romero}, \citenamefont {Coyle},
  \citenamefont {Paine}, \citenamefont {Soloviev}, \citenamefont {Scali},\ and\
  \citenamefont {Krompiec}}]{Buonaiuto2026QCQMC}%
  \BibitemOpen
  \bibfield  {author} {\bibinfo {author} {\bibfnamefont {G.}~\bibnamefont
  {Buonaiuto}}, \bibinfo {author} {\bibfnamefont {A.}~\bibnamefont
  {M{\'a}rquez~Romero}}, \bibinfo {author} {\bibfnamefont {B.}~\bibnamefont
  {Coyle}}, \bibinfo {author} {\bibfnamefont {A.~E.}\ \bibnamefont {Paine}},
  \bibinfo {author} {\bibfnamefont {V.~P.}\ \bibnamefont {Soloviev}}, \bibinfo
  {author} {\bibfnamefont {S.}~\bibnamefont {Scali}},\ and\ \bibinfo {author}
  {\bibfnamefont {M.}~\bibnamefont {Krompiec}},\ }\href
  {https://arxiv.org/abs/2603.25582} {\bibinfo {title} {A unified quantum
  computing quantum {M}onte {C}arlo framework through structured state
  preparation}} (\bibinfo {year} {2026}),\ \Eprint
  {https://arxiv.org/abs/2603.25582} {arXiv:2603.25582} \BibitemShut {NoStop}%
\bibitem [{\citenamefont {Chen}\ \emph {et~al.}(2021)\citenamefont {Chen},
  \citenamefont {Huang}, \citenamefont {Kueng},\ and\ \citenamefont
  {Tropp}}]{Chen2021Concentration}%
  \BibitemOpen
  \bibfield  {author} {\bibinfo {author} {\bibfnamefont {C.-F.}\ \bibnamefont
  {Chen}}, \bibinfo {author} {\bibfnamefont {H.-Y.}\ \bibnamefont {Huang}},
  \bibinfo {author} {\bibfnamefont {R.}~\bibnamefont {Kueng}},\ and\ \bibinfo
  {author} {\bibfnamefont {J.~A.}\ \bibnamefont {Tropp}},\ }\href
  {https://doi.org/10.1103/PRXQuantum.2.040305} {\bibfield  {journal} {\bibinfo
   {journal} {PRX Quantum}\ }\textbf {\bibinfo {volume} {2}},\ \bibinfo {pages}
  {040305} (\bibinfo {year} {2021})}\BibitemShut {NoStop}%
\bibitem [{\citenamefont {Childs}\ \emph {et~al.}(2018)\citenamefont {Childs},
  \citenamefont {Maslov}, \citenamefont {Nam}, \citenamefont {Ross},\ and\
  \citenamefont {Su}}]{Childs2018PNAS}%
  \BibitemOpen
  \bibfield  {author} {\bibinfo {author} {\bibfnamefont {A.~M.}\ \bibnamefont
  {Childs}}, \bibinfo {author} {\bibfnamefont {D.}~\bibnamefont {Maslov}},
  \bibinfo {author} {\bibfnamefont {Y.}~\bibnamefont {Nam}}, \bibinfo {author}
  {\bibfnamefont {N.~J.}\ \bibnamefont {Ross}},\ and\ \bibinfo {author}
  {\bibfnamefont {Y.}~\bibnamefont {Su}},\ }\href
  {https://doi.org/10.1073/pnas.1801723115} {\bibfield  {journal} {\bibinfo
  {journal} {Proc. Natl. Acad. Sci. U.S.A.}\ }\textbf {\bibinfo {volume}
  {115}},\ \bibinfo {pages} {9456} (\bibinfo {year} {2018})}\BibitemShut
  {NoStop}%
\bibitem [{\citenamefont {Kiss}\ \emph {et~al.}(2023)\citenamefont {Kiss},
  \citenamefont {Grossi},\ and\ \citenamefont {Roggero}}]{KGR2023}%
  \BibitemOpen
  \bibfield  {author} {\bibinfo {author} {\bibfnamefont {O.}~\bibnamefont
  {Kiss}}, \bibinfo {author} {\bibfnamefont {M.}~\bibnamefont {Grossi}},\ and\
  \bibinfo {author} {\bibfnamefont {A.}~\bibnamefont {Roggero}},\ }\href
  {https://doi.org/10.22331/q-2023-04-13-977} {\bibfield  {journal} {\bibinfo
  {journal} {Quantum}\ }\textbf {\bibinfo {volume} {7}},\ \bibinfo {pages}
  {977} (\bibinfo {year} {2023})}\BibitemShut {NoStop}%
\bibitem [{\citenamefont {Cugini}\ \emph {et~al.}(2026)\citenamefont {Cugini},
  \citenamefont {Atif},\ and\ \citenamefont {Suba\c{s}{\i}}}]{CAS2026}%
  \BibitemOpen
  \bibfield  {author} {\bibinfo {author} {\bibfnamefont {D.}~\bibnamefont
  {Cugini}}, \bibinfo {author} {\bibfnamefont {T.~A.}\ \bibnamefont {Atif}},\
  and\ \bibinfo {author} {\bibfnamefont {Y.}~\bibnamefont {Suba\c{s}{\i}}},\
  }\href {https://arxiv.org/abs/2603.13495} {\bibinfo {title} {Resource-optimal
  importance sampling for randomized quantum algorithms}} (\bibinfo {year}
  {2026}),\ \Eprint {https://arxiv.org/abs/2603.13495} {arXiv:2603.13495}
  \BibitemShut {NoStop}%
\bibitem [{\citenamefont {Lee}\ and\ \citenamefont {Choi}(2026)}]{Lee2026}%
  \BibitemOpen
  \bibfield  {author} {\bibinfo {author} {\bibfnamefont {S.}~\bibnamefont
  {Lee}}\ and\ \bibinfo {author} {\bibfnamefont {S.}~\bibnamefont {Choi}},\
  }\href {https://arxiv.org/abs/2604.26263} {\bibinfo {title} {{qSHIFT}: An
  adaptive sampling protocol for higher-order quantum simulation}} (\bibinfo
  {year} {2026}),\ \Eprint {https://arxiv.org/abs/2604.26263}
  {arXiv:2604.26263} \BibitemShut {NoStop}%
\bibitem [{\citenamefont {Mohammadipour}\ and\ \citenamefont
  {Li}(2026)}]{Mohammadipour2026}%
  \BibitemOpen
  \bibfield  {author} {\bibinfo {author} {\bibfnamefont {P.}~\bibnamefont
  {Mohammadipour}}\ and\ \bibinfo {author} {\bibfnamefont {X.}~\bibnamefont
  {Li}},\ }\href {https://arxiv.org/abs/2604.26865} {\bibinfo {title}
  {{MLMC-qDRIFT}: Multilevel variance reduction for randomized quantum
  {Hamiltonian} simulation}} (\bibinfo {year} {2026}),\ \Eprint
  {https://arxiv.org/abs/2604.26865} {arXiv:2604.26865} \BibitemShut {NoStop}%
\bibitem [{\citenamefont {Ouyang}\ \emph {et~al.}(2020)\citenamefont {Ouyang},
  \citenamefont {White},\ and\ \citenamefont {Campbell}}]{Ouyang2020}%
  \BibitemOpen
  \bibfield  {author} {\bibinfo {author} {\bibfnamefont {Y.}~\bibnamefont
  {Ouyang}}, \bibinfo {author} {\bibfnamefont {D.~R.}\ \bibnamefont {White}},\
  and\ \bibinfo {author} {\bibfnamefont {E.~T.}\ \bibnamefont {Campbell}},\
  }\href {https://doi.org/10.22331/q-2020-02-27-235} {\bibfield  {journal}
  {\bibinfo  {journal} {Quantum}\ }\textbf {\bibinfo {volume} {4}},\ \bibinfo
  {pages} {235} (\bibinfo {year} {2020})}\BibitemShut {NoStop}%
\bibitem [{\citenamefont {Jin}\ and\ \citenamefont
  {Li}(2025)}]{JinLi2021PartiallyRandom}%
  \BibitemOpen
  \bibfield  {author} {\bibinfo {author} {\bibfnamefont {S.}~\bibnamefont
  {Jin}}\ and\ \bibinfo {author} {\bibfnamefont {X.}~\bibnamefont {Li}},\
  }\href {https://doi.org/10.1007/s42967-023-00336-z} {\bibfield  {journal}
  {\bibinfo  {journal} {Commun. Appl. Math. Comput.}\ }\textbf {\bibinfo
  {volume} {7}},\ \bibinfo {pages} {442} (\bibinfo {year} {2025})}\BibitemShut
  {NoStop}%
\bibitem [{\citenamefont {Hagan}\ and\ \citenamefont
  {Wiebe}(2023)}]{Hagan2023Composite}%
  \BibitemOpen
  \bibfield  {author} {\bibinfo {author} {\bibfnamefont {M.}~\bibnamefont
  {Hagan}}\ and\ \bibinfo {author} {\bibfnamefont {N.}~\bibnamefont {Wiebe}},\
  }\href {https://doi.org/10.22331/q-2023-11-14-1181} {\bibfield  {journal}
  {\bibinfo  {journal} {Quantum}\ }\textbf {\bibinfo {volume} {7}},\ \bibinfo
  {pages} {1181} (\bibinfo {year} {2023})}\BibitemShut {NoStop}%
\bibitem [{\citenamefont {Pocrnic}\ \emph {et~al.}(2024)\citenamefont
  {Pocrnic}, \citenamefont {Hagan}, \citenamefont {Carrasquilla}, \citenamefont
  {Segal},\ and\ \citenamefont {Wiebe}}]{Pocrnic2024}%
  \BibitemOpen
  \bibfield  {author} {\bibinfo {author} {\bibfnamefont {M.}~\bibnamefont
  {Pocrnic}}, \bibinfo {author} {\bibfnamefont {M.}~\bibnamefont {Hagan}},
  \bibinfo {author} {\bibfnamefont {J.}~\bibnamefont {Carrasquilla}}, \bibinfo
  {author} {\bibfnamefont {D.}~\bibnamefont {Segal}},\ and\ \bibinfo {author}
  {\bibfnamefont {N.}~\bibnamefont {Wiebe}},\ }\href
  {https://doi.org/10.1103/PhysRevResearch.6.013224} {\bibfield  {journal}
  {\bibinfo  {journal} {Phys. Rev. Res.}\ }\textbf {\bibinfo {volume} {6}},\
  \bibinfo {pages} {013224} (\bibinfo {year} {2024})}\BibitemShut {NoStop}%
\bibitem [{\citenamefont {Paganelli}\ \emph {et~al.}(2026)\citenamefont
  {Paganelli}, \citenamefont {Grossi}, \citenamefont {Giachero}, \citenamefont
  {O'Brien},\ and\ \citenamefont {Kiss}}]{Paganelli2026When}%
  \BibitemOpen
  \bibfield  {author} {\bibinfo {author} {\bibfnamefont {F.}~\bibnamefont
  {Paganelli}}, \bibinfo {author} {\bibfnamefont {M.}~\bibnamefont {Grossi}},
  \bibinfo {author} {\bibfnamefont {A.}~\bibnamefont {Giachero}}, \bibinfo
  {author} {\bibfnamefont {T.~E.}\ \bibnamefont {O'Brien}},\ and\ \bibinfo
  {author} {\bibfnamefont {O.}~\bibnamefont {Kiss}},\ }\href
  {https://arxiv.org/abs/2604.07448} {\bibinfo {title} {When is randomization
  advantageous in quantum simulation?}} (\bibinfo {year} {2026}),\ \Eprint
  {https://arxiv.org/abs/2604.07448} {arXiv:2604.07448} \BibitemShut {NoStop}%
\bibitem [{\citenamefont {G{\"u}nther}\ \emph {et~al.}(2026)\citenamefont
  {G{\"u}nther}, \citenamefont {Witteveen}, \citenamefont {Schmidhuber},
  \citenamefont {Miller}, \citenamefont {Christandl},\ and\ \citenamefont
  {Harrow}}]{Gunther2026PartialRandom}%
  \BibitemOpen
  \bibfield  {author} {\bibinfo {author} {\bibfnamefont {J.}~\bibnamefont
  {G{\"u}nther}}, \bibinfo {author} {\bibfnamefont {F.}~\bibnamefont
  {Witteveen}}, \bibinfo {author} {\bibfnamefont {A.}~\bibnamefont
  {Schmidhuber}}, \bibinfo {author} {\bibfnamefont {M.}~\bibnamefont {Miller}},
  \bibinfo {author} {\bibfnamefont {M.}~\bibnamefont {Christandl}},\ and\
  \bibinfo {author} {\bibfnamefont {A.}~\bibnamefont {Harrow}},\ }\href
  {https://doi.org/10.1103/ynxb-p2xq} {\bibfield  {journal} {\bibinfo
  {journal} {PRX Quantum}\ }\textbf {\bibinfo {volume} {7}},\ \bibinfo {pages}
  {020332} (\bibinfo {year} {2026})}\BibitemShut {NoStop}%
\bibitem [{\citenamefont {Akahoshi}\ \emph {et~al.}(2024)\citenamefont
  {Akahoshi}, \citenamefont {Maruyama}, \citenamefont {Oshima}, \citenamefont
  {Sato},\ and\ \citenamefont {Fujii}}]{Akahoshi2023}%
  \BibitemOpen
  \bibfield  {author} {\bibinfo {author} {\bibfnamefont {Y.}~\bibnamefont
  {Akahoshi}}, \bibinfo {author} {\bibfnamefont {K.}~\bibnamefont {Maruyama}},
  \bibinfo {author} {\bibfnamefont {H.}~\bibnamefont {Oshima}}, \bibinfo
  {author} {\bibfnamefont {S.}~\bibnamefont {Sato}},\ and\ \bibinfo {author}
  {\bibfnamefont {K.}~\bibnamefont {Fujii}},\ }\href
  {https://doi.org/10.1103/PRXQuantum.5.010337} {\bibfield  {journal} {\bibinfo
   {journal} {PRX Quantum}\ }\textbf {\bibinfo {volume} {5}},\ \bibinfo {pages}
  {010337} (\bibinfo {year} {2024})}\BibitemShut {NoStop}%
\bibitem [{\citenamefont {Kanasugi}\ \emph {et~al.}(2026)\citenamefont
  {Kanasugi}, \citenamefont {Toshio}, \citenamefont {Maruyama},\ and\
  \citenamefont {Oshima}}]{Kanasugi2026EarlyFTQPE}%
  \BibitemOpen
  \bibfield  {author} {\bibinfo {author} {\bibfnamefont {S.}~\bibnamefont
  {Kanasugi}}, \bibinfo {author} {\bibfnamefont {R.}~\bibnamefont {Toshio}},
  \bibinfo {author} {\bibfnamefont {K.}~\bibnamefont {Maruyama}},\ and\
  \bibinfo {author} {\bibfnamefont {H.}~\bibnamefont {Oshima}},\ }\href
  {https://arxiv.org/abs/2603.22778} {\bibinfo {title} {Enabling chemically
  accurate quantum phase estimation in the early fault-tolerant regime}}
  (\bibinfo {year} {2026}),\ \Eprint {https://arxiv.org/abs/2603.22778}
  {arXiv:2603.22778} \BibitemShut {NoStop}%
\bibitem [{\citenamefont {Watson}(2024)}]{Watson2024RandomlyCompiled}%
  \BibitemOpen
  \bibfield  {author} {\bibinfo {author} {\bibfnamefont {J.~D.}\ \bibnamefont
  {Watson}},\ }\href {https://arxiv.org/abs/2411.04240} {\bibinfo {title}
  {Randomly compiled quantum simulation with exponentially reduced circuit
  depths}} (\bibinfo {year} {2024}),\ \Eprint
  {https://arxiv.org/abs/2411.04240} {arXiv:2411.04240} \BibitemShut {NoStop}%
\bibitem [{\citenamefont {Zhuk}\ \emph {et~al.}(2024)\citenamefont {Zhuk},
  \citenamefont {Robertson},\ and\ \citenamefont
  {Bravyi}}]{Zhuk2024DynamicMPF}%
  \BibitemOpen
  \bibfield  {author} {\bibinfo {author} {\bibfnamefont {S.}~\bibnamefont
  {Zhuk}}, \bibinfo {author} {\bibfnamefont {N.}~\bibnamefont {Robertson}},\
  and\ \bibinfo {author} {\bibfnamefont {S.}~\bibnamefont {Bravyi}},\ }\href
  {https://doi.org/10.1103/PhysRevResearch.6.033309} {\bibfield  {journal}
  {\bibinfo  {journal} {Phys. Rev. Res.}\ }\textbf {\bibinfo {volume} {6}},\
  \bibinfo {pages} {033309} (\bibinfo {year} {2024})}\BibitemShut {NoStop}%
\bibitem [{\citenamefont {Watson}\ and\ \citenamefont
  {Watkins}(2025)}]{Watson2024TrotterMitigation}%
  \BibitemOpen
  \bibfield  {author} {\bibinfo {author} {\bibfnamefont {J.~D.}\ \bibnamefont
  {Watson}}\ and\ \bibinfo {author} {\bibfnamefont {J.}~\bibnamefont
  {Watkins}},\ }\href {https://doi.org/10.1103/kw39-yxq5} {\bibfield  {journal}
  {\bibinfo  {journal} {PRX Quantum}\ }\textbf {\bibinfo {volume} {6}},\
  \bibinfo {pages} {030325} (\bibinfo {year} {2025})}\BibitemShut {NoStop}%
\bibitem [{\citenamefont {Kiumi}\ and\ \citenamefont
  {Koczor}(2025)}]{Kiumi2025TEPAI}%
  \BibitemOpen
  \bibfield  {author} {\bibinfo {author} {\bibfnamefont {C.}~\bibnamefont
  {Kiumi}}\ and\ \bibinfo {author} {\bibfnamefont {B.}~\bibnamefont {Koczor}},\
  }\href {https://doi.org/10.1088/2058-9565/ae1160} {\bibfield  {journal}
  {\bibinfo  {journal} {Quantum Sci. Technol.}\ }\textbf {\bibinfo {volume}
  {10}},\ \bibinfo {pages} {045071} (\bibinfo {year} {2025})}\BibitemShut
  {NoStop}%
\bibitem [{\citenamefont {Granet}\ and\ \citenamefont
  {Dreyer}(2024)}]{Granet2024Hamiltonian}%
  \BibitemOpen
  \bibfield  {author} {\bibinfo {author} {\bibfnamefont {E.}~\bibnamefont
  {Granet}}\ and\ \bibinfo {author} {\bibfnamefont {H.}~\bibnamefont
  {Dreyer}},\ }\href {https://doi.org/10.1038/s41534-024-00877-y} {\bibfield
  {journal} {\bibinfo  {journal} {npj Quantum Inf.}\ }\textbf {\bibinfo
  {volume} {10}},\ \bibinfo {pages} {82} (\bibinfo {year} {2024})}\BibitemShut
  {NoStop}%
\bibitem [{\citenamefont {Dai}\ \emph {et~al.}(2026)\citenamefont {Dai},
  \citenamefont {Hasselgren},\ and\ \citenamefont {Kiumi}}]{Dai2026structure}%
  \BibitemOpen
  \bibfield  {author} {\bibinfo {author} {\bibfnamefont {J.~W.}\ \bibnamefont
  {Dai}}, \bibinfo {author} {\bibfnamefont {F.}~\bibnamefont {Hasselgren}},\
  and\ \bibinfo {author} {\bibfnamefont {C.}~\bibnamefont {Kiumi}},\ }\href
  {https://arxiv.org/abs/2606.23544} {\bibinfo {title} {Structure-aware
  variance reduction for unbiased randomized {H}amiltonian simulation}}
  (\bibinfo {year} {2026}),\ \Eprint {https://arxiv.org/abs/2606.23544}
  {arXiv:2606.23544} \BibitemShut {NoStop}%
\bibitem [{\citenamefont {{\c{S}}ahino{\u{g}}lu}\ and\ \citenamefont
  {Somma}(2021)}]{Sahinoglu2021LowEnergy}%
  \BibitemOpen
  \bibfield  {author} {\bibinfo {author} {\bibfnamefont {B.}~\bibnamefont
  {{\c{S}}ahino{\u{g}}lu}}\ and\ \bibinfo {author} {\bibfnamefont {R.~D.}\
  \bibnamefont {Somma}},\ }\href {https://doi.org/10.1038/s41534-021-00451-w}
  {\bibfield  {journal} {\bibinfo  {journal} {npj Quantum Inf.}\ }\textbf
  {\bibinfo {volume} {7}},\ \bibinfo {pages} {119} (\bibinfo {year}
  {2021})}\BibitemShut {NoStop}%
\bibitem [{\citenamefont {Chen}\ and\ \citenamefont
  {Brand{\~a}o}(2024)}]{Chen2024AverageCase}%
  \BibitemOpen
  \bibfield  {author} {\bibinfo {author} {\bibfnamefont {C.-F.}\ \bibnamefont
  {Chen}}\ and\ \bibinfo {author} {\bibfnamefont {F.~G. S.~L.}\ \bibnamefont
  {Brand{\~a}o}},\ }\href {https://doi.org/10.1007/s00220-023-04912-5}
  {\bibfield  {journal} {\bibinfo  {journal} {Commun. Math. Phys.}\ }\textbf
  {\bibinfo {volume} {405}},\ \bibinfo {pages} {32} (\bibinfo {year}
  {2024})}\BibitemShut {NoStop}%
\bibitem [{\citenamefont {Yu}\ \emph {et~al.}(2025)\citenamefont {Yu},
  \citenamefont {Xu},\ and\ \citenamefont {Zhao}}]{Yu2025ObservableDriven}%
  \BibitemOpen
  \bibfield  {author} {\bibinfo {author} {\bibfnamefont {W.}~\bibnamefont
  {Yu}}, \bibinfo {author} {\bibfnamefont {J.}~\bibnamefont {Xu}},\ and\
  \bibinfo {author} {\bibfnamefont {Q.}~\bibnamefont {Zhao}},\ }\href
  {https://doi.org/10.1038/s42005-025-02260-5} {\bibfield  {journal} {\bibinfo
  {journal} {Commun. Phys.}\ }\textbf {\bibinfo {volume} {8}},\ \bibinfo
  {pages} {340} (\bibinfo {year} {2025})}\BibitemShut {NoStop}%
\bibitem [{\citenamefont {Li}(2024)}]{Li2024Principal}%
  \BibitemOpen
  \bibfield  {author} {\bibinfo {author} {\bibfnamefont {L.}~\bibnamefont
  {Li}},\ }\href {https://doi.org/10.1103/PhysRevA.110.062614} {\bibfield
  {journal} {\bibinfo  {journal} {Phys. Rev. A}\ }\textbf {\bibinfo {volume}
  {110}},\ \bibinfo {pages} {062614} (\bibinfo {year} {2024})}\BibitemShut
  {NoStop}%
\bibitem [{\citenamefont {Wu}\ \emph {et~al.}(2026)\citenamefont {Wu},
  \citenamefont {Fan},\ and\ \citenamefont {Zhang}}]{Wu2025fluctuation}%
  \BibitemOpen
  \bibfield  {author} {\bibinfo {author} {\bibfnamefont {Y.-X.}\ \bibnamefont
  {Wu}}, \bibinfo {author} {\bibfnamefont {Y.-Z.}\ \bibnamefont {Fan}},\ and\
  \bibinfo {author} {\bibfnamefont {D.-B.}\ \bibnamefont {Zhang}},\ }\href
  {https://doi.org/10.1103/52wr-1hys} {\bibfield  {journal} {\bibinfo
  {journal} {Phys. Rev. Appl.}\ }\textbf {\bibinfo {volume} {25}},\ \bibinfo
  {pages} {054034} (\bibinfo {year} {2026})}\BibitemShut {NoStop}%
\bibitem [{\citenamefont {Fan}\ \emph {et~al.}(2025)\citenamefont {Fan},
  \citenamefont {Wu},\ and\ \citenamefont {Zhang}}]{Fan2025adaptive}%
  \BibitemOpen
  \bibfield  {author} {\bibinfo {author} {\bibfnamefont {Y.-Z.}\ \bibnamefont
  {Fan}}, \bibinfo {author} {\bibfnamefont {Y.-X.}\ \bibnamefont {Wu}},\ and\
  \bibinfo {author} {\bibfnamefont {D.-B.}\ \bibnamefont {Zhang}},\ }\href
  {https://doi.org/10.1103/4nmq-twj8} {\bibfield  {journal} {\bibinfo
  {journal} {Phys. Rev. A}\ }\textbf {\bibinfo {volume} {112}},\ \bibinfo
  {pages} {062431} (\bibinfo {year} {2025})}\BibitemShut {NoStop}%
\bibitem [{\citenamefont {Kivlichan}\ \emph {et~al.}(2019)\citenamefont
  {Kivlichan}, \citenamefont {Granade},\ and\ \citenamefont
  {Wiebe}}]{Kivlichan2019PhaseEstimation}%
  \BibitemOpen
  \bibfield  {author} {\bibinfo {author} {\bibfnamefont {I.~D.}\ \bibnamefont
  {Kivlichan}}, \bibinfo {author} {\bibfnamefont {C.~E.}\ \bibnamefont
  {Granade}},\ and\ \bibinfo {author} {\bibfnamefont {N.}~\bibnamefont
  {Wiebe}},\ }\href {https://arxiv.org/abs/1907.10070} {\bibinfo {title} {Phase
  estimation with randomized {H}amiltonians}} (\bibinfo {year} {2019}),\
  \Eprint {https://arxiv.org/abs/1907.10070} {arXiv:1907.10070} \BibitemShut
  {NoStop}%
\bibitem [{\citenamefont {Poulin}\ \emph {et~al.}(2015)\citenamefont {Poulin},
  \citenamefont {Hastings}, \citenamefont {Wecker}, \citenamefont {Wiebe},
  \citenamefont {Doherty},\ and\ \citenamefont
  {Troyer}}]{Poulin2015TrotterStep}%
  \BibitemOpen
  \bibfield  {author} {\bibinfo {author} {\bibfnamefont {D.}~\bibnamefont
  {Poulin}}, \bibinfo {author} {\bibfnamefont {M.~B.}\ \bibnamefont
  {Hastings}}, \bibinfo {author} {\bibfnamefont {D.}~\bibnamefont {Wecker}},
  \bibinfo {author} {\bibfnamefont {N.}~\bibnamefont {Wiebe}}, \bibinfo
  {author} {\bibfnamefont {A.~C.}\ \bibnamefont {Doherty}},\ and\ \bibinfo
  {author} {\bibfnamefont {M.}~\bibnamefont {Troyer}},\ }\href
  {https://doi.org/10.26421/QIC15.5-6-1} {\bibfield  {journal} {\bibinfo
  {journal} {Quantum Inf. Comput.}\ }\textbf {\bibinfo {volume} {15}},\
  \bibinfo {pages} {361} (\bibinfo {year} {2015})}\BibitemShut {NoStop}%
\bibitem [{\citenamefont {Huang}\ \emph {et~al.}(2020)\citenamefont {Huang},
  \citenamefont {Kueng},\ and\ \citenamefont {Preskill}}]{Huang2020Shadows}%
  \BibitemOpen
  \bibfield  {author} {\bibinfo {author} {\bibfnamefont {H.-Y.}\ \bibnamefont
  {Huang}}, \bibinfo {author} {\bibfnamefont {R.}~\bibnamefont {Kueng}},\ and\
  \bibinfo {author} {\bibfnamefont {J.}~\bibnamefont {Preskill}},\ }\href
  {https://doi.org/10.1038/s41567-020-0932-7} {\bibfield  {journal} {\bibinfo
  {journal} {Nat. Phys.}\ }\textbf {\bibinfo {volume} {16}},\ \bibinfo {pages}
  {1050} (\bibinfo {year} {2020})}\BibitemShut {NoStop}%
\bibitem [{\citenamefont {Zhao}\ \emph {et~al.}(2021)\citenamefont {Zhao},
  \citenamefont {Rubin},\ and\ \citenamefont {Miyake}}]{Zhao2021Fermionic}%
  \BibitemOpen
  \bibfield  {author} {\bibinfo {author} {\bibfnamefont {A.}~\bibnamefont
  {Zhao}}, \bibinfo {author} {\bibfnamefont {N.~C.}\ \bibnamefont {Rubin}},\
  and\ \bibinfo {author} {\bibfnamefont {A.}~\bibnamefont {Miyake}},\ }\href
  {https://doi.org/10.1103/PhysRevLett.127.110504} {\bibfield  {journal}
  {\bibinfo  {journal} {Phys. Rev. Lett.}\ }\textbf {\bibinfo {volume} {127}},\
  \bibinfo {pages} {110504} (\bibinfo {year} {2021})}\BibitemShut {NoStop}%
\bibitem [{\citenamefont {Wan}\ \emph {et~al.}(2023)\citenamefont {Wan},
  \citenamefont {Huggins}, \citenamefont {Lee},\ and\ \citenamefont
  {Babbush}}]{Wan2023Matchgate}%
  \BibitemOpen
  \bibfield  {author} {\bibinfo {author} {\bibfnamefont {K.}~\bibnamefont
  {Wan}}, \bibinfo {author} {\bibfnamefont {W.~J.}\ \bibnamefont {Huggins}},
  \bibinfo {author} {\bibfnamefont {J.}~\bibnamefont {Lee}},\ and\ \bibinfo
  {author} {\bibfnamefont {R.}~\bibnamefont {Babbush}},\ }\href
  {https://doi.org/10.1007/s00220-023-04844-0} {\bibfield  {journal} {\bibinfo
  {journal} {Commun. Math. Phys.}\ }\textbf {\bibinfo {volume} {404}},\
  \bibinfo {pages} {629} (\bibinfo {year} {2023})}\BibitemShut {NoStop}%
\bibitem [{\citenamefont {Sakurai}\ and\ \citenamefont
  {Napolitano}(2020)}]{SakuraiNapolitano2020}%
  \BibitemOpen
  \bibfield  {author} {\bibinfo {author} {\bibfnamefont {J.~J.}\ \bibnamefont
  {Sakurai}}\ and\ \bibinfo {author} {\bibfnamefont {J.}~\bibnamefont
  {Napolitano}},\ }\href {https://doi.org/10.1017/9781108587280} {\emph
  {\bibinfo {title} {Modern Quantum Mechanics}}},\ \bibinfo {edition} {3rd}\
  ed.\ (\bibinfo  {publisher} {Cambridge University Press},\ \bibinfo {year}
  {2020})\BibitemShut {NoStop}%
\bibitem [{\citenamefont {Neyman}(1934)}]{Neyman1934}%
  \BibitemOpen
  \bibfield  {author} {\bibinfo {author} {\bibfnamefont {J.}~\bibnamefont
  {Neyman}},\ }\href {https://doi.org/10.2307/2342192} {\bibfield  {journal}
  {\bibinfo  {journal} {J. R. Stat. Soc.}\ }\textbf {\bibinfo {volume} {97}},\
  \bibinfo {pages} {558} (\bibinfo {year} {1934})}\BibitemShut {NoStop}%
\bibitem [{\citenamefont {Owen}(2013)}]{Owen2013}%
  \BibitemOpen
  \bibfield  {author} {\bibinfo {author} {\bibfnamefont {A.~B.}\ \bibnamefont
  {Owen}},\ }\href {https://artowen.su.domains/mc/} {\emph {\bibinfo {title}
  {Monte {Carlo} Theory, Methods and Examples}}}\ (\bibinfo  {publisher}
  {self-published},\ \bibinfo {year} {2013})\BibitemShut {NoStop}%
\bibitem [{\citenamefont {Wecker}\ \emph {et~al.}(2015)\citenamefont {Wecker},
  \citenamefont {Hastings},\ and\ \citenamefont {Troyer}}]{Wecker2015}%
  \BibitemOpen
  \bibfield  {author} {\bibinfo {author} {\bibfnamefont {D.}~\bibnamefont
  {Wecker}}, \bibinfo {author} {\bibfnamefont {M.~B.}\ \bibnamefont
  {Hastings}},\ and\ \bibinfo {author} {\bibfnamefont {M.}~\bibnamefont
  {Troyer}},\ }\href {https://doi.org/10.1103/PhysRevA.92.042303} {\bibfield
  {journal} {\bibinfo  {journal} {Phys. Rev. A}\ }\textbf {\bibinfo {volume}
  {92}},\ \bibinfo {pages} {042303} (\bibinfo {year} {2015})}\BibitemShut
  {NoStop}%
\bibitem [{\citenamefont {Crawford}\ \emph {et~al.}(2021)\citenamefont
  {Crawford}, \citenamefont {van Straaten}, \citenamefont {Wang}, \citenamefont
  {Parks}, \citenamefont {Campbell},\ and\ \citenamefont
  {Brierley}}]{Crawford2021}%
  \BibitemOpen
  \bibfield  {author} {\bibinfo {author} {\bibfnamefont {O.}~\bibnamefont
  {Crawford}}, \bibinfo {author} {\bibfnamefont {B.}~\bibnamefont {van
  Straaten}}, \bibinfo {author} {\bibfnamefont {D.}~\bibnamefont {Wang}},
  \bibinfo {author} {\bibfnamefont {T.}~\bibnamefont {Parks}}, \bibinfo
  {author} {\bibfnamefont {E.}~\bibnamefont {Campbell}},\ and\ \bibinfo
  {author} {\bibfnamefont {S.}~\bibnamefont {Brierley}},\ }\href
  {https://doi.org/10.22331/q-2021-01-20-385} {\bibfield  {journal} {\bibinfo
  {journal} {Quantum}\ }\textbf {\bibinfo {volume} {5}},\ \bibinfo {pages}
  {385} (\bibinfo {year} {2021})}\BibitemShut {NoStop}%
\bibitem [{\citenamefont {Arrasmith}\ \emph {et~al.}(2020)\citenamefont
  {Arrasmith}, \citenamefont {Cincio}, \citenamefont {Somma},\ and\
  \citenamefont {Coles}}]{Arrasmith2020}%
  \BibitemOpen
  \bibfield  {author} {\bibinfo {author} {\bibfnamefont {A.}~\bibnamefont
  {Arrasmith}}, \bibinfo {author} {\bibfnamefont {L.}~\bibnamefont {Cincio}},
  \bibinfo {author} {\bibfnamefont {R.~D.}\ \bibnamefont {Somma}},\ and\
  \bibinfo {author} {\bibfnamefont {P.~J.}\ \bibnamefont {Coles}},\ }\href
  {https://arxiv.org/abs/2004.06252} {\bibinfo {title} {Operator sampling for
  shot-frugal optimization in variational algorithms}} (\bibinfo {year}
  {2020}),\ \Eprint {https://arxiv.org/abs/2004.06252} {arXiv:2004.06252}
  \BibitemShut {NoStop}%
\bibitem [{\citenamefont {Scali}\ \emph {et~al.}(2026)\citenamefont {Scali},
  \citenamefont {Soloviev}, \citenamefont {M\'{a}rquez~Romero}, \citenamefont
  {Coyle}, \citenamefont {Buonaiuto}, \citenamefont {Paine}, \citenamefont
  {Fetherolf}, \citenamefont {D{\'\i}ez~Garc{\'\i}a}, \citenamefont
  {Krompiec},\ and\ \citenamefont {Kirsopp}}]{OpenQARP}%
  \BibitemOpen
  \bibfield  {author} {\bibinfo {author} {\bibfnamefont {S.}~\bibnamefont
  {Scali}}, \bibinfo {author} {\bibfnamefont {V.~P.}\ \bibnamefont {Soloviev}},
  \bibinfo {author} {\bibfnamefont {A.}~\bibnamefont {M\'{a}rquez~Romero}},
  \bibinfo {author} {\bibfnamefont {B.}~\bibnamefont {Coyle}}, \bibinfo
  {author} {\bibfnamefont {G.}~\bibnamefont {Buonaiuto}}, \bibinfo {author}
  {\bibfnamefont {A.}~\bibnamefont {Paine}}, \bibinfo {author} {\bibfnamefont
  {J.~H.}\ \bibnamefont {Fetherolf}}, \bibinfo {author} {\bibfnamefont
  {M.}~\bibnamefont {D{\'\i}ez~Garc{\'\i}a}}, \bibinfo {author} {\bibfnamefont
  {M.}~\bibnamefont {Krompiec}},\ and\ \bibinfo {author} {\bibfnamefont
  {J.}~\bibnamefont {Kirsopp}},\ }\href {https://arxiv.org/abs/2609.15697}
  {\bibinfo {title} {{OpenQARP}: a modular framework for quantum application
  research}} (\bibinfo {year} {2026}),\ \Eprint
  {https://arxiv.org/abs/2609.15697} {arXiv:2609.15697} \BibitemShut {NoStop}%
\bibitem [{\citenamefont {Sun}\ \emph {et~al.}(2020)\citenamefont {Sun},
  \citenamefont {Zhang}, \citenamefont {Banerjee}, \citenamefont {Bao},
  \citenamefont {Barbry}, \citenamefont {Blunt}, \citenamefont {Bogdanov},
  \citenamefont {Booth}, \citenamefont {Chen}, \citenamefont {Cui} \emph
  {et~al.}}]{Sun2020}%
  \BibitemOpen
  \bibfield  {author} {\bibinfo {author} {\bibfnamefont {Q.}~\bibnamefont
  {Sun}}, \bibinfo {author} {\bibfnamefont {X.}~\bibnamefont {Zhang}}, \bibinfo
  {author} {\bibfnamefont {S.}~\bibnamefont {Banerjee}}, \bibinfo {author}
  {\bibfnamefont {P.}~\bibnamefont {Bao}}, \bibinfo {author} {\bibfnamefont
  {M.}~\bibnamefont {Barbry}}, \bibinfo {author} {\bibfnamefont {N.~S.}\
  \bibnamefont {Blunt}}, \bibinfo {author} {\bibfnamefont {N.~A.}\ \bibnamefont
  {Bogdanov}}, \bibinfo {author} {\bibfnamefont {G.~H.}\ \bibnamefont {Booth}},
  \bibinfo {author} {\bibfnamefont {J.}~\bibnamefont {Chen}}, \bibinfo {author}
  {\bibfnamefont {Z.-H.}\ \bibnamefont {Cui}}, \emph {et~al.},\ }\href
  {https://doi.org/10.1063/5.0006074} {\bibfield  {journal} {\bibinfo
  {journal} {J. Chem. Phys.}\ }\textbf {\bibinfo {volume} {153}},\ \bibinfo
  {pages} {024109} (\bibinfo {year} {2020})}\BibitemShut {NoStop}%
\bibitem [{\citenamefont {Whitfield}\ \emph {et~al.}(2011)\citenamefont
  {Whitfield}, \citenamefont {Biamonte},\ and\ \citenamefont
  {Aspuru-Guzik}}]{Whitfield2011Simulation}%
  \BibitemOpen
  \bibfield  {author} {\bibinfo {author} {\bibfnamefont {J.~D.}\ \bibnamefont
  {Whitfield}}, \bibinfo {author} {\bibfnamefont {J.}~\bibnamefont
  {Biamonte}},\ and\ \bibinfo {author} {\bibfnamefont {A.}~\bibnamefont
  {Aspuru-Guzik}},\ }\href {https://doi.org/10.1080/00268976.2011.552441}
  {\bibfield  {journal} {\bibinfo  {journal} {Mol. Phys.}\ }\textbf {\bibinfo
  {volume} {109}},\ \bibinfo {pages} {735} (\bibinfo {year}
  {2011})}\BibitemShut {NoStop}%
\bibitem [{\citenamefont {Seeley}\ \emph {et~al.}(2012)\citenamefont {Seeley},
  \citenamefont {Richard},\ and\ \citenamefont
  {Love}}]{Seeley2012BravyiKitaev}%
  \BibitemOpen
  \bibfield  {author} {\bibinfo {author} {\bibfnamefont {J.~T.}\ \bibnamefont
  {Seeley}}, \bibinfo {author} {\bibfnamefont {M.~J.}\ \bibnamefont
  {Richard}},\ and\ \bibinfo {author} {\bibfnamefont {P.~J.}\ \bibnamefont
  {Love}},\ }\href {https://doi.org/10.1063/1.4768229} {\bibfield  {journal}
  {\bibinfo  {journal} {J. Chem. Phys.}\ }\textbf {\bibinfo {volume} {137}},\
  \bibinfo {pages} {224109} (\bibinfo {year} {2012})}\BibitemShut {NoStop}%
\bibitem [{\citenamefont {Schollw\"ock}(2011)}]{Schollwock2011}%
  \BibitemOpen
  \bibfield  {author} {\bibinfo {author} {\bibfnamefont {U.}~\bibnamefont
  {Schollw\"ock}},\ }\href {https://doi.org/10.1016/j.aop.2010.09.012}
  {\bibfield  {journal} {\bibinfo  {journal} {Ann. Phys.}\ }\textbf {\bibinfo
  {volume} {326}},\ \bibinfo {pages} {96} (\bibinfo {year} {2011})}\BibitemShut
  {NoStop}%
\bibitem [{\citenamefont {Szabo}\ and\ \citenamefont
  {Ostlund}(1996)}]{SzaboOstlund}%
  \BibitemOpen
  \bibfield  {author} {\bibinfo {author} {\bibfnamefont {A.}~\bibnamefont
  {Szabo}}\ and\ \bibinfo {author} {\bibfnamefont {N.~S.}\ \bibnamefont
  {Ostlund}},\ }\href
  {https://store.doverpublications.com/products/9780486691862} {\emph {\bibinfo
  {title} {Modern Quantum Chemistry: Introduction to Advanced Electronic
  Structure Theory}}}\ (\bibinfo  {publisher} {Dover Publications},\ \bibinfo
  {year} {1996})\BibitemShut {NoStop}%
\bibitem [{\citenamefont {White}(1992)}]{White1992}%
  \BibitemOpen
  \bibfield  {author} {\bibinfo {author} {\bibfnamefont {S.~R.}\ \bibnamefont
  {White}},\ }\href {https://doi.org/10.1103/PhysRevLett.69.2863} {\bibfield
  {journal} {\bibinfo  {journal} {Phys. Rev. Lett.}\ }\textbf {\bibinfo
  {volume} {69}},\ \bibinfo {pages} {2863} (\bibinfo {year}
  {1992})}\BibitemShut {NoStop}%
\bibitem [{\citenamefont {Or\'us}(2014)}]{Orus2014}%
  \BibitemOpen
  \bibfield  {author} {\bibinfo {author} {\bibfnamefont {R.}~\bibnamefont
  {Or\'us}},\ }\href {https://doi.org/10.1016/j.aop.2014.06.013} {\bibfield
  {journal} {\bibinfo  {journal} {Ann. Phys.}\ }\textbf {\bibinfo {volume}
  {349}},\ \bibinfo {pages} {117} (\bibinfo {year} {2014})}\BibitemShut
  {NoStop}%
\bibitem [{\citenamefont {D{\'\i}ez~Garc{\'\i}a}\ and\ \citenamefont
  {M{\'a}rquez~Romero}(2024)}]{Garcia2024TensorSurvey}%
  \BibitemOpen
  \bibfield  {author} {\bibinfo {author} {\bibfnamefont {M.}~\bibnamefont
  {D{\'\i}ez~Garc{\'\i}a}}\ and\ \bibinfo {author} {\bibfnamefont
  {A.}~\bibnamefont {M{\'a}rquez~Romero}},\ }\href
  {https://doi.org/10.1109/ACCESS.2024.3519676} {\bibfield  {journal} {\bibinfo
   {journal} {IEEE Access}\ }\textbf {\bibinfo {volume} {12}},\ \bibinfo
  {pages} {193212} (\bibinfo {year} {2024})}\BibitemShut {NoStop}%
\bibitem [{\citenamefont {Krompiec}\ \emph {et~al.}(2026)\citenamefont
  {Krompiec}, \citenamefont {Kirsopp}, \citenamefont {M\'{a}rquez~Romero},\
  and\ \citenamefont {Perez~Soloviev}}]{oopccd}%
  \BibitemOpen
  \bibfield  {author} {\bibinfo {author} {\bibfnamefont {M.}~\bibnamefont
  {Krompiec}}, \bibinfo {author} {\bibfnamefont {J.~J.~M.}\ \bibnamefont
  {Kirsopp}}, \bibinfo {author} {\bibfnamefont {A.}~\bibnamefont
  {M\'{a}rquez~Romero}},\ and\ \bibinfo {author} {\bibfnamefont
  {V.}~\bibnamefont {Perez~Soloviev}},\ }\href
  {https://doi.org/10.1021/acs.jctc.5c01452} {\bibfield  {journal} {\bibinfo
  {journal} {J. Chem. Theory Comput.}\ }\textbf {\bibinfo {volume} {22}},\
  \bibinfo {pages} {257} (\bibinfo {year} {2026})}\BibitemShut {NoStop}%
\bibitem [{\citenamefont {Aaronson}(2020)}]{Aaronson2020ShadowTomography}%
  \BibitemOpen
  \bibfield  {author} {\bibinfo {author} {\bibfnamefont {S.}~\bibnamefont
  {Aaronson}},\ }\href {https://doi.org/10.1137/18M120275X} {\bibfield
  {journal} {\bibinfo  {journal} {SIAM J. Comput.}\ }\textbf {\bibinfo {volume}
  {49}},\ \bibinfo {eid} {STOC18-368} (\bibinfo {year} {2020})}\BibitemShut
  {NoStop}%
\bibitem [{\citenamefont {Elben}\ \emph {et~al.}(2023)\citenamefont {Elben},
  \citenamefont {Flammia}, \citenamefont {Huang}, \citenamefont {Kueng},
  \citenamefont {Preskill}, \citenamefont {Vermersch},\ and\ \citenamefont
  {Zoller}}]{Elben2023Toolbox}%
  \BibitemOpen
  \bibfield  {author} {\bibinfo {author} {\bibfnamefont {A.}~\bibnamefont
  {Elben}}, \bibinfo {author} {\bibfnamefont {S.~T.}\ \bibnamefont {Flammia}},
  \bibinfo {author} {\bibfnamefont {H.-Y.}\ \bibnamefont {Huang}}, \bibinfo
  {author} {\bibfnamefont {R.}~\bibnamefont {Kueng}}, \bibinfo {author}
  {\bibfnamefont {J.}~\bibnamefont {Preskill}}, \bibinfo {author}
  {\bibfnamefont {B.}~\bibnamefont {Vermersch}},\ and\ \bibinfo {author}
  {\bibfnamefont {P.}~\bibnamefont {Zoller}},\ }\href
  {https://doi.org/10.1038/s42254-022-00535-2} {\bibfield  {journal} {\bibinfo
  {journal} {Nat. Rev. Phys.}\ }\textbf {\bibinfo {volume} {5}},\ \bibinfo
  {pages} {9} (\bibinfo {year} {2023})}\BibitemShut {NoStop}%
\bibitem [{\citenamefont {Huang}\ \emph {et~al.}(2021)\citenamefont {Huang},
  \citenamefont {Kueng},\ and\ \citenamefont
  {Preskill}}]{Huang2021Derandomization}%
  \BibitemOpen
  \bibfield  {author} {\bibinfo {author} {\bibfnamefont {H.-Y.}\ \bibnamefont
  {Huang}}, \bibinfo {author} {\bibfnamefont {R.}~\bibnamefont {Kueng}},\ and\
  \bibinfo {author} {\bibfnamefont {J.}~\bibnamefont {Preskill}},\ }\href
  {https://doi.org/10.1103/PhysRevLett.127.030503} {\bibfield  {journal}
  {\bibinfo  {journal} {Phys. Rev. Lett.}\ }\textbf {\bibinfo {volume} {127}},\
  \bibinfo {pages} {030503} (\bibinfo {year} {2021})}\BibitemShut {NoStop}%
\bibitem [{\citenamefont {Hadfield}\ \emph {et~al.}(2022)\citenamefont
  {Hadfield}, \citenamefont {Bravyi}, \citenamefont {Raymond},\ and\
  \citenamefont {Mezzacapo}}]{Hadfield2022LocallyBiased}%
  \BibitemOpen
  \bibfield  {author} {\bibinfo {author} {\bibfnamefont {C.}~\bibnamefont
  {Hadfield}}, \bibinfo {author} {\bibfnamefont {S.}~\bibnamefont {Bravyi}},
  \bibinfo {author} {\bibfnamefont {R.}~\bibnamefont {Raymond}},\ and\ \bibinfo
  {author} {\bibfnamefont {A.}~\bibnamefont {Mezzacapo}},\ }\href
  {https://doi.org/10.1007/s00220-022-04343-8} {\bibfield  {journal} {\bibinfo
  {journal} {Commun. Math. Phys.}\ }\textbf {\bibinfo {volume} {391}},\
  \bibinfo {pages} {951} (\bibinfo {year} {2022})}\BibitemShut {NoStop}%
\bibitem [{\citenamefont {Jordan}\ and\ \citenamefont
  {Wigner}(1928)}]{JordanWigner1928}%
  \BibitemOpen
  \bibfield  {author} {\bibinfo {author} {\bibfnamefont {P.}~\bibnamefont
  {Jordan}}\ and\ \bibinfo {author} {\bibfnamefont {E.}~\bibnamefont
  {Wigner}},\ }\href {https://doi.org/10.1007/BF01331938} {\bibfield  {journal}
  {\bibinfo  {journal} {Z. Phys.}\ }\textbf {\bibinfo {volume} {47}},\ \bibinfo
  {pages} {631} (\bibinfo {year} {1928})}\BibitemShut {NoStop}%
\bibitem [{DLMF()}]{DLMF}%
  \BibitemOpen
  DLMF,\ \href {https://dlmf.nist.gov/2.1} {\bibinfo {title} {{NIST Digital
  Library of Mathematical Functions}}},\ \bibinfo {howpublished} {Release 1.2.8
  of 2026-09-15, Section 2.1},\ \bibinfo {note} {{F.~W.~J. Olver}, A.~B. {Olde
  Daalhuis}, D.~W. Lozier, B.~I. Schneider, R.~F. Boisvert, C.~W. Clark, B.~R.
  Miller, B.~V. Saunders, H.~S. Cohl, and M.~A. McClain, eds.}\BibitemShut
  {Stop}%
\bibitem [{\citenamefont {Kingma}\ and\ \citenamefont
  {Ba}(2015)}]{Kingma2014Adam}%
  \BibitemOpen
  \bibfield  {author} {\bibinfo {author} {\bibfnamefont {D.~P.}\ \bibnamefont
  {Kingma}}\ and\ \bibinfo {author} {\bibfnamefont {J.}~\bibnamefont {Ba}},\
  }in\ \href {https://arxiv.org/abs/1412.6980} {\emph {\bibinfo {booktitle}
  {3rd International Conference on Learning Representations (ICLR)}}}\
  (\bibinfo {year} {2015})\ \Eprint {https://arxiv.org/abs/1412.6980}
  {arXiv:1412.6980} \BibitemShut {NoStop}%
\end{thebibliography}
\end{document}